# The density and peculiar velocity fields of nearby galaxies

Michael A. Strauss

*School of Natural Sciences, Institute for Advanced Study, Princeton, New Jersey 08540*

and

Jeffrey A. Willick

*Observatories of the Carnegie Institute of Washington, 813 Santa Barbara Street, Pasadena, California 91101-1292*

## Abstract

We review the quantitative science that can be and has been done with redshift and peculiar velocity surveys of galaxies in the nearby universe. After a brief background setting the cosmological context for this work, the first part of this review focuses on redshift surveys. The practical issues of how redshift surveys are carried out, and how one turns a distribution of galaxies into a smoothed density field, are discussed. Then follows a description of major redshift surveys that have been done, and the local cosmography out to 8,000 km s$^{-1}$ that they have mapped. We then discuss in some detail the various quantitative cosmological tests that can be carried out with redshift data. The second half of this review concentrates on peculiar velocity studies, beginning with a thorough review of existing techniques. After discussing the various biases which plague peculiar velocity work, we survey quantitative analyses done with peculiar velocity surveys alone, and finally with the combination of data from both redshift and peculiar velocity surveys. The data presented rule out the standard Cold Dark Matter model, although several variants of Cold Dark Matter with more power on large scales fare better. All the data are consistent with the hypothesis that the initial density field had a Gaussian distribution, although one cannot rule out broad classes of non-Gaussian models. Comparison of the peculiar velocity and density fields constrains the Cosmological Density Parameter. The results here are consistent with a flat universe with mild biasing of the galaxies relative to dark matter, although open universe models are by no means ruled out.

# Contents

# 1  Introduction

The last two decades have seen a tremendous growth in our awareness of the richness of the structure of the Universe on scales of galaxies and larger. Technical advances in spectrographs and detectors in both the visible and radio wavebands have allowed redshifts of large numbers of galaxies to be measured efficiently, giving us the opportunity to map their distribution for the first time. We are in the midst of a golden age of exploration. The redshift surveys of the last fifteen years have discovered the existence of superclusters, great voids, and coherent structures stretching across the entire volume covered by the surveys. But it is not just "geography" (or cosmography, as it is called in the astronomical context) that has been done: hidden in the statistics of the galaxy distribution is much important information about the structure of the Universe on large scales. Thus we can deduce much about the early history of the universe, and its global structure, from careful quantitative analyses of redshift surveys: this will be the central theme of this review.

The Hubble Law states that at distances much less than the Hubble radius, the expansion of the universe causes the recession velocity of a galaxy $cz$ to be proportional to its distance $r$:

$$cz = H_0 r, \tag{1}$$

where $H_0$ is the *Hubble Constant*, whose value remains uncertain by a factor of two; in astronomer's units, it is often written as $H_0 = 100h$ km s$^{-1}$ Mpc$^{-1}$, the quantity $h$ parameterizing our ignorance of its value. Mpc stands for megaparsecs, the common unit of distance for much extragalactic work: $1\,\text{Mpc} = 3.08 \times 10^{24}\,\text{cm}$. Thus in physicist's units, $H_0 = 3.25 \times 10^{-18}\,h$ sec, or $H_0^{-1} \approx 10^{10}\,h^{-1}$ yr. In practice, we will rarely be troubled by the uncertain value of $h$ in this review, because we will measure distances in units of km s$^{-1}$, wherein $H_0 \equiv 1$. The observational evidence for the linearity of Eq. (1) is reviewed in Peebles (1993) and Lauer & Postman (1992); cf. § 3.6 below.

At the low redshifts discussed in the majority of this review, relativistic effects are for the most part unimportant and the Hubble law is an excellent approximation. However, galaxies have motions above and beyond their Hubble velocities, deviations from the isotropic expansion that holds only in the theoretical idealization of a perfectly homogeneous universe. Thus Eq. (1) is modified to:

$$cz = H_0 r + \hat{\mathbf{r}} \cdot [\mathbf{v}(\mathbf{r}) - \mathbf{v}(\mathbf{0})], \tag{2}$$

where $\hat{\mathbf{r}}$ is the unit vector towards the galaxy in question, $\mathbf{v}(\mathbf{r})$ is the *peculiar velocity* at position $\mathbf{r}$, and $\mathbf{v}(\mathbf{0})$ is the peculiar velocity of the observer. The development over the past fifteen years of redshift-independent methods of measuring distances have allowed the measurement of the peculiar velocity field $\mathbf{v}(\mathbf{r})$. As we will see in detail, observations of the peculiar velocity field are of great importance; it is their study which occupies the second half of this review. In particular, under the hypothesis that structure formed by the gravitational growth of small perturbations on an initially uniform background, there is a direct relation between the density and velocity fields, which we will exploit to great advantage to put constraints on cosmological models.

Thus the main emphasis in this review will be on quantitative constraints that can be placed on cosmological models, from redshift and peculiar velocity surveys of the local Universe. By local, we mean at distances small compared with the horizon distance, or equivalently, at look-back times small relative to the age of the universe, or redshifts small compared to the speed of light. That is, for most of this review, we will not be concerned about the general relativistic generalization of Eq. (1), and can work in the Newtonian limit almost exclusively. Another definition of the local universe is that within which evolutionary effects in the galaxy properties can be assumed to be negligible. In practice, we will restrict ourselves to recession velocities below 20,000 km s$^{-1}$ ($z = 0.067$).

There are a number of major reviews and books that are relevant as background material for this review. The reader is assumed to be familiar with the basics of Big Bang cosmology, as reviewed in Peebles (1993) and Padmanabhan (1993). Somewhat more advanced material can be found in White (1991), Efstathiou (1991), and Suto (1993). Older texts that also discuss this material include Peebles (1971), Efstathiou & Silk (1983), Peebles (1980) and Weinberg (1972). A useful pocket summary of the field is found in Scott et al. (1994). We will many times skirt the issues of astroparticle physics in the cosmological context; Kolb & Turner (1990) is an excellent overview of the field. Redshift surveys are discussed in Geller & Huchra (1988) and Giovanelli & Haynes (1991), and peculiar velocity surveys are reviewed in Burstein (1990) and Dekel (1994). The subject can also be followed in the proceedings of many conferences, among them Brück, Coyne, & Longair (1982), Kolb et al. (1986), Madore & Tully (1986), Burbidge & Hewitt (1986), Kormendy & Knapp (1987), Audouze, Pellatan, & Szalay (1987), Rubin & Coyne (1988), van den Bergh & Pritchet (1988), Latham & da Costa (1991), Chincarini et al. (1993), and Bouchet & Lachiéze-Rey (1994).

We stress observational aspects of large-scale structure studies in this review. Not all aspects of the field can be covered. In particular, we only touch upon observations of the high-redshift universe, either in galaxies (Spinrad & Kron 1995) or in background radiation (Partridge 1994). We focus exclusively on galaxies as probes of large-scale structure, and do not discuss the distribution of clusters of galaxies (e.g., Bahcall 1988). See the review of Efstathiou (1994) for a more unified treatment of the large-scale distribution of galaxies and clusters. We also do not review evidence for dark matter on Galactic scales (Faber & Gallagher 1979; Trimble 1987; Binney & Tremaine 1987), or models for galaxy formation (e.g., White 1994). Finally, although peculiar velocities are a major focus of this paper, we do not review the closely related question of the determination of the Hubble Constant (cf. Jacoby et al. 1992).

This review is too long to be read in a single sitting, but the different chapters are largely independent. Chapter 2 covers the basics of Big Bang cosmology to the extent needed for the rest of the review, and can be skipped by the reader familiar with this material. Chapter 3 gives a history of redshift surveys, and discusses many of the practical issues needing to be addressed in their quantitative analysis. Chapter 4 is a qualitative tour of the structures we see within 8000 km s$^{-1}$, as revealed to us via redshift surveys. Those who are interested in quantitative results can skip ahead to Chapter 5, which details the various statistical measures of large-scale structure that have been drawn from redshift surveys.

We then shift the focus to peculiar velocity surveys. Chapter 6 parallels Chapter 3 and discusses the practical aspects of peculiar velocity measurements. We put special emphasis on the various biases which plague peculiar velocity work, and the methods necessary to correct them. Quantitative analyses of peculiar velocity surveys are discussed in Chapter 7, and analyses of redshift and peculiar velocity surveys together are covered in Chapter 8. Chapter 9 is a summary chapter.

Those readers who are familiar with much of the background material may thus find it most useful to read Chapters 5, 7, and 8, after skimming Chapters 3 and 6. We recommend that students new to the field read through the entire review, making sure that they understand the basic concepts introduced in Chapter 2.

## 2 Theoretical Background

For the majority of this review, we will work in the context of the Big Bang model. In order to set a common language and notation, we briefly review the relevant points of Big Bang cosmology to which we will find ourselves referring. The Big Bang model itself is discussed in § 2.1, in which we define the basic parameters of an homogeneous and isotropic expanding universe, and write down the equations for their evolution. § 2.2 gives the evolution equations for perturbations in such a homogeneous universe, and § 2.3 discusses the power spectrum of these perturbations, and introduces the concept of dark matter. § 2.4 discusses models for the relative distribution of galaxies and dark matter in the context of the biasing paradigm. § 2.5 summarizes the outstanding questions we would like to address with the observations.

### 2.1 The Big Bang Model and its Parameters

It is an observational fact that all galaxies (with the exception of galaxies in the Local Group and a few galaxies associated with the Virgo Cluster) have positive redshifts, and it is observed that redshifts are proportional to distance (§ 3.6). This is interpreted as due to the expansion of the Universe. The Cosmological Principle, as formulated originally by Einstein, states that on large enough scales (to be quantified below) the Universe is homogeneous and isotropic; this model together with the tenets of general relativity leads to the prediction that we do not live in a static Universe[1]. In particular, the Cosmological Principle implies that the covariant line element between two points is given by:

$$ds^2 = c^2 \, dt^2 - a^2(t) \, dl^2, \tag{3}$$

where $a(t)$ is the *scale factor*. Our observation that the universe is expanding means that $a(t)$ is an increasing function of time. The *comoving* distance $l$ between any two points taking part in the Hubble expansion is a constant; the *proper* distance between two points is the quantity $s$ at a given constant time $t$, i.e., such that $dt = 0$. Comparing with Eq. (1), we see immediately that

$$H_0 = \frac{1}{s}\frac{ds}{dt} = \left.\frac{\dot{a}}{a}\right|_{t=t_0}, \tag{4}$$

where the subscript 0 refers to values of quantities at the present time $t_0$. Indeed, the quantity $\dot{a}/a$ is not constant with time, and thus neither is the Hubble Constant; however, at any given time, it is independent of position and direction.

Consider two observers separated by a proper distance $\Delta l$ small compared with the distance to the horizon. By Hubble's law, they are moving apart from one another at a speed given by $v = \frac{\dot{a}}{a}\Delta l$. Now consider a plane wave of radiation, with definite wavelength $\lambda$ as measured by the first of the observers, traveling towards the second. The second observer will observe this radiation to be shifted by the first-order Doppler shift (because $v \ll c$) to a wavelength

$$\lambda(1 + v/c) = \lambda\left(1 + \frac{\dot{a}}{a}\Delta l/c\right). \tag{5}$$

---

[1] The introduction of a cosmological constant does allow solutions indicating a static Universe, but these solutions are unstable.

Thus, recognizing that the proper time required for the radiation to travel the distance $\Delta l$ is $\Delta t = \Delta l/c$, and taking the limit $\Delta l \to 0$, we find

$$\frac{\dot{\lambda}}{\lambda} = \frac{\dot{a}}{a} \tag{6}$$

or

$$\lambda(t) \propto a(t). \tag{7}$$

The *redshift* of a galaxy $z$ is defined as

$$z = \frac{\lambda(t) - \lambda_0}{\lambda_0}, \tag{8}$$

where $\lambda_0$ is the wavelength of a plane wave emitted by the galaxy at the time of emission (the rest wavelength), and $\lambda(t)$ is the wavelength of the plane wave at the present (the observed wavelength). Thus the redshift and the scale factor are directly linked:

$$1 + z = a_0/a(t). \tag{9}$$

At low redshifts, the recession velocity of a galaxy is simply given by $cz$. At high redshifts, this expression clearly breaks down, and one must go to general relativistic generalizations of it.

Under the assumption that the universe is homogeneous and isotropic, the line element can be expressed in spherical coordinates:

$$ds^2 = c^2\, dt^2 - a^2(t) \left[ \frac{dr^2}{1 - kr^2} + r^2 (d\theta^2 + \sin^2\theta\, d\phi^2) \right]. \tag{10}$$

This is called the Friedman-Robertson-Walker metric, after the men who first wrote it down. The quantity $k$ is the *curvature constant*, an integration constant of Einstein's equations[2]. Thus $k = 0$ corresponds to a Euclidean geometry for the three spatial dimensions; this is called a *flat* universe.

Dynamical equations for the scale factor $a$ follow from the metric, together with the equations of general relativity and the ideal fluid approximation[3]. One finds (e.g., Chapter 15 of Weinberg 1972) two equations:

$$\frac{\ddot{a}}{a} = -\frac{4}{3}\pi G \left( \rho + \frac{3p}{c^2} \right) + \frac{\Lambda}{3} \tag{11}$$

and

$$H^2 \equiv \left(\frac{\dot{a}}{a}\right)^2 = \frac{8}{3}\pi G \rho - \frac{kc^2}{a^2} + \frac{\Lambda}{3}. \tag{12}$$

In these equations, $G$ is Newton's gravitational constant, $\rho$ is the (non-relativistic) mass density of the universe, $p$ is the pressure, $H$ is the Hubble Constant (compare with Eq. 4), where the lack of a subscripted 0 indicates that it is a function of time, and $\Lambda$ is the Cosmological Constant.

---

[2] The curvature constant is sometimes written in the form $\pm \frac{1}{R^2}$ because in a closed universe, $a_0 R$ can be interpreted as the radius of the universe (cf. Eq. 5.9 of Peebles 1993).
[3] These equations may also be derived heuristically following a purely Newtonian argument (cf. Chapter 5 of Peebles 1993).

In the present universe, non-relativistic matter completely dominates the energy density of the universe, making the pressure term negligible; we drop it from now on. In this zero-pressure limit, mass conservation implies $\rho \propto a^{-3}(t)$. The wavelength of any relativistic species will be redshifted with the expansion of the universe; to the extent that the number of particles of this species is preserved, the energy density in relativistic species drops as $a^{-4}(t)$. The *Cosmic Microwave Background* (CMB), isotropic radiation with a pure black-body spectrum and a temperature of $T_0 = 2.735^\circ K$, represents such a relativistic species. The energy density in the CMB today is negligible relative to the density in non-relativistic matter, but because of its faster fall-off with $a$, it dominated the energy density of the universe for $z > 2.4 \times 10^4 \Omega_0 h^2$. The CMB will not be reviewed thoroughly here, although we will find ourselves referring to it often. See Efstathiou (1991), Peebles (1993), and Partridge (1994) for recent reviews.

Setting the Cosmological Constant $\Lambda$ to zero for the moment, and dropping the pressure term, Eq. (12) can be written:

$$\dot{a}^2(t) = \left(\frac{8}{3}\pi G \rho_0\right) \frac{1}{a(t)} - kc^2. \tag{13}$$

For $k$ negative or zero, the right hand side, and therefore $\dot{a}^2(t)$, stays positive-definite. We know that $\dot{a}(t)$ is positive now (the universe is observed to be expanding, not contracting) and thus $a(t)$ will continue to increase for all time. This is what we call the open or flat universe. However, if $k$ is positive, there will be some time in the future at which the second term on the right hand side of Eq. (13) balances the first term, and thus $\dot{a}^2(t)$ returns to zero. As $\ddot{a}$ is negative-definite (Eq. 11) this means that $\dot{a}(t)$ changes sign, and $a(t)$ begins to decrease, eventually reaching zero. This is a closed universe. The controlling factor deciding the fate of the universe is the density; the division between the open and closed universe happens at the critical density, whose current value is:

$$\rho_{\text{crit}} \equiv \frac{3H_0^2}{8\pi G}. \tag{14}$$

Thus we define the *Cosmological Density Parameter* $\Omega_0$ as

$$\Omega_0 \equiv \frac{\rho_0}{\rho_{\text{crit}}}, \tag{15}$$

whose value is less than unity for an open universe, greater than unity for a closed universe, and exactly zero for a flat universe. If we re-introduce the Cosmological constant, we get a flat universe ($k = 0$) in the case that $\Omega_0 + \Omega_\Lambda = 1$, where

$$\Omega_\Lambda \equiv \frac{\Lambda}{3H_0^2} \tag{16}$$

is the contribution to the cosmological density due to vacuum energy. Of course, in this case, the concepts of open and closed universes become more complicated, because $\ddot{a}$ can change sign. See Harrison (1981) for a complete inventory of cosmological models.

We can use this line of reasoning to peer into the past as well. As $\ddot{a}$ is negative-definite (for $\Omega_\Lambda = 0$, as we will assume) and $\dot{a}$ is positive-definite, then at some finite time $t$ in the past $a(t) = 0$. This is the origin of the concept of the Big Bang: the dynamical equations for $a(t)$ indicate that the universe was of infinitesimal extent some finite time in the past. Let us define that time $t = 0$. We can now ask for the age of the universe (Weinberg 1972):

$$t_0 = \begin{cases} \frac{1}{2}H_0^{-1}\Omega_0(\Omega_0-1)^{-3/2}\left[\cos^{-1}\left(\frac{2}{\Omega_0}-1\right) - \frac{2}{\Omega_0}(\Omega_0-1)^{1/2}\right] & \Omega_0 > 1, \\ \frac{2}{3}H_0^{-1} & \Omega_0 = 1, \\ \frac{1}{2}H_0^{-1}\Omega_0(1-\Omega_0)^{-3/2}\left[\cosh^{-1}\left(1-\frac{2}{\Omega_0}\right) - \frac{2}{\Omega_0}(1-\Omega_0)^{1/2}\right] & \Omega_0 < 1. \end{cases} \quad (17)$$

As we could have guessed from the units, the present age of the universe is proportional to $H_0^{-1}$, with a numerical coefficient which is a decreasing function of $\Omega_0$. The corresponding equations for $\Lambda \neq 0$ universes are non-analytic, and will not be presented here.

We need one further definition, the *acceleration parameter*[4], defined at the present time:

$$q_0 \equiv -\frac{\ddot{a}a}{\dot{a}^2} = \frac{\Omega_0}{2} - \Omega_\Lambda, \quad (18)$$

where the second equality follows directly from Equations 11 and 12. The acceleration parameter is a measure of relativistic effects in the relation between observables; thus, for example, the first-order corrections to Eq. (1) are proportional to $q_0$.

This completes our survey of the global structure of space-time in a homogeneous and isotropic universe. The cosmology is defined by the following numbers:
– $\Omega_0$, the cosmological density parameter;
– $H_0$, the Hubble constant;
– $\Lambda$, the cosmological constant;
– $q_0$, the acceleration parameter, and
– $t_0$, the present age of the universe.

These parameters are not independent of one another in the context of the Big Bang model, as we have seen, but each are measured or constrained by a variety of different observations. It is one of the important tests of Big Bang cosmology that one find consistency between the values found for the different parameters. We will put the greatest emphasis in this review on measurements of $\Omega_0$, because this is the quantity that can best be constrained with analyses of redshift and peculiar velocity surveys.

Standard Big Bang cosmology has built into it a number of apparent paradoxes. The first is often referred to as the horizon problem: the size of causally connected regions at the time of recombination subtends only a few degrees on the sky, leaving the large-scale isotropy observed in the CMB unexplained: this must be taken as an (arbitrary) initial condition. The second is the so-called flatness problem: $\Omega_0 = 1$ is an unstable solution to the evolution equations, in the sense that the fact that we observe $\Omega_0$ to be within an order of magnitude of unity today requires that it be very finely tuned in the early universe.

There is a class of cosmological models under the general rubric of the *inflationary paradigm*, which addresses these problems, and predicts tight constraints on the parameters of the Big Bang model. Inflation is reviewed thoroughly in Chapter 8 of Kolb & Turner (1990). In its simplest form, the inflationary model posits a period in the very early universe when a super-cooled phase transition caused the vacuum energy density to become dominant; Eq. (12) then implies that the scale factor grows exponentially. If this exponential expansion continues long enough for the curvature term to become exponentially small, the flatness problem is solved; the result is a universe that is globally flat (i.e., $k = 0$ or $\Omega_0 + \Omega_\Lambda = 1$). Moreover, as the observable universe inflated from a region that was causally connected and in thermal contact, the homogeneity and isotropy of the universe (the horizon problem) are explained. The inflationary model also predicts a scale-invariant spectrum of adiabatic density fluctuations[5], a subject to which we turn in § 2.3 after a discussion of gravitational instability theory.

---
[4] Some authors refer to $q_0$ as the deacceleration parameter.

The inflationary prediction of a flat universe is apparently in direct contradiction with measurements of the total mass density of the universe, at least until recently. Measurements of the luminosity density of the universe (cf., § 3.4) imply that the mass density of the universe in stars is only $\Omega \sim 0.01$ (§ 3.5). There is abundant evidence for non-luminous matter associated with galaxies and clusters of galaxies, whose total contribution is as much as $\Omega_0 \sim 0.2$ (cf., Faber & Gallagher 1979; Kormendy & Knapp 1987; Trimble 1987). The inflationary prediction of $\Omega_0 = 1$ in a universe without a cosmological constant requires that there be additional dark matter distributed on scales larger than that of clusters; indeed, as we shall see in detail later in this review, there is recent evidence from large-scale flows that $\Omega_0$ is larger than that inferred from dynamics within clusters. The material which makes up the dark matter, however, remains completely unknown. The physical properties of the dark matter greatly influences the distribution of matter on large scales, and thus redshift surveys of galaxies have the potential to constrain the form of dark matter. In order to explore these issues, we need a theory for the growth of structure in the presence of gravity.

*2.2 The Gravitational Instability Paradigm*

The Big Bang model as outlined in the previous section explicitly assumes a homogeneous and isotropic universe. We believe that the Cosmological Principle does in fact hold on the largest scales (for reasons that will be discussed in § 5.5), and that at early times the universe was very close to homogeneous. However, we observe structure all around us: from planets, to galaxies, to superclusters of galaxies, matter tends to aggregate and form structures, rather than distribute itself uniformly. One of the great questions facing cosmology is how this structure came to be. The widely accepted view is that small density fluctuations present in the beginning grew by gravitational instability into the structures that we see today. In this section, we briefly review the theory of gravitational instability in an expanding universe.

One starts by writing down the equations of mass continuity, force, and gravitation in an expanding universe, in proper coordinates (ignoring relativistic effects):

$$\frac{\partial \rho}{\partial t} + \nabla \cdot (\rho \mathbf{v}) = 0, \tag{19}$$

$$\frac{\partial \mathbf{v}}{\partial t} + (\mathbf{v} \cdot \nabla)\mathbf{v} + \nabla \phi = 0, \tag{20}$$

$$\nabla^2 \phi = 4\pi G \rho. \tag{21}$$

Here $\rho$ is the mass density field, $\mathbf{v}$ is the velocity field, $\phi$ is the gravitational potential, and we have dropped terms depending on pressure, which we assume are negligible. All spatial derivatives are with respect to proper distance. If we expand these equations to first order in all quantities measuring departure from uniformity, convert to comoving coordinates, and subtract the zeroth order solutions[6], the first two equations simplify to

$$\frac{\partial \delta}{\partial t} + \frac{1}{a} \nabla \cdot \mathbf{v} = 0, \tag{22}$$

$$\frac{\partial \mathbf{v}}{\partial t} + \frac{\dot{a}}{a} \mathbf{v} + \frac{1}{a} \nabla \phi = 0, \tag{23}$$

---

[5] There do exist variants on the standard inflationary model which predict other forms for the density fluctuation spectrum, and which predict noticeable curvature; see the references at the end of §8.5 of Kolb & Turner (1990).
[6] The subtraction of the zeroth order solution involves some subtleties having to do with the gravitational potential of a uniform universe; these are discussed further in §8 of Shu (1992).

where $\delta$ is the dimensionless density contrast,

$$\delta(\mathbf{r}) \equiv \frac{\rho(\mathbf{r}) - \rho_0}{\rho_0}, \tag{24}$$

and $\rho_0$ is the mean mass density. Taking the time derivative of the continuity equation and substituting into the divergence of the force equation yields, with the Poisson equation (Eq. 21):

$$\frac{\partial^2 \delta}{\partial t^2} + \frac{2\dot{a}}{a}\frac{\partial \delta}{\partial t} = 4\pi G \rho_0 \delta. \tag{25}$$

Because Eq. (25) is a second-order partial differential equation in time alone, we can separate the spatial and temporal dependences, and write:

$$\delta = A(\mathbf{x})D_1(t) + B(\mathbf{x})D_2(t), \tag{26}$$

where $D_1$ and $D_2$ are growing and decaying modes, respectively. We will be using the growing mode throughout this review; it is given by

$$D_1(t) = \frac{\dot{a}}{a} \int_0^a \dot{a}^{-3}\,\mathrm{d}a. \tag{27}$$

In the general case of a non-vanishing cosmological constant, this integral is not analytic. For $\Omega_\Lambda = 0$ models, however, analytic expressions for $D_1(t)$ and $D_2(t)$ can be obtained (§11 of Peebles 1980). A particularly simple solution exists in the special case of a flat universe, in which $a(t) \propto t^{2/3}$ (Eq. 17), and the right hand side of Eq. (25) is simply $\frac{3}{2}\left(\frac{\dot{a}}{a}\right)^2$, so we find:

$$\frac{\partial^2 \delta}{\partial t^2} + \frac{4}{3t}\frac{\partial \delta}{\partial t} = \frac{2}{3t^2}\delta, \tag{28}$$

which has an analytic solution in power laws [7] of $t$:

$$\delta(\mathbf{x}, t) = A(\mathbf{x})\,t^{2/3} + B(\mathbf{x})\,t^{-1}. \tag{29}$$

More generally, the solutions depend on $\Omega_0$; in particular, the growth is faster for increasing $\Omega_0$. For $\Omega_0 < 1$, the expansion of the universe (the drag term represented by $\frac{2\dot{a}}{a}\frac{\partial \delta}{\partial t}$ in Eq. 25) dominates the gravitational attraction of the matter, and the clustering freezes out (i.e., it stops growing) at $z \approx 1/\Omega_0 - 1$. A full discussion of these solutions, including the effects of pressure which is important in the early universe, will take us too far afield; cf. Chapter 9 of Kolb & Turner (1990) for more details.

If we wait until late times, the growing mode of Eq. (26) will dominate, and we can rewrite Eq. (22) as:

$$\nabla \cdot \mathbf{v} = -a\delta \frac{\dot{D}_1}{D_1} = -aH_0 f\delta, \tag{30}$$

where

$$f \equiv \frac{1}{H_0 D_1}\frac{\mathrm{d}D_1}{\mathrm{d}t} = \frac{1}{H_0 D_1}\frac{\mathrm{d}D_1}{\mathrm{d}a}\frac{\mathrm{d}a}{\mathrm{d}t} = \frac{\mathrm{d}\ln D_1}{\mathrm{d}\ln a}. \tag{31}$$

---

[7] Pressureless gravitational growth in a non-expanding universe is quite different: the drag term is not present, and perturbations grow exponentially.

We are still in comoving coordinates, as indicated by the $a$ on the right hand side of Eq. (30). The expression for $D_1$, and therefore for $f$, is a function of $\Omega_0$ and $\Lambda$. A good approximation in the general case is given by Lahav et al. (1991):

$$f(\Omega_0, \Lambda) = \Omega_0^{0.6} + \frac{\Omega_\Lambda}{70}\left(1 + \frac{1}{2}\Omega_0\right);\qquad(32)$$

other approximations can be found in Peebles (1984), Regös & Geller (1989), Lightman & Schechter (1990), Martel (1991), and Carroll, Press, & Turner (1992). Thus the influence of $\Lambda$ on dynamics at low redshift is minimal (Lahav et al. 1991). Eq. (30) can be inverted via the methods of electrostatics in the usual way to yield, after returning to proper coordinates [8]:

$$\mathbf{v}(\mathbf{r}) = \frac{H_0 f}{4\pi}\int d^3\mathbf{r}' \frac{\delta(\mathbf{r}')(\mathbf{r}' - \mathbf{r})}{|\mathbf{r}' - \mathbf{r}|^3}.\qquad(33)$$

Eq. (33) reveals the physical content of this first-order expansion: linear perturbation theory states that peculiar velocities are proportional to gravitational acceleration. We will find ourselves using Eqs. (30) and (33) throughout this paper. These equations can be generalized using higher-order perturbation theory; we will introduce these results when needed.

If we measure the quantity $\mathbf{r}$ in units of km s$^{-1}$, then $H_0 \equiv 1$, and we see that a comparison of the velocity field $\mathbf{v}(\mathbf{r})$ and density field $\delta(\mathbf{r})$ gives a direct measure of $f(\Omega_0, \Lambda)$. One of the central themes of this review will be exploiting this comparison to put constraints on $f$.

Linear theory makes the assumption that the change in comoving position of galaxies as the universe expands is negligible. Zel'dovich (1970) made an important extension of linear theory by assuming that the difference between the Lagrangian position $\mathbf{q}$ and Eulerian position $\mathbf{x}$ of a particle in a gravitating system is separable in space and time:

$$\mathbf{x}(\mathbf{q}) = \mathbf{q} + D_1(t)\psi(\mathbf{q}),\qquad(34)$$

where $D_1(t)$ is the growing mode in linear theory (Eq. 27) and $\psi$, which determines the amplitude of the velocity field, is proportional to the gradient of the gravitational potential. See Shandarin & Zel'dovich (1989) for a review of the full ramifications of this deceptively simple equation. This approximation is used as a basis for several of the non-linear schemes described in § 5.2.2.

2.3 *Power Spectra, Initial Conditions, and Dark Matter*

We have seen that initially small perturbations grow by gravitational instability. It remains to characterize the distribution with scale of those initial perturbations. Let us define the Fourier Transform $\tilde{\delta}(\mathbf{k})$ of the fractional density field $\delta(\mathbf{r})$ at some early time $t$ such that:

$$\delta(\mathbf{r}) = \frac{1}{(2\pi)^3}\int d^3\mathbf{k}\, \tilde{\delta}(\mathbf{k})\, e^{-i\mathbf{k}\cdot\mathbf{r}}.\qquad(35)$$

(Care needs to be taken when comparing the results of different authors, as there is inconsistency in the literature about where the factors of $2\pi$ go in the definition of the Fourier Transform.) Because of the isotropy assumed in the Cosmological Principle, the statistical properties of $\tilde{\delta}(\mathbf{k})$ are independent of the direction of $\hat{\mathbf{k}}$, and so it makes sense to define a *power spectrum* $P(k)$:

$$\left\langle \tilde{\delta}(\mathbf{k}_1)\tilde{\delta}^*(\mathbf{k}_2) \right\rangle = (2\pi)^3 P(k)\, \delta^D(\mathbf{k}_1 - \mathbf{k}_2), \tag{36}$$

where $\delta^D$ is a Dirac delta function, and the averaging on the left-hand side is over directions of $\mathbf{k}$. As Bertschinger (1992) makes clear, $P(k)$ is a power *spectral density*, and thus represents the power per unit volume in $k$-space. One often sees the power spectrum defined as $P(k) \equiv \left\langle |\tilde{\delta}(\mathbf{k})|^2 \right\rangle$, but this is incorrect (at least for the Fourier Transform convention we've adopted), as can be seen by comparing the units of the two expressions: the power spectrum has units of volume, as does $\tilde{\delta}$.

---
[8] One has the freedom to add an arbitrary divergence-free term to the right-hand side of Eq. (33). This term corresponds to the decaying solution, and thus will be negligible at the present.

The quantity $\tilde{\delta}(\mathbf{k})$ is complex, and thus $P(k)$ is a complete statistical description of the density field [9] only if the phases of $\tilde{\delta}(\mathbf{k})$ are random. This is a natural prediction of inflationary models, and is often called the *random-phase hypothesis*. By the Central Limit Theorem and Eq. (35), random phases imply that the one-point distribution function of $\delta(\mathbf{r})$ is Gaussian [10], so the random-phase hypothesis is often also referred to as the *Gaussian hypothesis*. Random phases can strictly hold only in the limit of very small perturbations: $\delta(\mathbf{r})$ cannot be smaller than $-1$ (Eq. 24), but has no upper bound, and therefore it develops a positive skewness (about which we will have much more to say in § 5.4) as perturbations grow by gravitational instability. Until the lower bound on $\delta$ becomes important, linear theory holds, and because gravitational growth of perturbations is independent of scale in linear theory, the *shape* of the power spectrum is independent of time.

One way to quantify the density fluctuation field is in terms of the mass fluctuations within a spherical window of radius $R$:

$$\begin{aligned}
\left\langle \left(\frac{\delta M}{M}\right)^2 \right\rangle \bigg|_R &= \left\langle \left(\int d^3\mathbf{r}\, W(|\mathbf{r}-\mathbf{x}|)\delta(\mathbf{r})\right)^2 \right\rangle_{\mathbf{x}} \\
&= \frac{1}{(2\pi)^6} \left\langle \int d^3\mathbf{r}\, d^3\mathbf{k}\, d^3\mathbf{k}'\, \tilde{\delta}(\mathbf{k})\tilde{\delta}^*(\mathbf{k}')\, e^{i\mathbf{r}\cdot(\mathbf{k}-\mathbf{k}')}\, W^2(|\mathbf{r}-\mathbf{x}|) \right\rangle_{\mathbf{x}} \\
&= \frac{1}{(2\pi)^3} \int d^3\mathbf{k}\, P(k)\widetilde{W}^2(kR) \tag{37}
\end{aligned}$$

where $W(R)$ is the *window function* used; $\widetilde{W}(kR)$ is its Fourier Transform. For a *tophat* window function, which is unity out to some radius $R$, and then drops to zero,

$$\widetilde{W}(x) = \frac{3 j_1(x)}{x} \tag{38}$$

where $j_1(x) \equiv (\sin x - x\cos x)/x^2$ is the first spherical Bessel function. This is very crudely a step function to $k \approx 1/R$. This implies that characteristic mass fluctuations on a scale $k = 1/R$ are given roughly by

$$\delta M/M \sim k^{3/2} P(k)^{1/2}. \tag{39}$$

It is interesting to compare the mass fluctuations within a sphere of radius $R$ with the corresponding fluctuations in the bulk flow velocity within the same sphere. Using Eq. (30) and following a derivation

---

[9] That is, the multi-variate distribution function of $\delta(\mathbf{k}_i), i = 1,\ldots,N$ is a multi-variate Gaussian with covariance matrix given by Eq. (36).

[10] There are classes of models whose initial conditions are explicitly non-Gaussian: among them are models with initial seeds of gravitational growth (primordial black holes, cosmic string loops and wakes, monopoles, textures), and with initial explosions. We will have little to say about these models in this review.

very similar to that in Eq. (37), one finds:

$$\langle v^2 \rangle \Big|_R = \frac{H_0^2 f^2}{2\pi^2} \int dk\, P(k) \widetilde{W}^2(kR), \qquad (40)$$

which differs from Eq. (37) with two fewer powers of $k$ in the integrand. This means that the peculiar velocity field on a given scale is sensitive to components of the power spectrum on larger scales than is the density field, and thus is a useful probe of the largest-scale power. We illustrate this in Fig. 1, which compares the cumulative contribution to the integrals in Eqs. (39) and (40) for wavenumbers smaller than $k$. This calculation used a top-hat window of radius 50 $h^{-1}$ Mpc, and assumed a standard Cold Dark Matter model (see below). The rms velocity includes contributions from much larger scales than do the rms mass fluctuations.

The simplest inflationary models predict a so-called *scale-invariant* power spectrum, given by

$$P(k) \propto k, \qquad (41)$$

so-called because the potential fluctuations on a scale $\ell$ at the time that $\ell$ is equal to the horizon size, are independent of $\ell$ for this spectrum (cf., Kolb & Turner 1990). This form was in fact predicted well before inflation had been suggested (Peebles & Yu 1970, Harrison 1970, Zel'dovich 1972), as it is unique among power-law power spectra in avoiding divergences at large and small $k$ in various observable quantities.

The pure power-law power spectrum thought to exist in the early universe is sometimes called the primordial power spectrum. However, the power spectrum that is directly observable is that which holds following the epoch at which the energy densities in relativistic and non-relativistic particles are equal. The equations of gravitational instability we worked out above hold only for pressureless (i.e., non-relativistic) matter; in the relativistic case, the pressure of the matter is such as to retard the growth of perturbations on all scales within the radius of the universe. Thus perturbations on scales smaller than the horizon (the scale over which the universe is causally connected, $\sim ct$) can only grow after non-relativistic matter becomes dominant, and this is thus a convenient time at which to characterize the power spectrum.

While perturbations cannot grow on scales smaller than the horizon during the radiation-dominated era, they can and do grow on scales larger than the horizon. After the epoch of matter-radiation equality, perturbations on sub-horizon scales start to grow, although at a different rate[11]. The combination of these two effects causes a bending in an originally pure power-law power spectrum on the scale of the horizon size at matter-radiation equality. The index of the power law decreases by four over this bend. Thus following matter-radiation equality, the power spectrum which is proportional to $k^1$ on large scales bends over to become $k^{-3}$ on small scales. One quantifies this in terms of a *transfer function*, whose square is the quantity by which the primordial power spectrum is multiplied to generate the final power spectrum.

The power spectrum we have described is generic for universes in which the dark matter is non-relativistic (or "cold") when a galaxy mass is contained within the horizon; its form depends principally on the horizon size at the epoch of matter-radiation equality, which scales as $1/\Omega_0 h^2$. The so-called *Standard Cold Dark Matter* model (CDM) goes further and invokes inflation to give $\Omega_0 = 1$ and Eq. (41) for the primordial power spectrum; the Hubble Constant is set to $H_0 = 50$ km s$^{-1}$ Mpc$^{-1}$ in order to get an age of the universe in concordance with the ages of the oldest globular clusters (§ 9.6). Although the CDM model has been very popular ever since it was first proposed (some of the early important papers include Peebles 1982, Blumenthal *et al.* 1984, and Davis *et al.* 1985), due to its naturalness, lack of free parameters, and successes in predicting many of the observed properties of galaxies, there exists no compelling model invoking particles or physical processes *known to exist* that predicts any form of CDM. Particle physicists have come up with a number of hypothetical particles which could constitute the CDM, axions and the lightest supersymmetric partner to the photon among the most popular at the moment (cf. Primack, Seckel, & Sadoulet 1988; Kolb & Turner 1990).

---

[11] This is true for the dark matter, which does not couple with the radiation. However, baryonic matter is still ionized at the epoch of matter-radiation equality, and thus is tightly coupled to the photons. Therefore, fluctuations in the baryonic component cannot grow until recombination. These considerations are of great importance in quantifying fluctuations in the CMB on small scales.

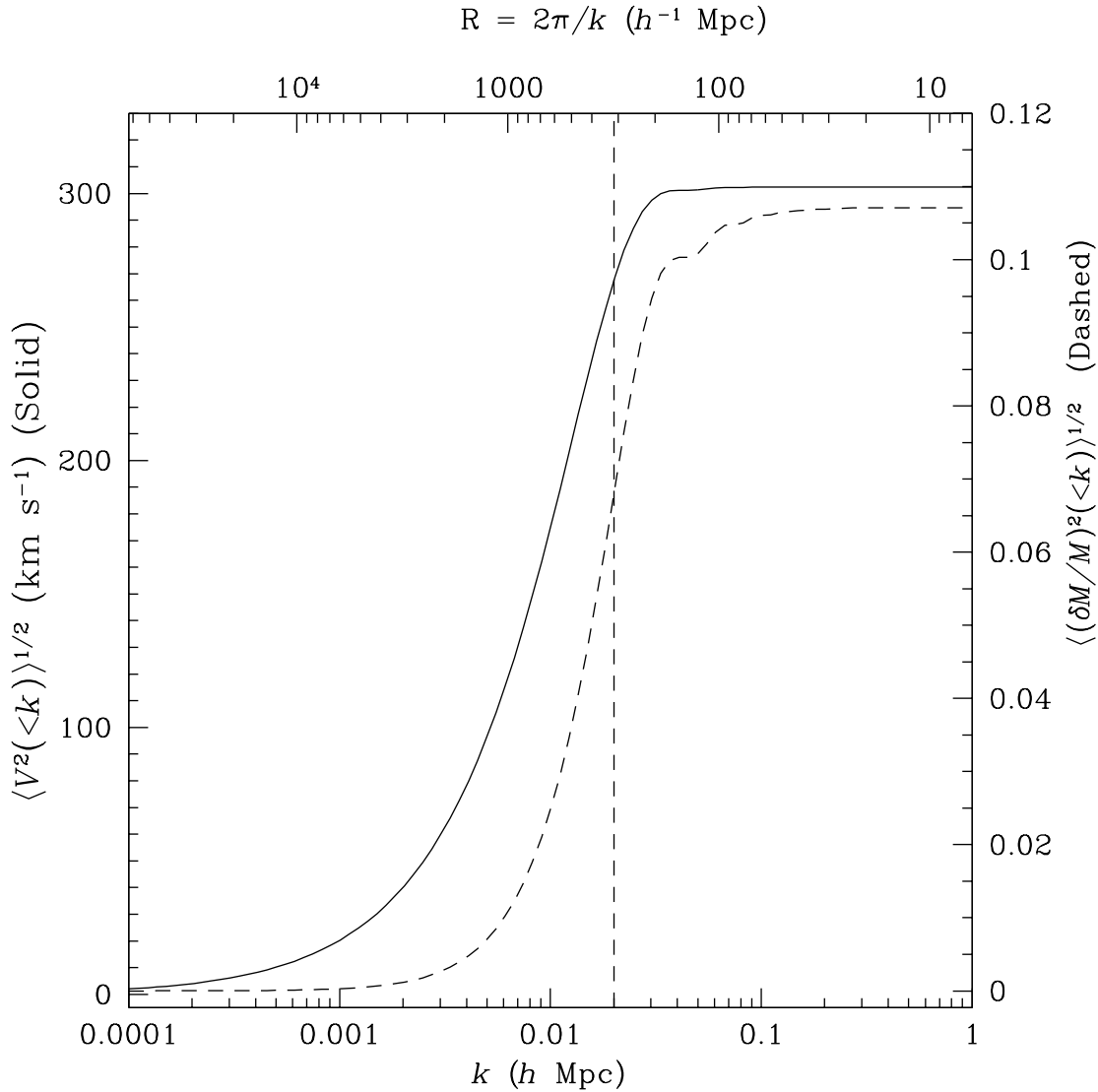

Fig. 1. The solid curve shows the square root of the contribution to the mean squared bulk flow within a tophat of radius 50 $h^{-1}$ Mpc as a function of the lower cutoff in $k$, for a standard CDM model. The dashed line is the square root of the contribution to the mean squared mass fluctuations (right-hand scale). Note that the velocity field includes much more contribution from small $k$ (large scales).

The power spectrum of the Cold Dark Matter model has been calculated numerically by a number of workers (Peebles 1982, 1984; Bond & Efstathiou 1984; Efstathiou, Bond, & White 1992), and fit to various functional forms; Efstathiou *et al.* (1992) give:

$$P(k) = \frac{Ak^n}{\left\{1 + \left[\left(\frac{6.4k}{\Gamma}\right) + \left(\frac{3.0k}{\Gamma}\right)^{3/2} + \left(\frac{1.7k}{\Gamma}\right)^2\right]^{1.13}\right\}^{2/1.13}}, \qquad (42)$$

where $A$ is a normalizing constant, $k$ is measured in units of $h\,\mathrm{Mpc}^{-1}$, $\Gamma \equiv \Omega_0 h$, and $n$ is the power-law index of the primordial power spectrum. The quantity $h$ is the Hubble constant in units of 100 km s$^{-1}$ Mpc$^{-1}$. Thus Standard CDM sets $\Gamma = 0.5$; models with smaller $\Gamma$ have more power on large scales. The quantity $\Gamma$ is inversely proportional to the Hubble radius at matter-radiation equality, and thus sets the scale at which the power spectrum bends over. This equation (and the equality between $\Gamma$ and $\Omega_0 h$) holds only in a universe with a negligible contribution of baryons to $\Omega_0$. The normalization for the case of $n = 1$ and $\Omega_0 = 1$ (standard CDM) can be written (Bunn, Scott, & White 1994; White, Scott, & Silk 1994):

$$A = \frac{6\pi^2}{5} \frac{\langle Q \rangle^2}{T_0^2} \eta_0^4 = 1.45 \times 10^{16} \frac{\langle Q \rangle^2}{T_0^2} \left( h^{-1} \mathrm{Mpc} \right)^4, \tag{43}$$

where $\eta_0$ is the conformal time at the present, $\langle Q \rangle$ is the measured quadrupole anisotropy of the CMB background, and $T_0$ is its measured temperature.

There is another physical effect that can come in. We have assumed that the dark matter is cold at the epoch when a galaxy mass is contained within the horizon. However, imagine that the dark matter is made up of neutrinos with a mass around 100 eV (already ruled out for $\nu_e$, but well within experimental bounds for $\nu_\mu$ and $\nu_\tau$). It can be shown from considerations of thermodynamic equilibrium in the early universe that the number density of neutrinos in this case would be enough to close the universe (i.e., $\Omega_\nu = 1$). Moreover, neutrino dark matter would be relativistic at a time when a galaxy mass was enclosed within the horizon, and thus in their free-streaming, would wash out any perturbations on scales comparable to the horizon scale at that epoch. The consequence is a power spectrum that cuts off exponentially on small scales. The evolution of large-scale structure in such a *Hot Dark Matter* model, or HDM, is very different from that in CDM. Because there is no power on the scale of galaxies in the HDM model, larger scale structures collapse first in pancakes, that then fragment later to form galaxies. This is often referred to as the *top-down* scenario. In CDM, on the other hand, there is power on galaxy scales, and galaxies form first, thereafter clustering into larger scales. Indeed, collapse is simultaneous for all scales for which $P(k) \propto k^{-3}$ (Eq. 39). This is referred to as hierarchical clustering, or the *bottom-up* scenario.

The mass perturbations predicted by inflation arise from perturbations in the gravitational potential, and thus relativistic and non-relativistic components fluctuate in tandem. These are called *adiabatic* fluctuations, in contrast with *isocurvature* fluctuations, in which the two fluctuate in opposite senses to give no net fluctuations in the mass density. The *Primordial Baryon Isocurvature* models (PBI), first suggested by Peebles (1987; cf. Bardeen, Bond, & Efstathiou 1987), invoke neither a hypothetical new particle (as in CDM) nor a hypothetical mass for a particle now believed to be massless (as the neutrino in HDM). Rather, baryons dominate the mass density of the universe. The PBI models characteristically show a strong peak in the power spectrum at the Jeans length at the epoch of matter-radiation equality.

The power spectra of a variety of models are shown in Fig. 2, based on a similar figure in Strauss *et al.* (1995). Shown are Standard CDM, HDM, and two variants of the PBI model; note the bump on large scales in the PBI models. Also shown are two variants of Standard CDM, one with $\Gamma = 0.3$ (LCDM, this increases the epoch of matter-radiation equality, and therefore the wavelength of the turnover in the power spectrum), and one which invokes somewhat non-standard inflationary models to assume a primordial power spectrum given by $P(k) \propto k^{0.7}$ (Tilted CDM, or TCDM). Finally, we show a hybrid model with a roughly 2 : 1 mixture of CDM and HDM (C+HDM). The parameters for these models are given in Table 1. We will discuss these models in detail in § 9.1.

All of these models are normalized to the anisotropies in the CMB on a scale of 10°, as measured by the *Cosmic Background Explorer* (COBE) (Smoot *et al.* 1992, Wright *et al.* 1992, Efstathiou, Bond, & White 1992), based on the assumption that the observed anisotropies are due to potential fluctuations at the surface of last scattering (the Sachs-Wolfe (1967) effect). Note that inflationary models (or for that matter, other competing models for the generation of fluctuations in the early universe) are not specific enough to predict the *amplitude* of the fluctuations; we simply do not know the relevant particle physics well enough. Thus COBE provides the only unambiguous way to normalize power spectra[12]. Indeed, in the pre-COBE days, the normalization of power spectra was a free parameter which was only weakly constrained by galaxy clustering (Eq. 52). A thorough discussion can be found in Efstathiou (1991) and Peebles (1993).

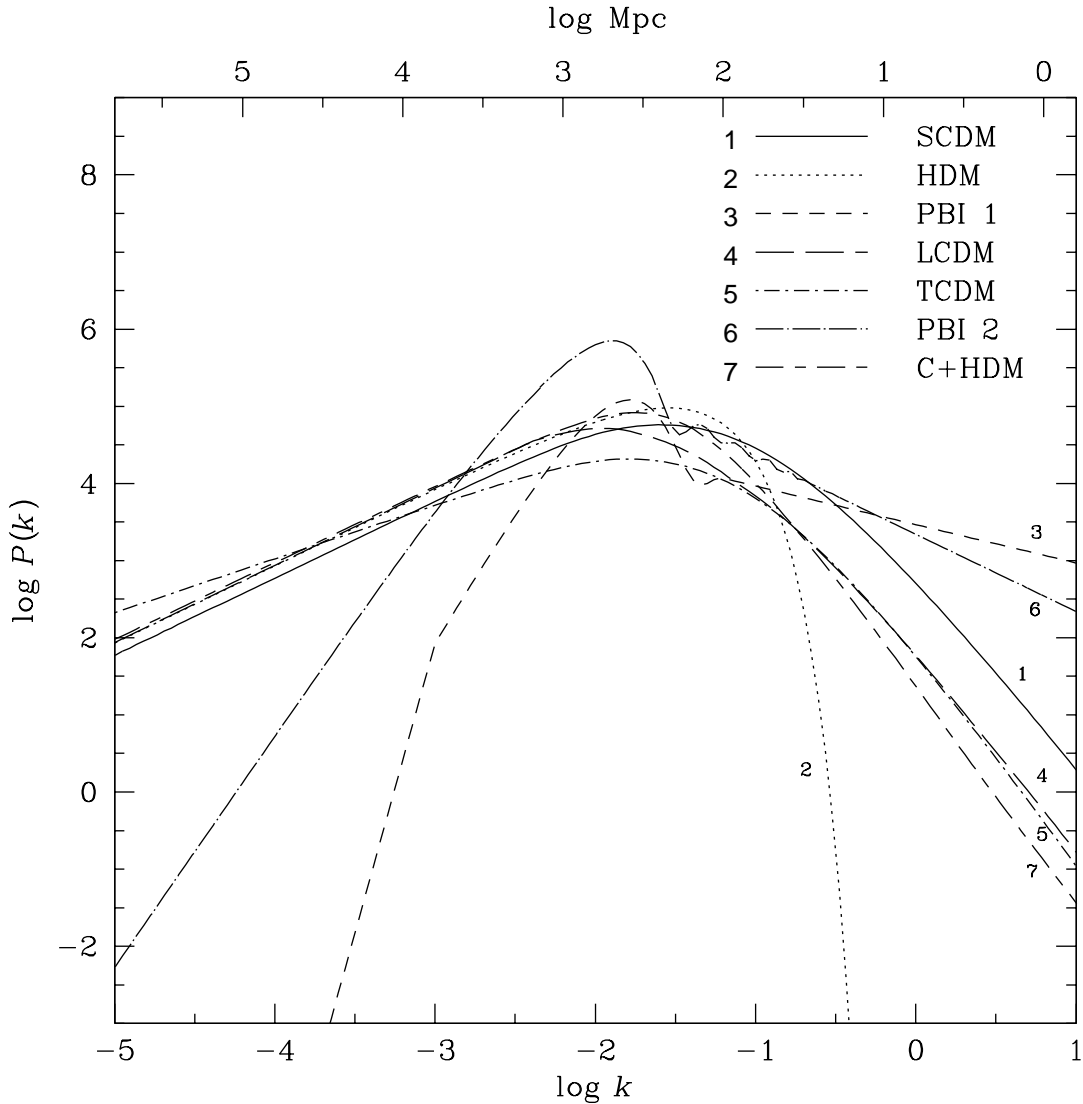

Fig. 2. The power spectra in seven models of large-scale structure: Standard CDM, Tilted CDM (Tilted CDM), $\Gamma = 0.3$ CDM (LCDM), HDM, Mixed Dark Matter (C+HDM), and two variants of the PBI model. All are normalized to the COBE observations.

---

[12] It has been pointed out (e.g., Davis *et al.* 1992) that in inflationary models that predict a deviation from the Harrison-Zel'dovich power spectrum, there is substantial generation of gravity waves, and thus the Sachs-Wolfe effect does not account for all the anisotropies seen by COBE.

Table 1: The Parameters of the Power Spectra

| Model | $\Omega_0{}^a$ | $\sigma_8{}^b$ | $H_0{}^c$ | $n^d$ or $m^e$ | $x^f$ |
|---|---|---|---|---|---|
| Standard CDM | 1.0 | 1.05 | 50 | 1.0 | 0.0 |
| HDM | 1.0 | 0.86 | 50 | 1.0 | 0.0 |
| PBI 1 | 0.2 | 0.90 | 80 | $-0.5$ | 0.1 |
| PBI 2 | 0.3 | 1.02 | 50 | $-1.0$ | 0.1 |
| LCDM | 0.3 | 0.67 | 67 | 1.0 | 0.0 |
| TCDM | 1.0 | 0.50 | 50 | 0.7 | 0.0 |
| C+HDM | $1.0^g$ | 0.67 | 50 | 1.0 | 0.0 |

$^a$ All models are spatially flat so that $\Lambda_0 = 1 - \Omega_0$.
$^b$ rms fraction density fluctuations within $8\,h^{-1}$ Mpc spheres.
$^c$ Hubble Constant in units of km s$^{-1}$ Mpc$^{-1}$.
$^d$ Adiabatic primordial spectral index.
$^e$ Isocurvature primordial spectral index.
$^f$ Ionization fraction.
$^g$ $\Omega_{CDM} = 0.7$, $\Omega_{HDM} = 0.3$.

*2.3.1 The Correlation Function*

The *autocorrelation function*, or, more commonly, the *correlation function* of the mass density field, is one of the most powerful methods of quantifying galaxy clustering. It is defined by

$$\xi(r) \equiv \langle \delta(\mathbf{x}) \delta(\mathbf{x} + \mathbf{r}) \rangle, \tag{44}$$

where the average is over position $\mathbf{x}$ and isotropy guarantees that $\xi(r)$ is independent of the direction of $\mathbf{r}$. Eq. (44) is only applicable for continuous density fields; for a point process such as the distribution of galaxies, we define $\xi(r)$ operationally as the mean excess number of galaxy pairs at separation $r$, over that expected for a pure Poisson distribution. That is, the mean number of galaxies in a spherical shell a distance $r$ from a given galaxy of thickness $\mathrm{d}r$ is given by

$$N(r) = 4\pi r^2 \, \mathrm{d}r \, n[1 + \xi(r)], \tag{45}$$

where $n$ is the mean number density of galaxies. If we expand the $\delta$'s in Eq. (44) in inverse Fourier Transforms, we find:

$$\begin{aligned} \xi(r) &= \left\langle \int \mathrm{d}^3\mathbf{k}\,\mathrm{d}^3\mathbf{k}'\, e^{i\mathbf{k}\cdot\mathbf{r}}\, e^{i\mathbf{x}\cdot(\mathbf{k}-\mathbf{k}')} \delta(\mathbf{k})\delta(\mathbf{k}') \right\rangle \\ &= \frac{1}{(2\pi)^3} \int \mathrm{d}^3\mathbf{k}\, P(k)\, e^{i\mathbf{k}\cdot\mathbf{r}}. \end{aligned} \tag{46}$$

Thus the correlation function and the power spectrum form a Fourier Transform pair. We discuss the application of $\xi(r)$ in much greater detail in § 5.1. Just as for the power spectrum, the correlation function gives a complete statistical description of the density field for a Gaussian field; all higher-order correlations are zero. Thus higher-order correlations are a measure of non-Gaussianity, as we discuss in § 5.4.

*2.4 The Relation Between the Mass and Galaxy Density Fields*

Our discussion thus far has been quite abstract; galaxies have been mentioned only in passing. However, it is galaxies that we observe directly, while we believe that most of the material of the universe is made up of dark matter; we need to make a connection between the galaxies and the mass density field. The simplest assumption, and one that was implicitly made until the last decade, is that the distribution of galaxies is a mirror of the distribution of dark matter. This could come about, for example, if each galaxy were surrounded by a halo of dark matter with total mass in proportion to the (visible) luminosity of a galaxy, with no extra component of dark matter either smoothly distributed, or lying in clumps with no associated visible matter. However, there is very little evidence that this is true, and some indirect evidence that it is false. For example, one could imagine that a component of the dark matter is distributed like that of the galaxy field smoothed on scales of 5 $h^{-1}$ Mpc, and does not follow the details of the galaxy distribution on smaller scales. If this were the case, one would get a misleading impression of the small-scale distribution of dark matter from studying galaxies. A specific model for this sort of effect was first suggested by Kaiser (1984), in which galaxies form only at the high-density peaks of the mass density field. The galaxy distribution is then said to be *biased* with respect to the mass distribution. The term biasing is used to refer to a number of different but related effects.

The original argument of Kaiser (1984) was concerned not with the biasing of galaxies relative to dark matter, but of clusters of galaxies relative to galaxies: clusters of galaxies are likely to be produced where the initial density contrast, smoothed on larger scales, is high, and therefore is likely to produce a higher than average density of other clusters. The result is a stronger correlation function (Eq. 44) of clusters than that of galaxies, as is indeed observed (Bahcall & Soneira 1983; Bahcall & West 1992). Thus this form of bias is known to exist.

It was then realized that a similar effect could cause galaxies to be biased relative to the underlying dark matter (Bardeen *et al.* 1986, Davis *et al.* 1985, Peacock & Heavens 1985), causing the clustering of the former to be stronger than that of the latter, and decreasing the apparent value of $\Omega_0$ from dynamical studies. The so-called *peaks biasing* model originally proposed by Kaiser (1984) makes a definite prediction for the relation between the correlation function of the galaxies and dark matter, at least on large scales:

$$\xi_{\text{galaxies}}(r) = b^2 \, \xi_{\text{dark matter}}(r), \tag{47}$$

where $b$ is the *biasing parameter*, which is (approximately) a constant, independent of scale, related to the threshold above which galaxies are presumed to have formed. In practice, we will often assume the much more specific *linear biasing model*, in which

$$\delta_{\text{galaxies}}(\mathbf{r}) = b \, \delta_{\text{dark matter}}(\mathbf{r}). \tag{48}$$

This is in fact the intuitive definition of biasing that most people have, in which a constant of proportionality relates the dark matter density field to that of the galaxies. Note that in practice, the galaxy distribution is a point process, not a continuous field, and thus the bias defined as above must always make reference to a smoothing length. Eq. (48) implies Eq. (47) through Eq. (44), although the converse is certainly not true. Furthermore, for $b > 1$ (as is usually assumed) Eq. (48) cannot be strictly true, because both $\delta_{\text{galaxies}}$ and $\delta_{\text{dark matter}}$ are limited below by $-1$, and thus one occasionally sees models like:

$$1 + \delta_{\text{galaxies}}(\mathbf{r}) = [1 + \delta_{\text{dark matter}}(\mathbf{r})]^b \,, \tag{49}$$

which gets around this problem. Weinberg (1994; cf., Coles 1993) has shown empirically that in a wide range of models which invoke *local* biasing, i.e., in which

$$\delta_{\text{galaxies}}(\mathbf{r}) = f\left[\delta_{\text{dark matter}}(\mathbf{r})\right], \tag{50}$$

the ratio of the fluctuations $\sigma(R)$ in the galaxies and dark matter is independent of scale $R$ on scales larger than $\approx 5\,h^{-1}$ Mpc. This of course will not hold in *non-local* biasing models, in which galaxy formation is influenced by events many Mpc away (Babul & White 1991; Bower *et al.* 1993); in these schemes, the ratio of the amplitude of dark matter and galaxy fluctuations can be a strong function of scale.

Any given prescription relating $\delta_{\text{galaxies}}$ to $\delta_{\text{dark matter}}$ is expected to hold only in the mean; at any given point in space, there will be fluctuations around this mean in the relation between the two. A complete model of biasing must specify the distribution of these fluctuations.

To the extent that the linear biasing model holds, comparisons of peculiar velocities and gravity via Eq. (30) will not be able to constrain the quantity $f$, but rather the combination

$$\beta \equiv \frac{f(\Omega_0, \Lambda)}{b}. \tag{51}$$

Thus our ignorance about the relative distribution of galaxies and dark matter translates directly into ignorance of the value of $\Omega_0$. Note in particular that for $b > 1$, as is usually assumed, $\Omega_0 = 1$ models can be reconciled with observed low values of $\beta$. Thus biasing is often invoked to explain why some dynamical observations imply a value of $\Omega_0$ substantially below the inflationary prediction of unity (§ 8.1).

An early review of the biasing model and specific physical models which predict biasing can be found in Dekel & Rees (1987); recent work has found biasing of the galaxies relative to the dark matter in numerical simulations of galaxy formation and the growth of structure in an expanding universe (e.g., White et al. 1987; Gelb & Bertschinger 1994a; Cen & Ostriker 1992b; Katz, Hernquist, & Weinberg 1992).

If galaxies can be biased with respect to the underlying dark matter, different populations of galaxies can in principle be biased with respect to one another. We know that this holds on small scales for elliptical and spiral galaxies: although elliptical galaxies make up roughly 15% of the population of galaxies in the field, they are by far the dominant population in the cores of rich clusters (Hubble 1936, Oemler 1974, Dressler 1980ab, Dressler 1984, Postman & Geller 1984, Whitmore, Gilmore, & Jones 1993). This is reflected in a steeper correlation function of ellipticals relative to spirals (Davis & Geller 1976, Giovanelli, Haynes, & Chincarini 1986). We refer to this as a form of *relative biasing* of two galaxy populations, and will discuss further observational evidence for this below.

Finally, the term biasing is occasionally (mis)-used to refer to the normalization of the power spectrum. It is observed that the variance of optically selected galaxy number counts is approximately unity within spheres of radius $8\,h^{-1}$ Mpc (Eq. 37, Davis & Peebles 1983b) and thus the bias can be defined as

$$\frac{1}{b^2} = \frac{1}{(2\pi)^3} \int d^3\mathbf{k}\, P(k)\widetilde{W}^2(kR), \tag{52}$$

at $R = 8\,h^{-1}$ Mpc, with $W$ given by Eq. (38). This definition is consistent with that in Eq. (48) only to the extent that the variance in galaxy counts at $R = 8\,h^{-1}$ Mpc is actually unity, whereas in fact it varies from sample to sample. Thus we will shun this definition of $b$ in this paper.

## 2.5 Outstanding Questions

This finishes our overview of the theoretical framework within which we will interpret the results of redshift and peculiar velocity surveys. As a form of summary, we list the issues we would like to address with observational data:

(i) The Friedman-Robertson-Walker metric is characterized by a number of parameters: $\Omega_0, H_0, \Lambda, q_0$, and $t_0$, which, although not independent in the model, are constrained observationally by quite different methods. Redshift and peculiar velocity surveys have the most power to constrain $\Omega_0$ (mostly through Eq. 30), and that is what we will concentrate on for this review.

(ii) The gravitational instability model is compelling, but needs to be tested observationally. Linear theory and gravitational instability imply a unique relation between the mass and velocity fields (Eq. 30), furnishing one such test.

(iii) The power spectrum following the epoch of radiation-matter equality is of paramount importance for determining the evolution of large-scale structure in our universe. Moreover, as we have seen, it can give vital clues as to the form and amount of the dark matter. We will attempt to constrain it with observations.
(iv) The power spectrum discussed above is a complete statistical description of the mass fluctuations in the early universe only to the extent that the Fourier modes had random phases. We will see that there do exist direct observational tests of this from the nearby galaxy distribution.
(v) The physical processes by which galaxies formed are very poorly understood. Redshift and peculiar velocity surveys can put constraints on the relative distribution of galaxies and dark matter; this relative distribution ultimately must be predicted by galaxy formation models.
(vi) At the end, we wish to step back and ask whether the various observational constraints we have found allow a coherent picture consistent with the theoretical framework. That is, are our observations consistent with the Big Bang model, with structure forming via gravitational instability? To what extent can we rule out competing models?

At this point, we are ready to start addressing these questions with observations of the galaxy distribution and velocity field at low redshifts.

## 3   Redshift Surveys: Setting the Quantitative Groundwork

In this chapter, we discuss the history of redshift surveys, and the basic methods required to do quantitative work with them. § 3.1 introduces the main quantities that differentiate one redshift survey from another, and § 3.2 summarizes the history of redshift surveys. § 3.3 discusses the methods used to measure redshifts with optical and radio telescopes. Flux-limited surveys have a number density of objects that drops as a function of distance from the Earth; this is quantified with the selection function, as discussed in § 3.4. The selection function is closely related to the luminosity function, whose astrophysical implications are discussed in § 3.5 and § 3.6. With the selection function in hand, the density field of galaxies can be determined, using methods discussed in § 3.7. One of the difficulties in determining the density field is the lack of redshift surveys at low Galactic latitudes. Methods to correct for this are the topic of § 3.8.

### 3.1   The Variety of Redshift Surveys

In this review, we will concentrate on redshift surveys of well-defined samples of galaxies. By well-defined, we mean those in which the selection criteria are quantifiable and reproduceable (at least in a statistical way), for without this, it is impossible to do quantitative analyses with them. In practice, this usually means that a sample is defined as limited by some photometric property, usually received flux or diameter in some band. A sample may also have secondary selection criteria as well, such as galaxy morphology, color, or surface brightness. Unfortunately, the data one has available to define a sample are rarely of very high photometric accuracy, and thus the limits are always approximate to some extent. A redshift survey sample is thus defined by several factors:
(i) The region of sky covered.
(ii) The photometric quantities with respect to which the sample is defined, and the errors thereof.
(iii) The limits on these quantities.
(iv) The fraction of the galaxies meeting the selection criteria for which redshifts are measured.
From these parameters, several other characteristics of the survey follow, including the total number of galaxies included, the number density of objects surveyed (i.e., its sparseness), and some measure of a typical redshift (the "depth") in the survey.

Several authors, including Geller & Huchra (1988), Giovanelli & Haynes (1992) and Borgani (1994) have given lists of major redshift surveys of galaxies. Redshift surveys fall into several categories:
- One-dimensional ("pencil-beam") surveys, which go deep over a very small region of sky. Given the slow spectrographs of twenty years ago, these were all that were available in the early days of redshift surveys (Gregory & Thompson 1978; Kirshner et al. 1978; 1981). Recent pencil-beam surveys push to very faint magnitudes and therefore high redshifts (Broadhurst et al. 1988, 1990; Colless et al. 1990), especially with the advent of multi-object spectrographs. These surveys are especially useful for studying galaxy evolution.
- Two-dimensional surveys, in which a long thin strip on the sky is covered. The most influential of these is the CfA2 "slice" survey (de Lapparent, Geller, & Huchra 1986) to be discussed below.
- Three-dimensional surveys, in which a substantial solid angle on the sky is covered. This is the major emphasis of this review.
- Targeted surveys of special kinds of objects. These include surveys of clusters of galaxies, active and unusual galaxies of various kinds, and pairs and groups of galaxies. Recent cluster redshift surveys include those of Huchra et al. (1990b), Zabludoff, Huchra, & Geller (1990), Huchra et al. (1992), Guzzo et al. (1992), Lauer & Postman (1994), and Dalton et al. (1994); see Bahcall (1988) and Bahcall & West (1992) for recent reviews of large-scale structure as traced by galaxy clusters. Redshift surveys of active and unusual galaxies are typically carried out not for studies of large-scale structure, but rather to learn about the properties of the galaxies themselves; we will not discuss them further.
- "Blind" redshift surveys carried out in the 21 cm line of neutral hydrogen (HI). These are not pointed at galaxies or galaxy candidates as defined in other passbands, but survey the sky over a range of redshifts to learn about the HI content of the universe (Kerr & Henning 1987; Weinberg et al. 1991; Hoffman, Lu, & Salpeter 1992; Szomoru et al. 1994a; Briggs 1994).

A redshift survey requires a predefined sample of targets, and thus is often defined by catalogs of galaxies detected in photographic surveys of the sky. There have been heroic efforts to define complete galaxy samples from photographic sky surveys prepared on Schmidt plates, but because a telescope at any given location on earth can only see a fraction of the sky, galaxy samples tend to be restricted to either the Northern or Southern celestial hemisphere; moreover, dust extinction at low Galactic latitudes from our own Galaxy means that galaxy samples are woefully incomplete at low latitudes (the *zone of avoidance*.) The most important large galaxy catalogs from which redshift survey samples have been drawn include:
- The Catalog of Galaxies and Clusters of Galaxies (CGCG), prepared by Zwicky and collaborators (1961–1968) from blue plates taken at the Palomar 18″ Schmidt telescope. It covers the entire Northern Sky ($\delta > -2.5°$) except for regions at very low Galactic latitudes, and claims completeness to blue magnitudes $\leq 15.5$. It contains $\approx 30{,}700$ entries.
- The Uppsala General Catalog of Galaxies (UGC) prepared by Nilson (1973) from the Palomar Observatory Sky Survey plates taken on the Palomar 48″ Schmidt Telescope. It claims completeness for galaxies with diameters measured (by eye) on the blue plates greater than $1'$, although Hudson & Lynden-Bell (1991) show that the survey is incomplete below $1.65'$. The survey includes 12,940 objects.
- The European Southern Observatory Galaxy Catalog (ESO) prepared by Lauberts (1982), from the ESO-Schmidt plates of the Southern Sky taken with the 48″ Schmidt telescope at Siding Springs. It contains 16,160 galaxies with blue diameters greater than $1'$ and with $\delta < -17.5°$, although Hudson & Lynden-Bell (1991) claim incompleteness below $1.3'$. Lauberts & Valentijn (1989) have done photo-densitometer scans of most of the ESO catalog, resulting in a uniform set of diameters and magnitudes.
- The Catalogue of Principal Galaxies (Paturel et al. 1989) consists of a compilation of data for 73,197 galaxies drawn from existing catalogs.
- The Third Reference Catalogue of Bright Galaxies (RC3; de Vaucouleurs et al. 1991) is a major compilation of existing catalogs. It contains 23,022 galaxies with photometric data put onto a common basis.

The RC3 is three volumes long, and probably represents the last comprehensive compilation of galaxy data to be published in hard-cover. Major galaxy databases are now maintained on-line in computer readable formats, and include the redshift compilation of Huchra *et al.* (1992), consisting of $\sim 57,000$ entries with $\sim 52,000$ redshifts, and the NASA/IPAC Extragalactic Database (NED) available via Internet (Helou *et al.* 1991), containing data on 286,674 galaxies and galaxy candidates (as of May 1994), most without redshifts. The largest single galaxy catalog in existence is not yet publicly available, and contains $2 \times 10^6$ galaxies brighter than $b = 20.5$ measured by photodensitometer scans (the APM machine) of the ESO-Schmidt plates over 1.3 ster in the Southern Galactic Cap (Maddox *et al.* 1990abc).

*3.2 History of Redshift Surveys*

We summarize here some of the major redshift surveys, and give a brief history of the field. Our emphasis is on redshift surveys of nearby galaxies covering substantial areas of sky, and thus we do not include all pencil-beam surveys and those of unusual types of objects.

Large-scale structure was recognized from two-dimensional maps of the galaxy distribution very early on; see for example Charlier (1922; quoted in Chapter 3 of Peebles 1993), and Chapter 3 of Hubble (1936). de Vaucouleurs (1948) recognized the existence of the Local Supercluster from two-dimensional data, although it took many years before his discovery was appreciated by the general astronomical community. When Shane & Wirtanen (1967) published their deep galaxy counts from Lick astrograph plates over a substantial area of the Northern Sky, detailed quantitative analyses of the galaxy distribution could be done (Groth & Peebles 1977; Fry & Peebles 1978); more importantly, the resulting map (Seldner *et al.* 1977, cf. Figure 3.9 of Peebles 1993) showed the community the rich structure that could be found in the galaxy distribution. However, these sky maps lacked the third dimension which is provided by redshift measurement.

The first redshift surveys carried out to study the large-scale distribution of galaxies were those of Gregory & Thompson (1978a, 1978b, 1984) and Gregory, Thompson, & Tifft (1981), who did pencil beam surveys towards Coma, Perseus, and other clusters, and started the process of identifying superclusters. The first redshift survey to cover a large fraction of the sky was that of the Revised Shapley-Ames Galaxy Catalog (Sandage & Tammann 1981), consisting of all galaxies with $B_T < 13.2$ and with Galactic latitude $|b| > 30°$. The sample contains 1191 galaxies covering 6.29 ster, and has a median redshift of 1500 km s$^{-1}$. Major analysis papers of this dataset include Davis, Geller & Huchra (1978), Sandage, Tammann, & Yahil (1979), and Yahil *et al.* (1980). Much of the early work is reviewed in Oort (1983).

Kirshner, Oemler, & Shectman (1978), and Kirshner *et al.* (1981; 1983; 1987) carried out a series of pencil beam surveys, with the original purpose of studying the galaxy luminosity function. Three such pencil beams in the region of the constellation Boötes found a common gap in the redshift distribution. This so-called Boötes void is now known to have a diameter of 6000 km s$^{-1}$ (Kirshner *et al.* 1987; Strauss & Huchra 1988; Dey, Strauss, & Huchra 1990). At the time it was first discovered, such vast regions of space substantially underdense relative to the mean were a great surprise to the community, although a few workers, notably Einasto and colleagues (Einasto, Joeveer, & Saar 1980), had been pointing out the existence of filamentary structure and voids in galaxy maps for some time.

The Center for Astrophysics (CfA) redshift survey (Huchra *et al.* 1983) includes all galaxies in the CGCG with $m_z \leq 14.5$ mag in the region of sky defined by $\delta > -2.5°$ and $b < -30°$, and $\delta > 0°$ and $b > +40°$. The sample contains 2417 galaxies over 2.67 ster, with a median redshift of 3300 km s$^{-1}$. This survey showed a galaxy distribution in which voids and filaments were ubiquitous, although none of the voids seen matched that in Boötes void discovered by Kirshner *et al.* (1981). Major early analysis papers of the CfA survey include Davis *et al.* (1980), Davis & Huchra (1982), Davis *et al.* (1982), and Davis

& Peebles (1983b), but this workhorse of redshift surveys has been analyzed in detail in dozens of other papers by many authors over the years.

Fisher & Tully (1981) used the Green Bank and Effelsburg radio telescopes to carry out the first systematic redshift survey of galaxies using the H I 21 cm line. Their aim was to include all galaxies with redshifts less than 3000 km s$^{-1}$, and thus their selection criteria are somewhat ill-defined. Their original survey included data for 1787 galaxies; it has since been extended to 2367 galaxies with $cz < 3000$ km s$^{-1}$, and published in tabular and graphical form by Tully & Fisher (1987) and Tully (1987).

The Southern Sky Redshift Survey (SSRS; da Costa et al. 1988, 1991) consists of all galaxies in the ESO catalog with B band diameters (corrected to face-on) greater than 1.9', and with $\delta < -17.5°$ and $|b| > 30°$. The sample consists of 1963 galaxies covering 1.96 ster; 12% do not have redshifts (these are largely galaxies with very low central surface brightnesses). The median redshift is 4000 km s$^{-1}$. Analysis papers include Davis et al. (1988), Park, Gott, & da Costa (1992), Maurogordato, Schaeffer, & da Costa (1992), and Lachiéze-Rey, da Costa, & Maurogordato (1993).

Giovanelli & Haynes have used the Arecibo radio telescope to carry out an H I survey of galaxies with $\delta \geq 0°$ and $b < 10°$ from the CGCG and UGC catalogs; early-type galaxies without detectable H I have redshifts measured optically. Their sample contains $\approx 5000$ galaxies, and is centered on the Pisces-Perseus supercluster. Reviews of their work include Giovanelli & Haynes (1988) and Giovanelli & Haynes (1991).

In 1986, de Lapparent, Geller & Huchra published their first 6° wide slice of a deeper extension of the original CfA survey. The galaxy distribution in this two-dimensional slice is shown in Fig. 3, from the data in Huchra et al. (1990a). Redshift is the radial coordinate, and right ascension is the angular coordinate; declination is suppressed. This figure, which is often compared to a slice through a network of bubbles, shows clearly that voids like that in Boötes are ubiquitous, and that galaxies are aligned along sheet-like structures. The structure stretching across the entire slice at $cz \approx 8000$ km s$^{-1}$ has been called the Great Wall (Geller & Huchra 1989), and is the largest structure known to exist (cf., Di Nella & Paturel 1994). Geller, Huchra, and collaborators are extending the original CfA survey to $m_Z = 15.5$, over the full area covered by the CGCG. The survey is nearing completion, and will contain roughly 15,500 galaxies at high latitudes and another 3000 in regions near the Galactic plane. The median redshift is $\approx 7500$ km s$^{-1}$. The results from this survey are reported in de Lapparent et al. (1986a, 1988, 1989, 1991), Huchra et al. (1990a), Vogeley et al. (1991, 1992, 1994), Ramella et al. (1992), and Park et al. (1994).

The *Infrared Astronomical Satellite* (*IRAS*) flew in 1983, and did a full-sky survey with $\sim 1'$ resolution in four broad bands centered at 12, 25, 60, and 100$\mu$m. Because infrared radiation is not impeded by Galactic extinction, galaxy samples selected from the *IRAS* database are well-suited for full-sky redshift surveys. They do have the drawback, however, that early-type galaxies have very little dust or star formation, and thus are not represented in *IRAS* galaxy samples.

The first major redshift survey of *IRAS* galaxies (Soifer et al. 1987) consisted of the 324 galaxies with $f_{60} > 5.4$ Jy with $|b| > 30°$ in the Northern sky. Strauss and collaborators carried out a survey covering 11.06 ster (88% of the sky, missing only regions at very low Galactic latitudes and areas not surveyed by *IRAS*) and including all 2658 galaxies to $f_{60} = 1.936$ Jy. Fisher et al. (1995) followed this up with a deeper survey over the same region of sky to $f_{60} = 1.2$ Jy, for a total of 5339 galaxies; redshifts are 99.6% complete. The median redshift of this survey is 5800 km s$^{-1}$. Scientific results from this survey include Strauss et al. (1990, 1992abc), Yahil et al. (1991), Fisher et al. (1992, 1993, 1994abcd), Bouchet et al. (1993), and Dekel et al. (1993).

A parallel effort led by Rowan-Robinson consisted of a sparse sample of one in six *IRAS* galaxies to a flux limit of $f_{60} = 0.6$ Jy. The selection criteria differed in detail from the *IRAS* surveys mentioned before, and covered a solid angle of 10.3 ster with 2184 galaxies. This survey is deeper but sparser than the complete *IRAS* surveys, with a median redshift of 8400 km s$^{-1}$. This survey is nicknamed QDOT, after the institutions of the investigators. Major papers include Rowan-Robinson et al. (1990), Efstathiou et al.

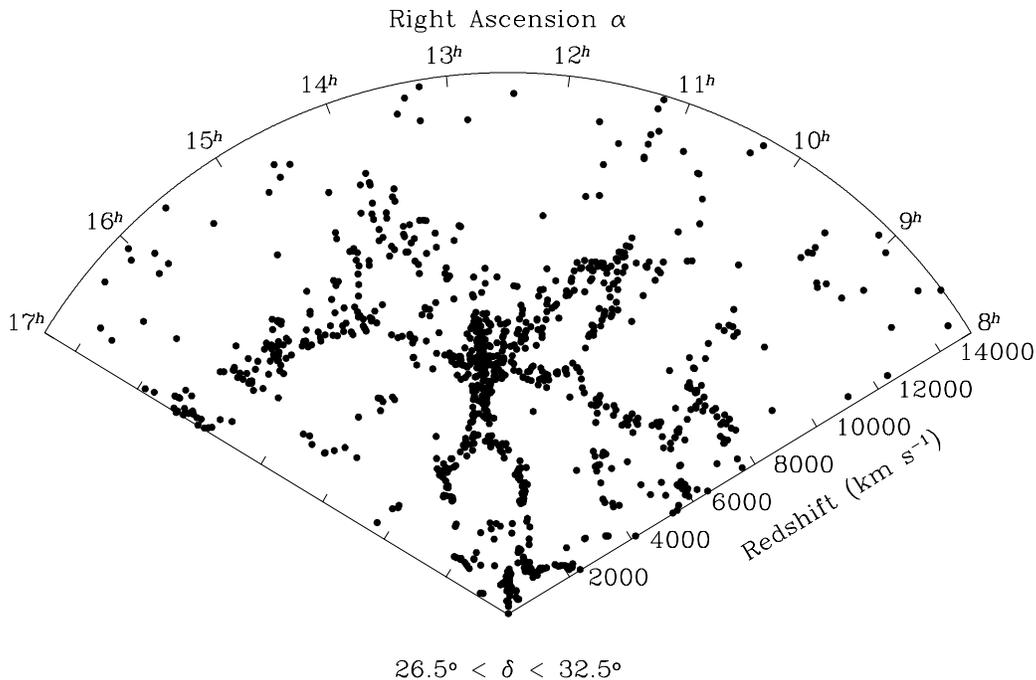

Fig. 3. The galaxy distribution from the survey of de Lapparent *et al.* (1986a), using the data of Huchra *et al.* (1990a). Redshift is plotted in the radial direction, and right ascension is the angular coordinate. Declination is suppressed. The prominent structure in the center of the figure is the Coma cluster.

(1990), Saunders *et al.* (1990), Saunders *et al.* (1991), Kaiser *et al.* (1991), and Lawrence *et al.* (1994).

Loveday *et al.* (1992a, b, 1994) have carried out a redshift survey following up the APM photometric survey of Maddox *et al.* (1990abc). They measured redshifts for one in twenty of galaxies brighter than $b_J = 17.15$ over 1.3 steradians centered on the Southern Galactic Cap. The sample contains 1787 galaxies with a median redshift of 15,200 km s$^{-1}$.

Although not a redshift survey itself, the effort of Hudson (1993ab, 1994ab) should also be mentioned. He calculated the redshift incompleteness (using the compilation of Huchra *et al.* 1992) of the UGC and ESO catalogs as a function of diameter and position on the sky, and statistically corrected the redshift maps accordingly, allowing him to recreate the full optical galaxy density field, which he used for comparison to peculiar velocity studies (§ 8.1).

There are also a number of redshift surveys in progress or in the advanced planning stages (see the volume edited by Maddox & Aragón-Salamanca 1995 for papers on these and other surveys in progress):
– Strauss and collaborators have completed a redshift survey of optically selected galaxies (the ORS) to

roughly the CfA depth from the UGC and ESO galaxy catalogs. In the declination strip not covered by either of these two catalogs ($-17.5° < \delta < -2.5°$) they use the recently completed diameter-limited *Extension to the Southern Galaxy Catalog* (Corwin & Skiff 1994). The sample is limited by Galactic extinction to $|b| > 20°$, covers 8.09 ster, and contains 8457 galaxies, of which redshifts are available for 8286. The median redshift is 3900 km s$^{-1}$. The sample and the resulting galaxy distribution are presented in Santiago et al. (1995ab) and Santiago (1994).

– Freudling, da Costa, & Pellegrini (1994) have compiled a similar optically selected sample of galaxies based on the CfA and SSRS surveys, and the *Morphological Catalog of Galaxies* (MCG) in the declination strip not covered by either survey (Huchra et al. 1993). Their zone of avoidance is more extensive than that of the ORS. Preliminary results can be found in Freudling & da Costa (1994).

– The collaboration which carried out the QDOT *IRAS* redshift survey is in the process of extending their survey to include *every* galaxy brighter than 0.595 Jy at 60$\mu$m within their surveyed area. The survey should be completed in fall 1994, and will contain $\approx 15,000$ galaxies. A parallel effort is using the *IRAS* Faint Source Survey to push to 0.2 Jy at 60$\mu$m over a much smaller area of sky (cf. Lonsdale et al. 1990).

– da Costa and collaborators are carrying out a redshift survey in the Southern Hemisphere of comparable scope to the CfA2 in the North (although over a somewhat smaller solid angle). As there exists no galaxy sample equivalent to the CGCG in the South, the galaxy selection and photometry is being carried out from photo-densitometer scans of photographic plates. The sample consists of 3600 galaxies over 1.13 ster. First results are in da Costa (1994) and da Costa et al. (1994a, b).

– Geller & Huchra are measuring redshifts for all galaxies to $r = 16.5$ in a strip $1° \times 100°$ centered on the original CfA "slice" (de Lapparent et al. 1986a). The sample includes 2400 galaxies with a median redshift of 20,000 km s$^{-1}$, and is called the "Century Survey".

– The team that carried out the Boötes Void survey discussed above is conducting an extensive redshift survey from the Las Campanas Du Pont 2.5m telescope, using a multi-object spectrograph with 100 fibers. Their sample is limited in $R$ band CCD magnitude in the range $15.0 < R < 17.7$, using photometry they obtain themselves on the Las Campanas 40″ telescope. Their survey area is 1000 square degrees in a series of constant declination strips, and they expect to include 25,000 galaxies in their completed survey. The median redshift is $\sim 30,000$ km s$^{-1}$. Preliminary results are discussed in Shectman et al. (1992), and the PhD. theses of Tucker (1994) and Lin (1995).

The more ambitious surveys mentioned here are going deeper than existing catalogs of galaxies extend. Thus workers are being forced to define galaxy samples themselves. This turns out to be a boon (although a great deal more work) as now the galaxy selection criteria can be fine-tuned for the purposes of redshift survey work and large-scale structure studies. In addition, the modern automated methods of galaxy selection and photometry yield more reliable catalogs with much more robust magnitudes than those measured by eye. Two further redshift surveys in the advanced planning stages are discussed in a final concluding section (§ 9.7).

*3.3 The Measurement of Galaxy Redshifts*

The majority of the redshift surveys discussed above have redshifts measured in the optical part of the spectrum. This is done typically with telescopes of aperture between one and five meters, attached to low- or medium resolution ($R \equiv \frac{\lambda}{\Delta \lambda} \sim 1000$) spectrographs. At much higher resolution, one resolves the internal motions of the galaxies themselves; at much lower resolution, individual spectral features begin to blend with one another.

The optical spectra of the vast majority of galaxies fall into one of several types:
– A pure absorption line spectrum characteristic of G and K stars. This type of spectrum is associated with quiescent early-type galaxies without a recent history of star formation.

- An absorption-line spectrum showing features characteristic of A and F stars (in particular, Hydrogen Balmer line absorption) in addition to lines from G and K stars, and emission lines characteristic of H II regions of ionized gas. Such spectra are often associated with late-type galaxies with star formation. In some extreme cases (the so-called extragalactic H II regions), almost no stellar continuum at all is seen, but the emission lines are very strong.
- Galaxies harboring an active nucleus, in which strong emission lines of high ionization species are seen. Strongly active galaxies represent only a few percent of the galaxy population, but are the focus of intense study; see Osterbrock (1989; 1991) for detailed reviews.

The measurement of redshifts of low-surface brightness galaxies is always difficult. If the object is too faint to measure a continuum with enough signal-to-noise ratio to detect absorption lines, one can either "pray for H$\alpha$", hoping that an H II region will fall on the slit, or observe these objects in the radio, looking for 21 cm transition of H I; late-type low-surface brightness galaxies are often rich in H I, and surveys of these objects have been carried out with radio telescopes (Bothun et al. 1985, Schneider, Thuan, & Mangum 1992).

The measurement of emission-line redshifts is straightforward, and is usually done by fitting multiple Gaussians to the lines that are seen. Absorption-line redshifts require more subtlety: the absorption lines are typically of lower signal-to-noise ratio, and the features that are seen often consist of blends of lines, making it impossible to assign the feature to a unique redshift. A redshift is measured by comparing the spectrum of the galaxy in question to that of a star of known (small) radial velocity, typically a K0-K3 giant taken with the same instrumental setup. Heuristically, one slides the spectrum of the stellar template back and forth until it matches the spectrum of the galaxy, the amount of the shift being a measure of the redshift. In practice, this is done by taking cross-correlating the spectra of the galaxy and the stellar template in either Fourier or real space, and fitting the peak with a smooth function (Sargent et al. 1977; Tonry & Davis 1979; Heavens 1993). Using such techniques redshifts of nearby galaxies are typically measured with $\sim 50$ km s$^{-1}$ accuracy (e.g., Huchra et al. 1983).

In the days when blue-sensitive photographic plates were the detecting element in spectrographs, spectra for redshifts of nearby galaxies were often taken in the blue region of the spectrum covering the region around 4000Å, in order to detect the Calcium H and K lines. Modern CCD detectors are more sensitive in the red, and spectra are typically taken centered on the 5175Å Mg b feature, or, for galaxies that are likely to have strong emission lines, centered on H$\alpha$ and the [N II] doublet at $\sim 6550$Å. CCD detectors on most modern spectrographs have of the order of 800 pixels along the dispersion direction, with roughly two pixels per dispersion element, implying a spectral coverage of 2000Å for $R = 1000$. As CCDs get larger, it is becoming possible to cover more of the visible part of the spectrum, and double spectrographs which use a dichroic to split the light into red and blue halves, each going to a separate camera, are now operating on the Palomar 5m, Lick 3m, ARC 3.5m, Keck 10m, and other telescopes.

Most of the redshift surveys listed above have been carried out by observing one object at a time. However, now that multi-object spectrographs exist on many of the world's largest telescopes, it is possible to obtain spectra for many galaxies in a single exposure. In practice, the size of the telescope determines the faintness of the galaxies for which one can obtain redshifts in a reasonable exposure time, and this faintness limit in turn determines the mean number density of galaxies on the sky. Thus one designs a multi-object spectrograph with a number of fibers and field of view with these considerations in mind. This is the approach taken by the Kirshner et al. Las Campanas survey and the Sloan Digital Sky Survey (§ 9.7), among others.

Redshift surveys are also carried out in the radio, where one looks for the H I 21 cm line in emission. As elliptical galaxies tend to be gas-poor, such surveys are limited to later-type galaxies. The largest such redshift surveys have been done at Arecibo Observatory in Puerto Rico, and with the 140-foot and the late 300-foot telescopes at Green Bank, West Virginia. These surveys are carried out at much higher resolution than in the optical, typically $R = 30,000$. Thus radio redshifts are typically measured to much higher accuracy than optical redshifts, with quoted errors of 5–10 km s$^{-1}$. The combination of loss of sensitivity, and the fact that the 21 cm line is redshifted into an unprotected band, means that radio surveys are unable to probe redshifts much beyond 10,000 km s$^{-1}$, although the Giant Metrewave Radio Telescope near Puna, India, will be able to get around these problems with its enormous collecting area and relative isolation. In some of the older H I redshift literature, the redshift was not defined by Eq. (8), but by

$$z = \frac{\nu_0 - \nu(t)}{\nu_0}, \qquad (53)$$

which agrees with Eq. (8) only for infinitesimal $z$. The reader of the older literature must be aware of this potential for confusion, although modern workers consistently use the "optical definition" of $z$, Eq. (8).

Redshifts are measured, by necessity, on a telescope attached to the Earth. The Earth takes part in many motions: it is rotating on its own axis (0.3 km s$^{-1}$ at the equator), and it is orbiting around the Sun (30 km s$^{-1}$). For extragalactic work, the former correction is negligible, but redshifts are usually published with the correction to the heliocentric frame. However, the Sun is in orbit around the center of the Milky Way ($\approx 225$ km s$^{-1}$), the Milky Way is falling towards our nearest large companion, M31 at 119 km s$^{-1}$ (Binney & Tremaine 1987), and the whole Local Group of galaxies takes part in the larger-scale velocity field which we will discuss in detail below. Because motions on scales smaller than that of the Local Group are very non-linear, we will not include them in our models, but rather refer to redshifts relative to the barycenter of the Local Group. Estimates for the correction from the heliocentric to Local Group barycentric frame have been given by Yahil, Tammann, & Sandage (1977), de Vaucouleurs, de Vaucouleurs, & Corwin (1976) and Lynden-Bell & Lahav (1988). These three determinations are consistent with one another; for example, Yahil et al. quote the motion of the sun relative to the barycenter as 307 km s$^{-1}$ towards Galactic coordinates $l = 105°$, $b = -7°$.

*3.4 Determination of the Luminosity and Selection Functions*

Redshift surveys have the property that the number density of sample objects is a decreasing function of redshift. We quantify this in terms of a *selection function* $\phi(r)$, defined as the fraction of galaxies at distance $r$ which meet the sample selection criteria. In order to use the galaxies in a sample as tracers of the general galaxy distribution, we must correct for the fact that the galaxies represent an increasingly smaller proportion of the parent population at larger and larger distances. We do this by giving sample objects weights proportional to $1/\phi(r)$ in order to account for the galaxies below the flux limit at distance $r$. The selection function is closely related to the *luminosity function* $\Phi(L)$, which gives the number density of galaxies of luminosity $L$, per unit luminosity. The relationship between $\phi(r)$ and $\Phi(L)$ is most easily seen in the case of a survey in which the sample is defined by a minimum energy flux $f_{\min}$; the derivation to follow may be trivially modified for a survey limited by apparent diameter or other photometric quantity, or for direction-dependent selection (as in the case of samples affected by Galactic extinction; cf., Santiago et al. 1995b). In the latter case, one generalizes the selection function to a quantity depending on both distance and direction on the sky.

The integral of $\Phi(L)$ over all $L$ gives the total number density of galaxies, $n$. In practice, the luminosity function is poorly constrained at the faint end, and so we cut off the integral at a lower limit $L_s \equiv 4\pi r_s^2 f_{\min}$ for some small $r_s$ (typically 500 km s$^{-1}$), and reject galaxies with lower luminosities. At any distance $r > r_s$, we can only observe galaxies with luminosities greater than

$$L_{min} = 4\pi r^2 f_{\min}. \qquad (54)$$

Consequently, the ratio of observable to total galaxies at distance $r$ is

$$\phi(r) = \frac{\int_{4\pi r^2 f_{\min}}^{\infty} \Phi(L)\,dL}{\int_{L_s}^{\infty} \Phi(L)\,dL}. \tag{55}$$

At distances closer than $r_s$, we set $\phi(r) \equiv 1$. Note that the luminosity function is a property of the galaxies themselves, while the selection function depends on the photometric limits of the sample in question.

The use of the selection function to weight galaxies in large-scale structure studies makes the assumption that the luminosity function is *universal*, independent of local density. This assumption means that galaxies of any luminosity are equally good tracers of the large-scale galaxy distribution (cf. § 5.10), or equivalently, that that bivariate distribution of galaxies in position and luminosity is a separable function.

There are several methods for measuring the luminosity or selection function from redshift surveys. The simplest is called the $1/V_{\max}$ method, due to Schmidt (1968): the luminosity of a galaxy in the sample determines a maximum distance to which it could be placed and still remain within the sample:

$$r_{\max} = r_i \left(\frac{f_i}{f_{\min}}\right)^{\frac{1}{2}}, \tag{56}$$

where $r_i$ and $f_i$ are the measured distance and flux of the source. We can then define a maximum volume, $V_{\max} = \omega r_{\max}^3/3$, where $\omega$ is the solid angle covered by the survey. The estimator for the luminosity function is then

$$\Phi(L)\Delta L = \sum \frac{1}{V_{\max}}, \tag{57}$$

where the sum is over all galaxies with measured luminosities between $L$ and $L + \Delta L$. This method is easily generalized to the case of a variable flux limit, caused, for example, by Galactic extinction. However, it makes the assumption that the mean underlying number density of galaxies is uniform, without any density inhomogeneities. This is a dangerous assumption, for we shall see that the inhomogeneities can be substantial.

There is an extensive literature of methods to derive selection functions in density-independent ways (Lynden-Bell 1971, Turner 1979, Kirshner *et al.* 1978, Sandage, Tammann, & Yahil 1979, Davis & Huchra 1982, Nicoll & Segal 1982, Efstathiou, Ellis, & Peterson 1988, Choloniewski 1986, Binggeli *et al.* 1988, and Yahil *et al.* 1991). We present here our own favorite, following the development of Yahil *et al.* (1991). We phrase the problem in terms of likelihoods: given that a galaxy $i$ is at distance $r_i$, what is the probability that it fall in the interval $L < L_i < L + dL$? This probability is simply the luminosity function $\Phi(L_i)$, normalized by the integral over all luminosities it could have at that distance, given the flux limit of the survey:

$$F(L_i|r_i)\,dL = \frac{\Phi(L_i)\,dL}{\int_{L_{\min}(r_i)}^{\infty} \Phi(L)\,dL}, \tag{58}$$

where $L_{\min}(r)$ was defined in Eq. (54). Note that as written, $F(L_i|r_i)$ is a probability *density*, per unit luminosity. It will be convenient to rephrase Eq. (58) in terms of the selection function:

$$F(L_i|r_i) \propto -\frac{1}{\phi(r_i)} \frac{\partial \phi(r)}{\partial r}\bigg|_{r=r_{\max i}} \tag{59}$$

where $r_{\max}$ was defined in Eq. (56). This is the likelihood for a single galaxy, given the selection function; the likelihood for the entire sample is then the product of $F$ over all the galaxies in the sample. We then solve for the selection function by maximizing the likelihood with respect to the parameters in terms of which we model $\phi(r)$. Note that we need not worry about the constant of proportionality in Eq. (59), as it is independent of the selection function itself, and therefore is irrelevant when maximizing with respect to $\phi$.

Note that the likelihood function is indeed independent of the underlying density distribution, as it is a conditional probability for the luminosities, *given* the distances to each object. In particular, this method is independent of the mean density of the sample, which drops out of the ratio in Eq. (58). Thus we need an independent way to define the mean density. Davis & Huchra (1982) discuss various methods for defining the mean density of a sample given the selection function. In general, the mean density is given by a weighted sum over the sample:

$$n = \frac{\sum_{\text{galaxies } i} w_i}{\int dV \phi w}. \tag{60}$$

In the presence of density inhomogeneities, the variance in $n$ is minimized with the weights:

$$w_i = \frac{1}{1 + n J_3 \phi(r_i)}, \tag{61}$$

where

$$J_3 = 4\pi \int r^2 \xi(r) \, dr \tag{62}$$

is a measure of the density fluctuations on the scale of the survey. Because of the presence of $n$ in the expression for the weights, one has to calculate $n$ iteratively. In practice, one often uses the simpler estimator, given by $w_i = 1/\phi(r_i)$, thus:

$$n = \frac{1}{V} \sum_{\text{galaxies } i} \frac{1}{\phi(r_i)}. \tag{63}$$

In principle, one sums over all the galaxies of the survey, in which case $V$ is the volume out to the most distant object of the sample. In practice, this would be horribly noisy, and one sums out to a radius beyond which the sample becomes so sparse that the shot noise of Eq. (63) dominates.

Given a value of $n$, one can define the luminosity function in terms of the selection function. One often sees the luminosity function expressed in logarithmic intervals of luminosity, so that:

$$\hat{\Phi}(L) \equiv \ln 10\, L \Phi(L) = -1.15 n \left( r \frac{\partial \phi}{\partial r} \right) \bigg|_{r = r_{\max}(L)}. \tag{64}$$

How in practice does one maximize the likelihood with respect to an unknown function? One approach is to use a parameterized form for the selection function, and maximize with respect to those parameters (Sandage *et al.* 1979, Yahil *et al.* 1991). It is easier to work with a parameterized form for $\phi$ rather than $\Phi$ because it is always easier to differentiate than to integrate. Yahil *et al.* (1991) suggest a simple generic form for $\phi$ with three parameters, which is applicable to a variety of situations:

$$\phi(r) = \left( \frac{r}{r_s} \right)^{-2\alpha} \left( \frac{r_*^2 + r^2}{r_*^2 + r_s^2} \right)^{-\beta}. \tag{65}$$

Alternatively, one can write the luminosity function in the form of a series of steps:

$$\Phi(L) = \Phi_k; \quad L_k < L < L_{k+1}; \quad k = 1, \ldots, N. \tag{66}$$

The steps $L_k$ are typically logarithmically spaced, and the aim is to solve for the parameters $\Phi_k$. The selection function can then be written

$$\phi(r) = \frac{\sum_{i=1}^{N} \Phi_i H_i[L_{\min}(r)]}{\sum_{i=1}^{N} \Phi_i} \qquad (67)$$

where $L_{\min}(r)$ was defined above and

$$H_i(L) = \begin{cases} 0, & L > L_{i+1} \\ \frac{L_{i+1} - L}{L_{i+1} - L_i}, & L_i < L < L_{i+1} \\ 1, & L < L_i \end{cases} \qquad (68)$$

Setting the derivative of the likelihood function (Eq. 58) with respect to the $\Phi_i$ to zero yields a series of implicit equations for the $\Phi$'s which are solved by iteration (Nicoll & Segal 1982; Efstathiou et al. 1988). This method has two drawbacks (Koranyi & Strauss 1994). First, because the luminosity function is taken to be constant over a finite interval, the selection function has discontinuous first derivatives, and in fact is biased upwards; this makes further analyses based on the selection function misleading. This can be fixed by fitting a smooth curve through the steps and calculating the selection function from this. Alternatively, one can use a generalization of Eq. (66), in which one linearly interpolates the luminosity function between steps. Second and more serious, when the number of steps becomes small, the luminosity function is biased downwards, especially at the faint end, an effect which requires a large number of steps (40 or more) to mitigate. We will discuss one effect of this bias in § 3.6.

Maximum-likelihood methods in general, and the ones we have described here in particular, have the disadvantage that there is no explicit measure of goodness-of-fit. Thus when fitting for a parameterized form, one can be sure to get the best values of the parameters out, but one can never know whether the form itself is adequate. Sandage et al. (1979) and Yahil et al. (1991) have developed a simple *a posteriori* method to test how well a given selection function fits the sample. One can compare the observed distribution of luminosities in a redshift sample with that expected, given the distances and a model for the luminosity function. Eq. (58) is the probability distribution function of luminosity for a galaxy at a given distance, with a sharp cutoff at $L_{\min}(r)$. One can carry out this comparison for any well-defined subset of the data (defined by redshift, position on the sky, morphological type, local density, etc.) that does not have explicit selection on luminosity. Fig. 4 illustrates this for the *IRAS* 1.2 Jy survey: in each panel, the smooth curve is the predicted luminosity distribution, while the histogram is that observed; the agreement is excellent. The different panels show the luminosity distribution in different redshift shells, indicating that there are no systematic errors as a function of redshift. Actually, it remains unclear how useful this test is; experiments show that the luminosity function must be very seriously in error before this test shows disagreement between observed and predicted luminosity distributions.

As discussed in § 3.1, the photometric data on which a redshift survey is based are not always of the highest quality. This has two effects: objects will be scattered in and out of the sample by flux errors, and errors in fluxes will cause luminosities to be in error. Because there are many more faint than bright galaxies, the net effect of errors is to cause a systematic *over-estimation* of luminosities. It is straightforward to show that constant fractional errors in the fluxes result in a derived luminosity function that is a convolution of the true underlying luminosity function with the flux error distribution. Monte-Carlo experiments by Santiago et al. (1995b) show that despite this effect, the density field derived from redshift survey data is quite robust to flux errors, if they are proportional to the fluxes themselves. That is, the bias in the selection function caused by flux errors is very nearly cancelled by the bias in galaxy counts caused by objects scattering across the flux limit. The density field is more robust than is the luminosity function

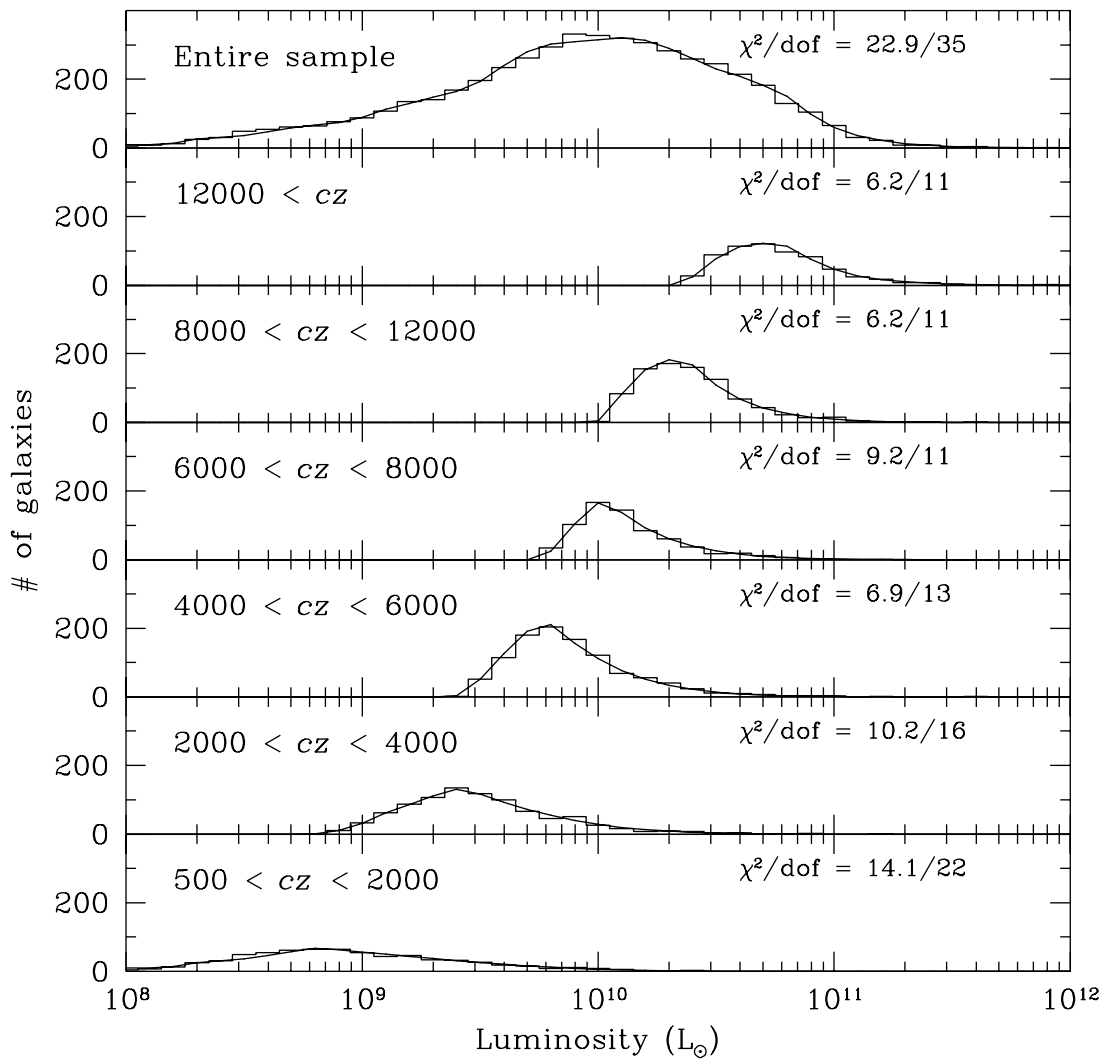

Fig. 4. The luminosity distribution (histogram) of the *IRAS* 1.2 Jy sample in different redshift ranges, together with the predicted distributions, given the luminosity function. The values of $\chi^2/dof$ given in each panel refer to the difference between the two curves, with errors given by Poisson statistics.

itself to flux errors, as long as the selection function is measured from the same dataset.

A comprehensive review of determinations of the luminosity function is given in Binggeli *et al.* (1988), cf. Felten (1977) for earlier work. Modern determinations of the luminosity functions of galaxies in optically selected bands include Efstathiou *et al.* (1988), Loveday *et al.* (1992b, 1994), de Lapparent *et al.* (1989), and Marzke, Huchra, & Geller (1994), while the *IRAS* luminosity function has been discussed by Saunders *et al.* (1990), Yahil *et al.* (1991), Spinoglio & Malkan 1989, and Soifer *et al.* (1989). The optical luminosity function is often fit to a form originally suggested by Schechter (1976):

$$\Phi(L) = \Phi_0 L^{-\alpha} e^{-L/L_*}, \qquad (69)$$

with free parameters $\alpha$ and $L_*$; the sharp exponential cutoff means that galaxies with luminosities well above $L_*$ (the "knee" of the luminosity function) are quite rare. The 60$\mu$m luminosity function of *IRAS* galaxies does not show such a sharp cutoff, but rather is well-fit by two power-laws (cf. Eq. 65); there is therefore a substantial population of ultraluminous *IRAS* galaxies, whose properties are the subject of much research (e.g., Sanders et al. 1989). This also means that the selection function of *IRAS* galaxies is not as steep as that of optically selected galaxies, and thus *IRAS* redshift surveys include a much more extensive tail of high-redshift galaxies than does an optically selected sample with the same median redshift (Figure 4 of Santiago et al. 1995a).

## 3.5 Luminosity Functions: Scientific Results

The determination of the optical luminosity function of galaxies in optical bands has only been done adequately in the photographic *B*-band, in which the comprehensive galaxy catalogs have been compiled. Indeed, much remains to be learned about the distribution of galaxy properties in quantities other than luminosity. For example, Binggeli et al. (1988) stress that the luminosity function of galaxies is a strong function of galaxy morphology (cf. Loveday et al. 1992b, 1994), and although some work has been done on the diameter function of galaxies (Lahav et al. 1988; Maia & da Costa 1990; Hudson & Lynden-Bell 1991), the bivariate distribution of diameters and luminosities has barely been explored (Choloniewski 1985; Sodré & Lahav 1993). Even less is known about the bivariate distribution of luminosities and galaxy colors. Proper analyses of these basic properties of galaxies will require large samples of galaxies with excellent photometry in a variety of bands. This is one of the main scientific goals of the Sloan Digital Sky Survey. The galaxy luminosity function is a basic datum that any detailed model of galaxy formation must match; theoretical papers addressing this include White et al. (1987), Schaeffer & Silk (1988), White & Frenk (1991), Cen & Ostriker (1992a), Cole et al. (1994a), and White (1994).

The measured luminosity function becomes very uncertain at the faint end (e.g., Efstathiou et al. 1988). The faint-end slope of the luminosity function has important ramifications for a number of issues in astrophysics. First, number counts of faint galaxies show an excess over the number predicted by simple extrapolation of the galaxy luminosity function today (Tyson 1988; Cowie et al. 1989; Maddox et al. 1990d; Ellis 1993); if there is a population of galaxies that were bright in the past but whose surface brightnesses dimmed substantially as star formation ceased, they could be hidden today in the faint end of the luminosity function (McGaugh 1994). As photographic and CCD surveys push to ever lower surface brightness levels, new populations of galaxies are appearing (Bothun et al. 1987; Schombert & Bothun 1988; Schombert et al. 1992; Dalcanton 1994), including many galaxies of quite substantial luminosities.

Understanding the low-luminosity population is also important for understanding the evolution of metals (i.e., elements beyond Helium in the periodic table) in the universe. Cowie (1988) shows that the integrated surface brightness of galaxies in the sky is directly proportional to the total metal abundance:

$$S_\nu = \frac{c}{4\pi} \epsilon_{\nu 0} \rho Z, \tag{70}$$

where $\rho Z$ is the volume density of metals in the universe, and $\epsilon_{\nu 0}$ is the light emitted per unit frequency $\nu$ per unit mass of metals returned to the interstellar medium. Quantifying the full galaxy population and therefore the total surface brightness of the night sky due to galaxies (Spinrad & Stone 1978, Toller 1990) will thus give insights into the production of metals and the chemical evolution of the universe.

The numbers of low-luminosity galaxies is also vitally important for quantifying the luminosity density of the universe. The total emissivity of optical light per unit volume in the universe is given by $\int L\Phi(L)dL$, which diverges if the faint-end logarithmic slope of $\Phi(L)$ is steeper than $\alpha = 2$ (Eq. 69)[13]. We observe a dark night sky, which says that the luminosity density of the universe does not in fact diverge at the faint end, thus either the faint-end slope is somewhat shallower than $\alpha = 2$, or there is a cut-off at some point. The observational situation is still very much in a state of flux; this is an active area of research.

---

[13] Note that some workers use the luminosity function per unit *log* luminosity, for which the definition of $\alpha$ differs by unity from that here. See Eq. (64).

The optical luminosity density of the universe can be turned into a mass density of the stars if one assumes a mass-to-light ratio for the stars. Alternatively, Loveday et al. (1988) calculate the mass-to-light ratio necessary to close the universe (i.e., such that $\Omega_0 = 1$), given the measured luminosity density, as $1580 \pm 190 h$ in solar units. For comparison, standard spectral synthesis models give mass to blue luminosity ratios of $4 - 7$.

The recent interest in low-luminosity and low surface brightness galaxies reminds us that the galaxy samples which form our basis for redshift surveys are likely to suffer from incompleteness (Disney 1976; McGaugh 1994). Even a survey limited at bright magnitudes or large diameters will miss a population of low-surface brightness galaxies, some of which are quite luminous (Bothun et al. 1987), although the numbers of luminous low-surface brightness galaxies seems to be quite small. There is also a bias against finding compact galaxies of very high surface brightness; these galaxies can be difficult to distinguish from stars (which are much more numerous than galaxies at all Galactic latitudes until one pushes to 22nd magnitude and fainter) and thus will be missed from the survey. The extreme examples of compact galaxies are the quasars: 3C 273 is bright enough to satisfy the selection criteria of the CfA redshift survey, but appears absolutely stellar on photographic plates; the host galaxy is apparent only with deep exposures under superb conditions (Hutchings & Neff 1991; Bahcall, Kirhakos, & Schneider 1994).

*3.6 Testing the Hubble Law with Redshift Surveys*

The observational evidence for the Hubble law (Eq. 1) for nearby galaxies is mostly based on distance indicator relations for individual galaxies. These usually take the form of *standard candles*, whereby the absolute luminosity of the galaxy or some well-defined part of it (e.g., Cepheid variables) are assumed to be known *a priori* (§ 6.1). Observation of its apparent luminosity gives the distance via the inverse-square law. This can then be compared with the redshift as a check of Eq. (1). The history of this approach is summarized in Chapter 5 of Peebles (1993); see Lauer & Postman (1992) for a recent test of the Hubble law. There has been no strong evidence for deviations from Eq. (1) beyond that expected from peculiar motions (Eq. 2). Nevertheless, one could imagine generalizing Eq. (1) to the form

$$cz = H_p r^p; \qquad (71)$$

Segal et al. (1993, and references therein) suggest a so-called Chronometric Cosmology, in which $p = 2$ at low redshifts. Segal et al. (1993) suggest an intriguing measure of $p$ from redshift surveys. Rather than using distance indicators to measure the distances to individual galaxies, one can use the luminosity function of galaxies as a distance indicator, assuming the luminosity function to be universal. In fact, luminosity function fitting has been used in the past to measure distances to galaxy clusters (Schechter 1976; Schechter & Press 1976; Gudehus 1989); here we wish to use it to test the Hubble law. Imagine that one interprets the results of a redshift survey using Eq. (71) with the incorrect value for $p$. One would expect that the luminosity function determined from subsets of the sample in different redshift ranges would not be consistent with one another; agreement would be found only for the correct value for $p$, and this then has the potential for distinguishing between different cosmological models. Segal et al. (1993) use the *IRAS* 1.936 Jy redshift survey (Strauss et al. 1992b) to claim that the data are consistent with $p = 2$ and inconsistent with the conventional $p = 1$. However, the luminosity function in different redshift shells is a much less powerful statistic than Segal et al. claim. In the limit of thin redshift shells, it is just measuring the *flux* distribution of the sample, and in any redshift survey, the majority of the sources are close to the flux limit, independent of cosmology. Koranyi & Strauss (1994) show that this distribution is almost independent of $p$, if one self-consistently derives the luminosity function from the sample for each value of $p$. Moreover, because Segal et al. use the step-wise luminosity function method described above, their luminosity function is systematically in error, the error being worse for smaller $p$ than for larger.

Therefore, their luminosity function for $p = 2$ is a better fit to the data than is $p = 1$, and they erroneously conclude that the Hubble law is ruled out by the data.

Although the luminosity function is a poor discriminant of models, one can still use redshift surveys alone to put constraints on $p$. In a homogeneous universe, the number of galaxies in a redshift survey with redshift between $r$ and $r + \mathrm{d}r$ is $n\phi(r)\omega r^2 \, \mathrm{d}r$. The upper panel of Fig. 5 compares the observed histogram of redshifts of galaxies in logarithmic bins from the 1.2 Jy *IRAS* survey with that predicted for $p = 1, 2$, and 3. Note that this is an independent test from that of the density-independent luminosity function diagnostic described above. In the $p \neq 1$ models, the predicted and observed redshift histograms are in bad disagreement. The resulting density relative to the mean in shells (the ratio of the histogram to the smooth curves) is shown in the lower panel. In the standard cosmology, the almost full-sky coverage of the *IRAS* survey averages over most density structures, meaning that the sample is close to mean density at all redshifts (cf. de Lapparent et al. 1988), although the overdensity associated with the Local Supercluster is apparent at $cz = 1000$ km s$^{-1}$. However, for the $p = 2$ and $p = 3$ cosmologies, one must argue that the Local Group lies in the middle of a void, surrounded by a vast spherical shell at the mean density between 4000 and 10,000 km s$^{-1}$, which is surrounded by another spherical void extending to the horizon.

## 3.7 The Smoothed Density Field

We have seen above that the number density of galaxies in a flux limited redshift survey is a decreasing function of distance. We correct for this by assigning each galaxy a weight given by the inverse of the selection function $\phi(\mathbf{r})$. In order to define the extent of the structures traced by the galaxies, we would like to define a continuous density field. This is done by smoothing (cf., Eq. 24):

$$\delta(\mathbf{r}) = \frac{1}{n} \sum_{\text{galaxies } i} \frac{W(|\mathbf{r} - \mathbf{r}_i|/r_{\text{smooth}})}{\phi(r_i)} - 1, \tag{72}$$

where the window function $W$ is normalized to unit integral:

$$\int \mathrm{d}^3\mathbf{r} \, W(r/r_{\text{smooth}}) = 1. \tag{73}$$

Common window functions used are top-hat, parabolic, and Gaussian:

$$r_{\text{smooth}}^3 W(x) = \begin{cases} \frac{3}{4\pi}, & x < 1 \quad \text{tophat} \\ \frac{15}{8\pi}(1 - x^2), & x < 1 \quad \text{parabolic} \\ \left(\frac{2}{\pi}\right)^{\frac{1}{2}} e^{-\frac{1}{2}x^2} & \text{Gaussian}. \end{cases} \tag{74}$$

All that remains is to choose a smoothing radius $r_{\text{smooth}}$. Because the mean interparticle spacing of a redshift survey is an increasing function of distance from the origin (indeed, it is given by $[4\pi n\phi(r)/3]^{-1/3}$) a smoothing length which shows full detail at small distances will contain increasingly fewer particles at larger distances, with corresponding ever-increasing shot noise. For this reason, it is common to choose a smoothing length which grows with radius, proportional to the mean interparticle spacing. This has the disadvantage of not being amenable to Fourier Transform techniques for carrying out the smoothing in Eq. (72); more important, the smoothed density field has different statistical properties at low and high redshift. Nevertheless, it is useful for qualitative and cosmographical description of the structures that are seen, and in some sense shows the maximum amount of information in the redshift survey.

The smoothed density field will be subject to errors from a variety of sources. These include:

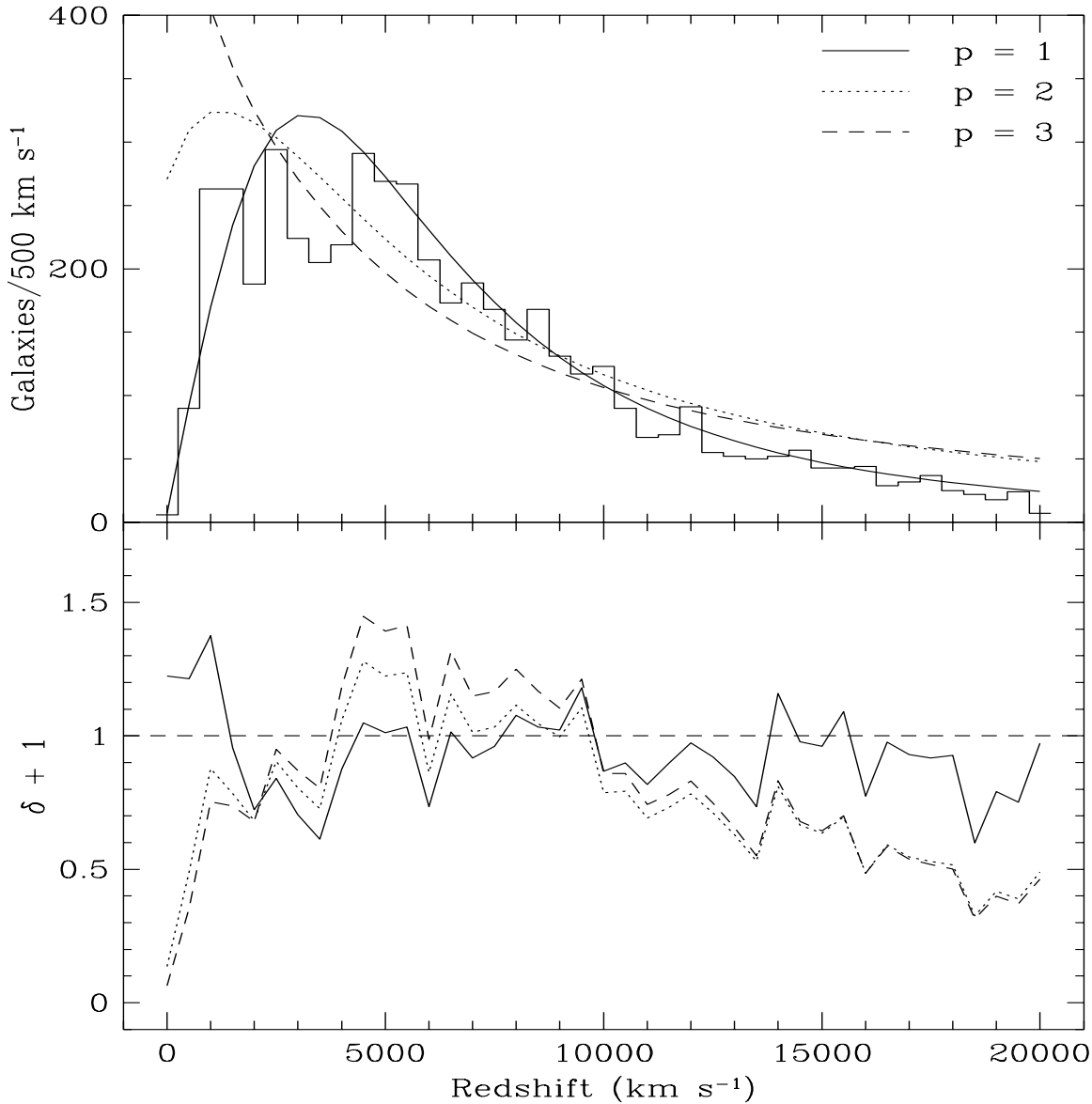

Fig. 5. The upper panel shows the redshift distribution of the *IRAS* 1.2 Jy sample, together with the expected distributions in a homogeneous universe, for various redshift-distance relations. The lower panel shows the ratio between the observed and expected distributions, which is the radial density field. Note that $p = 1$ (the Hubble Law) remains close to the mean density at all radii.

- Shot noise. The galaxy distribution is a point process, and one is always limited by the finite number of galaxies within a smoothing length. The assumption of a universal luminosity function implies that the subset of galaxies luminous enough to enter a flux-limited sample at a given redshift are fair tracers of the full population of galaxies. Under this assumption, the uncertainty in the density field due to shot noise is simply

$$\left\langle \delta(\mathbf{r})^2 \right\rangle^{1/2} = \frac{1}{n} \left[ \sum_{\text{galaxies } i} \left( \frac{W(|\mathbf{r} - \mathbf{r}_i|)}{\phi(r_i)} \right)^2 \right]^{1/2}, \qquad (75)$$

because the mean value of $\delta$ is zero by definition. There is a further contribution to the shot noise in the *mass* density field which depends on the variance in the mass-to-light ratio of galaxies, but this term is important only for nearby galaxies in typical redshift surveys, simply because $1/\phi$ grows so quickly (cf.

Appendix A of Strauss *et al.* 1992c).
- The physically interesting quantity in most applications is the density field in real space, while redshifts are the quantities that are measured. One can assume a model for the velocity field to map from one to the other (as we will discuss in detail in § 5.9). Ignoring the problem altogether causes systematic errors in the density field (Kaiser 1987). In practice, quantifying this source of error requires a specific model for the velocity field, although linear theory makes specific predictions for this distortion (§ 5.2.1).
- The derived density field depends on the assumed selection function; a small error in the selection function causes a *systematic* error in the density field, which for simple flux-limited surveys, will be a function of redshift. This source of error can only be quantified with Monte-Carlo simulations. As mentioned above, Santiago *et al.* (1995b) show that photometry errors are largely compensated for in the selection function, in such a way that the derived density field is quite robust. One other way to get around this problem is to define a *volume-limited subsample* of galaxies from a flux-limited survey. This is done by choosing all galaxies in the sample with distances less than some value $r_{\max}$, whose luminosities are such that they would enter the sample even if placed at $r_{\max}$. For a sample with a uniform flux limit, this means that one chooses those objects with luminosities greater than $L_{\min}(r_{\max})$. This has the additional advantage that it is robust to luminosity-dependent clustering (non-universal luminosity function). Of course, volume-limiting a sample means throwing out the majority of the objects.
- There exists no redshift survey with complete sky coverage; there are always regions of the sky uncovered by the survey. The *IRAS* survey of Strauss *et al.* (1992b) and Fisher *et al.* (1995) is the survey with the most complete sky coverage, 88%. These must be corrected for in Eq. (72). We discuss approaches to this problem in § 3.8.

One can filter the data to minimize the shot noise. The Wiener Filter (cf. Press *et al.* 1992, §13.3) minimizes the *variance* between the measured and true density fields. Let $\delta(\mathbf{r})$ be the true underlying fractional density field of galaxies, and let $\mathcal{D}(\mathbf{r})$ be the measured density field, with shot noise $N$ included:

$$\mathcal{D}(\mathbf{r}) = \delta(\mathbf{r}) + N; \tag{76}$$

We assume for simplicity that the shot noise is independent of position. Let us define a filter $F$ such that the difference between the filtered density field $F\mathcal{D}$ and the true density field $\delta$ is minimized. In practice, we will work with Fourier Transforms, and write:

$$\text{variance} = \int d^3\mathbf{r}\, [\delta(\mathbf{r}) - F(\mathbf{r})\mathcal{D}(\mathbf{r})]^2 = \int d^3\mathbf{k}\, \left[\tilde{\delta}(\mathbf{k}) - \tilde{F}(\mathbf{k})\tilde{\mathcal{D}}(\mathbf{k})\right]^2, \tag{77}$$

by Parseval's Theorem. Substituting in Eq. (76) and minimizing with respect to the unknown function $\tilde{F}(\mathbf{k})$, one finds the Wiener filter:

$$\tilde{F}(\mathbf{k}) = \frac{\left\langle \tilde{\delta}^2(\mathbf{k}) \right\rangle}{\left\langle \tilde{\delta}^2(\mathbf{k}) \right\rangle + N^2}, \tag{78}$$

where $\left\langle \tilde{\delta}^2(\mathbf{k}) \right\rangle$ is proportional to the *power spectrum* $P(k)$ of the underlying density field[14]. The Wiener filter requires a model for the underlying power spectrum. We will discuss methods for estimating the power spectrum from redshift survey data in § 5.3. Note that the Wiener filter is defined in $k$-space, because in real space, the values of $\delta(\mathbf{r})$ at different $\mathbf{r}$ are correlated. Indeed, the Wiener filter can be derived in real space, in which case it is a matrix which takes this correlation directly into account (Zaroubi *et al.* 1994).

---

[14] Eq. (78) holds only in the situation in which the noise is assumed independent of position. In the more realistic case of radially dependent noise, Eq. (78) generalizes to a two-dimensional matrix; cf., Fisher *et al.* (1994d) for details.

The number of elements in the matrix is the square of the number of points at which one wishes to define the density field, which becomes a very non-trivial computational problem.

In a flux-limited redshift survey with a constant smoothing length, the noise term is an increasing function of $r$. In this case, one can either apply a Wiener filter in $\mathbf{r}$ space, as described above, or define a series of Wiener-filtered density fields at a range of values of $N^2$, and interpolate between these at each point.

The Wiener filter approaches unity when the signal-to-noise ratio is high. At low signal-to-noise ratio, it approaches zero; when there is no information on the underlying density field, the filter returns the most likely value for the density, namely its mean. Thus the Wiener filter has the disadvantage of biasing the contrast in the density field downward. The expectation value of the square of the filtered field is

$$\left\langle \widetilde{F}^2(\mathbf{k})\widetilde{\mathcal{D}}^2(\mathbf{k}) \right\rangle = \left( \frac{\left\langle \tilde{\delta}^2(\mathbf{k}) \right\rangle}{\left\langle \tilde{\delta}^2(\mathbf{k}) \right\rangle + N^2} \right)^2 \left\langle \widetilde{\mathcal{D}}^2(\mathbf{k}) \right\rangle$$

$$= \left( \frac{\left\langle \tilde{\delta}^2(\mathbf{k}) \right\rangle}{\left\langle \tilde{\delta}^2(\mathbf{k}) \right\rangle + N^2} \right)^2 \left( \left\langle \tilde{\delta}^2(\mathbf{k}) \right\rangle + N^2 \right)$$

$$= \widetilde{F}(\mathbf{k}) \left\langle \tilde{\delta}^2(\mathbf{k}) \right\rangle \leq \left\langle \tilde{\delta}^2(\mathbf{k}) \right\rangle, \tag{79}$$

because $\widetilde{F}(\mathbf{k}) \leq 1$ always. For comparison, the expectation value of the square of the unfiltered field is $\left\langle \tilde{\delta}^2(\mathbf{k}) \right\rangle + N^2 \geq P(\mathbf{k})$. Thus the Wiener filter *over-corrects* the density field for shot noise. In the case of a redshift survey for a constant smoothing length, the signal-to-noise ratio is a decreasing function of $r$, and thus the contrast of structures decreases with $r$, just as we had in the variable smoothing mentioned above. Thus the Wiener filter offers a natural way to invoke variable smoothing. For some problems, however, this lack of statistical similarity between the nearby and far-away parts of a redshift survey can be a drawback. In this case, one can define an alternative filter, the *power-preserving filter* (Yahil et al. 1994):

$$\widetilde{F}_{PP}(\mathbf{k}) = \left( \frac{\left\langle \tilde{\delta}^2(\mathbf{k}) \right\rangle}{\left\langle \tilde{\delta}^2(\mathbf{k}) \right\rangle + N^2} \right)^{1/2}, \tag{80}$$

just the square root of the Wiener filter. This is just one of a entire family of generalizations of the Wiener filter, as discussed in, e.g., Andrews & Hunt (1977). If we substitute the power-preserving filter into Eq. (79), we find $\left\langle F_{PP}^2(\mathbf{k})\widetilde{\mathcal{D}}^2(\mathbf{k}) \right\rangle = \left\langle \tilde{\delta}^2(\mathbf{k}) \right\rangle$, thus the name power-preserving. Of course, this filter does not share the minimum variance property of the Wiener filter, but it is easy to show that for signal-to-noise ratios greater than one, the root-mean-square difference between the filtered and true density fields is at most 10% larger for the power-preserving filter than for the Wiener filter. As with the Wiener filter, Eq. (80) only holds in when the noise is independent of position; Yahil et al. (1994) interpolate between maps filtered with Eq. (80) at different noise levels for the flux-limited case of noise increasing as a function of distance.

Although the power-preserving filter indeed preserves the second moment of the density distribution function, it does not do a good job of preserving higher-order moments at low signal-to-noise ratio (in particular, the skewness gets exaggerated), with the consequence that the filtered maps show sharper peaks in regions of low signal-to-noise ratio than in high. Work is ongoing to quantify this effect and correct for it.

## 3.8 Filling in the Galactic Plane

With the advent of galaxy surveys covering all of the sky outside of the Galactic plane, there has been a great deal of recent work expended on "finishing the job", that is, either extrapolating the density field at high latitudes to lower latitudes, or actually surveying in the zone of avoidance itself (Balkowski & Kraan-Korteweg 1994). There are two strong motivations for this. The first of these is cosmography: We would like to have a complete map of of the structures of the local universe. The two largest superclusters in our immediate neighborhood, the Pisces-Perseus Supercluster and the Hydra-Centaurus Supercluster, both lie at low Galactic latitudes, and in both cases, there are overdensities on the opposite side of the plane (Camelopardalis and Pavo-Indus-Telescopium, respectively) to which they are plausibly physically connected.

The second reason for mapping the galaxy distribution at low latitudes is for dynamical studies. Peculiar velocities and densities are related in linear theory by Eq. (33); if we wish to use this equation as a test of gravitational instability theory or to measure $\Omega_0$, we need as complete a map of the density field as possible.

Simply leaving unsurveyed regions unfilled causes systematic errors in any dynamical modeling. The absence of galaxies does not correspond to absence of structure; indeed, the unfilled regions will act as a void with negative $\delta$, from which Eq. (33) predicts a systematic outflow. Filling the unfilled regions of a survey with the average density of galaxies corrects for this. The next order of approximation involves interpolating the density field from higher-latitude regions. The *IRAS* 1.936 Jy redshift survey has a low-latitude excluded zone that is only 10° wide; for this, Yahil *et al.* (1991) advocate a linear interpolation of the density field across the Galactic plane. Lynden-Bell, Lahav, & Burstein (1989) have a 30° wide excluded zone at low Galactic latitudes in their survey of optically selected galaxies, and an additional 15° wide zone in the gap between the areas of sky covered by the UGC and ESO catalogs. They use a cloning procedure, in which galaxies are duplicated from adjacent high-latitude regions into the excluded zones. A more elaborate interpolation scheme was used by Scharf *et al.* (1992) who expanded the angular distribution of *IRAS* galaxies in spherical harmonics, and extrapolated them into the excluded zone. They found a prominent overdensity in Puppis (which had been recognized independently by Kraan-Korteweg & Huchtmeier 1992 and Yamada *et al.* 1993), which Lahav *et al.* (1993b) was able to show was a poor cluster at roughly the distance of Virgo. Lahav *et al.* (1994) advocate filtering the spherical harmonics with a Weiner filter to reduce the noise; they show with $N$-body simulations that the resulting reconstruction is quite robust for excluded zones less than 30° wide.

The observed velocity field at high latitudes depends on the density field at low latitudes. Using a density field reconstruction called POTENT (§ 7.5), Kolatt, Dekel, & Lahav (1994) have reconstructed the velocity field in the Zone of Avoidance. We will compare the results of this reconstruction to the density field of galaxies in § 8.

An alternative approach to the various interpolation schemes discussed here is to survey the low-latitude sky directly. It is difficult to do quantitative work here, as the extinction is large and very variable. In regions that the extinction is not too strong (say, less than $A_B = 0.5$ mag), one can correct for extinction statistically in the selection function, following the methods of Santiago *et al.* (1995b); the fewer galaxies in regions of high extinction are simply given greater weight to compensate. However, this process introduces a much larger shot noise. Various ongoing low-latitude searches for galaxies in the optical bands, and redshift surveys thereof, are summarized in the volume edited by Balkowski & Kraan-Korteweg (1994).

One can also select galaxies in wavebands that are less affected than the optical by Galactic extinction. Galaxy catalogs selected at $60\mu$m from the *IRAS* database have been used for redshift survey work. The principal limitations here are that any infrared color scheme that selects galaxies from stars and other Galactic objects becomes severely contaminated at low Galactic latitudes by emission nebulae and infrared cirrus (which has the same infrared colors as quiescent galaxies). At very low Galactic latitudes, the very high source density starts causing systematic errors in the *IRAS* fluxes. Finally, it becomes increasingly difficult to optically identify the galaxies in order to obtain redshifts when the optical extinction becomes too high. Despite these problems, the redshift surveys of Strauss et al. (1992b), Fisher et al. (1995), and Lawrence et al. (1994) extend down to within 5° of the Galactic plane. Low-latitude galaxy catalogs selected from the *IRAS* survey have also been published by Ichikawa & Nishida (1989) and Yamada et al. (1993).

Neutral hydrogen 21 cm emission is also unaffected by Galactic extinction. Lu et al. (1990) have done H I surveys of *IRAS* sources, although Strauss et al. (1990) show that optical surveys are more efficient. But 21 cm surveys can also be done blindly, independent of any other catalog. Kerr & Henning (1987) used the late Green Bank 300 ft telescope to point to 1900 points; they discovered 16 previously uncatalogued galaxies. Kraan-Korteweg et al. (1994b) announce the discovery in H I of a new member of the group of which Maffei 1, Maffei 2, and IC 342 is a part. This is the first published result of a dedicated survey of the Northern Galactic plane ($|b| < 5°$) with the 25m Dwingeloo radio telescope, which is surveying out to a redshift of 4000 km s$^{-1}$ with a resolution of 4 km s$^{-1}$ and a beam size of 0.5° (FWHM).

Sky surveys at $K$ band ($2.2\mu$m) are planned by both American and European teams. The extinction in this waveband is appreciably smaller than that in the optical bands, meaning that galaxy catalogs selected from this database will be able to penetrate to much lower Galactic latitudes. The Two Micron All Sky Survey (2MASS; Kleinmann 1992) will indeed cover the entire sky, and thus the galaxy sample selected from this survey will be ideal for measuring the dipole moment of the galaxy distribution (cf., § 5.7). The Deep Extragalactic Near Infrared Survey (DENIS; Deul 1992), is planned only for the Southern Hemisphere, but will have superior angular resolution to the 2MASS survey.

## 4   Redshift Surveys: A Cosmographical Tour

As discussed in the Introduction, much of the work of redshift surveys has been cosmographical in origin, in which workers have explored the variety of structures that are traced by the large-scale galaxy distribution, and have defined the structures that are visible. We have traced out the history of the process by which structures have been explored above. Until the late 1970's, the mental image of the galaxy distribution that a majority of workers in the field had was an approximately uniform distribution, with randomly placed clusters embedded in it. When the distribution of galaxies on the sky from the Shane-Wirtanen (1967) counts were published by Seldner et al. (1977), people started becoming aware of the richness of the structures that the galaxies traced. The discovery of the Boötes Void by Kirshner et al. (1981), and the publication of the redshift maps of the CfA1 survey (Davis et al. 1982), and the Pisces-Perseus region (Giovanelli et al. 1986) brought the message of the Lick maps home, although it was the publication of the CfA2 slice (de Lapparent et al. 1986a; Fig. 3 above) that really captured the community's (and the public's) imagination. The CfA2 survey, now extended over a volume ten times that of the first slice, has been pivotal in characterizing the variety of structures in which galaxies are found. The galaxy distribution shows structures on all scales that our surveys have probed, from pairs and small groups of galaxies on $1\,h^{-1}$ Mpc scales, to the Great Wall (Geller & Huchra 1989) which has an extent of at least 150 $h^{-1}$ Mpc.

Much of this review is devoted to a quantitative study of the structures that are seen. Before launching into this, we would like to briefly familiarize ourselves with the specific features in the Galaxy distribution. The process of mapping structures and giving them names has been carried out recently by Tully (1987b), who emphasizes the structures within 3000 km s$^{-1}$; Pellegrini et al. (1990), who trace structures in the Southern Hemisphere, Saunders et al. (1991), which is the best reference for the nature of galaxy structures on very large scales, Giovanelli & Haynes (1991), which shows the distribution of galaxies projected on the sky, Strauss et al. (1992a), Hudson (1993a), Santiago et al. (1995a), and Fisher et al. (1995). The best existing survey for tracing the full extent of the local structures (at least within 10,000 km s$^{-1}$) is the 1.2 Jy *IRAS* redshift survey of Fisher *et al.* (1995), given its close to full sky coverage and moderate sampling. Because the sampling is not very dense, it is not ideal for defining individual structures on the smallest scales; see the references above for more detailed mapping of individual structures.

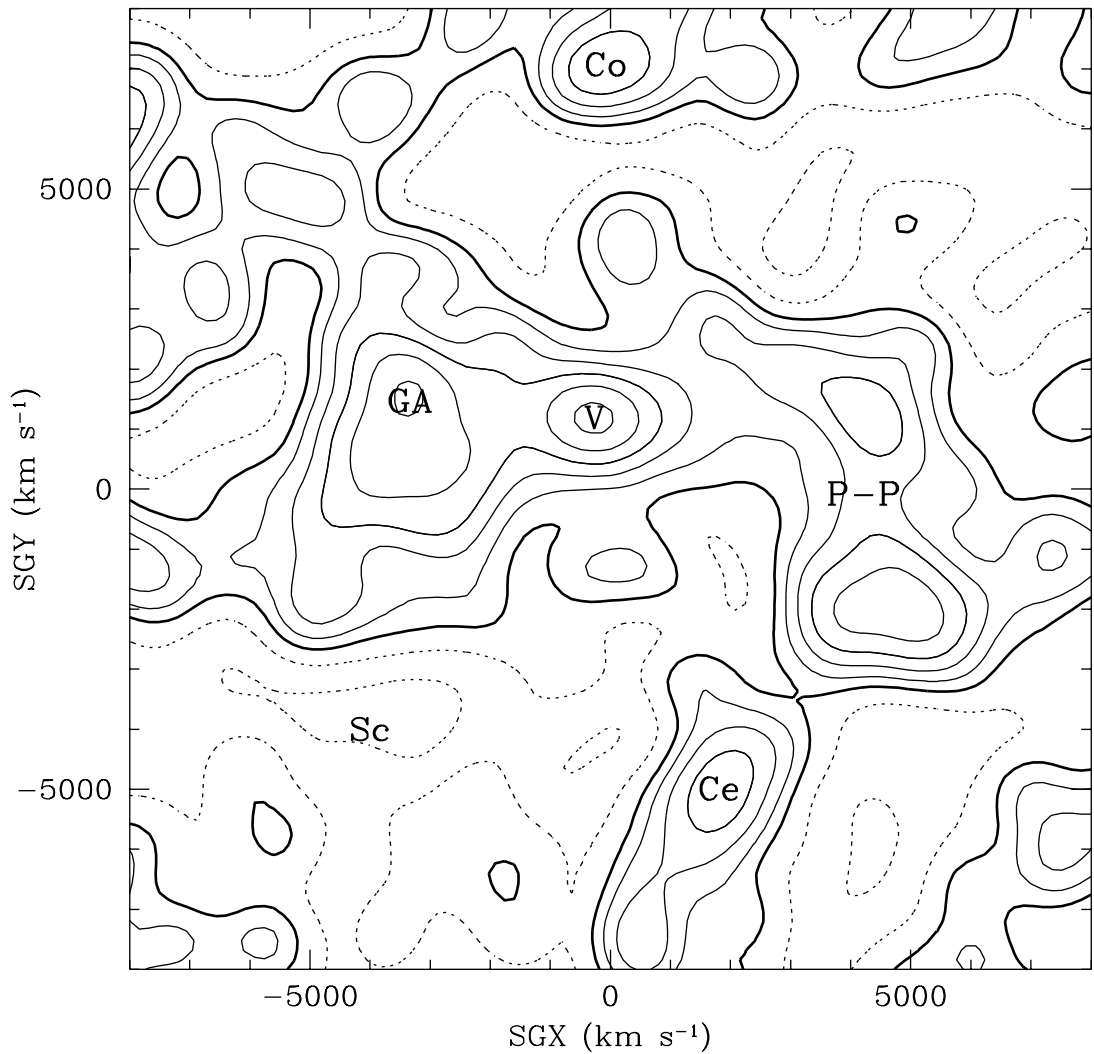

Fig. 6. The density field of *IRAS* galaxies in the Supergalactic plane. A 500 km s$^{-1}$ Gaussian smoothing has been applied. Prominent structures are labeled: V=Virgo, GA=Great Attractor, P-P=Perseus-Pisces, Co=Coma-A1367 Supercluster, Sc = Sculptor Void, and Ce=Cetus Wall.

The density field of the 1.2 Jy *IRAS* survey, with a Gaussian smoothing of $r_{\rm smooth} = 500$ km s$^{-1}$ and with the power-preserving filter applied (§ 3.7), is shown in a series of slices in Fig. 6 through 8. The density field is that obtained by a self-consistent correction for peculiar velocities, as detailed in § 5.9 below, with $\beta = 1$. The slices shown in these figures are made parallel to the principal planes defined in *Supergalactic* coordinates. de Vaucouleurs (1948) pointed out that distribution of nearby galaxies ($cz < 3000$ km s$^{-1}$) is largely confined to a planar structure, which he called the *Supergalactic* plane, or the plane of the Local Supercluster. The Supergalactic plane is almost perpendicular to the Galactic plane.

The slice through the Supergalactic plane (Supergalactic $Z = 0$) is shown in Fig. 6. The contours are of the galaxy density field $\delta$ (Eq. 24). The heavy contour is at the mean density, $\delta = 0$, while the dashed contours are at $\delta = -1/3$ and $-2/3$. The solid contours represent positive values of $\delta$ and are logarithmically spaced in $1 + \delta$, with three contours corresponding to an increase of a factor of two. One's first visual impression from these maps is the irregularity of the shapes of the largest structures. One sees immediately that modeling superclusters as uniform spheres can be misleading! One also notes the rough symmetry between the high and low-density regions. If the distribution function of the primordial density field is symmetric between under-dense and overdense regions, then linear theory predicts this symmetry to be preserved. However, because $\delta$ is has a strict lower limit of $-1$, while there is no corresponding upper limit, we expect a positive skewness to develop for late enough times and small enough smoothing scales that $\langle \delta^2 \rangle$ approaches unity (§ 5.4). This is evident in the maps here, although there is still a rough balance between the volume of space occupied by over- and under-dense regions. One can do statistics with the *topology* of the isodensity contours; we will discuss this in § 5.6.

In this figure, the Local Group sits at the origin. Some of the prominent structures are labeled with identifying flags. The nearest substantial overdensity is found at $X = -250$ km s$^{-1}$, $Y = 1150$ km s$^{-1}$; this the Virgo Cluster[15] (V), the nearest large cluster of galaxies. The Ursa Major cluster, a somewhat more diffuse and spiral-rich cluster, is too close to Virgo to be resolved as a separate structure with this smoothing. The traditional definition of the Local Supercluster refers to it as a flattened structure in the Supergalactic plane of extent $\approx 2000$ km s$^{-1}$, with the Virgo cluster at its center (indeed, one often hears the Local Supercluster referred to as the Virgo Supercluster), but as this figure shows, the Local Supercluster is far from an isolated structure. In particular, at this smoothing, it joins with the *Great Attractor* (GA), which is the large extended structure centered at $Z = -3400$ km s$^{-1}$, $Y = 1500$ km s$^{-1}$. The Great Attractor is often referred to as two separate superclusters, the *Hydra-Centaurus* (positive supergalactic Y) and *Pavo-Indus-Telescopium* (negative supergalactic Y) Superclusters, although that division is somewhat artificial, being imposed by the zone of avoidance. The *IRAS* survey clearly shows the two superclusters to be contiguous (the effective smoothing length is greater than the width of the excluded zone). The Great Attractor was first named as such by Dressler (1987b), when peculiar velocity surveys showed a convergence in the galaxy velocity field towards a point corresponding to the Hydra-Centaurus supercluster (Lilje, Yahil, & Jones 1986; Lynden-Bell *et al.* 1988) as we discuss in § 7.1.2 below. Subsequent redshift surveys (cf. Strauss & Davis 1988; Dressler 1988; 1991) mapped the full extent of the overdensity of galaxies associated with the Great Attractor. See Lynden-Bell, Lahav, & Burstein (1989) for a brief history of definitions of the Great Attractor. In the context of redshift surveys, we define the Great Attractor as the extended overdensity of galaxies that dominates the left hand side of Fig. 6.

On the opposite side of the sky, there is an extended chain of galaxies with two distinct density peaks, at $X = 4300$ km s$^{-1}$, $Y = 1300$ km s$^{-1}$, and $X = 4750$ km s$^{-1}$, $Y = -2000$ km s$^{-1}$. This is the *Perseus-Pisces* Supercluster (P-P), the Northern part of which is sometimes referred to as the Camelopardalis Supercluster. These contour plots do not give justice to the remarkable tightness and coherence of the galaxy distribution here; Giovanelli *et al.* (1986) show that the structure lies nearly in the plane of the sky, and displays a remarkably filamentary structure, with rich clusters embedded in it. There is a void in the foreground of the Perseus-Pisces supercluster centered at $X = 2000$ km s$^{-1}$, $Y = -1400$ km s$^{-1}$. There is also a void beyond the Virgo cluster, which stretches to the *Coma-A1367* supercluster (Co), centered at $X = 0, Y = 7100$ km s$^{-1}$. This supercluster is embedded in the Great Wall (Fig. 3), although the Great Wall itself lies beyond the boundaries of this figure for the most part. The void centered at $X = -4000$ km s$^{-1}$, $Y = -4000$ km s$^{-1}$ was discovered by da Costa *et al.* (1988); Dekel & Rees (1994) refer to it as the Sculptor Void (Sc). Finally, the radially directed structure centered at $X = 1800$ km s$^{-1}$, $Y = 5000$ km s$^{-1}$ is the Cetus Wall (Ce), discovered by the SSRS.

Fig. 7a is a slice at $Z = 3000$ km s$^{-1}$, and shows that the region above the Supergalactic plane is marked by an extensive void. This void extends almost to the Local Group, and is apparent in the distribution of galaxies very near us as the Local Void (cf. Tully 1987b). The overdensity extending across the top of the figure is a piece of the Great Wall. Below the Supergalactic plane (Fig. 7b $Z = -3000$ km s$^{-1}$), one sees the extensions of the Great Attractor (in the form of the Pavo-Indus-Telescopium supercluster, P-I-T) and the Pisces-Perseus Superclusters.

---

[15] It is standard practice to name clusters and superclusters after the constellations in which they are found.

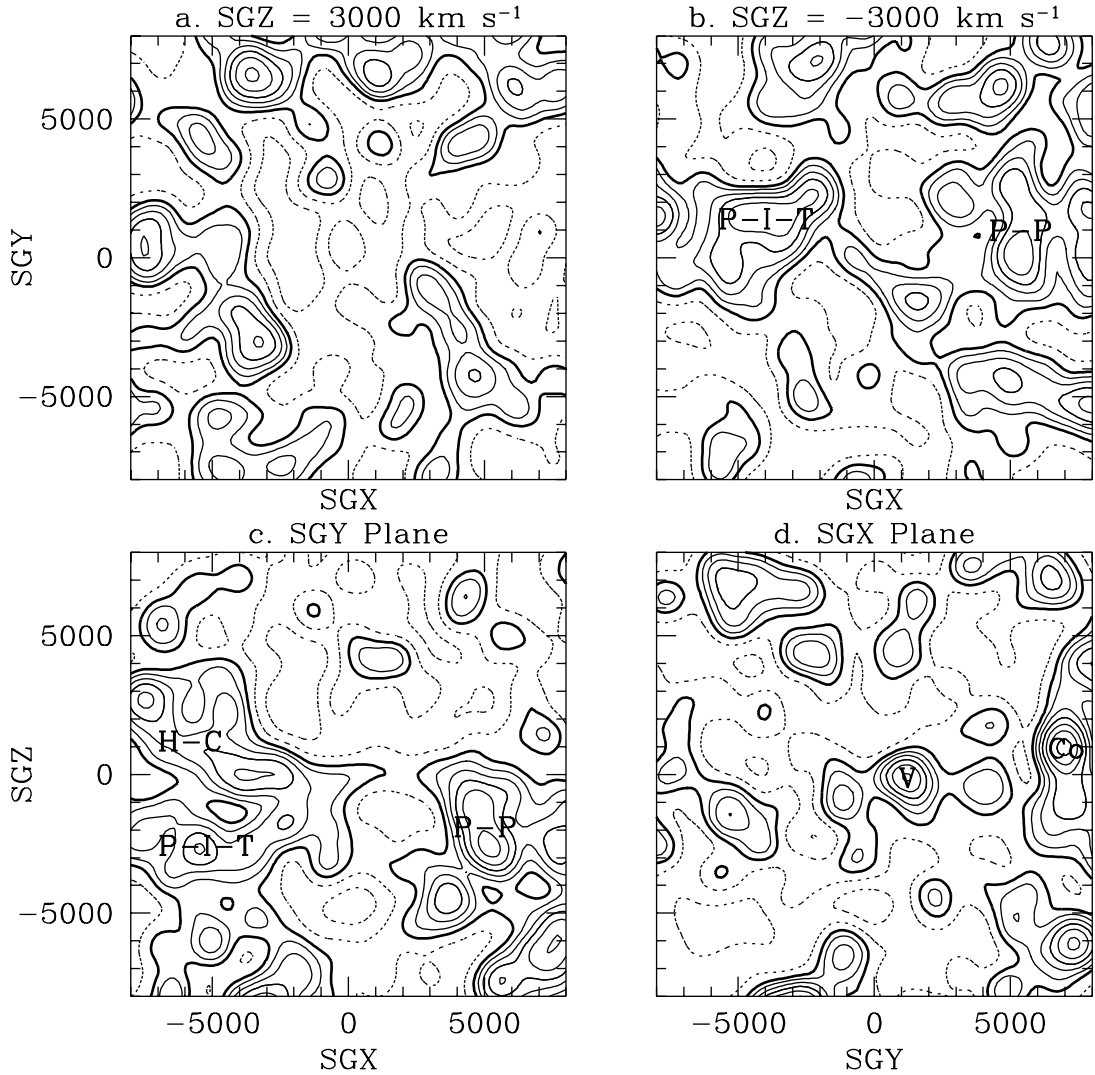

Fig. 7. The density field in various slices in Supergalactic coordinates a: $Z = +3000$ km s$^{-1}$. b: $Z = -3000$ km s$^{-1}$. c: $Y = 0$. d: $X = 0$. Prominent structures marked include H-C=Hydra-Centaurus, and P-I-T=Pavo-Indus-Telescopium.

Panel c of Fig. 7 is the $Y = 0$ slice. This slice lies almost in the Galactic plane, and cuts cleanly through the Great Attractor (negative $X$) and the Northern part of the Pisces-Perseus Supercluster (positive $X$). The Great Attractor is seen to be bimodal in this cut, and we label the Hydra-Centaurus (H-C) and P-I-T superclusters separately. The extent of the void above the Supergalactic plane (Fig. 7a) is now clear; it occupies the entire upper part of this figure. A slice at $X = 0$ (Fig. 7d) passes through the Virgo cluster, and the Coma-A1367 superclusters. The Great Wall is apparent at $Y = 7000$ km s$^{-1}$ in this slice.

The slice at $Y = 3000$ km s$^{-1}$ (Fig. 8a) shows no dramatic structures, and is largely dominated by the void that lies between the Local and Coma-A1367 superclusters. The slice at $Y = -3000$ km s$^{-1}$ (Fig. 8b) cuts through the southern (i.e., smaller $Y$) part of the Pisces-Perseus supercluster, and shows it to be extended in the Supergalactic $Z$ direction. The region south of the Great Attractor is dominated by the Sculptor Void (Sc).

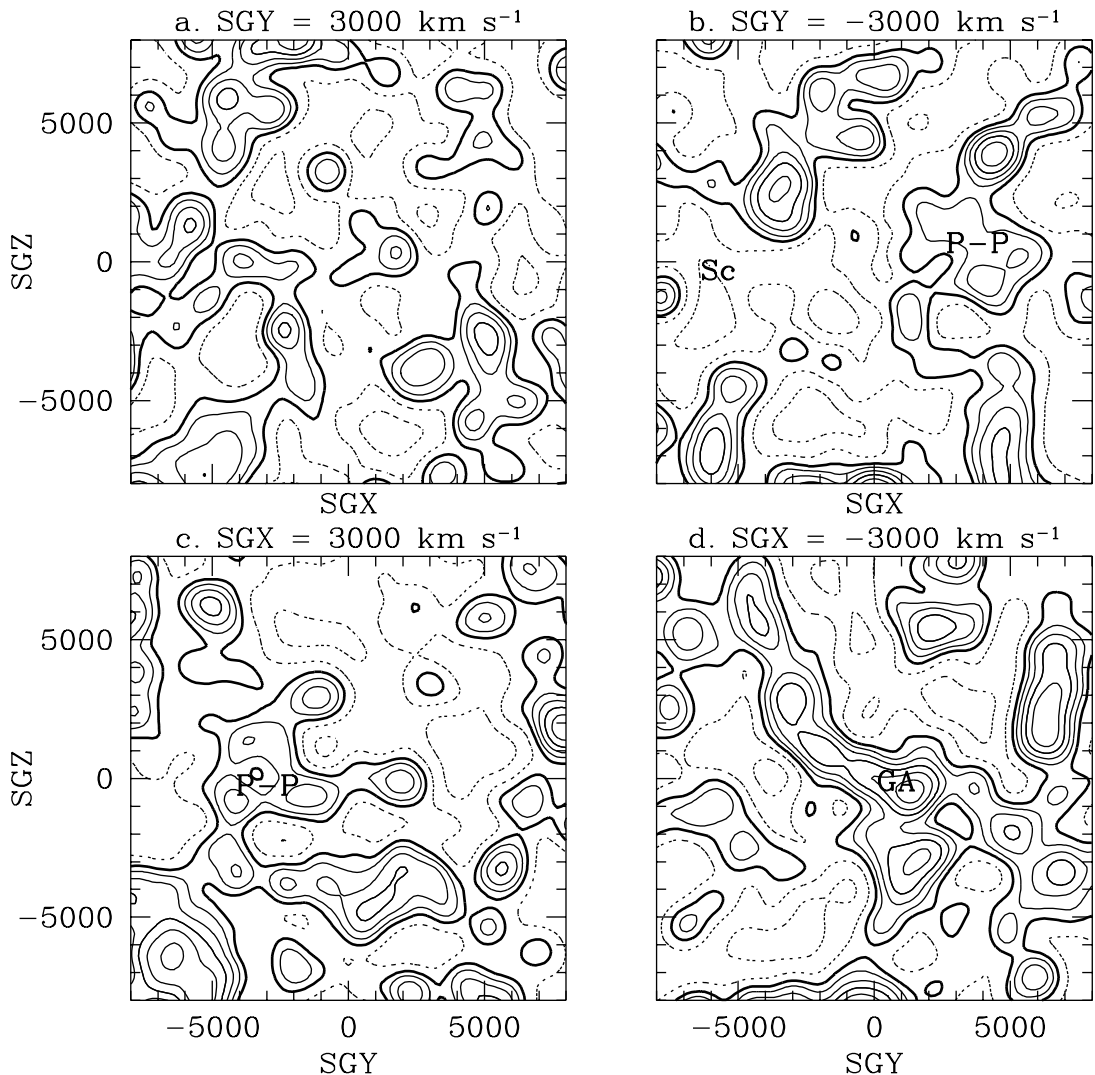

Fig. 8. The density field in various slices in Supergalactic coordinates a: $Y = +3000$ km s$^{-1}$. b: $Y = -3000$ km s$^{-1}$. a: $X = +3000$ km s$^{-1}$. b: $X = -3000$ km s$^{-1}$.

$X = 3000$ km s$^{-1}$ (Fig. 8c) slices through the Pisces-Perseus supercluster, showing it to be multi-modal. Indeed, there is a void between the Perseus-Pisces Supercluster in the Supergalactic plane, and its counterpart at $Z = -3000$ (Fig. 8b). Finally, the slice at $X = -3000$ (Fig. 8d) cuts through the Great Attractor. The bimodality apparent in the $Y = 0$ slice is apparent here as well. The Sculptor Void appears here, bracketed by a remarkable wall of galaxies extending from the Great Attractor to the upper left hand corner of the figure.

In this brief overview of the galaxy distribution as seen by *IRAS*, we have not been comprehensive: we have not endeavored to give every structure that is visible a name, nor have we given justice to the detailed mapping that has been done by a large number of people (see the references above). Moreover, with our 500 km s$^{-1}$ Gaussian smoothing, we have washed out some of the more remarkable structures that are apparent in the redshift data: the very thin walls and filaments, and the tight clusters. However, the main emphasis in this survey is on quantitative and statistical analyses of the galaxy distribution, and thus we now move beyond cosmography, to see what quantitative science we can do with the data presented here.

## 5 Redshift Surveys: Galaxy Clustering

This chapter will discuss quantitative measures of galaxy clustering, and how we might use the results to put constraints on cosmological models for structure formation. Much of the background material has been introduced earlier in § 2, although we will find ourselves introducing new concepts as we go along. We are not exhaustive in this section, and do not attempt to describe every statistic that has been used as a measure of galaxy clustering. In particular, we do not survey the work done with multi-fractal measures or percolation methods; these approaches have been thoroughly reviewed in this journal by Borgani (1994).

We start with a discussion of the two-point correlation function $\xi(r)$ in § 5.1. Before going onto the power spectrum in § 5.3, we discuss the effects of redshift space distortions and non-linear effects in § 5.2. Non-linear effects give non-zero high-order correlations, which we discuss in § 5.4 and § 5.5. Topological measures of large-scale structure are discussed in § 5.6.

The dipole moment of the galaxy distribution is closely related to the motion of the Local Group (§ 5.7). One can expand the density distribution in higher-order multipoles as well, as discuss in § 5.8. We return to the subject of redshift-space distortions with methods to correct for them in § 5.9. We finish this chapter with a discussion of the relative distribution of different types of galaxies, in § 5.10.

### 5.1 The Two-Point Correlation Function

The two-point correlation function was introduced in Eq. (44) as the autocorrelation function of the (continuous) density field. As it is the Fourier Transform of the power spectrum $P(k)$ (Eq. 46), it also gives a complete statistical description of the density field to the extent that the phases are random. Although one can always define a smoothed density field as described in § 3.7, and then apply Eq. (44), the resulting correlation function would be cut off on scales smaller than the smoothing length. Instead, we simply apply Eq. (45), which we rewrite as follows: the joint probability that galaxies be found at positions $\mathbf{r}_1$ and $\mathbf{r}_2$ within the infinitesimal volumes $dV_1$ and $dV_2$ is

$$dP = n^2 \, dV_1 \, dV_2 \, \phi(\mathbf{r}_1) \phi(\mathbf{r}_2) \left[1 + \xi(r_{12})\right], \tag{81}$$

where $r_{12} = |\mathbf{r}_1 - \mathbf{r}_2|$. In the case of a volume-limited sample ($\phi \equiv 1$) we thus find

$$\xi(r) = \frac{N(r)}{4\pi r^2 n^2 V} - 1, \tag{82}$$

where $N(r) \, dr$ is the number of pairs found in the sample with separations between $r$ and $r + dr$, and $4\pi r^2 n^2 V$ is the number expected in a uniform distribution of galaxies. This expression is appropriate only for the case of a volume-limited sample of galaxies, and ignores edge effects. In practice then, one does the following (Davis & Peebles 1983b): one generates on the computer a sample of points with no intrinsic clustering, but with the same selection criteria as those of the real galaxies. Thus the mock catalog matches the true catalog in the solid angle coverage and in the selection function. One then counts pairs with separation between $r$ and $r + dr$ both in the real data (call this quantity $N_{DD}(r)$, with $D$ standing for data), and between the real data and the mock catalog ($N_{DR}(r)$, with $R$ standing for random). The correlation function is then estimated as:

$$\xi(r) = \frac{N_{DD}(r)}{N_{DR}(r)} \frac{n_R}{n_D} - 1, \tag{83}$$

where $n_D$ and $n_R$ are the mean number densities of galaxies in the data and random samples. One usually makes the number of random galaxies much larger than that of the real sample so that additional shot noise is not introduced. In general, one can weight the pair counts any way one wants:

$$N_{DD}(r) = \sum_{ij} w(\mathbf{r}_i, r) w(\mathbf{r}_j, r), \tag{84}$$

with a similar equation for $N_{RR}$, where the sum is over pairs of galaxies at positions $\mathbf{r}_i$ and $\mathbf{r}_j$ such that $r < |\mathbf{r}_i - \mathbf{r}_j| < r + dr$. For flux-limited samples, weighting by the inverse of the product of selection functions for the two galaxies gives equal volume weighting. Saunders, Rowan-Robinson, & Lawrence (1992) show that the variance in $\xi(r)$ is minimized for the weights:

$$w(r_i, r) = \frac{1}{1 + 4\pi n J_3(r) \phi(r_i)}, \tag{85}$$

where $J_3(r)$ was defined above (Eq. 62). Note that Eq. (85) bears a resemblance to the minimum variance weights for the mean density, Eq. (61).

The accuracy of Eq. (83) is limited by the accuracy of the estimate of the mean number density of galaxies; one cannot measure $\xi(r)$ on scales beyond that on which it falls below the fractional uncertainty in the mean density of the sample. The mean density is uncertain due to the possibility of large-scale structure on the scale of the survey itself. That is, one never knows the extent to which a given volume is a fair sample of the universe. The rms fluctuations in the density on the scale of a survey are given by Eq. (37), but of course, in order to estimate this, we need to know $P(k)$, which is what we are trying to find in the first place. However, there exist estimators of $\xi(r)$ whose sensitivity to uncertainties in the mean density is appreciably weaker than that of Eq. (83) (Landy & Szalay 1993; Hamilton 1993b). In particular, Hamilton (1993b) shows that the fractional error in the estimator:

$$\xi(r) = \frac{N_{DD}(r) N_{RR}(r)}{N_{DR}^2} - 1 \tag{86}$$

is proportional to the *square* of the fractional error in the mean density. This advantage of Eq. (86) over Eq. (83) actually only holds for weighting $w(r_i) \neq 1/\phi(r_i)$; otherwise, the two estimators give quite similar results. In any case, the differences between the two become apparent only on very large scales, where the correlation function is quite weak.

There has been a great deal of confusion over the proper estimation of the error in the correlation function. There are two sorts of statistical error one could try to calculate:
(i) The difference between the true correlation function of galaxies in the volume surveyed, and that measured. This error arises from the finite number of galaxies in the sample.
(ii) Even if one could measure the correlation function in the volume surveyed without error, it will differ from the true correlation function of the entire universe of galaxies, due to power on scales larger than the sample.

Many of the analyses in the literature consider only one of these two effects. The problem is inherently difficult because the variance in the two-point correlation function necessarily depends on the three- and four-point correlation functions (to be discussed in § 5.4). In addition, there exists strong covariance between the estimate of the correlation function on different scales, which of course is strongest for estimates at two closely spaced scales.

Peebles (1973) and Kaiser (1986) calculate the shot noise contribution to the correlation function errors in terms of a cluster model: if we think of the galaxy distribution as a smooth field with clusters embedded, the number of galaxies associated with each cluster is $1 + 4\pi n J_3$ (we restrict ourselves for the moment to the case of a volume-limited sample). The number of independent pairs of galaxies at a given separation is the number of observed pairs, $N_{DD}$, divided by the number associated with the clusters. That is, because galaxies are clustered, a majority of the galaxy pairs are redundant. The resulting error in the correlation function is then just given by Poisson statistics:

$$\sigma(\xi) = \frac{1 + 4\pi n J_3}{N_{DD}^{1/2}} \tag{87}$$

Kaiser (1986) notes that $N_{DD} \propto n^2$, and thus for $4\pi n J_3 \gg 1$, $\sigma(\xi)$ is independent of $n$. Thus he argues that given a finite amount of telescope time, one wants to sparse-sample to the level that $4\pi n J_3 \approx 1$ (i.e., roughly one galaxy per cluster), maximizing the volume covered to minimize the effect of power on scales larger than the sample. This is a meaningful strategy, *if in fact one's primary motivation is to measure $\xi(r)$ on large scales*. This is the motivation behind the 1-in-6 sampling of the QDOT survey (Lawrence et al. 1994), and the 1-in-20 sampling of the APM survey (Loveday et al. 1992ab).

Ling, Frenk, & Barrow (1986) argue that the effect of shot noise can best be estimated by making bootstrap realizations of a given sample, and calculating the scatter in the determination of $\xi(r)$ from each of these. Although the mean $\xi(r)$ over the bootstraps is unbiased, this results in an *overestimation* of the errors, as shown by Mo, Jing, & Börner (1992) and Fisher et al. (1994a). Fisher et al. (1994a) describe a brute-force way to make realistic estimates of the correlation function error covariance matrix: make a series of independent $N$-body realizations of a given sample, and calculate the scatter of the estimates of $\xi(r)$ from each of these. Of course, this estimate will be only as good as the power spectrum assumed for the simulation itself, although it is probably not terribly sensitive to the details. Existing redshift surveys are not large enough to be able to afford to split the surveys up into pieces and compute errors from the variance in the estimated correlation function in each, although Hamilton (1993b) describes a practical method to estimate correlation function errors from one's dataset itself, by considering the contribution that each subvolume in a sample makes to the correlations.

The two-point correlation function can be applied not only to redshift data, but also to surveys containing only angular data. The *angular* correlation function can be defined in analogy with Eq. (81): the joint probability that galaxies be found at *angular* positions $\hat{\mathbf{r}}_1$ and $\hat{\mathbf{r}}_2$ within the infinitesimal solid angles $d\omega_1$ and $d\omega_2$ is

$$dP = \mathcal{N}^2 \, d\omega_1 \, d\omega_2 \left[1 + w(\theta_{12})\right], \tag{88}$$

where $\theta_{12}$ is the angle between $\hat{\mathbf{r}}_1$ and $\hat{\mathbf{r}}_2$, and $\mathcal{N}$ is the number density of galaxies on the sky. Groth & Peebles (1977) calculated the angular correlation function of the Shane-Wirtanen Lick galaxy counts (cf., the beginning of § 4); they found that $w(\theta)$ is well fit by a power law of slope 0.77 for scales smaller than about 2°, with a sharp break on larger scales (cf, the challenge to their results by Geller, Kurtz, & de Lapparent (1984) and de Lapparent, Kurtz, & Geller (1986) over the issue of plate matching). More recently, the angular correlation function of the APM galaxies has been calculated by Maddox et al. (1990a), who reproduce the Groth & Peebles (1977) results on small scales; however, the APM correlation function breaks on a somewhat larger scale than that of the Lick counts. The APM data are of higher photometric accuracy than the Lick data, and because the catalog was generated automatically it is immune from the inevitable systematic effects that counting galaxies by eye entails. Other recent determinations of the angular correlation function include Picard (1991) and Collins, Nichol, & Lumsden (1992). Bernstein (1994) has carried out a detailed analytic analysis of the error in the angular correlation function (using the estimator of Landy & Szalay 1993), including the covariance terms, and using the hierarchical hypothesis (Eq. 120) to include the effects of three-point and four-point correlations. The resulting expressions are too complicated to reproduce here, but do an excellent job of matching the errors measured from Monte-Carlo simulations.

The angular correlation function $w(\theta)$ for a flux-limited sample is related to the spatial correlation function by Limber's (1953) equation (cf. Rubin 1954):

$$w(\theta) = \frac{\int d^3\mathbf{r}_1 \, d^3\mathbf{r}_2 \, \phi(r_1)\phi(r_2)\xi(r_{12})}{\left[\int d^3\mathbf{r}\phi(r)\right]^2}, \tag{89}$$

where $\phi$ is the selection function, $r_{12} \equiv |\mathbf{r}_1 - \mathbf{r}_2|$, and $\theta$ is the angle between $\mathbf{r}_1$ and $\mathbf{r}_2$. It is straightforward to show from Eq. (89) that a power-law spatial correlation function of logarithmic slope $\gamma$ corresponds to an angular correlation function with logarithmic slope $\gamma - 1$. Thus we expect the spatial correlation function to be a power law with slope $\gamma = 1.77$, at least on small scales. The spatial correlation function has been determined for essentially all the large redshift surveys discussed above in § 3.1; important papers include Davis & Peebles (1983b), Bean et al. (1983), Shanks et al. (1983), Davis et al. (1988), de Lapparent, Geller, & Huchra (1988), Strauss et al. (1992a), Fisher et al. (1994a), Moore et al. (1994), and Loveday et al. (1994). Because of the much smaller number of galaxies included in redshift surveys than in the angular catalogs, the spatial correlation function is determined with lower accuracy on large scales. However, these studies and others have demonstrated convincingly that the spatial correlation function is indeed a power-law on small scales, with a break at approximately 2000 km s$^{-1}$. The much quoted relation of Davis & Peebles (1983b),

$$\xi(r) = \left(\frac{r}{5.4 \, h^{-1} \, \text{Mpc}}\right)^{-1.8}, \tag{90}$$

is consistent with the observed power-law behavior of $w(\theta)$. Eq. (90) also implies that the rms galaxy fluctuations within spheres of radius 8 $h^{-1}$ Mpc are unity (Eq. 37)[16]. This is a scale below which the clustering is clearly strongly non-linear, and much of the formalism developed in § 2.2 becomes irrelevant.

A primordial power spectrum of power law slope greater than zero implies that $P(0) = 0$. It then follows from Eq. (46) that the volume integral of the spatial correlation function over all of space must be zero, meaning that the correlation function must go negative at some point. Given a power spectrum, for example, that of Standard Cold Dark Matter, one can use Eq. (46) to predict that the correlation function goes negative on scales above 33 $h^{-1}$ Mpc, and reaches a minimum at 46 $h^{-1}$ Mpc with an amplitude of $-1.5 \times 10^{-3}$. This is too small an effect to have been measured in any existing galaxy sample. Indeed, if one defines the mean density of a sample from the sample itself (as is usually done), the integral of the correlation function *over the volume of the survey* is forced to zero. This effect tends to bias the correlation function low, at least when it is estimated for small volumes.

The power-law nature of the correlation function has prompted some workers to suggest that galaxies follow a fractal distribution, with no preferred scale (e.g., Coleman & Pietronero 1992). In such a model, it would not be possible to define a mean density of the universe; it would be a function of the scale on which one measured it. However, the correlation function is defined in terms of $\delta$, which has the mean density subtracted already. The correlation function of $1 + \delta$ predictably is *not* scale-free (e.g., Guzzo et al. 1991; Calzetti, Giavalisco, & Meiksin 1992). Peebles (1993) shows that the observed scaling of the angular correlation function with depth rules out simple fractal models (cf. Davis *et al.* 1988). More complicated, multi-fractal models have been proposed; they are reviewed thoroughly in Borgani (1994).

---

[16] Eq. (90) has been corrected for redshift space distortions; see § 5.2.1.

*5.2 Distortions in the Clustering Statistics*

The next logical topic to discuss would be the determination of the power spectrum of the galaxy distribution. Before we do so, however, we outline the principal effects which cause the observed correlation function and power spectrum to differ from those which held following the epoch of radiation-matter equality, extrapolated via linear theory to the present. These are:

(i) The galaxy distribution may be biased with respect to the underlying mass distribution. If linear biasing holds (Eq. 48), the correlation function and power spectra of galaxies differ from that of the underlying dark matter by a constant factor $b^2$; the shape of these functions is unaffected. However, if the biasing is a (*a priori* unknown) function of scale, then the power spectrum of galaxies gives us no information of that of dark matter.
(ii) We measure redshifts, not distances of galaxies. As Eq. (2) shows, the two differ from one another because of peculiar velocities.
(iii) Although in linear theory, the shape of the power spectrum does not change as clustering grows (simply because the growth rate is independent of scale), this is not true on scales that have gone non-linear, on which the rms mass fluctuations are not much less than unity.

We will discuss the redshift space distortions and the non-linear effects here.

*5.2.1 Redshift Space Distortions*

The effect of peculiar velocities on the shape of structures can be understood heuristically by imagining the gravitational influence of a rich cluster of galaxies. On small scales, within the virialized cluster itself, galaxies have peculiar velocities of 1000 km s$^{-1}$ or more, which causes a characteristic stretching of the redshift space distribution along the line of sight. This is called the "*Finger of God*", which points directly at the origin in a redshift pie diagram; the Finger of God associated with the Coma cluster is apparent in Fig. 3. Thus a compact configuration of galaxies is stretched out along the line of sight, greatly reducing the correlations: *the small-scale velocity dispersion of galaxies causes the correlation function to be underestimated in redshift space.*

On larger scales, a different effect operates: galaxies outside the cluster itself feel the gravitational influence of the cluster, and thus have peculiar velocities falling into it. A galaxy on the far side of the cluster will thus have a negative radial peculiar velocity, and appear closer to us in redshift space than in real space, while a galaxy on the near side will have a positive peculiar velocity, and appear further away from us. Thus the gravitational influence of a cluster causes a compression of structures, thus *enhancing* the correlation function.

These two effects can be quantified. The effect of the large-scale motions can be calculated via linear theory, as was first done in the present context by Kaiser (1987; cf. Sargent & Turner 1977). In brief, one calculates the change in the Jacobian of the volume element in going from real to redshift space. The result is that in linear theory, both the power spectrum and correlation function are enhanced in redshift relative to their real space counterparts by a factor:

$$K(\beta) = \left(1 + \frac{2}{3}\beta + \frac{1}{5}\beta^2\right), \tag{91}$$

where $\beta$ was defined above in Eq. (51). For $\beta = 1$, $K = 1.87$, so this is not a small effect. This calculation is done in the "distant observer approximation", in which the volume surveyed is supposed to subtend a small angle from the point of view of the observer. For further discussion of this approximation, see Cole,

Fisher, & Weinberg (1994), and Zaroubi & Hoffman (1994).

One way to disentangle the effects of peculiar velocities from true spatial correlations is to divide the vector separating any two galaxies into components in the plane of the sky ($r_p$ in the notation of Fisher et al. 1994a), and along the line of sight ($\pi$). The effects of redshift space distortions are purely radial, and thus the correlation function projected onto the plane of the sky is a measure of the real space correlation function. In practice, then, we measure $\xi_s$ as a function of both $r_p$ and $\pi$. The subscript $s$ reminds that this is a quantity measured in redshift space. The projection of $\xi_s(r_p, \pi)$ onto the $r_p$ axis yields a quantity closely related to the angular correlation function (Davis & Peebles 1983b):

$$w_p(r_p) \equiv 2 \int_0^\infty d\pi\, \xi_s(r_p, \pi) = 2 \int_0^\infty dy\, \xi_r \left[(r_p^2 + y^2)^{1/2}\right], \qquad (92)$$

where here $\xi_r$ is the desired real space correlation function, as indicated by the subscript $r$. For a power-law correlation function $\xi_r(r) = (r/r_0)^{-\gamma}$, the integral can be done analytically, yielding

$$w_p(r_p) = r_p\, \xi_r(r_p) \frac{\Gamma(1/2)\Gamma[(\gamma-1)/2]}{\Gamma(\gamma/2)}. \qquad (93)$$

This is the approach that Davis & Peebles (1983b) took to find the result in Eq. (90); Fisher et al. (1994a) used this method to find

$$\xi_r(r) = \left(\frac{r}{3.76\, h^{-1}\, \mathrm{Mpc}}\right)^{-1.66} \qquad (94)$$

for *IRAS* galaxies. Saunders, Rowan-Robinson, & Lawrence (1992) measured $\xi_r(r)$ for *IRAS* galaxies using a related approach: they cross-correlated the QDOT redshift survey with its parent 2D catalog (Rowan-Robinson et al. 1991) to suppress the redshift spacing distortions; they find results in excellent agreement with Eq. (94) (cf. Loveday et al. 1994 for a measurement of the APM correlation function using the same technique). The significance of the discrepancy in the amplitudes and slopes between the *IRAS* (Eq. 94) and optical (Eq. 90) correlation functions will be discussed in § 5.10.

If we could measure the Kaiser effect of Eq. (91) directly, we could constrain the parameter $\beta$. The difficulty is that Eq. (91) is only valid on large scales where linear theory is valid, and where the competing effect of small-scale velocity dispersion is unimportant. But of course, the correlations are small and difficult to measure on these large scales. Gramann, Cen, & Bahcall (1994) and Brainerd & Villumsen (1994) point out that if the small-scale velocity dispersion were as large as predicted by standard CDM, then the Kaiser effect would be swamped by the suppression of the correlation function due to the velocity dispersion until one gets to truly enormous scales. Nevertheless, several groups have attempted to measure the Kaiser effect directly from redshift survey data. Fry & Gaztañaga (1993) compared the correlation function measured in redshift space for various redshift surveys to the angular correlation function of the same samples (which are free from redshift space distortions). They found $\beta = 0.53 \pm 0.15$ for the CfA survey, $\beta = 1.10 \pm 0.16$ for the SSRS, and $\beta = 0.84 \pm 0.45$ for the *IRAS* 1.936 Jy survey. Alternatively, one looks for the anisotropy of $\xi_s(r_p, \pi)$ in the radial and transverse directions. Hamilton (1992) defines the angular moments of the correlation function as

$$\xi_\ell(r) \equiv \frac{1}{2} \int_{-1}^{1} \xi_s(r_p, \pi)(1 + 2\ell) P_\ell(\mu)\, d\mu, \qquad (95)$$

where $P_\ell$ is the $\ell^{\mathrm{th}}$ Legendre polynomial and $\mu$ is the cosine of the angle between the line of sight and the redshift separation vector. He then shows that:

$$\frac{\xi_2(r)}{-\xi_0(r) + \frac{3}{r^3}\int_0^r \xi_0(s)s^2\,\mathrm{d}s} = \frac{\frac{4}{3}\beta + \frac{4}{7}\beta^2}{1 + \frac{2}{3}\beta + \frac{1}{5}\beta^2} \tag{96}$$

in the linear regime. Hamilton (1993a) applies this to the *IRAS* 1.936 Jy sample to find $\beta = 0.69^{+0.28}_{-0.24}$.

Fisher et al. (1994b) took a somewhat different approach, including the effects of both the small-scale velocity dispersion and large-scale Kaiser effect, and fitting directly to $\xi_s(r_p, \pi)$. Following Peebles (1980), the relation between the real and redshift space correlation functions can be written as

$$1 + \xi_s(r_p, \pi) = \int \mathrm{d}w_3 f(w_3|r) \left(1 + \xi_r\left\{\left[r_p^2 + (\pi - w_3)^2\right]^{1/2}\right\}\right), \tag{97}$$

where the integral is over possible values of the radial peculiar velocity difference $w_3$, and $f(w_3|r)$ is the distribution function of $w_3$ at a given separation $r$. An exponential distribution is a good approximation to the distribution function found in $N$-body simulations; Fisher et al. (1994b) show that it gives a better fit than does a Gaussian to the real data. Taking the shape of the first and second moments of $f$ as a function of $r$ from $N$-body models, Fisher et al. were able to reduce the model-fitting to two free parameters: the amplitude of the second moment $\sigma$ (i.e., the pairwise velocity dispersion) at $r = 100$ km s$^{-1}$, and the amplitude of the first moment $v_{12}$ (i.e., the mean pairwise streaming of galaxies) at $r = 1000$ km s$^{-1}$. They show, in analogy with Eq. (91), that linear theory predicts the following form for the first moment:

$$v_{12}(r) = \frac{2\beta H_0}{r^2[1 + \xi_r(r)]} \int_0^r \mathrm{d}r' r'^2 \xi_r(r'). \tag{98}$$

Thus the measurement of $v_{12} = 109^{+64}_{-47}$ km s$^{-1}$ at 1000 km s$^{-1}$, together with the real-space correlation function (Eq. 94), directly yields a value for $\beta$; they find $\beta = 0.45^{+0.27}_{-0.18}$. Although this statistic is shown to be unbiased with the help of $N$-body simulations, its power is limited by the volume of the sample. We assume that any anisotropy measured in the sample is due to redshift-space distortions, but the real space correlation function will be isotropic only to the extent that the sample includes enough volume to average over the orientation of elongated superclusters.

It is not *a priori* obvious that the approach of writing the effect of the redshift space distortions as a convolution with the velocity distribution function (Eq. 97) is consistent with linear theory in the form of Eq. (91); linear theory predicts covariance between the velocity and density fields that is not included in Eq. (97). However, Fisher (1995) has been able to reproduce Eq. (91) by expanding Eq. (97) to second order, assuming that $f(w|r)$ is Gaussian, with mean value and dispersion as given by linear theory.

The Fisher et al. (1994b) analysis also measures the distortions on non-linear scales to derive the pair-wise velocity dispersion at 100 km s$^{-1}$, $\sigma = 317^{+40}_{-49}$ km s$^{-1}$. This is to be compared with the Davis & Peebles (1983b) value of $340 \pm 40$ km s$^{-1}$ from the CfA survey, also measured by looking at redshift space distortions[17]. The predicted values of $\sigma$ for different power spectra (as calculated with $N$-body simulations) are quite different, so this quantity is of great interest for constraining models. High resolution dissipationless simulations of Standard CDM indicate $\sigma \sim 1000$ km s$^{-1}$ for the dark matter (e.g., Davis et al. 1985; Gelb & Bertschinger 1994b), far in excess of what is observed. However, this number has been controversial: it is the velocity dispersions of galaxies, not dark matter particles, which are observed. The velocity dispersion of *halos* of dark matter particles in simulations (Brainerd & Villumsen 1993; Gelb & Bertschinger 1994) are smaller than that of the dark matter itself, although these simulations suffer from the so-called over-merging problem, in which groups of galaxies merge in enormous supergalaxies, which have no counterparts in the real world. Because $\sigma$ is weighted by pairs of galaxies, such over-merging tends to cause the estimate of $\sigma$ to be biased low. Hydrodynamical simulations of CDM, which tend to avoid the over-merging problem, radiatively dissipate some of the energy that would otherwise go into galaxy motions, and in fact find a lower value for $\sigma$ (e.g., Cen & Ostriker 1993), although still not low enough to match the observed value above. Weinberg (1994) shows that the velocity dispersion is very sensitive to the details of the biasing model used to define galaxies from the distribution of dark matter.

---

[17] A recent reanalysis of the CfA sample by Davis (private communication) finds $\sigma = 380$ km s$^{-1}$ after correction of a small error in the original code.

Velocity dispersion analyses of other redshift surveys (Mo *et al.* 1994) show a larger $\sigma$ for some samples, although the pair-weighting nature of $\sigma$ makes it quite sensitive to a few rare clusters in a survey volume (Zurek *et al.* 1994). What is needed to settle this controversy is a new statistic which is less weighted by the rare high velocity-dispersion clusters, and thus more strongly reflects the velocity dispersion in the field.

The small-scale velocity dispersion can also be used to apply the *Cosmic Virial Theorem* (Peebles 1976ab; Peebles 1980): if one assumes statistical equilibrium of clustering on small scales, and takes the continuum limit, one can show that

$$\sigma^2(r) = \frac{9 H_0^2 \Omega_0}{4\pi b\, \xi_r(r)} \int\limits_r^\infty \frac{\mathrm{d}r}{r} \int \mathrm{d}^3 \mathbf{z}\, \frac{\mathbf{r}\cdot\mathbf{z}}{z^3} \zeta(r, z, |\mathbf{r} - \mathbf{z}|), \qquad (99)$$

where $\zeta$ is the three-point correlation function, to be discussed in § 5.4. This equation can be simplified assuming a hierarchical model for the three-point correlation function (Eq. 118), and a power-law form for the two-point correlation function (Eq. 75.14 of Peebles 1980). In practice, the application of this equation is hampered by our poor knowledge of the three-point correlation function. More importantly, the linear biasing model assumed in Eq. (99) is questionable at best on these very inhomogeneous scales. Finally, Carlberg, Couchman, & Thomas (1990) argue that galaxies may not be fair tracers of the *velocity field* on small scales, a form of *velocity bias*. The Cosmic Virial Theorem has been used to argue for a small value of $\Omega_0 = 0.2\,\mathrm{e}^{\pm 0.4}$ (Davis & Peebles 1983b), but if dark matter is not clustered with galaxies on the very small scales of 100 km s$^{-1}$, Eq. (99) will return an underestimated value of $\Omega_0$ (Bartlett & Blanchard 1994).

Cole, Fisher, & Weinberg (1994) take a parallel approach to Hamilton (1992; 1993a), based on the power spectrum. Just as we can separate the vector between two galaxies into parallel and perpendicular components, we can separate a wave vector $\mathbf{k}$ into parallel and perpendicular components, *if we work with a subsample of a survey with small total opening angle with respect to the observer*, the distant observer approximation discussed above. In analogy with Eq. (95), they define angular moments of the power spectrum:

$$\mathcal{P}_\ell(k) = \frac{2\ell + 1}{2} \int\limits_{-1}^{1} \mathrm{d}\mu\, P(k, \mu) P_\ell(\mu), \qquad (100)$$

and show that the rational expression in Eq. (96) can be expressed as $\mathcal{P}_2(k)/\mathcal{P}_0(k)$. Applying this to the 1.2 Jy *IRAS* redshift survey, they find $\beta = 0.35 \pm 0.05$, although they emphasize that non-linear effects cause this to be a lower limit. A recent re-analysis by Cole, Fisher, & Weinberg (1995) parameterizes the effects of non-linearity in the velocity field by including a small-scale velocity dispersion in their model, in analogy to the analysis by Fisher *et al.* (1994b). They find $\beta = 0.52 \pm 0.13$ for the *IRAS* 1.2 Jy sample, and $\beta = 0.54 \pm 0.3$ for the QDOT survey.

As in all these methods, non-linear effects have the potential to break the degeneracy between $\Omega_0$ and $b$. Non-linear effects make the effective value of $\beta$ as derived from this method grow as a function of scale until the linear regime is reached; the scale at which the curve asymptotes is thus a measure of the strength of the *mass* clustering, and can be used to put constraints on the bias parameter. Unfortunately, existing redshift surveys are not extensive enough to measure this effect with confidence.

### 5.2.2 Non-linear Effects

In linear perturbation theory, the real space density field differs from the primordial density field only by a universal scaling factor. However, this is no longer true when non-linear effects become important. There has been a great deal of work in recent years on non-linear extensions to Eq. (30) (Bernardeau 1992a; Gramann 1993a; Nusser et al. 1991; Giavalisco et al. 1993; Mancinelli & Yahil 1994; cf. Mancinelli et al. 1994 for a comparison of these techniques), and various approximate non-linear schemes to bridge the gap between linear theory and $N$-body simulations (Peebles 1989a, 1990, 1994; Weinberg 1991; Matarrese et al. 1992; Brainerd, Scherrer, & Villumsen 1993; Bagla & Padmanabhan 1994). Here we wish to concentrate on methods to take redshift surveys back in time to their initial conditions. Weinberg (1989; 1991) adopts the assumption that the initial density distribution function is Gaussian, and notes that the rank order of densities is likely to be preserved even as non-linear effects skew the distribution function. Thus he applies a technique called *Gaussianization*, whereby the rank order of the densities at different points is conserved, but the densities are reassigned to fit a Gaussian form. The details of this method depend on assumptions about galaxy biasing and the power spectrum. The idea is to apply the Gaussianization technique to redshift survey data, measure the power spectrum of the resulting density field, and then evolve the resulting initial conditions forward in time again using an $N$-body code. To the extent that the assumed and measured power spectra match, and the final results agree with the original data, one has demonstrated consistency with the input model. Weinberg (1989) applied this technique on a volume-limited subsample of the Pisces-Perseus survey of Giovanelli & Haynes (1988). The power spectrum was consistent with that of standard CDM. Unbiased models did not work, not reproducing the filamentary structure of the real data; $b = 2$ was a better match to the real data. Most importantly, the analysis tests, and finds consistency, with the assumptions of Gaussian initial conditions and gravitational instability.

Nusser & Dekel (1992) have developed a time machine to take the observed density field (as derived from a redshift survey or the POTENT method; cf. § 7.5 below) back in time. The equations of motion allow a decaying mode (Eq. 26), which gets amplified if one simply reverses the density evolution equations. Nusser & Dekel (1992) instead start with the Zel'dovich equation (Eq. 34), which when expressed in Eulerian coordinates yields a first-order differential equation for the velocity potential which only allows a growing mode:

$$\frac{\partial \varphi_v}{\partial D_1} - \frac{1}{2}\left(\nabla^2 \varphi_v\right)^2 = 0, \qquad (101)$$

where $\varphi_v$ is the potential associated with the scaled velocity field $v \equiv \mathbf{v}/a\dot{D}_1$. This *Zel'dovich-Bernoulli* equation can be integrated backwards in time from observations of the density fields; $N$-body tests show the results to reproduce the initial conditions better than does linear theory.

Gramann (1993a) shows that consideration of the continuity equation in the context of the Zel'dovich equation yields a correction term $C_g$ to the right hand side of Eq. (101), given by:

$$\nabla^2 C_g \equiv \sum_{i \neq j} \frac{\partial^2 \varphi_g}{\partial x_i^2}\frac{\partial^2 \varphi_g}{\partial x_j^2} - \left(\frac{\partial^2 \varphi_g}{\partial x_i \partial x_j}\right)^2, \qquad (102)$$

where $\varphi_g$ is the gravitational potential. $N$-body tests show this equation reproduces the non-linear evolution better than does Eq. (101), although this approach has not been applied to redshift surveys yet.

Nusser & Dekel (1993) have used the Zel'dovich approximation in another version of their time machine. Assuming laminar flow, one can write down the eigenvalues of the space derivatives of the initial velocity field in Lagrangian space in terms of the eigenvalues of the space derivatives of the observed velocity field in Eulerian space $\partial v_i/\partial x_i$. Assuming further that linear theory holds in the initial conditions (as it should), one can derive the initial density field via Eq. (30); the final result is

$$\delta_{initial} = -D \sum_{i=1}^{3} \frac{\partial v_i/\partial x_i}{1 - (\partial v_i/\partial x_i)/(H_0 \Omega_0^{0.6})}, \qquad (103)$$

where $D$ is the time dependence of the growing mode of gravitational instability (Eq. 27). They use the *IRAS* 1.936 Jy redshift survey and the methods of § 5.9 to generate the predicted velocity field and thus the initial density field, from which they determine the initial density distribution function. They find it to be accurately Gaussian. Application of their technique to the observed velocity field is described in § 7.5.1.

Another approach to non-linear gravitational evolution was taken by Peebles (1989a; 1990), and amplified by Giavalisco et al. (1993). One can derive the *exact* equations of motion for a multi-body gravitating system by finding the stationary points of the action $S$:

$$S \equiv \int_0^{t_0} L \, dt = \int_0^{t_0} dt \sum_i \left[\frac{1}{2} m_i a^2 \dot{\mathbf{x}}_i^2 - m\phi(\mathbf{x}_i)\right], \qquad (104)$$

where the sum is over the particles in the system, $m_i$ are their masses, $\mathbf{x}_i$ are their comoving positions, and $\phi$ is the gravitational potential. Giavalisco et al. then expand the positions $\mathbf{x}_i$ in a Taylor series, of which the Zel'dovich equation (Eq. 34) is the first two terms:

$$\mathbf{x}_i = \mathbf{x}_{i,0} + \sum_j [D(t) - D_0]^j \, \mathbf{C}_{j,i}, \qquad (105)$$

where the $\mathbf{C}_{j,i}$ are coefficients to be determined. Setting the derivative of the action $S$ with respect to the $\mathbf{C}_{j,i}$ yields the set of equations

$$\int_0^{t_0} dt \, [D(t) - D_0]^j \left[-\frac{d}{dt}(a^2 \dot{\mathbf{x}}_i) + a\nabla \phi(\mathbf{x}_i)\right] = 0, \qquad (106)$$

which can be solved for the unknowns $\mathbf{C}_{j,i}$. This method is exact except in regions in which multi-streaming has occurred, that is, where a single point in Eulerian space corresponds to more than one point in Lagrangian space. Giavalisco et al. (1993) have tested this approach against spherical infall models, and show that it converges very quickly. Peebles (1989a; 1990) has used this method in the analysis of the dynamics of the Local Group, and the first attempts to extend this method to redshift surveys can be found in Peebles (1994), and Shaya, Peebles, & Tully (1994).

Hamilton et al. (1991) have taken an empirical approach to non-linear evolution of the power spectrum. Considerations of galaxy conservation within a radius $r_0$ fixed in Lagrangian space around a galaxy in an $\Omega_0 = 1$ universe yields the hypothesis that the quantity $a^2 J_3(r_0)/r_0^3$ is invariant with time, where $J_3$ is given by Eq. (62). Tests with $N$-body models show this in fact to be the case, and that this quantity is independent of the initial power spectrum, allowing the initial correlation function to be read off that measured. Using this method on the *IRAS*, CfA, and APM correlation functions allowed them to reproduce the initial correlation function, which they found to be best fit by a model invoking a mix of cold and hot dark matter. Extensions of their method can be found in Peacock & Dodds (1994), discussed further below, and Mo, Jain, & White (1995).

## 5.3 The Power Spectrum

In principle, the correlation function should contain all the information about the power spectrum, given that the two are a Fourier Transform pair (Eq. 46). However, there are two strong reasons to calculate the power spectrum directly from redshift surveys:

(i) Determination of the correlation function on large scales depends on the calculation of the *difference* of two numbers close to unity (Eq. 83), while errors in the mean density of a sample affect the amplitude but not the shape of the power spectrum.

(ii) The error analysis and covariance between the determination of the power spectrum on different scales is more straightforward than for the correlation function.

The power spectrum can be calculated from a galaxy redshift survey as follows. The unsmoothed density field is given by a sum over Dirac delta functions:

$$\delta(\mathbf{r}) = \frac{1}{nV} \sum_i \frac{\delta^D(\mathbf{r}_i)}{\phi(r_i)} - 1. \qquad (107)$$

Taking the Fourier Transform of this yields

$$\hat{\delta}_{\mathbf{k}} = \frac{1}{nV} \sum_i \frac{1}{\phi(r_i)} e^{i\mathbf{k}\cdot\mathbf{r}_i} - W(\mathbf{k}), \qquad (108)$$

where

$$W(\mathbf{k}) \equiv \frac{1}{V} \int_V d^3\mathbf{r}\, e^{i\mathbf{k}\cdot\mathbf{r}}. \qquad (109)$$

Our estimator of the power spectrum is then

$$\Pi(\mathbf{k}) \equiv V \hat{\delta}_{\mathbf{k}} \hat{\delta}_{\mathbf{k}}^*, \qquad (110)$$

where the factor of $V$ on the right hand side gets the units right. Several lines of algebra (Fisher *et al.* 1993) show that the expectation value of this estimator is given by

$$\langle \Pi(\mathbf{k}) \rangle = \int d^3\mathbf{k}'\, P(k') G(\mathbf{k}-\mathbf{k}') + \frac{1}{nV} \int d^3\mathbf{r} \frac{1}{\phi(r)}, \qquad (111)$$

where

$$G(\mathbf{k}-\mathbf{k}') \equiv \frac{V}{(2\pi)^3} |W(\mathbf{k}-\mathbf{k}')|^2. \qquad (112)$$

Thus the power spectrum estimator is given by the true power spectrum convolved with an expression involving the Fourier Transform of the volume, plus a shot noise term. In the limit of an infinitely large volume, $G$ approaches a Dirac delta function.

The power spectrum as so defined is a function of the direction of $\mathbf{k}$. In practice, one averages $\langle \Pi(\mathbf{k}) \rangle$ over $4\pi$ steradians in $k$-space. Fisher *et al.* (1993) introduce the trick of measuring the power spectrum within cylinders embedded within the survey volume, whose long axis of length $2R$ is parallel to the vector $\mathbf{k}$. If one then chooses $kR = n\pi$ with $n$ a positive integer, the window function $W$ vanishes (Eq. 109). This has two benefits: $\delta(\mathbf{k})$, and therefore the power spectrum now scale exactly with mean density, and thus errors in the mean density affect only the *amplitude*, and not the shape, of the power spectrum. In addition, the values of the power spectrum at different values of **k** are uncorrelated; there is no covariance between them.

Feldman, Kaiser, & Peacock (1994) have taken a slightly different approach. They include a weight function in Eq. (108), and derive an expression for the variance in the power spectrum estimator, assuming that the error distribution of $P(k)$ is exponential (which follows from a Gaussian distribution of $\delta(\mathbf{r})$). Minimizing the ratio of this variance to $P^2$ gives the optimum weight function for galaxy $i$:

$$w_i = \frac{1}{1 + n\phi(r_i)P(k)}, \tag{113}$$

which is of a similar form to the optimum weight for the mean density (Eq. 61) and the correlation function (Eq. 85). With this weight function, the variance in the estimate of the power spectrum is given by

$$\sigma^2[P(k)] = \frac{(2\pi)^3}{V_k \int d^3\mathbf{r}\,[n w \phi(r)]^2}, \tag{114}$$

where $V_k$ is the volume in $k$-space occupied by the bin in question. This expression assumes that the bins are spaced far enough apart that the covariance is negligible (this happens roughly for separations $\Delta k > 2\pi/R$, where $R$ is the characteristic dimension of the volume surveyed). They also derive an expression for the off-diagonal terms of the covariance matrix; see their paper for details.

The power spectrum of galaxies has been calculated for a number of redshift surveys (Baumgart & Fry 1991; Peacock & Nicholson 1991; Park, Gott, & da Costa 1992, Vogeley *et al.* 1992; Fisher *et al.* 1993, Feldman *et al.* 1994; Park *et al.* 1994; da Costa *et al.* 1994b; Lin 1995). Although there is reasonable agreement in the literature now about the shape of the power spectrum on small scales, its amplitude, especially on large scales, remains uncertain. The data are consistent with a slope of $P(k) \propto k^{-1.4}$ on small scales, and several authors (e.g., da Costa *et al.* 1994b) show an abrupt change of slope at $2\pi/k = 50\,h^{-1}$ Mpc. All theoretical power spectra show a turnover on scales of 100 $h^{-1}$ Mpc or more (Fig. 2); this has not yet been seen unequivocally in the data. A number of authors (Peacock 1991; Torres, Fabbri, & Ruffini 1994; Kashlinsky 1992; Branchini, Guzzo, & Valdarnini 1994; Padmanabhan & Narasimha 1993) have derived empirical power spectra to fit the various data sets. The most thorough of these analyses is that of Peacock & Dodds (1994), who have combined a number of the above datasets, together with the real-space correlation function (from the angular correlation function; Baugh & Efstathiou 1993). They correct each for the Kaiser effect (Eq. 91) on large scales, and the effects of small-scale velocity dispersion, following Peacock (1991). Moreover, they correct for non-linear effects using the approach of Hamilton *et al.* (1991) (§ 5.2.2), extending the formalism to the $\Omega_0 \neq 1$ case. They combine the different samples, asking for consistency while adjusting five free parameters: four bias values (for Abell clusters, radio galaxies, optical galaxies, and *IRAS* galaxies), plus $\Omega_0$, which determines the strength of the Kaiser effect (Eq. 91). They find that $\Omega_0^{0.6}/b_{IRAS} = 1.0 \pm 0.2$. The small errors on this number imply that the redshift space distortion is unambiguously detected, largely based on the comparison with the real-space correlation function of Baugh & Efstathiou (1993). The constraint on $b_{IRAS}$ separately is less strong, but is consistent with unity. The ratios of the various bias factors are:

$$b_{Abell} : b_{radio} : b_{optical} : b_{IRAS} = 4.5 : 1.9 : 1.3 : 1. \tag{115}$$

Their results on the reconstructed linear power spectrum are shown in Fig. 9. The points are means of the power spectrum from the various data sets, for $\Omega_0 = b_{IRAS} = 1$. The two curves are standard CDM (dashed), and $\Gamma = 0.25$ CDM, normalized to the power spectrum implied by the CMB anisotropies as measured by the COBE satellite (indicated by the box on the left-hand side of the figure). The data are clearly far more consistent with the $\Gamma = 0.25$ CDM model than with standard CDM, both in amplitude and in shape (indeed, Peacock & Dodds come to this conclusion without consideration of the normalization afforded by the COBE data). This result is consistent with the conclusions of a number of workers in the field; we discuss the issues further in § 9.1.

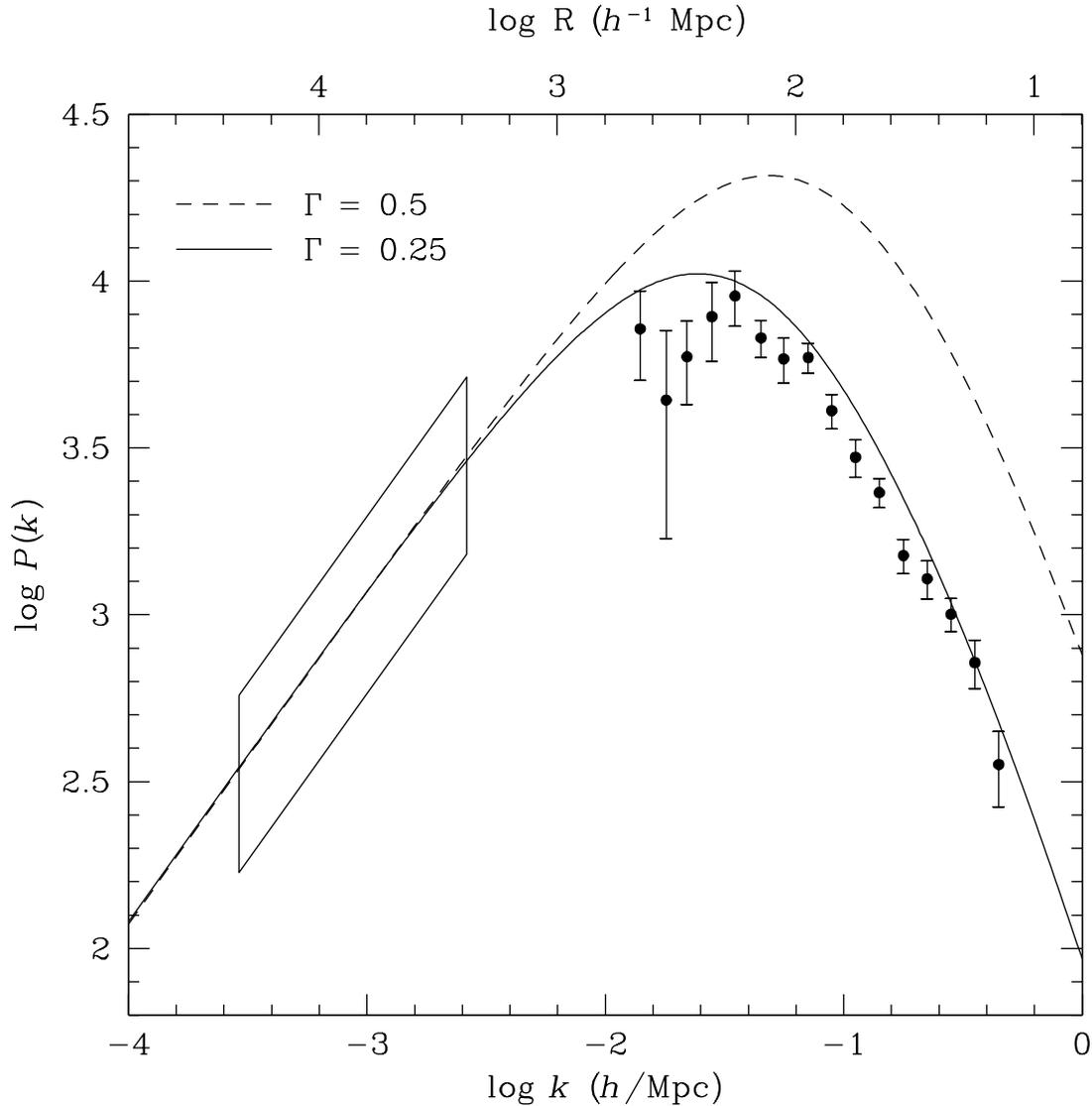

Fig. 9. The power spectrum as derived from a variety of redshift surveys, after correction for non-linear effects, redshift distortions, and relative biases; from Peacock & Dodds (1994). The two curves show the Standard CDM power spectrum, and that of CDM with $\Gamma = 0.25$. Both are normalized to the COBE fluctuations, shown as the box on the left-hand side of the figure.

The second moment of the density distribution function is directly related to the power spectrum via Eq. (37). It can be calculated directly from redshift surveys as the second moment of the count distribution function (after correction for shot noise; cf., Peebles 1980; Saunders et al. 1991), and thus represents another handle on the power spectrum itself. This has been done by Efstathiou et al. (1990), Saunders et al. (1991), Loveday et al. (1992a), Bouchet et al. (1993), and Moore et al. (1994), among others; the results they find are consistent with those shown in Fig. 9. A compilation of second moment results for *IRAS* galaxies is shown in Fisher et al. (1994a). It makes the qualitative point that the variance drops as the scale increases, as is required by the Cosmological Principle, and follows for any power spectrum with $n > -3$ (Eq. 39).

## 5.4 Higher-Order Statistics

As we have mentioned above, the power spectrum, or its Fourier Transform, the two-point correlation function, is a complete statistical description of the density field $\delta$ only to the extent that the phases of the Fourier modes of $\delta$ are random, implying that the one-point distribution function of the density field is Gaussian. Even if this condition holds for the density field in the early universe (as is predicted by inflationary models), it begins to break down as soon as non-linear effects start to develop. We discussed theoretical approaches to this problem in § 5.2.2. In this and the following section, we discuss methods of measuring these non-linear effects from the data.

One can obviously extend the definition of the two-point correlation function to higher order. We can define the three-point correlation function for the continuous density field as

$$\zeta(r, z, |\mathbf{r} - \mathbf{z}|) = \langle \delta(\mathbf{x})\delta(\mathbf{x} + \mathbf{r})\delta(\mathbf{x} + \mathbf{z}) \rangle. \tag{116}$$

However, the practical definition in terms of the point distribution is more complicated, because of the need to correct for the contribution due to the fact that galaxies have a two-point correlation function. In analogy to Eq. (81), the probability of finding a galaxy at distinct positions $\mathbf{r}_1, \mathbf{r}_2$, and $\mathbf{r}_3$ within volume elements $dV_1$, $dV_2$, and $dV_3$ is:

$$\begin{aligned} dP = {} & n^3 \, dV_1 \, dV_2 \, dV_3 \phi(\mathbf{r}_1)\phi(\mathbf{r}_2)\phi(\mathbf{r}_3) \\ & \times [1 + \xi(r_a) + \xi(r_b) + \xi(r_c) + \zeta(r_a, r_b, r_c)], \end{aligned} \tag{117}$$

where $r_a, r_b$ and $r_c$ are the sides of the triangle defined by the three points. One can immediately see the difficulty in measuring $\zeta$, as it requires subtracting four terms from the triple counts. In addition, the three-point correlation function is now a function of three numbers, not just one. The problem only gets worse for higher-order correlation functions; the equivalent expression to Eq. (117) for the four-point correlation function has fifteen terms on the right hand side (Eq. 35.1 of Peebles 1980). Despite these difficulties, the three- and four-point correlation functions have been measured for the Shane-Wirtanen counts (Fry & Peebles 1978; Fry 1983; Szapudi, Szalay, & Boschan 1992), the CfA redshift survey (Bonometto & Sharp 1980; Gaztañaga 1992), and the *IRAS* samples (Meiksin, Szapudi, & Szalay 1992; Bouchet *et al.* 1993). These samples have measured non-zero three- and four-point correlation functions on small scales, indicating that the phases are indeed not random. More importantly, they have shown that the three- and four-point correlation functions display a certain symmetry with respect to the two-point correlation function:

$$\zeta(r_a, r_b, r_c) = Q\left[\xi(r_a)\xi(r_b) + \xi(r_a)\xi(r_c) + \xi(r_b)\xi(r_c)\right] \tag{118}$$

where $Q$ is independent of scale and triangle configuration, to the level that the data can distinguish these things. In particular, there are no "loop terms", proportional to $\xi(r_a)\xi(r_b)\xi(r_c)$, which puts strong constraints on the form of biasing (e.g., Szalay 1988). A similar expression holds for the four-point correlation function. It has therefore been hypothesized that Eq. (118) can be generalized to the $N^{\text{th}}$ order correlation function, indicated as $\xi_N$ (not to be confused with the $\xi_\ell$ of Eq. 95!). Balian & Schaeffer (1989) assume that the $N$-point correlation function shows scale invariance:

$$\xi_N(\lambda \mathbf{r}_1, \ldots, \lambda \mathbf{r}_N) = \lambda^{-\gamma(N-1)} \xi_N(\mathbf{r}_1, \ldots, \mathbf{r}_N), \tag{119}$$

for any $\lambda$. They show that this allows one to write

$$\overline{\xi_N}(V) = S_N \, \overline{\xi_2}^{N-1}(V), \tag{120}$$

where the constants $S_N$ uniquely define the hierarchical scaling, and the correlation functions averaged over a sphere are given by:

$$\overline{\xi_N}(V) \equiv \frac{1}{V^N} \int_V d^3\mathbf{r}_1\, d^3\mathbf{r}_2 \ldots d^3\mathbf{r}_N\, \xi_N(\mathbf{r}_1, \mathbf{r}_2, \ldots, \mathbf{r}_N). \qquad (121)$$

This volume-averaging integrates over the shape information in the high-order correlation function. Although there is much that can be learned from the dependence of the high-order correlations on the angles between the $N$ points (Suto & Matsubara 1994; Fry 1994), the volume-averaged statistic is much more robust for small datasets. Moreover, the $\overline{\xi_N}(V)$ are equal to the irreducible $N^{\rm th}$-order moments of the density distribution function: the skewness is $\overline{\xi_3}(V) = \langle \delta^3 \rangle$, the kurtosis is $\overline{\xi_4}(V) = \langle \delta^4 \rangle - 3 \langle \delta^2 \rangle^2$, and so on (Peebles 1980). For a power-law correlation function, the relation between $\xi(r)$ and its volume average can be calculated analytically (Peebles & Groth 1976):

$$\overline{\xi_2} = \frac{72 \xi_2}{(3-\gamma)(4-\gamma)(6-\gamma)\, 2^\gamma}. \qquad (122)$$

There has been a great deal of interest in recent years to calculate the hierarchy of $S_N$, both from the theoretical and observational sides. The use of volume averaging mitigates the need to calculate the $N$-point correlation function with its dependence on $N(N-1)/2$ separations. In practice, one calculates the moments $\mu_N$ of the galaxy count distribution function, which are then corrected for shot-noise effects following § 36 of Peebles (1980; cf. Szapudi & Szalay 1993; Gaztañaga & Yokohama 1993) to yield $\overline{\xi_N}$. This only works for volume-limited samples, or angular data; flux-limited samples require a more elaborate correction (Saunders *et al.* 1991). One can then compare the observations with the predicted scaling, Eq. (120). As we will see momentarily, this scaling holds remarkably well, giving support to the scale invariant hypothesis (Eq. 119).

However, Eq. (119), or equivalently, Eq. (120), seems to have been pulled out of a hat. Before we show the observational evidence for them, let us discuss their theoretical motivation. For initial conditions with random phases, all $S_N$ for $N \geq 3$ are zero initially. However, as clustering grows and becomes non-linear, the density distribution function becomes non-Gaussian, and higher-order moments become non-zero. Peebles (1980) first calculated the skewness (i.e., third moment) of the *unsmoothed* density field in second-order perturbation theory for an $\Omega_0 = 1$ universe, and showed that

$$\overline{\xi_3} = \frac{34}{7} \overline{\xi_2}^2, \qquad (123)$$

in agreement with Eq. (120). However, observationally, we are always limited to the smoothed density field, for which the quantity $S_3$ depends on the power spectrum (Juszkiewicz, Bouchet, & Colombi 1993):

$$\begin{aligned}
S_3 &= \int \frac{d^3\mathbf{k}\, d^3\mathbf{k}'}{\sigma^4 (2\pi)^6} P(k) P(k') W(k) W(k') W(|\mathbf{k} - \mathbf{k}'|) \\
&\quad \times \left[ \frac{34}{7} - 6\mu \frac{k}{k'} + \frac{8}{7} P_2(\mu) \right] \\
&= \frac{34}{7} - (\gamma_1 + 3), \qquad (124)
\end{aligned}$$

where $\mu$ is the cosine of the angle between $\mathbf{k}$ and $\mathbf{k}'$, $W$ is the Fourier Transform of the window (cf. Eq. 38),

and
$$\gamma_n \equiv \frac{\mathrm{d}^n \log P(k)}{\mathrm{d} \log k^n}, \qquad (125)$$

where the derivative is calculated at the smoothing scale. Eq. (124) is valid for $-3 \leq \gamma_1 < 1$. The calculations get more difficult for increasing $N$; calculation of $S_N$ requires the application of $N-1$-order perturbation theory. One can show straightforwardly, however, that in every case, the scaling of Eq. (120) holds (Fry 1984ab; Bernardeau 1992b). In a mathematical tour-de-force, Bernardeau (1994b) has set up a formalism for calculating the $S_N$ for all $N$ for a tophat window function, and presents expressions up to $N = 7$ as a function of the $\gamma_i$, $i = 1, \ldots, N$. It turns out that it is more difficult mathematically to do the calculations for Gaussian smoothing, although the cases $N = 3$ (Juszkiewicz et al. 1993) and $N = 4$ (Łokas et al. 1994) have analytic solutions. Analogous calculations have been done for the non-linear evolution of the power spectrum by Juszkiewicz (1981), Makino, Sasaki, & Suto (1992), Jain & Bertschinger (1994) and others. Feldman et al. (1994) examine the cumulative distribution function of $|\delta^2(k)|$, and show it to be accurately exponential, as expected in a Gaussian field. However, the expected distribution in the mildly non-linear regime in the presence of shot noise has not yet been calculated.

The results quoted thus far are for an $\Omega_0 = 1$ universe in real space. The dependence of the $S_N$ on $\Omega_0$ is extremely weak (Bouchet et al. 1992, 1994), and is also insensitive to the transformation from real to redshift space. Moreover, it can be shown that the scaling relations Eq. (120), continue to hold under arbitrary local biasing transformations (Eq. 50), although the values of the $S_N$ themselves change (Fry & Gaztañaga 1993; Juszkiewicz et al. 1994; Fry 1994). Biasing models which are non-local, in which the probability that a galaxy be formed at a given point is a function of events removed by tens of Mpc from that point, have been invoked to explain the mismatch of the observed power spectrum with Standard CDM (e.g., Babul & White 1991; Bower et al. 1993). However, such models break the scale-invariant hierarchy by adding loop terms (e.g., Szalay 1988), and Frieman & Gaztañaga (1994) have used the excellent agreement with the scale-invariant predictions (e.g., Fig. 10) to rule out a wide class of these models.

The calculation of $S_3$ and $S_4$ has been done for the CfA and SSRS samples (Gaztañaga 1992) and the *IRAS* 1.2 Jy sample (Bouchet et al. 1993), as well as for various angular catalogs (Szapudi, Szalay, & Boschan 1992; Meiksin, Szapudi, & Szalay 1992; Gaztañaga 1994; Szapudi et al. 1995). All these authors have found beautiful agreement with the predicted scaling relation: the results from the *IRAS* survey are shown in Fig. 10. Gaztañaga (1994) finds that the value of $S_3$ varies slightly with scale; this is expected by Eq. (124) if the power spectrum is not a pure power law. Inserting the power spectrum for the APM counts of Baugh & Efstathiou (1993) in Eq. (124) gives beautiful agreement with the observed values, implying that the biasing is very weak. This conclusion depends on the biasing being linear; non-linear biasing can mimic absence of biasing in the dependence of $S_3$ on scale.

Thus the scaling relations Eq. (120) were originally hypothesized on largely aesthetic grounds. They were found to be predicted by perturbation theory assuming Gaussian initial conditions, and growth of structure via gravitational instability. Indeed, calculations of the skewness in initially non-Gaussian models (Fry & Scherrer 1994; Bouchet et al. 1994) show that the leading behavior goes like $\overline{\xi_2}^{3/2}$, rather than $\overline{\xi_2}^{2}$ as observed. May we therefore conclude that we can rule out non-Gaussian models? Unfortunately, the answer is no. First, the $\overline{\xi_2}^{3/2}$ term decays with time, and at late times, may be negligible. Moreover, as many have quipped, referring to non-Gaussian models is a little like referring to non-elephant animals; the range of possible non-Gaussian models is vast. Weinberg & Cole (1992) set up a series of non-Gaussian models by skewing the density distribution function of an initially Gaussian model, but the resulting non-Gaussianity exists only on the smoothing scale on which $\delta$ is defined. On scales appreciably larger than this, the Central Limit Theorem guarantees that the distribution is Gaussian again, and the scaling laws between the various moments will continue to hold. Thus one needs to examine each specific non-Gaussian model in turn, ask for its predictions for the scaling either using analytic techniques or $N$-body simulations (Moscardini *et al.* 1991; Weinberg & Cole 1992), and compare with the data. This process has not been carried out in detail at this writing.

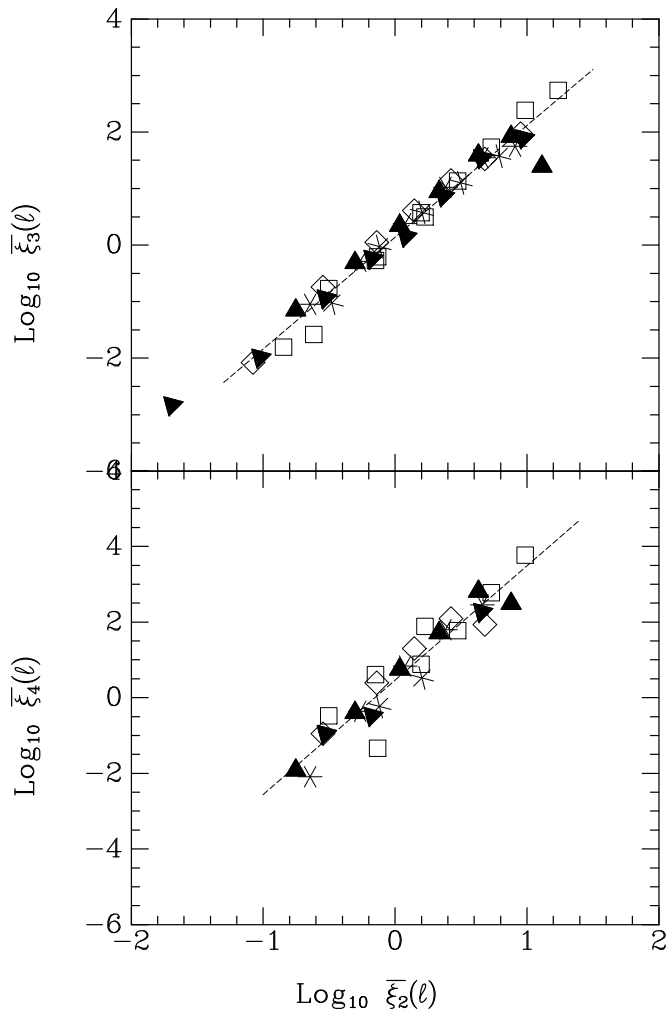

Fig. 10. The skewness (upper panel) and the kurtosis (lower panel) of the *IRAS* density field for various smoothing lengths, as a function of the variance. The plots are logarithmic. The lines drawn are least-square fits, with slopes $1.96 \pm 0.06$, and $3.03 \pm 0.18$, respectively. This figure is taken from Bouchet *et al.* (1993).

   Finally, Fig. 10 shows that the scaling relation between the moments predicted by second-order perturbation theory holds well into the highly non-linear regime, where it has no right to hold (although this has been a working hypothesis for the closure of the so-called BBGKY equations; cf., Davis & Peebles 1977). Similar behavior has been seen in $N$-body simulations (e.g., Juszkiewicz *et al.* 1994). There is controversy about the effect of redshift space distortions in these analyses: Lahav *et al.* (1993a), Suto & Matsubara (1994), and Matsubara & Suto (1994) argue on the basis of $N$-body simulations that redshift space distortions make the $S_N$ closer to constant than in real space in the non-linear regime, a conclusion supported by the analytic calculations of Matsubara (1994a), and the observations of the Pisces-Perseus region by Ghigna *et al.* (1994). However, Fry & Gaztañaga (1993) find that the $S_N$ are remarkably constant in both redshift and real space on small scales, in a variety of redshift surveys. In any case, there exists no analytic argument as to why hierarchical scaling should hold into the non-linear regime.

## 5.5 The Density Distribution Function and Counts in Cells

One of the striking features of the galaxy distribution is the presence of voids as much as 6000 km s$^{-1}$ in diameter. The statistical tools that we have presented thus far do not clearly indicate their presence; we see no feature in the correlation function on such scales. Thus we look for a statistic that is more specifically oriented to describing the visible structures that we see. One such statistic is the *void probability function*. Imagine laying down a series of spheres of radius $r$ randomly within a large volume populated with galaxies. Define the void probability function $P_0(r)$ as the fraction of those spheres which contain no galaxies. In the absence of clustering, Poisson statistics yields

$$P_0(r) = e^{-nV}, \qquad (126)$$

where $n$ is the mean density of galaxies and $V = 4\pi r^3/3$ is the volume of a sphere. In the clustered case, $P_0$ depends on the whole hierarchy of correlation functions, as shown by White (1979):

$$P_0(r) = \exp\left[\sum_{N=1}^{\infty} \frac{(-nV)^N}{N!} \overline{\xi_N}(r)\right], \qquad (127)$$

where the volume-averaged correlation functions were introduced in Eq. (121). Thus the void probability function is a complementary statistic to the correlation functions. One can compute not only $P_0$, but also the probability of observing $N$ galaxies within a sphere; it is related to $P_0$ as:

$$P_N(r) = \frac{(-n)^N}{N!} \frac{\partial^N P_0(n,r)}{\partial n^N}. \qquad (128)$$

The void probability function is clearly a strong function of the sparseness of a given sample, and thus masks to a certain extent the underlying galaxy distribution. One way around this is to define a sampling independent quantity $\chi$:

$$\chi(r) = -\frac{1}{nV} \ln P_0(r), \qquad (129)$$

so that a Poisson distribution gives $\chi = 1$ (Eq. 126). If the hierarchical hypothesis Eq. (120) holds, then Eq. (127) implies that $\chi$ is a universal function, independent of the sampling; indeed, this is observed for the *IRAS* galaxies (Bouchet et al. 1993, but see Vogeley et al. 1991). $\chi(r)$ is a smooth monotonically decreasing curve with no features; no particular scale is picked out.

The void probability function is potentially a useful discriminant of cosmological models. However, Weinberg & Cole (1992) and Little & Weinberg (1994) found that the void probability function is insensitive to the power spectrum or the density parameter, and is more sensitive to the details of the biasing scheme than to the bias value itself.

Under the scale-invariant hypothesis (Eq. 119), one can make quite detailed predictions for the form of the $P_N$ (Balian & Schaeffer 1989). Indeed, the density distribution function is given by (cf. Bernardeau & Kofman 1994):

$$f(\delta) = \frac{1}{2\pi i \sigma^2} \int_{-i\infty}^{+i\infty} dy \exp\left[\sum_{p=1}^{\infty} S_p(\sigma) \frac{(-1)^{p-1} y^p}{\sigma^2 p!} + \frac{(1+\delta)y}{\sigma^2}\right], \qquad (130)$$

which is an exact expression to the extent that the $S_p$ are exact[18]. The $P_N$ follow from this after convolving with a Poisson distribution to include the effects of shot noise.

The various predictions developed by Balian & Schaeffer (1989) based on the scale-invariant hypothesis have been checked in $N$-body simulations (Bouchet et al. 1991; Bouchet & Hernquist 1992), although very dense sampling is required to test the full suite of predictions. The $P_N$ have been derived observationally for various data sets (Alimi, Blanchard, & Schaeffer 1990; Maurogordato, Schaeffer, & da Costa 1992; Lahav & Saslaw 1992; Bouchet et al. 1993). The latter authors compare the observed counts in cells with various models, and find that a range of models (including those of Carruthers & Shih 1983; Saslaw & Hamilton 1984; Coles & Jones 1991) become degenerate at the sparse sampling of existing surveys, and the data cannot distinguish between them.

Another approach to the density distribution function (which is just $P_N(r)$ at constant $r$) was introduced by Juszkiewicz et al. (1994). On large scales, where the second moment of the distribution function $\sigma^2 \equiv \langle \delta^2 \rangle$ is small, the deviation of the distribution function from a Gaussian is expected to be small. Thus it makes sense to expand the distribution function in orthogonal polynomials relative to the Gaussian. The Edgeworth expansion does this:

$$f(x) = \frac{1}{(2\pi)^{1/2}} e^{-\frac{1}{2}x^2}$$
$$\times \left[ 1 + \frac{1}{6} S_3 H_3(x) \sigma + \left( \frac{1}{24} S_4 H_4(x) + \frac{1}{72} S_3^2 H_6(x) \right) \sigma^2 + \cdots \right], \qquad (131)$$

where the $H_N(x)$ are the Hermite polynomials, and $x = \delta/\sigma$. This is found to give an excellent fit to the distribution function in $N$-body models for small $\sigma$, although for $\sigma \geq 0.5$ the Edgeworth expansion starts going unphysically negative at moderate values of $x$. Maximum-likelihood calculations of $S_N$ using fits of Eq. (131) to the observed $P_N$ may be more robust than calculation of the moments directly. Indeed, the results of the moments method are heavily weighted by the tails of the distribution. This is dangerous when working within a finite volume, because one is sensitive to the rare dense clusters (Colombi, Bouchet, & Schaeffer 1994ab).

We motivated this section by pointing out that standard correlation statistics do a poor job of quantifying the largest scale features that are apparent to the eye in redshift maps. The void probability function goes part of the way in filling this need, although it is not as discriminating a statistic between different cosmological models as was hoped. There have been a number of papers discussing various statistics to capture the largest-scale features apparent in the redshift maps (Tully 1986, 1987a; Broadhurst et al. 1990; Babul & Starkman 1992), although again the robustness of these statistics, and their discriminatory power, have been questioned (Postman et al. 1989; Kaiser & Peacock 1991). One of the most successful statistics to describe large-scale structure in a way complementary to correlation functions uses concepts from topology, to which we now turn.

### 5.6  Topology and Related Issues

What is the mental picture we should have of the topology of the galaxy distribution? Is it a uniform sea of galaxies punctuated by rich clusters embedded in it, like meatballs in a bowl of spaghetti? Dramatic voids are what catch the eye in Fig. 3; would a better picture be a uniform distribution with voids scooped out of it, like a piece of swiss cheese? If the density distribution is Gaussian, then there should in fact be a topological symmetry between underdense and overdense regions, like a piece of sponge. Motivated by these considerations, Gott, Melott, & Dickinson (1986), who are responsible for these food analogies, suggested measuring the topology of the galaxy isodensity surfaces. In particular, given a surface in three-space, one can define the principal radii of curvature $a_1$ and $a_2$ at every point. By the Gauss-Bonnet theorem, the integral of the Gaussian curvature $K \equiv 1/a_1 a_2$ over the surface is given by

$$C \equiv \int K \, dA = 4\pi(1-g), \tag{132}$$

where $g$ is the genus number of the surface (the number of holes minus the number of disjoint pieces, plus 1). Thus measurements of the Gaussian curvature give the genus of the surface. A plot of genus of the isodensity surface as a function of density thus tells us the change in the topology at different contrast levels. What do we expect in the Gaussian model? At very high density contrasts, the isodensity contours will surround isolated clusters, and thus the genus will be negative. Similarly, for $\delta$ close to $-1$, the isodensity contours will surround isolated voids, and again the genus will be negative. The mean isodensity contour will be multiply connected and sponge-like, and thus have a positive genus. One can calculate analytically the genus number in the Gaussian case (Doroshkevich 1970; Bardeen et al. 1986; Hamilton, Gott, & Weinberg 1987; cf. Coles 1988 for specific non-Gaussian models):

$$g - 1 = \frac{V}{4\pi^2}\left(\frac{\langle k^2 \rangle}{3}\right)^{3/2}(1-\nu^2)\,e^{-\frac{1}{2}\nu^2}, \tag{133}$$

where $V$ is the volume of the survey, $\nu = \delta/\sigma$ is the level of the density in units of the rms of the density field,

$$\langle k^2 \rangle \equiv \frac{\int d^3\mathbf{k}\, k^2 W(k) P(k)}{\int d^3\mathbf{k}\, W(k) P(k)} \tag{134}$$

is the second moment of the smoothed power spectrum, and $W(k)$ is the Fourier Transform of the smoothing window. Measurements of the genus as a function of $\nu$ thus characterize the general topology of the density field. In particular, we can test the Gaussian hypothesis by comparing the observed form to Eq. (133). To the extent that the observed genus curve is well-fit by the Gaussian form, the amplitude of the curve is a measure of the shape of the power spectrum at the wavelength of the smoothing. The *amplitude* of the power spectrum cancels out of Eq. (134). Gott and collaborators use a volume-weighting technique to reduce the sensitivity of the topology statistic to non-linear evolution and to separate the topological information from that carried by the density distribution function. Calculations of the genus curve from redshift surveys have been carried out by Gott et al. (1989), Moore et al. (1992), and Vogeley et al. (1994), using a tessellation technique for measuring the genus number (Gott et al. 1986; cf. Weinberg 1988 for the source code). Unfortunately, the volume of existing surveys is small, and thus for smoothing lengths in the linear regime, the maximum genus levels are $\sim 20$. The results show a slight "meatball" shift relative to the Gaussian case; that is, the overdense contours show larger values of $g$ than Eq. (133) would predict. This is in the sense expected from non-linear evolution (Matsubara 1994b). The amplitude of the genus curve as a function of smoothing scale is in rough agreement with that predicted by CDM, although Moore et al. (1992) find some evidence for power in excess of CDM predictions on large scales. Statistical errors of the measurement of genus are usually calculated using bootstrap techniques, although these suffer from the same drawback as bootstraps for correlation functions (§ 5.1). At the moment, there is no rigorous error analysis of the genus statistic, nor any calculation of the effects of shot noise (which will tend to make the distribution function look more Gaussian). Very recently, Matsubara (1995) has studied the effect of redshift space distortions on the genus statistic.

---

[18] That is, if one wants a result accurate to $N^{\text{th}}$ order, one needs the $S_p$, $p < N$ calculated to $N^{\text{th}}$ order as well; it is not adequate to calculate the $S_p$ to lowest non-vanishing order. The higher-order corrections to the $S_p$ have not yet been calculated.

A related statistic was invented by Ryden (1988): the area of the isodensity surfaces. For a Gaussian field, one again expects a symmetric function, which again peaks at the mean density. The area is given by

$$A = \frac{2V}{\pi} \left(\frac{\langle k^2 \rangle}{3}\right)^{1/2} e^{-\frac{1}{2}\nu^2} . \tag{135}$$

Ryden et al. (1989) invented a clever technique to measure this statistic, involving counting how often skewers put randomly through the survey volume intersect the isodensity surface, and applied this to the CfA survey and the Giovanelli & Haynes Perseus-Pisces survey. At 1200 km s$^{-1}$ Gaussian smoothing, the results closely matched the linear theory predictions. At 600 km s$^{-1}$ smoothing, however, the data showed stronger deviations from Gaussianity than did any of the models examined. However, their models were probably not evolved forward to become sufficiently non-linear on 600 km s$^{-1}$ scales; this remains a problem for further investigation.

## 5.7 The Dipole

One of the early motivations behind redshift surveys of the full sky was to apply Eq. (33) to the Local Group. The CMB shows a dipole anisotropy of amplitude $\delta T/T \approx 10^{-3}$ (Kogut et al. 1993, and references therein), which is interpreted as a Doppler effect due to the motion of the earth relative to the rest frame of the last scattering surface. When transformed to the barycenter of the Local Group following Yahil et al. (1977), this motion is $627 \pm 22$ km s$^{-1}$ towards $l = 276 \pm 3°$, $b = +30 \pm 3°$ (Galactic coordinates). Given a full-sky redshift survey, a comparison of this motion with the dipole moment of the galaxy distribution is a direct measure of $\beta$. In fact, because both gravity and received light obey the inverse-square law, if one assumes a constant mass-to-light ratio for the galaxies, there is a direct proportionality between the peculiar velocity and the ratio of the dipole and monopole moments of the light distribution, allowing Eq. (33) to be applied using angular data only (Gott & Gunn 1973). The angular dipole moment of the galaxy distribution has been measured by a number of authors using a variety of galaxy catalogs (Meiksin & Davis 1986; Yahil, Walker, & Rowan-Robinson 1986; Villumsen & Strauss 1987; Lahav 1987; Harmon, Lahav, & Meurs 1987; Lahav, Rowan-Robinson, & Lynden-Bell 1988; Plionis 1988; Lynden-Bell, Lahav, & Burstein 1989; Kaiser & Lahav 1989; Scharf et al. 1992). The first impressive result of these analyses is that the vector direction of the light dipole agrees with that of the CMB dipole to within $10 - 30°$, depending on the specific sample and analysis used. There is greater disagreement in the amplitude of the dipole, with results varying from $\beta = 0.3$ to $\beta = 1.2$. Rather than discuss the details of this here, let us move to the application of Eq. (33) using redshift surveys.

The first measurement of the gravitational dipole from a redshift survey was by Davis & Huchra (1982), using a combination of the CfA and Revised Shapley-Ames (Sandage & Tammann 1981) surveys. Given the limited sky coverage of their sample, they were able only to measure the component of the acceleration towards the Galactic poles, which they compared to measured Virgocentric infall (§ 8.1.1), to find values of $\beta$[19] ranging from 0.38 to 0.74, depending on exactly what assumptions were made. More recently, Pellegrini & da Costa (1990) combined redshift survey data from several different surveys, and carried out a similar comparison to the Virgocentric infall; they found $\beta$ in the range 0.24 to 0.56. Further progress had to await the completion of redshift surveys of the entire sky. The dipole moment of the *IRAS* QDOT redshift surveys was calculated by Rowan-Robinson et al. (1990), who found convergence of the dipole only beyond 10,000 km s$^{-1}$. The amplitude of the dipole implied $\beta = 0.82 \pm 0.15$. Hudson (1993b) used his reconstruction of the optical galaxy density field to calculate the galaxy dipole; assuming that the dipole converges within 8000 km s$^{-1}$ allowed him to conclude that $\beta = 0.80^{+0.21}_{-0.13}$.

---

[19] This paper was written before the concept of biasing was formulated, and so the results are quoted in terms of $\Omega_0$, not $\beta$.

We cannot simply calculate the right hand side of Eq. (33) given a redshift survey. The quantity that we do calculate is the dipole moment of the galaxy distribution:

$$\text{Dipole} = \frac{\beta}{4\pi n} \sum_i \frac{W(r_i)\,\hat{\mathbf{r}}_i}{\phi(r_i)\,r_i^2}. \tag{136}$$

This differs from the dipole integral because of a number of effects, which we need to quantify:
(i) Shot noise, which of course gets worse as the sample gets sparser at greater distances. In addition, Eq. (136) assumes a constant mass per galaxy, independent of the luminosity. It is straightforward to recast Eq. (136) in terms of luminosities, which is more appropriate for a luminosity measured in the optical than the *IRAS* bands (cf., Hudson 1993b). Appendix A of Strauss *et al.* (1992c) shows that the shot noise is a more important effect than mass-to-light ratio for all but the nearest galaxies.
(ii) The integral extends over all of space, while the galaxy dipole is necessarily measured through a window $W(r)$, with cutoffs at both large and small distances, to minimize shot noise. Juszkiewicz, Vittorio, & Wyse (1990), Lahav, Kaiser, & Hoffman (1990), and Peacock (1992) point out that the contributions to the dipole due to material outside of the window can be substantial.
(iii) Eq. (33) assumes linear theory, while non-linear effects also come into play on small scales.
(iv) No redshift survey truly covers the whole sky. We discussed this issue in § 3.8.
(v) As Kaiser (1987) pointed out, redshift space distortions cause systematic errors in the derived dipole. In particular, if redshifts are corrected for a value of the Local Group peculiar velocity in error by an amount **v**, the derived dipole will have a spurious contribution (the *rocket effect*) given by

$$\frac{\beta \mathbf{v}}{3} \int \frac{dr}{r} \left(2 + \frac{d \ln \phi}{d \ln r}\right). \tag{137}$$

Strauss *et al.* (1992c) develop a maximum likelihood analysis that allows them to take the first three of these effects into account. A power spectrum must be assumed in the analysis in order to quantify the effect of density fluctuations outside the assumed window. Indeed, one can include parameters of the power spectrum in the maximum likelihood analysis, although in practice, the constraints one can put on models are not very strong. The rocket effect is minimized using self-consistent solutions for the velocity field (cf. § 5.9). The *IRAS* data imply $\beta = 0.55^{+0.20}_{-0.12}$, where the angle between the acceleration and velocity vectors is $18 - 25°$, depending on which self-consistent velocity field is used.

In models without large amounts of power on large scales, one expects the dipole to converge within the volumes probed by redshift surveys; that is, the contribution to the dipole on large scales should be negligible. The Strauss *et al.* results are consistent with the dipole converging within 4000 km s$^{-1}$, although this depends on which self-consistent density field is used, and how one corrects for the Kaiser effect. In addition, there is a substantial dipole moment contributed by galaxies between 17,000 and 20,000 km s$^{-1}$, aligned with the low-redshift dipole, although the sample is so sparse at those redshifts as to make this significant at only the 2 $\sigma$ level. If this large additional contribution to the dipole is found at higher significance level with deeper redshift surveys, this will imply large amounts of power on large scales, and a smaller value of $\beta$ than inferred above.

*5.8 Spherical Harmonics*

An approach to quantifying the galaxy density field complementary to correlation functions uses the method of spherical harmonics. The Fourier components of the density field that are used in the power spectrum are orthonormal within a cube. Spherical harmonics are an orthonormal set of functions on a sphere, and thus are especially appropriate for full-sky samples. They offer a natural way to smooth the data; if one expands to a given order in $l$, the smoothing length is an increasing function of distance from the observer, mimicking the drop-off in the sampling in a flux-limited sample.

Spherical harmonics are simply the generalization to higher order of the dipole analysis discussed in the previous section. They have been used for many years to describe the distribution of galaxies in the celestial sphere, when redshift information was unavailable (Peebles 1973; Peebles & Hauser 1973, 1974; Fabbri & Natale 1989; Scharf et al. 1992). One can express the galaxy density $\sigma(\theta, \phi)$ on the sky in spherical harmonics as

$$\sigma(\theta, \phi) = \sum_{l=0} \sum_{m=-l}^{m=+l} c_{lm} Y_{lm}(\theta, \phi), \qquad (138)$$

where the coefficients of the observed dataset are given by

$$c_{lm} = \sum_i Y_{lm}^*(\theta_i, \phi_i) - \frac{N}{\omega_{sample}} \int_{\omega_{sample}} d\omega Y_{lm}^*(\omega). \qquad (139)$$

The sum is over the $N$ galaxies in the sample, and the integral is over the solid angle subtended by the sample $\omega_{sample}$. For an exactly full-sky sample, the second term vanishes.

The spherical harmonics can be related to the power spectrum straightforwardly. Scharf et al. (1992) show that the expectation value of the square of the coefficients in a Gaussian model with power spectrum $P(k)$ is given by:

$$\left\langle |c_{lm}|^2 \right\rangle = \sum_{l',m'(l \neq 0)} |W_{ll',mm'}|^2 \left\langle |a_{l'm'}|^2 \right\rangle, \qquad (140)$$

where

$$W_{ll'}^{mm'} \equiv \int_{\omega_{sample}} d\omega Y_{lm}(\omega) Y_{l'm'}^*(\omega) \qquad (141)$$

is a tensor which couples together different modes in the case of incomplete sky coverage, and the spherical harmonics in the case of complete sky coverage are given by

$$\left\langle |a_l^m|^2 \right\rangle = \frac{2}{\pi} \int k^2 P(k) \, dk \left| \int r^2 \phi(r) j_l(kr) \, dr \right|^2 + \frac{N}{\omega_{sample}}. \qquad (142)$$

Scharf et al. (1992) used the angular distribution of the *IRAS* 1.936 Jy sample to put limits on the power spectrum, parameterized by an amplitude and a shape parameter, $\Gamma \equiv \Omega_0 h$ (Eq. 42). A maximum likelihood analysis yields $\Omega_0 h = 0.25$, $\sigma_8 = 0.8$, with remarkably tight error bars.

Scharf & Lahav (1993) have extended this analysis using the redshift information of the 1.936 Jy sample as well. Much the same formalism is used, with the addition of a redshift weighting factor $f(r)$ in the expansion of the density field (Eq. 138). In analogy to the dipole analysis described in the previous section, the growth of all the multipole moments with redshift is used as a diagnostic of the power spectrum. The resulting constraints on the power spectrum are similar to those which Scharf et al. (1992) found from the angular data alone, although Scharf & Lahav point out that redshift space distortions are a non-negligible effect.

Fisher, Scharf, & Lahav (1994) have calculated these redshift space distortions in linear theory. They find that the rms spherical harmonic amplitudes for a full sky redshift survey (i.e., $\omega_{sample} = 4\pi$), as measured in redshift space, is

$$\left\langle |a_{lm}|^2 \right\rangle = \frac{2}{\pi} \int k^2 P(k) \, \mathrm{d}k$$

$$\times \left| \int r^2 \phi(r) f(r) j_l(kr) \, \mathrm{d}r + \frac{\beta}{k} \int r^2 \phi(r) \frac{\mathrm{d}f}{\mathrm{d}r} \left[ j_l'(kr) - \frac{1}{3}\delta_{l1} \right] \mathrm{d}r \right|^2 ; \tag{143}$$

compare with Eq. (142). When measured for a real data set, there is of course an additional term due to shot noise. The $\beta$ dependence of Eq. (143) means that one can use the redshift distortions to measure $\beta$, in exact analogy to the analyses of redshift space distortions of the correlation function and power spectrum. Of course, the shape and amplitude of the power spectrum also come into play; using a generic CDM spectrum (Eq. 42), Fisher (1994) finds $\beta = 0.94 \pm 0.17$ and $\Gamma = 0.17 \pm 0.05$. A similar analysis by Heavens & Taylor (1994) yields $\beta = 1.1 \pm 0.3$.

*5.9 Recovering the Real Space Density Field*

As we have emphasized throughout this review, redshifts are not equivalent to distances; the two differ due to peculiar velocities. However, to the extent that peculiar velocities are due to gravity, and that linear theory holds, we can use Eq. (33) to estimate these peculiar velocities. Of course, this requires knowledge of the density field, of which have a distorted view due to peculiar velocities. Thus we look for a self-consistent solution to the density and velocity field, given redshifts and positions for a flux-limited redshift survey of galaxies, and assuming a value of $\beta$. In practice, this is doable only for full-sky redshift surveys, for which the integral in Eq. (33) can be carried out.

Yahil *et al.* (1991) describe an iterative technique to find this optimum solution. We describe it here in its latest incarnation (Willick *et al.* 1995d), as applied to the *IRAS* 1.2 Jy sample:

(i) Galaxies are initially placed at their redshift space positions. The selection function is calculated by the methods of § 3.4. The 10° wide zone of avoidance centered on the Galactic plane is filled with galaxies at a density interpolated from higher-density regions (§ 3.8).

(ii) A smoothed density field is defined by assigning the galaxies to a grid, weighting by $1/\phi$, and smoothing with a Gaussian using FFT's. The resulting density field is filtered with the power-preserving filter (§ 3.7), taking into account the increased shot noise as a function of distance from the origin.

(iii) Eq. (30) is solved for the velocity field, again using FFT techniques.

(iv) Given the density field $\delta(r)$ and velocity field $\mathbf{v}(r)$ along the line of sight $\hat{r}$ to a given galaxy with Local Group redshift $cz$ and flux $f$, one can calculate the probability $P(r|cz, f) \, \mathrm{d}r$ that it have distance between $r$ and $r + \mathrm{d}r$:

$$P(r|cz, f) \, \mathrm{d}r \propto [\delta(r) + 1] \, r^2 \, \mathrm{d}r \, \Phi(4\pi r^2 f)$$
$$\times \exp\left\{ -\frac{1}{2} \left[ \frac{cz - \{r + \hat{\mathbf{r}} \cdot [\mathbf{v}(\mathbf{r}) - \mathbf{v}(0)]\}}{\sigma_v} \right]^2 \right\}, \tag{144}$$

where $\Phi$ is the galaxy luminosity function, $\mathbf{v}(0)$ is the peculiar velocity at the origin, and $\sigma_v$ is a measure of the small-scale velocity dispersion not included in the smooth velocity field model. This method is inspired by the analysis discussed in § 8.1.3 below. If the velocity field model were perfect, the exponential would be replaced by a delta function at $cz = r + \hat{\mathbf{r}} \cdot [\mathbf{v}(r) - \mathbf{v}(0)]$; the Gaussian included here parameterizes our ignorance about the velocity field on scales smaller than the smoothing length.

In the presence of clusters, there are regions in which the peculiar velocity changes with distance fast enough that one finds *triple-valued zones*, in which a given redshift can correspond to three distances. This is illustrated in Fig. 11, which shows the relation between redshift and distance along a line of sight which intersects a large cluster. Infall into the cluster causes the dramatic S-curve. A galaxy at a redshift of 1200 km s$^{-1}$ could lie at any of the three distances. The probability function of Eq. (144) is shown as well; it correctly parcels the probability among these three solutions. In particular, the nearest crossing point is most strongly favored, given the luminosity function weighting.

(v) The density field is recalculated using the distribution function above and it and the velocity field are output. The value of $\beta$ is incremented on each iteration so that the effects of peculiar velocities grow gradually, the selection function is recalculated, and the iterations start again at step (iii) above.

This is the method used to generate Figs. 6-8 above. This basic approach of an iterated solution to the velocity field via Eq. (33) has been used by Yahil *et al.* (1991), Strauss *et al.* (1992c), Hudson (1993ab), and Freudling *et al.* (1994).

Nusser & Davis (1994a) take a different approach to the problem. They point out that the difference between the redshift and real space position of galaxies is directly related to the displacement of a galaxy from $t = 0$ to the present (the Zel'dovich approximation, Eq. 34). Conservation of galaxies, and the assumption that the velocity field is irrotational, allows them to write down a Jacobian for the transformation from the initial conditions in real space to the final configuration in redshift space, which yields a differential equation for the velocity potential $\Phi$. Expanding $\Phi$ and the density field $\delta(s)$ in redshift space on a given shell in spherical harmonics yields the equation:

$$\frac{1}{s^2}\frac{\mathrm{d}}{\mathrm{d}s}\left(s^2\frac{\mathrm{d}\Phi_{lm}}{\mathrm{d}s}\right) - \frac{1}{1+\beta}\frac{l(l+1)\Phi_{lm}}{s^2} = \frac{\beta}{1+\beta}\left(\delta_{lm} - \frac{1}{s}\frac{\mathrm{d}\ln\phi}{\mathrm{d}\ln s}\frac{\mathrm{d}\Phi_{lm}}{\mathrm{d}s}\right), \qquad (145)$$

where $\phi$ is the selection function, and $s$ is the redshift coordinate. This equation can be integrated for $\Phi_{lm}$ using standard numerical techniques, and the radial velocity field can then be derived by differentiation. This method has the advantage that it does not require iteration (every term in Eq. (145) is in redshift space), but because it requires a one-to-one correspondence between real and redshift space, it does not allow the existence of triple-valued zones.

A similar approach is taken by Fisher *et al.* (1994d). They decompose the density field with spherical harmonics and spherical Bessel functions for the radial component:

$$\delta(\mathbf{r}) = \sum_{l=0}^{l_{max}} \sum_{m=-l}^{+l} \sum_{n=1}^{n_{max}(l)} C_{ln} j_l(k_n r) Y_{lm}(\hat{\mathbf{r}}). \qquad (146)$$

They use the fact that for a survey with full-sky coverage, redshift space distortions couple only modes with a given $l$ but with different $n$'s, but there is no coupling between different angular modes. The coupling matrix is analytic, and its inverse allows the real space $\delta$ to be calculated from the redshift space $\delta$, one spherical harmonic at a time. The velocity field then follows directly from Eq. (33). In practice, Fisher *et al.* (1994d) use the Wiener filter to suppress shot noise. This method also does not require iteration. Fisher *et al.* (1994d) test the three methods presented here with the aid of a mock *IRAS* 1.2 Jy redshift survey drawn from an $N$-body simulation; all three give residual rms errors in the radial peculiar velocity field in the Local Group frame for galaxies within 6000 km s$^{-1}$ of $\approx 200$ km s$^{-1}$. These comparisons have only been made at the positions of the galaxies used in the analysis; there is a need to compare the full velocity and density fields on a uniform grid.

Other approaches to real space reconstruction of redshift surveys include Kaiser *et al.* (1991), Taylor & Rowan-Robinson (1994), Gramann (1993b), & Tegmark & Bromley (1994).

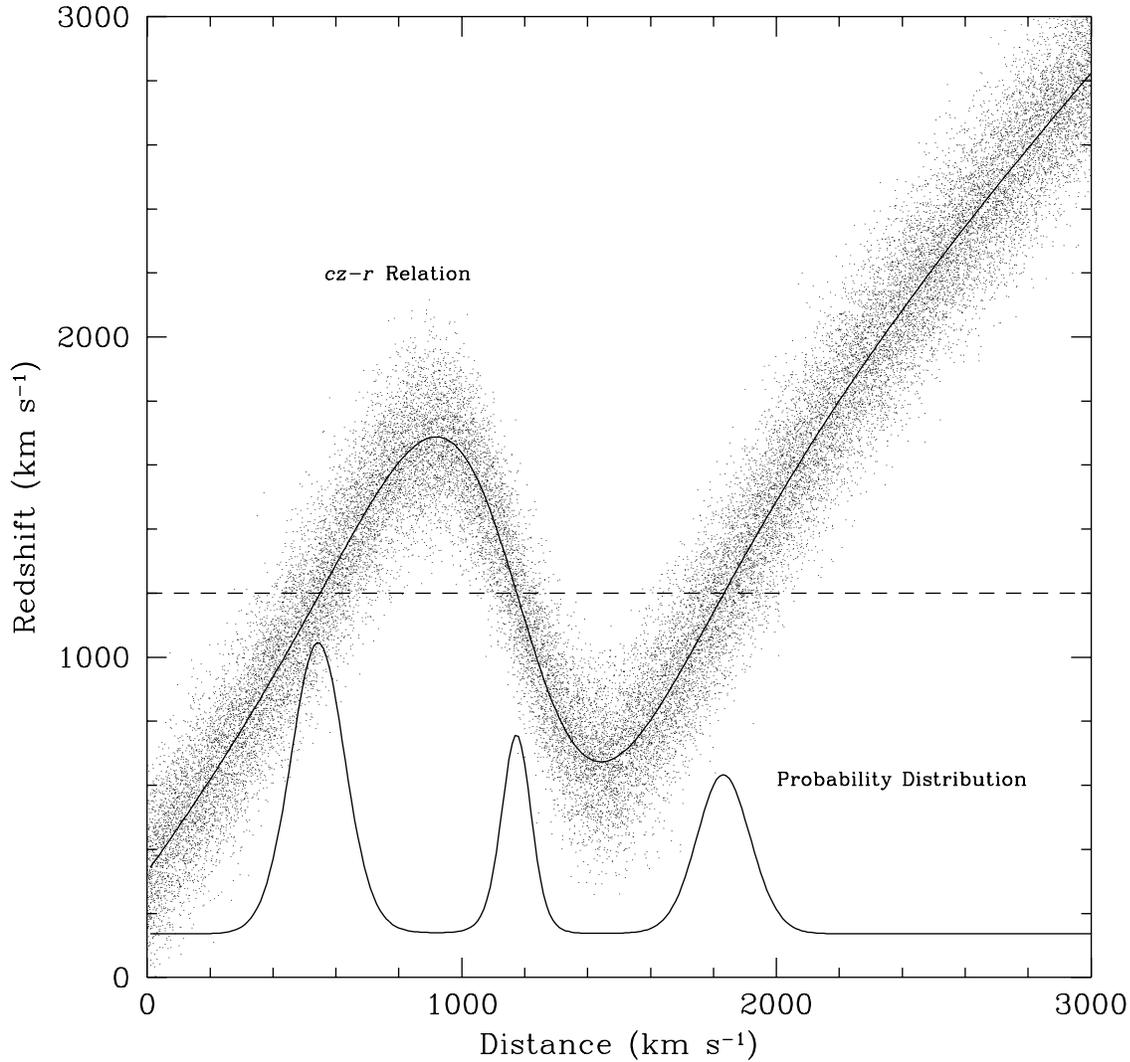

Fig. 11. An illustration of a triple-valued zone. The S-shaped curve shows the relation between redshift and distance along the line of sight to a cluster. A galaxy with a redshift of 1200 km s$^{-1}$ can lie at three distinct distances. When the small-scale noise inherent in any velocity field model, as given by the scattered points, is taken into account, the distribution function of distances for a given redshift gets smoothed out, as shown as the three-peaked curve at the bottom.

*5.10 Clustering of Different Types of Galaxies*

Galaxy formation is a very poorly understood process. Our difficulties in understanding it stem both from our ignorance of the nature of dark matter, which presumably forms the potential wells within which the baryons that eventually form the stars of the galaxy fall, and the extreme complexity of the diverse hydrodynamic and stellar dynamic effects that become important as gas begins to radiate and stars form. In the context of this review, our principal concern is to gain some understanding of the relative distribution of the dark matter and galaxies, because it is the sum of the two which gravitates, while we can observe only the latter directly. In § 2.4, we introduced the concept of biasing, and discussed the various contexts in which we might imagine that the distributions of galaxies and dark matter might differ. Here we ask how we might find observational evidence for biasing.

We observe that galaxies come in a variety of types, defined morphologically: from elliptical galaxies, through lenticulars, and then spirals both barred and unbarred, and finally irregulars and dwarfs of various sorts. If galaxies as a whole are biased relative to the dark matter, in the sense that the large-scale distributions of the two differ on large scales, one would expect that the process of galaxy formation caused the distribution of each of the galaxy types to be biased with respect to one another. Again, because galaxy formation is a poorly understood process, we do not have an accepted model for the Hubble sequence of morphological types, and we cannot be much more specific than this at this stage of our understanding. Nevertheless, this offers a well-posed observational problem: if we can measure biasing of one galaxy type relative to another, we have the potential of constraining biasing models.

One form of relative biasing has been known about since the time of Hubble: the cores of rich clusters are preferentially rich in elliptical galaxies. This was put on a firm quantitative basis by Dressler (1980ab; 1984) and Postman & Geller (1984), who showed that the relative fraction of elliptical galaxies rose from its mean of $\sim 15\%$ in the field starting at overdensities of $\delta \sim 200$, to nearly unity in the highest-density regions of clusters. This means that redshift maps of elliptical galaxies in redshift surveys look qualitatively very different from those of spiral galaxies: the clusters are much more prominent in the former (cf. Giovanelli et al. 1986; Huchra et al. 1990a). This dramatic segregation of ellipticals and spirals does not extend into the field, but subtle relative biasing on large scales between the two have not yet been ruled out.

A number of workers have looked for such effects in the correlation statistics (Davis, Geller, & Huchra 1978; Giovanelli, Haynes, & Chincarini 1986; Santiago & da Costa 1990; Einasto 1991). One of the more powerful statistics for this purpose is the *cross-correlation* function $\xi_{12}(r)$ for two populations of galaxies 1 and 2. The number of galaxies of type 2 within a shell a distance $r$ from a galaxy of type 1 is $n_2 \, dV[1+\xi_{12}(r)]$. One can compare the cross-correlation function with the auto-correlation function of either galaxy type; the ratio of the two is a measure of the relative bias of the two types of galaxies. This approach is especially appropriate when one of the two types of galaxies is quite a bit rarer than the other; the cross-correlation function is much more robust than the auto-correlation function of the rarer sample. The literature on searches for relative biases in different populations is vast: in addition to the morphology surveys discussed above, people have looked for segregation as a function of surface brightness (Davis & Djorgovski 1985; Bothun et al. 1986; Mo & Lahav 1993; Mo et al. 1994), luminosity (Hamilton 1988; Davis et al. 1988; Valls-Gabaud, Alimi, & Blanchard 1989; Thuan, Gott & Schneider 1987; Eder et al. 1989; Börner, Mo, & Zhou 1989; Salzer, Hanson, & Gavazzi 1990; Bouchet et al. 1993; Park et al. 1994; Loveday et al. 1994, Marzke et al. 1994), emission-line properties (Salzer et al. 1988), and even mass (White, Tully, & Davis 1988). In addition, there have been comparisons of the distributions of galaxies selected in different wavebands, including *IRAS* vs. optical (Babul & Postman 1990; Lahav, Nemiroff, & Piran 1990; Strauss et al. 1992a), and radio vs. optical (Shaver 1991; Peacock & Nicholson 1991; Mo, Peacock, & Xia 1993).

The results of these various studies are often contradictory, but can be summarized as follows: on small scales, the correlation functions of late-type, lower surface brightness galaxies are weaker than that of early type galaxies by a factor of 1.5 to 2, depending on the exact sample used. There is also evidence that the later-type galaxies show a shallower slope. There is a similar relation between the *IRAS* and optical correlation functions (Eqs. 90 and 94); this is not surprising, given that *IRAS* galaxies tend to be late-type spirals. There is a weak dependence of the correlation strength on luminosity in both optical and *IRAS* bands, in the sense that more luminous galaxies show stronger correlation. This is worrisome, as it is a violation of the universal luminosity function assumption (§ 3.4). In particular, it means that clustering statistics derived from flux-limited samples will have systematic errors. The correlation function on small scales is heavily weighted by pairs of galaxies nearby, where the sampling is higher and thus there are more pairs. However, the nearby objects have lower luminosity in the mean than those galaxies further away, and if the correlations of the former are indeed weaker, the derived slope of the correlation function will be too shallow. With these effects in mind, some workers (e.g., Park et al. 1994) have restricted themselves to volume-limited samples in calculating clustering statistics from redshift surveys.

One of the most striking features of the observed galaxy distribution is the presence of voids. Models with little power on small scales have galaxies forming from the fragmentation of pancakes, and thus naturally predict voids. Explosion models (Ikeuchi 1981; Ostriker & Cowie 1981) naturally evacuate large regions of space in which galaxies will not form. Alternatively, models that exhibit strong biasing predict that very few galaxies form in underdense regions, naturally creating voids. However, one would expect that in these various scenarios, those few galaxies within voids should have different physical properties from galaxies in the denser regions (Dekel & Silk 1986; Hoffman, Silk, & Wyse 1992; Brainerd & Villumsen 1992). A number of workers have compared the redshift maps of bright galaxies with dwarf galaxies (Eder et al. 1989; Thuan, Gott, & Schneider 1987), low surface brightness galaxies (Bothun et al. 1986; Mo et al. 1994), emission-line galaxies (Salzer et al. 1988), HI-rich galaxies (Weinberg et al. 1991), and *IRAS* galaxies (Babul & Postman 1990; Strauss et al. 1992a); no distinct population of galaxies that "fills the voids" has yet been found. Alternatively, a number of workers have looked for distinguishing physical properties of those galaxies in voids and in more normal environments (Hoffman, Lu, & Salpeter 1992; Szomoru et al. 1994b); no strong effects are seen, in contrast to the situation in clusters, where dramatic differences in mean galaxy properties are seen as a function of local density. Peebles (1989; 1993) argues that this lack of physical differences between void galaxies and those in mean density environments is a strong failing of the biasing model. Santiago & Strauss (1992) do a point-by-point comparison of the density fields as traced by different galaxy types in the CfA survey; the differences they see between ellipticals and spirals are statistically significant, with spirals being *over-represented* relative to ellipticals in the intermediate-density region ($\delta \approx 2$) around the Virgo cluster. Confirmation of this result will require larger samples with more accurate Hubble types; this is one of the principal motivations of the Optical Redshift Survey (Santiago 1993; Santiago et al. 1995a, b).

# 6 Peculiar Velocity Fields: Techniques of Measurement and Analysis

We have learned a great deal about the large-scale structure of the Universe from the redshift surveys discussed in preceding chapters. In these surveys we have generally taken the measured redshift $cz$ as a distance indicator, enabling us to construct three-dimensional maps of the distribution of galaxies. We have also gone one step beyond this, combining theory with observations to use the redshift maps as a means of studying the origin of large scale structure, and the value of the cosmological density parameter $\Omega_0$. These analyses have been premised on the gravitational instability paradigm, according to which structure grows due to gravitational amplification of initial density perturbations (§ 2.2). But redshift measurements do not, by themselves, lead to a conclusive test of the gravitational instability paradigm. Such a test requires the additional information we gain from what we broadly refer to as *peculiar velocity surveys*.

The limitations of redshift surveys stem from their inability to separate the two contributions to a galaxy's observed redshift: the *cosmological* component associated with the expansion of the Universe, and the *peculiar* component associated with the galaxy's proper motion with respect to the local rest frame (see Eq. 2). Such a separation is possible only if, in addition to the redshift $cz$, one has a redshift-independent distance measurement $d$. Peculiar velocity surveys are based on data sets consisting of redshifts and redshift-independent distance measurements, or "redshift-distance samples." From these two measurements one readily obtains the radial component of each galaxy's peculiar velocity,

$$u(d) = cz - H_0 d, \tag{147}$$

where we have used Eq. (2) and defined

$$u(r) = \hat{\mathbf{r}} \cdot [\mathbf{v}(\mathbf{r}) - \mathbf{v}(\mathbf{0})] . \tag{148}$$

Note it is only the radial component of the peculiar velocity that we can measure. We discuss an indirect means of reconstructing the transverse components of the peculiar velocity field in § 7.5 below.

The gravitational instability paradigm requires that the linear peculiar velocity and density fluctuation fields be related to one another according to the local expression Eq. (30), or the global expression Eq. (33). Those formulations of the velocity-density relation contain the Hubble constant $H_0$. In actual peculiar velocity analyses, explicit dependence on the Hubble constant is avoided, by working in a system of units in which distances are expressed as velocities—*i.e.*, in which the Hubble constant is defined to be unity. Eqs. (30) and (33) also involve actual *mass* density fluctuations, whereas redshift surveys provide galaxy density fluctuations only. We thus rewrite the two equations as they appear in an actual analysis:

$$\nabla \cdot \mathbf{v} = -\beta \delta_g \qquad (149)$$

and

$$\mathbf{v}(\mathbf{r}) = \frac{\beta}{4\pi} \int d^3\mathbf{r}' \frac{\delta_g(\mathbf{r}')(\mathbf{r}' - \mathbf{r})}{|\mathbf{r}' - \mathbf{r}|^3}, \qquad (150)$$

where $\beta = f(\Omega_0, \Lambda)/b$ (Eq. 51) and $b$ is the biasing parameter. It is understood that the derivatives in Eq. (149) and the integral in Eq. (150) are with respect to spatial variables expressed in km s$^{-1}$.

A proper analysis of redshift-distance samples will yield $\mathbf{v}(\mathbf{r})$, while we derive $\delta_g(\mathbf{r})$ from redshift survey data. We can thus test the validity of the gravitational instability paradigm by seeing whether these fields obey Eqs. (149) and (150)[20]. Furthermore, to the extent that the data are consistent with the gravitational instability picture, the ratio of the galaxy density and velocity divergence fields gives a direct measure of $\beta$. Indeed, with data of sufficient quality and quantity, we might hope eventually to go beyond the linear velocity-density relations, and thus break the degeneracy between $\Omega_0$ and $b$.

While confirmation of the gravitational instability picture and determination of $\Omega_0$ are their most important goals, peculiar velocity surveys serve other purposes as well. One is to characterize the mass distribution on very large scales. Because of shot noise, existing wide-angle redshift surveys cannot accurately measure number density fluctuations at distances much greater than $\sim 15{,}000$ km s$^{-1}$. However, large-amplitude ($\gtrsim 500$ km s$^{-1}$) coherent peculiar velocities on very large scales can be detected at such distances with relatively modest samples (*e.g.*, Lauer & Postman 1994; Dekel 1994). Velocity perturbations drop off more slowly with increasing scale than do density perturbations, as we saw in Fig. 1; mass density fluctuations of a few percent on a 20,000 km s$^{-1}$ scale—undetectable in present redshift surveys—can give rise to detectable bulk motions if $\Omega_0$ is not much less than unity. For this reason, peculiar velocity measurements constitute one of the best methods of measuring mass density fluctuations on very large scales; this quality makes them a powerful discriminant of rival cosmogonic theories (§ 7.1).

A related virtue of peculiar velocity surveys is their capacity to reveal mass fluctuations independently of bias. Through redshift surveys we can map the number density fluctuations $\delta_g$, and derive the mass density field $\delta$ through an *assumed* model of bias. But if the relationship between luminous and dark mass is, in reality, far different from the linear model we usually adopt—or worse, if that relationship has large variance (Cen & Ostriker 1992b) or is stochastic—redshift surveys may in fact tell us little about mass fluctuations. If the gravitational instability picture is valid, however, peculiar velocities uniquely reflect the mass density field. This point has been emphasized in recent years by Dekel, Bertschinger, and coworkers (*e.g.*, Bertschinger & Dekel 1989; Dekel *et al.* 1990; Dekel 1994), who have developed a method to reconstruct the mass density field from peculiar velocity measurements (§ 7.5).

---

[20] Babul *et al.* (1994) point out that in certain scenarios of *non-gravitational* growth of structure, Eqs. (149) and (150) might apparently be satisfied, yielding an incorrect value of $\beta$ (§ 8.2).

The use of redshift-distance samples as described above appears straightforward in principle. In reality, redshift-independent distances are far too inaccurate for Eq. (147) to be applied as written to real data. Much of the complexity (and controversy) surrounding peculiar velocity surveys arises from the steps needed to account for the consequences of distance measurement errors. In what follows we describe how workers in the field have attempted to deal with this complexity, and discuss the issues that remain unresolved. Still, it is worth emphasizing at the outset our optimism about the program of comparing the peculiar velocity and galaxy density fields in order to estimate $\beta$. Much progress has been made to date using existing large, fairly complete redshift-distance samples, and still larger and improved redshift-distance samples will soon become available. Redshift survey data yielding $\delta_g(\mathbf{r})$ for the local region are similarly improving. We consider it likely that within a decade or so, peculiar velocity analyses will have provided not only convincing substantiation of the gravitational instability picture, but also a believable estimate of $\beta$, and quite possibly, further insights into the subtle relationship between luminous and dark mass.

We begin this chapter with a discussion of distance indicator relations, with emphasis on the Tully-Fisher and $D_n$-$\sigma$ relations for spiral and elliptical galaxies, respectively. § 6.2 briefly discusses the evidence we have that these distance indicator relations are universal, and § 6.3 covers other distance indicator relations that are likely to become important in coming years. The rest of the chapter is devoted to the types of analyses one does with peculiar velocity data, emphasizing the inherent pitfalls and biases. This is first done at a qualitative level in § 6.4, followed by a mathematical treatment in § 6.5. The scientific results of peculiar velocity work are discussed in Chapters 7 and 8.

*6.1 Galaxian Distance Indicator Relations*

We obtain redshift-independent galaxy distances by means of *distance-indicator relations*, or DIs. The essence of a DI is an empirical correlation between two intrinsic properties of a galaxy, one of which is distance-dependent (e.g., diameter) and one of which is distance-independent (e.g., surface brightness).[21] Thus, for example, surface brightness might be known to correlate with absolute diameter; a comparison of the measured angular diameter with the inferred absolute diameter, obtained from this correlation, yields a distance estimate. In some cases the distance-independent property is simply the identification of an object as belonging to a particular class, such as Type 1a Supernovae (§ 6.3.3); in that case a DI may be called a "standard candle" or a "standard ruler," according to whether the distance-dependent property is luminosity or diameter.

Long before their use in peculiar velocity surveys, DIs were employed in the quest to measure the Hubble constant $H_0$ and the acceleration parameter $q_0$ (*e.g.*, Weinberg 1972). Determination of $H_0$ requires *absolute* calibration of a DI, which involves referencing DI measurements to a set of of fiducial galaxies whose distances in Mpc are known. For peculiar velocity work, absolute calibration is unnecessary; the DI measurements are instead referenced to a fiducial set whose distances in km s$^{-1}$ are known. (We discuss how this is done in practice in § 7.2.) In this review, we specifically do not discuss efforts to absolutely calibrate DIs and apply them toward determination of $H_0$; this subject has been recently reviewed by Jacoby et al. (1992). Estimation of $q_0$ does not require absolute calibration, but does entail observations of galaxies at substantial ($z \gtrsim 0.2$) redshifts. The peculiar velocity surveys we will discuss are strictly confined to $z \lesssim 0.06$ and thus have no bearing on the $q_0$ problem.

While many empirical correlations between distance-independent and distance-dependent galaxian properties exist, not all are equally useful as DIs. In what follows, we describe in detail those DIs which have been most fruitfully applied to the study of peculiar velocities, as well as some likely to be so employed in the future.

---

[21] Often the "distance-independent" quantity is in fact weakly distance-dependent. An example is surface brightness, which falls off with increasing redshift as $(1 + z)^{-4}$. However, as in that example, it is usually the case that the weak distance-dependence is either exactly modeled in terms of the redshift alone, and thus is easy to compensate for.

Both elliptical and spiral galaxies show correlations between galaxian luminosity and a relevant measure of internal velocity dispersion. In § 6.1.1, we discuss this correlation for spirals, the Tully-Fisher (TF) relation; § 6.1.2 discusses the corresponding correlation for ellipticals.

*6.1.1 The Tully-Fisher Relation for Spiral Galaxies*

Ordinary spiral galaxies are, for practical purposes, the most suitable objects to use as tracers of the peculiar velocity field in the local Universe. They are more numerous and more uniformly distributed than their luminous counterparts, ordinary elliptical galaxies (which tend to congregate in the cores of dense clusters). They are also bright enough to be carefully studied at fairly large distances, unlike the still more numerous but faint dwarf galaxies. As a result, the TF relation, which is the main DI for spirals, has been the workhorse in peculiar velocity surveys to date.

Since spiral galaxies are flattened systems supported by rotation, the relevant measure of internal velocity for Tully-Fisher is rotation velocity. Because spiral galaxies have "flat" rotation curves, their rotation velocities are well-defined, as we discuss below. In practice, the luminosity-rotation velocity correlation is well modeled as a power law,

$$L(v_{\rm rot}) \propto v_{\rm rot}^{\alpha}. \tag{151}$$

Using the conventional notation of astronomers, which we will follow here, this equation may be rewritten

$$M(\eta) = A - b\eta, \tag{152}$$

where $M = {\rm const.} - 2.5 \log L$ is the *absolute magnitude*, and

$$\eta \equiv \log(2v_{\rm rot}) - 2.5, \tag{153}$$

where $v_{\rm rot}$ is measured in km s$^{-1}$, is a convenient logarithmic measure of the rotation velocity. The normalization of $\eta$ is such that its value lies roughly between $\pm 0.3$ for most spirals. The quantities $A$ and $b$ are known as the "zeropoint" and "slope" of the TF relation; the slope is found to lie in the range $b \sim$ 5–10, depending on the details of how $v_{\rm rot}$ and $M$ are measured. Thus, the power-law exponent of the luminosity-rotation velocity relation is $\alpha \sim$ 2–4. The *apparent magnitude* $m$ of a galaxy is the observed quantity; it is related to the absolute magnitude via the inverse-square law:

$$m = M + 5 \log r \equiv M + \mu(r), \tag{154}$$

which defines the *distance modulus* $\mu(r)$.

The measurement of rotation velocities in TF studies has been carried out mainly through analysis of H I 21 cm profiles, especially prior to 1990. These profiles are derived from spatially unresolved data, and therefore do not give rotation velocity *per se,* but rather total 21 cm intensity as a function of velocity. This has led to a variety of efforts to define a 21 cm profile width which most faithfully reflects rotation velocity (*e.g.*, Tully & Fouqué 1985; Aaronson *et al.* 1986; Bicay & Giovanelli 1986ab). In recent years some workers have employed *optical* measures of rotation velocity, measured from the H$\alpha$ emission line (Dressler & Faber 1990b; Courteau 1992; Mathewson *et al.* 1992ab; Mathewson & Ford 1994), using a long slit oriented along the major axis of the galaxy. Such measurements have become increasingly useful as TF studies have been extended to the Southern Hemisphere, where fewer radio telescopes exist, and to larger distances, where confusion and sensitivity problems become severe in the radio. An alternative is to employ imaging Fabry-Perot spectroscopy in the H$\alpha$ line (Schommer *et al.* 1993), which allows one to map out the rotation velocity field in two-dimensional detail. The optical, unlike the H I, measurements are spatially resolved. Because 21 cm and optical measures of rotation can differ systematically (Tully & Fouqué 1985; Pierce & Tully 1988), it cannot be assumed that the slope or zeropoint of the H I and optical TF relations will be the same; recent work on combining the two types of data have addressed this issue (Mathewson *et al.* 1992b; Willick *et al.* 1995b).

Regardless of how velocity width is measured, the raw measurement differs from the true width (which enters into the TF relation) because of projection on the sky. The true width is estimated by dividing the raw value by the sine of the estimated inclination angle of the galaxy, obtained from the ellipticity of its optical image.

The existence of a correlation between optical luminosity and H I velocity width had been recognized since at least the early 1970s (see, for example, Bottinelli *et al.* 1971; Balkowski *et al.* 1973; Sandage & Tammann 1976), but Tully & Fisher (1977) were the first to characterize this correlation by the linear relation (Eq. 152), finding a slope $b \sim 6$. Their work brought its potential for use as an important cosmological tool to the attention of the astronomical community. Early work with the TF relation used blue-bandpass (B) measurements of luminosity, which were by far the most readily available in the late 1970s. It was recognized early on, however, that such measurements were subject to extinction of starlight by dust in both the source galaxy and in our own Galaxy. These effects represented systematic uncertainties in early applications of the TF relation.

An important advance was made in the late 1970's and early 1980's through the work of Aaronson and collaborators (*e.g.*, Aaronson, Huchra, & Mould 1979; Aaronson *et al.* 1982ab). They recognized that long-wavelength photometry would suffer less from internal and Galactic extinction than B band photographic or photoelectric data. The Aaronson group undertook a program of photoelectric H band ($\lambda = 1.6\mu$m) photometry of spiral galaxies in the Local Supercluster. These measurements were made through apertures whose size was dictated by instrumental constraints, and were subsequently scaled to a fiducial aperture roughly one-third the optical size of the galaxy. This awkward procedure was a drawback of the method, but the Aaronson group nonetheless found that the new infrared TF relation (IRTF) was tighter than the older short-wavelength version. They also found that the IRTF required little or no correction for internal and Galactic extinction, and that its slope $b$ was almost exactly 10, corresponding to a power-law exponent $\alpha = 4$.

The $v_{\rm rot}^4$ dependence of the IRTF was seen as favorable for two reasons. First, as we discuss further below, it drops out naturally from a simple theoretical "derivation" of the TF relation based on considerations of centrifugal equilibrium. Second, a $v^4$ power law also described the analogous *Faber-Jackson* relation for elliptical galaxies (§ 6.1.2). Since elliptical galaxies generally lack ongoing star formation, their radiant output in the blue is produced by an old stellar population. The blue-light luminosities of spiral galaxies, by contrast, may in some cases be dominated by the hot, massive stars which result from recent bursts of star formation. Aaronson *et al.* (1979) suggested that the H band photometry measured the light of the old stellar population, presumably the most faithful tracer of the mass, and therefore correlated better with the gravitational potential—and through it, the observed rotation velocity. The agreement in slope with that of the Faber-Jackson relation seemed to confirm this notion.

Attempts to explain the TF relation in terms of the composition and dynamics of spiral galaxies begin with the equation of centrifugal equilibrium,

$$v_{\rm rot}^2 \propto \frac{M}{R}. \tag{155}$$

If one then assumes a universal mass-to-light ratio $(M/L)$, and a constant mean surface brightness $\bar{I}$ for

spirals, so that

$$L \propto \bar{I} R^2, \qquad (156)$$

one obtains

$$v_{\rm rot}^2 \propto \frac{L}{R} \propto \frac{L}{\sqrt{L/\bar{I}}}, \qquad (157)$$

or

$$L \propto v_{\rm rot}^4, \qquad (158)$$

precisely the exponent characterizing the IRTF. Aaronson *et al.* (1979) cited this consistency as evidence that the near-infrared TF relation was more fundamental than shorter wavelength versions.

While this explanation of the TF slope is not unreasonable, plausible "derivations" of quite different slopes have been presented. For example, Sandage & Tammann (1976) began with Eq. (155), and also invoked the presumed constancy of $(M/L)$. They then supposed that the characteristic radius $R$ at which the equation is to be interpreted is independent of luminosity, on average. This yields $L \propto v_{\rm rot}^2$, corresponding to a TF slope $b = 5$, similar to the earliest estimates for the B bandpass (Tully & Fisher 1977; Bottinelli *et al.* 1983). Moreover, Burstein (1982) has argued on the basis of the observed slope of the TF relation and the relation between surface brightness and rotation velocity, that the assumption of a uniform mass-to-light ratio cannot hold true. The notion that one can straightforwardly derive the TF slope from first principles is now, in fact, considered dubious. It has become clear that the TF slope has more to do with technical issues of measurement than with the constancy of mass-to-light ratios or surface brightnesses. In particular, the size of the aperture and the photometric bandpass used materially affect the observed slope (Bottinelli *et al.* 1983; Pierce & Tully 1988; Willick 1991; Jacoby *et al.* 1992; cf. Fig. 12). Bernstein *et al.* (1994), for example, found an H band TF slope of $\sim 6$, rather than the canonical 10, when they used large apertures which encompassed essentially all of the light from a galaxy. Moreover, the assumption that spirals possess constant $(M/L)$ in H band light, has been challenged. Bothun (1986), for example, has pointed out that small variations in heavy-element abundances can greatly alter the infrared energy output of a stellar population, in which case the *age* of a galaxy should affect $(M/L)$. Pierce (1988,1991) has shown that, as measured in I band light, $(M/L)$ ratios show strong variations with luminosity. In short, it appears that poorly understood issues pertaining to the nature and spatial distribution of stellar populations will be central to an eventual understanding of the TF relation.

Beyond this, there is a fatal flaw in any simple model of the TF relation based on the equation of centrifugal support. Spiral galaxies possess roughly exponential light profiles (Freeman 1970) with a characteristic scale length. However, they have rotation curves which rise quickly from zero to a characteristic rotation speed that is observed to be remarkably constant out to many exponential scale lengths (*e.g.*, Rubin *et al.* 1980, 1982, 1985). This constancy of the rotation speed is one of the main pieces of evidence for the existence of dark halos in spiral galaxies (Faber & Gallagher 1979). Indeed, it is only because of such flat rotation curves that a well-defined rotation velocity of spiral galaxies exists. Thus the characteristic scale of the mass distribution in spiral galaxies is appreciably larger than is that of the light, although the two were equated in Eqs. (155)–(157). Moreover, the TF relation applies equally well to all galaxies of a given rotation speed, although their exponential scale lengths, as well as their bulge-to-disk ratios and morphologies, may vary considerably (Han 1991; Willick 1991). Evidently, there is a complex interplay between luminous and dark mass which determines the final state of spiral galaxies. While some progress has been made toward understanding these effects (*e.g.*, Blumenthal *et al.* 1986), our understanding of the physical basis of the TF relation remains schematic at best. The origin of the TF relation is shrouded in the mysteries of spiral galaxy formation.

With the advent of CCD detectors in the 1980s, highly accurate galaxy photometry has become routine. Two-dimensional digital detectors such as CCDs permit one to tailor the photometric aperture to the galaxy image in a natural way, in contrast to the somewhat arbitrary scalings required in fixed-aperture photoelectric photometry. CCDs are very sensitive in the R ($\lambda \sim 0.7 \mu m$) and I ($\lambda \sim 0.8 \mu m$) bandpasses, and most recent TF studies have been carried out in one of these bands (Bothun & Mould 1987; Pierce & Tully 1988; Willick 1991; Han 1991; Courteau 1992; Mathewson et al. 1992ab; Roth 1993; Mould et al. 1993; Mathewson & Ford 1994; Haynes et al. 1995). Dust extinction is less of a problem in R or I than in B, though not negligible as it is at $1.6 \mu m$ (a detailed discussion of models of internal extinction is given by Giovanelli et al. 1994). With the advent of imaging detectors sensitive at $\sim 2 \mu m$, a few workers are returning to the near infrared for TF work (Peletier & Willner 1993; Bernstein et al. 1994).

In Fig. 12 we plot B, R, and I TF relations obtained from the large data set of Mathewson et al. (1992b), and an H band TF relation from the Aaronson et al. (1982b) data set, as updated by Tormen & Burstein (1994, 1995). In each case a Hubble flow model of distances has been used; thus, the diagrams contain extra scatter due to the inaccuracy of the distance assignments, as discussed below. Nonetheless, the relation is well defined in all of the bandpasses, although the scatter in the R, I and H bandpasses is noticeably smaller than in the B bandpass. The B and R band data are photoelectric measurements taken from the RC3 and Lauberts & Valentijn (1989) catalogs, respectively, as compiled by D. Burstein (private communication). The slopes of the fitted TF relations are indicated in the figure; these TF fits are corrected for sample selection bias as described in § 6.5. As can be seen, the slope increases with increasing wavelength. The very large change from I to H is due only partially to wavelength increase; it stems mainly from the small apertures within which the H band magnitudes are measured (the I band magnitudes are "total" magnitudes) and the different definitions of velocity width used by Mathewson et al. vs. Aaronson et al.In all cases, the absolute magnitudes are expressed in the units suitable to peculiar velocity work, in which distances are measured in km s$^{-1}$.

An important and unresolved issue is the rms scatter of galaxies around the mean TF relation, usually quoted in magnitude units and denoted $\sigma$. The earliest estimates of the TF scatter were $\sigma \sim 0.5$–0.6 mag in B (Tully & Fisher 1977). The Aaronson group later found a somewhat tighter correlation, $\sigma \simeq 0.45$–0.50 mag, in the near infrared (e.g., Aaronson et al. 1986). However, Sandage, Tammann, and coworkers were critical of these relatively low estimates, arguing that sample selection effects (§ 6.4) were introducing gross biases into estimates of the scatter. By enlarging the data sets studied by the Aaronson group in the Virgo cluster, Sandage (1988) and Kraan-Korteweg, Cameron, & Tammann (1988) derived values ranging from $\sim 0.7$–1.0 mag, although Fukugita, Okamura, & Yasuda (1993) explain much of this increased scatter as due to appreciable depth effects in the cluster. More recently, Federspiel, Sandage, & Tammann (1994) have claimed that while the most luminous galaxies may have $\sigma = 0.4$ mag, the TF scatter for a typical galaxy is $\sim 0.6$–0.7 mag.

On the other hand, recent years have seen ever-decreasing estimates of the TF scatter by groups employing CCD photometry. Bothun & Mould (1987) suggested that an appropriate adjustment of the photometric aperture with rotation velocity width could reduce the scatter of the I band TF relation to $\sim 0.20$–0.25 mag. Pierce & Tully (1988) *observed* a scatter in the nearby Ursa Major cluster of $\sim 0.30$ mag, and calculated that the contribution of true cluster depth was such that the true scatter was 0.25 mag. Willick (1990,1991) studied a sample of relatively distant clusters and found an R band scatter of $\sim 0.30$ mag. Courteau (1992) identified a region of "quiet Hubble flow" within a larger field sample in which the R band TF scatter was 0.32 mag. Schommer et al. (1993) found a scatter of 0.25–0.3 mag for their I band Fabry-Perot sample (see above). Very recently, Bernstein et al. (1994) have studied a carefully selected I band CCD sample in the Coma Supercluster region which exhibits an apparent TF scatter of $\sim 0.10$ mag.

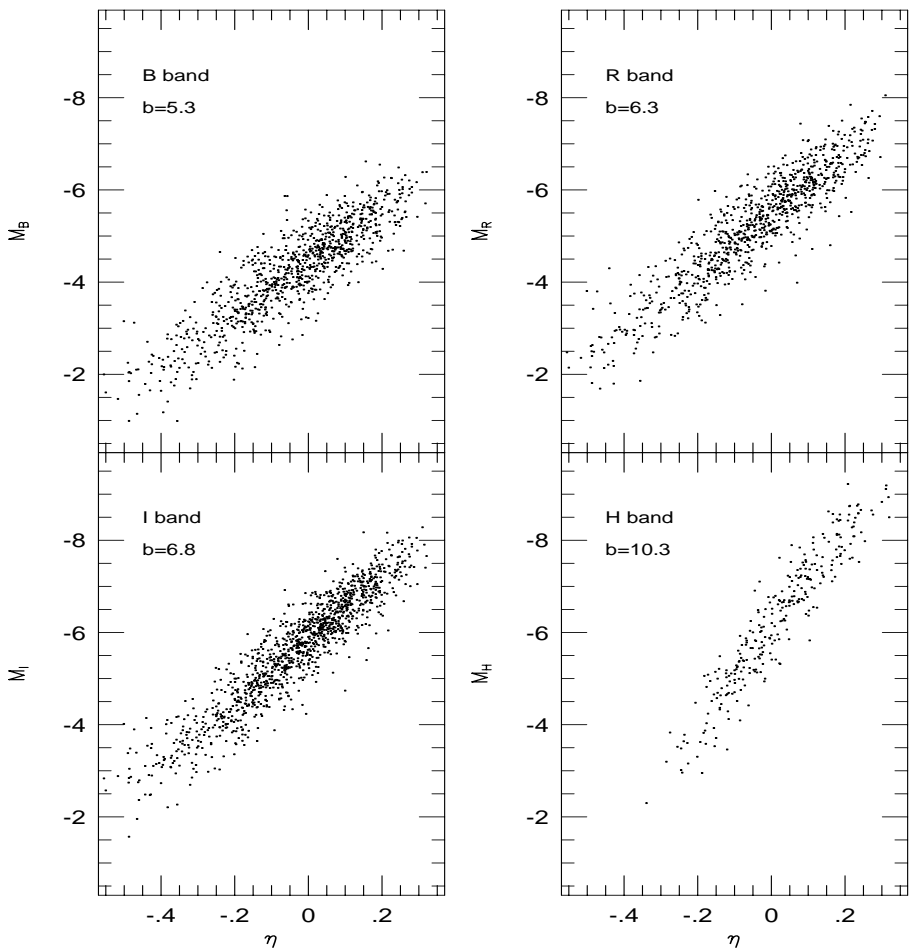

Fig. 12. TF relations in the B, R, I, and H bandpasses. The B, R, and I TF relations are obtained from the data set of Mathewson *et al.* (1992b). The B and R magnitudes for the Mathewson galaxies are photoelectric measurements obtained from the RC3 and Lauberts-Valentijn (1989) catalogs respectively; the I band magnitudes are from the CCD photometry, and the velocity widths from a combination of H I and optical measurements, carried out by the Mathewson group. The H band magnitudes are from the local sample of Aaronson *et al.* (1982b), as updated by Tormen & Burstein (1994, 1995). In each case the plotted absolute magnitudes are computed as $M = m - 5\log(cz)$, where $cz$ is the observed redshift, and are thus expressed in the units appropriate to peculiar velocity studies. Further details are given in the text.

The observed TF scatter comes from three sources: intrinsic or "cosmic" scatter, photometric and velocity width measurement errors, and incorrect distance assignments. Cosmic scatter is the quantity of greatest interest astrophysically; because of uncertainties in the other two sources of scatter, however, it is not well-constrained at present. Photometric errors *per se* are very small ($\lesssim 0.05$ mag) when CCDs are used. However, the extinction corrections made to the apparent magnitudes, while accurate in the mean, may be incorrect by $\sim 0.1$ mag in individual cases due to variations in the dust content of spiral galaxies. Raw velocity width measurements contain $\sim 5\%$ errors, and comparable errors are incurred in making the inclination correction; resulting errors in the width parameter $\eta$ must then be multiplied by the TF slope. Overall, measurement errors can account for $\sim 0.15$–$0.30$ mag, depending on bandpass, inclination, and quality of the H I or optical width determination. Cosmic scatter and measurement errors, added in quadrature, are what the quantity $\sigma$ is meant to signify. However, when a TF sample is used to estimate scatter, galaxies must be assigned distances; errors in these assignments contribute to the observed scatter. Minimizing this last contribution, and subtracting it from the observed scatter, is a tricky but necessary part of estimating the true TF scatter.

The observed value of the TF scatter is important not only in determining the size of random errors, but also of systematic errors due to statistical bias effects (§ 6.4). The discrepancy among estimates of the TF scatter discussed above probably arises from a combination of two factors: small number statistics, and different approaches to the problem of distance assignment errors. Large TF residuals that one author attributes to an erroneous distance model, such as the misassignment of a galaxy to a cluster, another might attribute to true scatter. In the first case the galaxy in question would be eliminated from the analysis and a relatively small scatter obtained, while in the second it is included and a higher value results. In many cases such a decision is subjective, and may depend on the worker's own biases with regard to the true scatter. As larger and more complete samples are studied, the importance of such subjective factors will diminish, and it is likely that scatter estimates will converge. A recent study using a data set consisting of over 3000 galaxies culled from five samples (Willick *et al.* 1995a,b,c; cf. § 7.2) finds the CCD R and I band scatters to be in the range $0.35$–$0.40$ mag, and the H band scatter to be $\sim 0.45$ mag. A partly overlapping sample of $\sim 2500$ spirals analyzed by Mathewson & Ford (1994) estimated the I band scatter at $0.42$ mag. Still, the issue is far from closed; claims of very small ($\sim 0.25$ mag) scatter, large TF data sets continue to appear in the literature (*e.g.*, Haynes *et al.* 1995).

*6.1.2  The Faber-Jackson and $D_n$-$\sigma$ Relations for Elliptical Galaxies*

The elliptical galaxy DI first used in peculiar velocity studies was the Faber-Jackson (FJ) relation (Faber & Jackson 1976). Like the TF relation, FJ expresses a power-law relationship between luminosity and internal velocity. Unlike spirals, luminous ellipticals do not rotate (Kormendy & Djorgovski 1989), and the relevant measure of internal velocity is the rms dispersion of radial stellar velocities in the central regions of the galaxy, $\sigma_e$. Thus the FJ relation is expressed as

$$L \propto \sigma_e^\alpha \,. \tag{159}$$

Unlike the case with TF, early work with FJ (Faber & Jackson 1976; Schechter 1980; Tonry & Davis 1981) produced reasonable agreement that the exponent of the power law was $\alpha \simeq 4$, corresponding to the infrared TF slope found by Aaronson and collaborators (see above). However, while early estimates of the TF scatter were $\sim 0.45$ mag, the scatter of the FJ relation was found to be roughly twice that. For this reason, although FJ was used with some success in studies of the Virgocentric motion of nearby galaxies (Schechter 1980; Tonry & Davis 1981), it did not seem to hold great promise as a DI.

The situation changed dramatically when two independent groups discovered that the addition of a third parameter significantly improved the accuracy of the FJ relation (Djorgovski & Davis 1987; Dressler *et al.* 1987b [22]). Working with newly acquired photometric and kinematic data for large samples of ellipticals, each group found that elliptical galaxies populate a planar region in a three dimensional parameter space. One expression of this *fundamental plane* describes the connection between the effective radius $R_e$ and the

---

[22] The members of this second collaboration were D. Burstein, R. Davies, A. Dressler, S. Faber, D. Lynden-Bell, R. Terlevich, and G. Wegner. As a group they became widely known among astronomers as the "7 Samurai" (hereafter 7S).

average stellar velocity dispersion and surface brightness interior to that radius:

$$R_e \propto \sigma_e^\alpha I_e^{-\beta} ; \tag{160}$$

the dual power law maps to a plane when logarithmic quantities are considered. The two groups both found that $\alpha \simeq 1.4$, $\beta \simeq 0.9$ (Faber et al. 1987). When using the fundamental plane as a DI, the 7S group preferred to incorporate the surface brightness and radius terms into a single photometric diameter they called $D_n$, the diameter interior to which the average surface brightness attained a fiducial value. In a subsequent study (Lynden-Bell et al. 1988a), the group provided a final calibration of the new DI:

$$D_n \propto \sigma_e^\gamma , \tag{161}$$

where the exponent $\gamma \simeq 1.20 \pm 0.1$. In this form the modified FJ relation is known as the $D_n$-$\sigma$ relation. Dressler (1987a) has shown that the $D_n$-$\sigma$ relation also holds for the bulges of early-type spirals, although this has not been exploited beyond his initial paper.

Lynden-Bell et al. (1988a) found that the scatter in the $D_n$-$\sigma$ relation was $\simeq 0.10$ dex in $\log D_n$, corresponding to about 23% accuracy as a distance predictor. Thus, $D_n$-$\sigma$ was found to be about twice as accurate as FJ, and compared favorably with TF. Indeed, the 7S group used the $D_n$-$\sigma$ distances to make one of the earliest inferences of very large-scale deviation from Hubble flow (Dressler et al. 1987a), as we discuss in greater detail in § 7.1. It is interesting to note, however, that while a fundamental plane led to a tighter DI, it also implied less insight into elliptical galaxy formation. For, while a one-dimensional relationship (such as TF) requires special conditions to be imposed at galaxy formation, a fundamental plane is simply a reflection of gravitational equilibrium, with the additional (albeit ill-understood; cf. Renzini & Ciotti 1993) assumption that $(M/L)$ be a function only of the other parameters in the relation (Faber et al. 1987; Gunn 1988). Thus the fundamental plane relations tell us little more than that elliptical galaxies are gravitationally bound structures, which is hardly surprising.

## 6.2 *Universality of the Distance Indicator Relations*

The measurement of redshift-independent distances by means of DIs such as TF and $D_n$-$\sigma$ assumes that the galaxian properties they embody are *universal, i.e.*, independent of spatial location or local environment. If, say, the zeropoint of the TF relation varied systematically from one place to another, then the distances inferred from it would carry a position-dependent error, giving rise to spurious peculiar velocity measurements. Recognizing this, some astronomers have questioned the reality of detections of large-scale departures from Hubble flow (§ 7.1)—or as one theorist asked in the provocative title to a paper on the subject, "Is Cosmic Drift a Cosmic Myth?" (Silk 1989).

Silk (1989) suggested, on the basis of plausible (but speculative) scenarios of galaxy formation, that both the TF and $D_n$-$\sigma$ relations would show a systematic cluster vs. field offset. Because the DIs are typically calibrated using cluster observations, the measured peculiar velocities of field galaxies (which dominate most analyses of large-scale motions) might then be invalid. de Carvalho & Djorgovski (1992) carried out a comparison of the properties of field and cluster ellipticals. They argued that as measured by the relationship between several distance-independent indices (surface brightness, color, metallicity, and velocity dispersion) the field and cluster galaxies differed systematically. The sense of these differences was that field galaxies were bluer, had higher surface brightness, and were more metal-poor than their field counterparts of similar velocity dispersion. They proposed that these differences would follow logically from star-formation histories if the field ellipticals were typically younger than cluster ellipticals. A related study was carried out by Guzmàn and coworkers (Guzmàn 1994; Guzmàn & Lucey 1993; Guzmàn, Lucey, & Bower 1993; cf., Gregg 1992), who also concluded that metallicity influences the fundamental plane relations (§ 6.1.2). Guzmàn (1994) explicitly argued that this "age effect" induces a significant difference between $D_n$-$\sigma$ distances for cluster and field galaxies, and that with appropriate corrections, the large peculiar velocities obtained from $D_n$-$\sigma$ data (§ 7.1) vanish.

There are, however, good reasons to view the results of the previous paragraph with considerable caution. The study by de Carvalho & Djorgovski (1992) mixed two data sets, those of Djorgovski & Davis (1987) and of the 7S group (Faber et al. 1989), which might well differ systematically from one another. Indeed, it is primarily the Djorgovski & Davis (1987) data set which manifests a clear cluster/field offset as measured by the distance-independent indices; no significant effect is found in the 7S data set. The work of Guzmàn and coworkers has been questioned on a variety of grounds (D. Burstein, private communication), in particular the use of an erroneous metallicity-velocity dispersion correlation and outdated stellar population models. Finally, it should be noted that the 7S group tested for metallicity and color dependences of the $D_n$-$\sigma$ relations and found them to be statistically insignificant (Lynden-Bell et al. 1988b). The 7S group recognized the existence of a fraction ($\sim$ 10–20%) of the total elliptical population in which stellar population anomalies might be present, but showed that these objects did not strongly affect their conclusions about the systematics of the flows (Burstein et al. 1990).

Another means of assessing whether measured peculiar velocities could be due to DI non-universality is to compare the velocity field in a given region of space obtained from different DIs. Kolatt & Dekel (1994) compared the peculiar velocity fields reconstructed by the POTENT technique (§ 7.5) from spiral (TF) and elliptical ($D_n$-$\sigma$) data separately, and found no statistically meaningful differences between the two. It is unlikely that zeropoint gradients would affect TF and $D_n$-$\sigma$ in precisely the same way. Another consistency test derives from a comparison of $D_n$-$\sigma$ distances with distances obtained for the same galaxies using the Surface Brightness Fluctuation (SBF) technique (Dressler 1994; see § 6.3.1). This comparison has validated $D_n$-$\sigma$ peculiar velocity estimates for $\sim$ 15 ellipticals in the Southern sky. Since the SBF technique is thought to be highly accurate, and involves measurements entirely unrelated to $D_n$-$\sigma$, these results represent another encouraging (albeit preliminary) indication of $D_n$-$\sigma$ universality.

In summary, then, there is no compelling evidence at present for substantial spatial or environmental variations in the TF or $D_n$-$\sigma$ relations. While there exist some indications that metallicity and/or age effects enter into the fundamental plane relations for a fraction of elliptical galaxies, these effects apparently do not influence the distance estimates obtained from $D_n$-$\sigma$ in a systematic way. As more data are acquired in the coming years, using a multiplicity of DIs, intercomparisons will permit more sensitive tests of DI universality.

### 6.3 Beyond TF and $D_n$-$\sigma$: A Look to the Future?

While the TF and $D_n$-$\sigma$ methods have been the workhorses in peculiar velocity surveys, there a number of other promising distance indicators likely to play a role in coming years.

#### 6.3.1 Surface Brightness Fluctuations

Over the last several years Tonry and collaborators (Tonry & Schneider 1988; Tonry, Ahjar, & Luppino 1989, 1990; Tonry & Schechter 1990; Tonry 1991) have begun to exploit an old idea first discussed in detail by Baum (1955): a sufficiently high resolution image of an elliptical galaxy will reveal the discrete nature of the ultimate sources of luminosity—billions of individual stars. A finite number of stars are included in a single pixel of an image; this number is subject to Poisson fluctuations, and therefore so is the pixel intensity. The farther away the galaxy is, the greater is the number of stars per pixel, and the smaller the Poisson fluctuations; thus the fluctuations are a measure of the distance of the galaxy. Tonry measures the *surface brightness fluctuation* (SBF) which characterizes an image. The mean SBF can in turn be related to the luminosity of the giant branch stars which dominate the light, which depends weakly on metallicity.

The technique is applied to ellipticals rather than spirals because the latter are multicomponent, often dusty systems in which the purely Poisson fluctuations cannot easily be isolated.

The uncertainty in SBF distance estimates varies in inverse proportion with the resolution of the images obtained. For galaxies within $\sim 3000$ km s$^{-1}$, SBF distances are estimated to have $\sim 5\%$ accuracy under the best observing conditions. Ground-based resolution is limited by microturbulence in the atmosphere ("seeing"). In collaboration with Tonry, Dressler (1994) has undertaken a program to obtain SBF distances for ellipticals studied with the $D_n$-$\sigma$ method using CCD images acquired at the du Pont Telescope in Chile, where the seeing is often $\lesssim 1''$. This program promises to yield accurate distances for nearby ellipticals, and may significantly improve our knowledge of the local peculiar velocity field. Extension of the SBF technique to distances of $\sim 5000$ km s$^{-1}$ and beyond will require either a dedicated program of space-based imaging, or the use of adaptive optics technology for enhancing ground-based resolution.

*6.3.2 BCG L-$\alpha$ relation*

The DIs discussed thus far apply to ordinary spiral or elliptical galaxies. There is also a class of extraordinary galaxies, the most luminous objects in the centers of rich clusters (called "brightest cluster galaxies" or BCGs), which can be used as DIs. BCGs were originally of interest because they are easily identified out to fairly large ($z \simeq 0.2$) redshifts. As a result, they were thought to be ideal tools for studying problems such as the linearity of the Hubble expansion (Sandage 1972; Sandage & Hardy 1973; Gunn & Oke 1975)[23]. In addition, because they are highly luminous objects and are unique in their environment, their study is not subject to the statistical biases associated with sample incompleteness we discuss in § 6.4.

The studies just mentioned considered BCGs as standard candles, *i.e.*, they assumed their luminosities were independent of any other galaxian property. Later, Hoessel (1980) showed that BCG luminosity correlates with the shape of the luminosity profile. Specifically, Hoessel defined a metric luminosity $L_m$ as the total light interior to the metric radius $r_m = 10\,h^{-1}$ kpc, and a luminosity shape parameter

$$\alpha = \left. \frac{d \log L}{d \log r} \right|_{r_m}, \tag{162}$$

and found a linear correlation between $L_m$ and $\alpha$. Recently, the $L$-$\alpha$ relation has been exploited by Lauer & Postman (1994; LP), and Postman & Lauer (1995), who studied the BCGs in the 119 Abell (1958) and Abell, Corwin, & Olowin (1989) clusters with redshifts less than 15,000 km s$^{-1}$. LP found a quadratic $L$-$\alpha$ relation with an equivalent distance error of $\sim 16\%$, comparable to the TF scatter. The BCG $L$-$\alpha$ relation is perhaps the most promising technique for extending peculiar velocity studies to much larger scales ($z \gtrsim 0.1$) than is currently possible, although it requires further study to test its universality.

*6.3.3 Possible Methods of Future Peculiar Velocity Work*

Two other DIs have recently been the focus of intense observational work, but thus far in the context of the distance scale (*i.e.*, Hubble constant determination) rather than the velocity field. The first of these is fitting the planetary nebula luminosity function (PNLF; Jacoby *et al.* 1989; Ciardullo *et al.* 1989ab; Jacoby *et al.* 1990; Ciardullo, Jacoby, & Harris 1991). "Planetary nebulae" are the luminescent ejectae of dying stars soon to become white dwarfs. Planetary nebulae may be detected and photometered in nearby ($cz \lesssim 1500$ km s$^{-1}$) elliptical galaxies using narrow-band imaging in the light of twice-ionized oxygen at $\lambda = 5007$ Å. The PNLF in this emission line is observed to have a universal form, with a sharp cutoff at the highest luminosities (Jacoby 1989; Ciardullo *et al.* 1989a). Fitting the observed distribution of planetary nebula brightnesses to the universal PNLF yields the distance to a galaxy with ∼ 5% accuracy. The PNLF and SBF methods yield consistent distances for Virgo Cluster galaxies (Ciardullo, Jacoby, & Tonry 1993). At distances much greater than that of Virgo it is quite difficult to reliably identify even luminous planetary nebulae using present techniques (Jacoby et al. 1992). As a result, the usefulness of the PNLF method for peculiar velocity work is uncertain.

---

[23] However, it was discovered that evolutionary effects are equally as important as those due to cosmology. See Sandage (1988) for a review.

A DI which has been around for some time, but which is currently undergoing a kind of renaissance, is the mean absolute luminosity of Type Ia Supernovae (Sne Ia). Of all the DIs discussed here, Sne Ia have the best-understood physical basis. It is generally agreed that Sne Ia result from thermonuclear disruption of a carbon-oxygen white dwarf that has accreted enough mass from a companion star to approach the Chandrasekhar mass of ∼ $1.4 M_\odot$ (e.g., Wheeler & Harkness 1990 and references therein). The radioactive decay of $^{56}$Ni and $^{56}$Co following the explosion power the subsequent observable light curve. Theoretical models suggest that the mass of the nickel ejectae is in the range 0.6–$0.8 M_\odot$ (Arnett et al. 1985); when combined with the observed rise time to maximum light of ∼ 17 days, an absolute luminosity of $M_B \simeq -19.6$ mag is predicted.

There has been some controversy over their intrinsic scatter as standard candles. The discovery that the peak luminosity of SN 1a correlates with their rate of decline (Pskovskii 1977, 1984; Phillips 1993; Hamuy et al. 1995, but see Tammann & Sandage 1994) may tighten up the scatter considerably, to as small as 0.1–0.2 mag (Hamuy et al. 1995; Reiss, Press, & Kirshner 1995).

Sne Ia have been used almost exclusively for Hubble constant determination (e.g., Saha et al. 1994a,b). Indeed, their idiosyncratic nature as DIs—one must wait for a supernova to appear in a galaxy in order to estimate its distance, and their occurrence is both unpredictable and rare—limits their utility for peculiar velocity work. However, they can be used as DIs to extremely large distances with relatively little loss of accuracy (e.g., Perlmutter et al. 1993), and may thus prove useful in extensions of peculiar velocity analyses to large distances.

Type II supernovae, which mark the end-point of the life of a massive star, can also be used as a distance indicator (Schmidt, Kirshner, & Eastman 1992; 1994). Spectroscopic observations yield the velocity of the ejecta and the temperature of the outer layers, from which the absolute brightness can be inferred with the use of model atmospheres. Comparison with the apparent brightness yields a distance. This method requires extensive model calculations and detailed observations, and has thus far been applied only to a handful of objects.

*6.4 Statistical Bias and Methods of Peculiar Velocity Analysis*

With the possible exception of the SBF technique, the DIs used in peculiar velocity surveys are not very accurate. The TF and $D_n$-$\sigma$ relations, for example, predict galaxy distances with only ∼ 20% accuracy. In the volume within which we hope to study the peculiar velocity field in reasonable detail, typical galaxies may lie at distances of ∼ 3000 km s$^{-1}$. The rms peculiar velocity error for such a galaxy is thus on the order of 600 km s$^{-1}$, which also happens to be the amplitude of typical peculiar velocities. At interesting distances, then, we cannot measure with any precision the peculiar velocity of a single galaxy. Meaningful analyses must use statistical techniques applied to large samples.

At first glance, this might not be considered a major problem. There are, after all, thousands of galaxies in the local volume for which we have obtained, or soon hope to obtain, distance indicator data. We might expect that by virtue of $\sqrt{N}$ statistics alone the signal-to-noise ratio of any statistical analysis could be made rather high. However, this expectation is not realized in practice. The analysis of DI data is instead subject to statistical bias effects; these effects result in random errors dropping more slowly than $1/\sqrt{N}$, and also in the possibility of large systematic errors if the biases are not properly corrected for. While there are a number of bias effects, all originate in a coupling of the DI scatter with external influences on the makeup and spatial distribution of the redshift-distance sample. The goal of this section is to explore the nature and consequences of this coupling in a variety of circumstances.

A peculiar velocity analysis has, logically if not sequentially, two steps. The first is *calibration* of the DI being used. If we assume for definiteness that it is the TF relation, calibration means determination of the zeropoint $A$, slope $b$, and scatter $\sigma$ which quantify the $M$–$\eta$ relationship (§ 6.1.1). The second step in the analysis is the inference[24] of distances and thus peculiar velocities for sample objects. Each step is potentially subject to biases. Those which enter into the calibration phase, if not corrected for, can lead to erroneous determinations of any or all of the parameters $A$, $b$, or $\sigma$. This incorrect TF relation will produce incorrect distances and peculiar velocities. Even with a properly calibrated TF relation, however, uncorrected biases in the distance-inference phase have the same result.

The separation of bias effects into "calibration" and "inferred-distance" problems (Willick 1994) is useful in emphasizing the different ways bias can affect peculiar velocity analyses. However, it does not distinguish the root causes of bias. These lie not in the goal but in the underlying assumptions of the analytic approach, as we will clarify in § 6.4.1. With one set of assumptions, an analysis will be subject to *selection* bias, which is intimately related (as the name implies) to observational selection criteria such as a flux limit. With the second set of assumptions, *Malmquist* bias will result. As defined here (but see next paragraph), Malmquist bias is not fundamentally related to a flux limit or any other selection criteria. It occurs because the true distance of a galaxy cannot be estimated from DI information alone, but requires knowledge of the actual line of sight density distribution as well. Selection bias can affect either DI calibration or peculiar velocity measurement—or both, as they often are carried out simultaneously (§ 6.4.3). Malmquist bias typically enters only when a calibrated DI is used to infer peculiar velocities; approaches to DI calibration which are subject to Malmquist bias are rarely used in practice and will not concern us here.

We organize our discussion mainly around the selection/Malmquist dichotomy, but we note that this distinction has rarely been maintained in the existing literature, where much confusion reigns. One source of confusion is the ambiguity surrounding the term "Malmquist bias," which historically has meant several different things. The term originated with the work of Malmquist (1924), who showed that flux limited samples were subject to a luminosity bias—one which in fact resembles our selection bias more closely than our Malmquist bias. Thus, early discussions of DI bias (Sandage, Tammann, & Yahil 1979; Teerikorpi 1984,1987; Aaronson *et al.* 1986) typically refer to bias effects associated with flux limits as "Malmquist biases." However, the distinct biases related to the line of sight galaxy density distribution were also referred to as "Malmquist" effects by Lynden-Bell *et al.* (1988a), who were the first to discuss them quantitatively in the extragalactic context. The latter usage has gained widespread acceptance, and it is the one we adopt here. Even so the impression has persisted in some quarters that Malmquist bias thus defined relates to a flux limit. One goal of the discussion to follow is to dispel this and related misperceptions brought about by conflicting uses of this catch-all phrase.

*6.4.1  Selection vs. Malmquist Bias – Method I vs. Method II*

Because of DI errors, data from a large number of galaxies are needed if the estimated peculiar velocity at a given point **r** is to be of value. Thus, in devising a method of analysis we must ask the question, "for what particular set of objects should data be averaged[25] to estimate peculiar velocity at position **r**?". There are two methodologically orthogonal answers to this question:

(i) the set of objects which *redshift space information* suggests are "near" **r**;

---

[24] We use the term "inference," rather than the stronger "prediction" (as is used, for example, by Burstein [1990]), to underscore the fact that redshift-independent distances are merely statistical estimators, obtained from more basic measurements, of true distances—and not very accurate ones at that.

[25] Here we use "averaged" in its widest sense, including smoothing or modeling techniques.

(ii) the set of objects which *the DI information* suggests are "near" **r**.

A method based on the first answer will be subject to selection bias; one based on the second answer will experience Malmquist bias. This holds true whether our goals are DI calibration, peculiar velocity measurement, or both.

It will prove useful to have terms describing methods of peculiar velocity analysis, in addition to terms describing the types of bias to which they are subject. When redshift-space information is used as the *a priori* indicator of true proximity, we will describe the analysis as belonging to "Method II." When DI information is used as the *a priori* indicator of true proximity, we will describe the analysis as belonging to "Method I." This terminology was originally introduced by Faber & Burstein (1988), although their usage was somewhat less general than ours. In § 6.4.2 and § 6.4.3, we will show, using prototypical examples, why Method II is subject to selection bias and Method I to Malmquist bias, and clarify how these biases affect the analysis. Later, in § 6.5 we will show how to quantify and correct for these biases. We will assume throughout that the DI in question is the TF relation, but the concepts apply equally to DIs such as $D_n$-$\sigma$.

*6.4.2 Bias in a Cluster Sample*

Suppose that a flux-limited Tully-Fisher sample consists of spiral galaxies thought to belong to an isolated rich cluster. A reasonable assumption to adopt in analyzing such a sample is that its members all lie at the same distance. This is an example of a Method II analysis: redshift-space information—from which we infer cluster membership—is treated as the *a priori* indicator of distance; the TF information has not entered into the assumption in any way. Let us further suppose that our goal is to calibrate the TF relation. Using the common distance assumption we may write the expected apparent magnitude of a cluster galaxy with velocity width parameter $\eta$ as $\langle m(\eta) \rangle = A - b\eta + \mu$ (see Eq. 152), where $\mu = 5 \log r_c$ is the cluster distance modulus; for now suppose that the cluster distance $r_c$, in km s$^{-1}$, is independently known. We also expect that the observed apparent magnitudes will scatter about $\langle m(\eta) \rangle$ with rms dispersion $\sigma$. These considerations seem to suggest that linear regression of $m$ on $\eta$ for the cluster sample will correctly give the TF slope and zeropoint (after subtracting the distance modulus), and that the scatter about that fit will yield $\sigma$.

However, because we have made *a priori* distance assignments based on redshift-space information, the above exercise will be subject to selection bias. How, specifically, does the bias enter in? Let $m_\ell$ denote the magnitude equivalent of the flux limit. Fig. 13 illustrates a Monte-Carlo simulation of a Tully-Fisher sample at a common distance, with a scatter of 0.35 mag, and a magnitude selection at $m_\ell = 16$. The dashed line is the input TF relation. Consider a subset of sample galaxies with similar and relatively large velocity width parameters $\simeq \eta_1 = 0.2$ such that they are typically much brighter than the flux limit, i.e., $\langle m(\eta_1) \rangle \ll m_\ell$. Their observed apparent magnitudes $m$ will be distributed uniformly about $\langle m(\eta_1) \rangle$ with rms dispersion $\sigma$. Now consider a second subset with relatively small width parameters $\simeq \eta_2 = -0.2$ such that $\langle m(\eta_2) \rangle \approx m_\ell$. They will exhibit a range of apparent magnitudes $m \leq \langle m(\eta_2) \rangle$. However, *none will have* $m > \langle m(\eta_2) \rangle$. Such galaxies exist in the cluster, but are "missing" from the sample because they are fainter than the magnitude limit. The mean apparent magnitude of the second subset will therefore be *brighter* than $\langle m(\eta_2) \rangle$, as the solid line in the figure shows. This is one definition of bias: the mean apparent magnitude does not conform to the TF prediction. A linear fit to the points which does not take into account the magnitude limit will be biased; as illustrated by the stars in Fig. 13: the slope of the fitted line is flatter (by 2 in this example), and the fitted zeropoint is brighter (by 0.3 mag in this example) than the true TF relation (the dashed line).

Finally, the scatter about the fit will not equal $\sigma$: because the faint part of the magnitude distribution at $\eta_2$ is cut off, the points cluster more tightly around the mean value. The flat slope, bright zeropoint, and reduced scatter of the fit are all characteristic of how selection bias affects DI calibration.

The calibration procedure itself was not, however, responsible for bias in the above example. Suppose that the TF parameters were known at the outset, and that instead it was the cluster distance modulus $\mu$ whose value was sought. We could still make the assumption that the cluster galaxies lie at the same distance. We would then seek to determine $\mu$ by fitting a line of known slope $-b$ to the TF data. The intercept of that line would (apparently) be $A+\mu$. Yet even though $A$ and $b$ are known, the fitted line would still be displaced from the correct (TF-predicted) one. Near the magnitude limit, the observed apparent magnitudes are still brighter in the mean than $\langle m(\eta) \rangle$. The derived value of $\mu$ will thus be biased small; when combined with the cluster redshift, this will result in a too-positive radial peculiar velocity. These effects are, independently of calibration issues, also characteristic of selection bias.

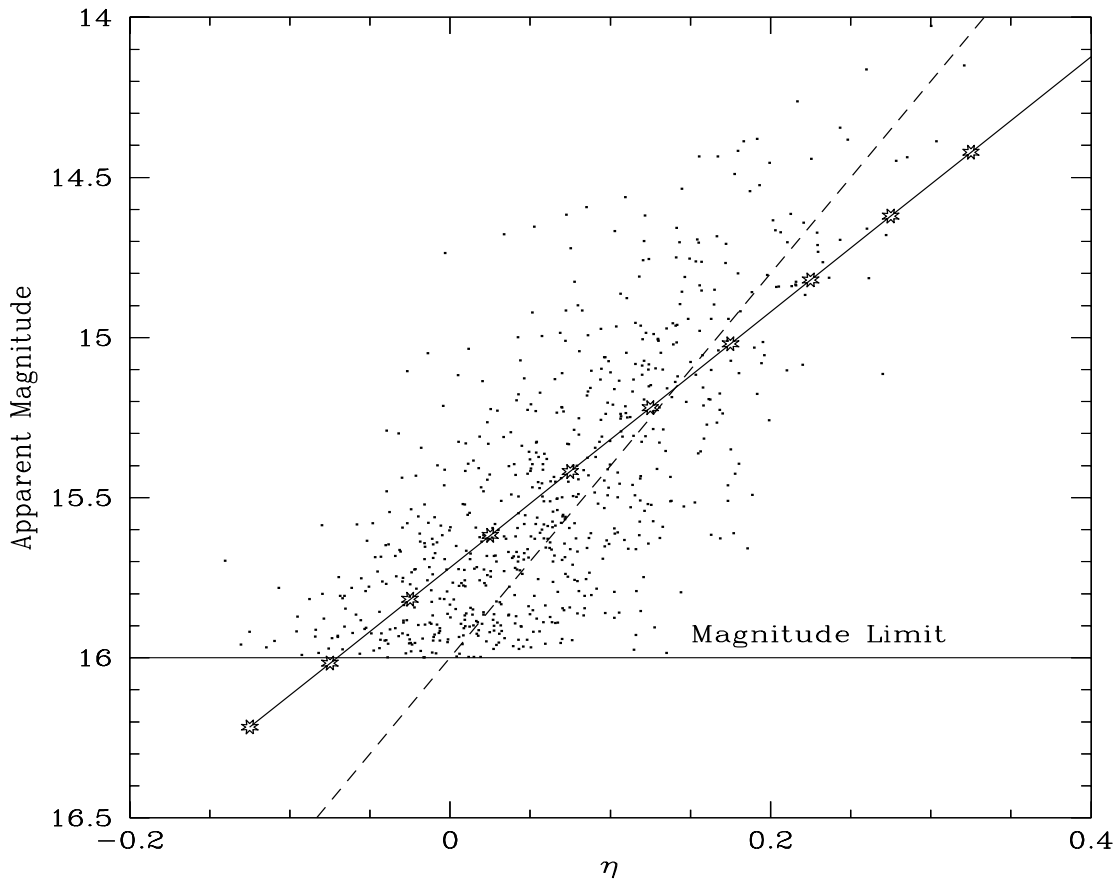

Fig. 13. A Monte-Carlo simulation illustrating the effects of a magnitude limit on the derived TF relation. Points were assumed to be at a common distance, and assigned a TF relation (indicated by the dashed line) with a scatter of 0.35 mag. The sample was limited at apparent magnitude 16. The stars indicate the best-fit linear TF relation which does not take the magnitude limit into account; it is shallower, with a brighter zeropoint.

Central to the Method II analyses just described was the assumption that the cluster galaxies lie at the same distance. This assumption, though plausible, is not self-evident; the galaxies might conceivably be spread out over a range of distances. Following this line of reasoning we might, if our TF relation were independently calibrated, neglect redshift space data initially and assign distances directly from the observables $m$ and $\eta$. These *inferred distances* $d$ are derived by forming a distance modulus $\mu(m,\eta) = m - M(\eta)$ and then taking $d = 10^{0.2\mu(m,\eta)}$. Because of DI errors, the true distance $r$ can differ appreciably from $d$. Still, a Method I analysis of the cluster would assume that only galaxies with similar values of $d$—not all galaxies in the sample—may be treated as equidistant. The redshifts of galaxies with similar values of $d$ would then be averaged together to obtain a mean value $\langle cz(d) \rangle$, and a corresponding inferred peculiar velocity $u(d) = \langle cz(d) \rangle - d$.

Because we have now made *a priori* distance assignments based on the DI information, this analysis will be subject to Malmquist bias. The specific manifestations of this bias are that $d$, $\langle cz(d) \rangle$, and and $u(d)$ are not equal to the average true distance, redshift, and radial peculiar velocity, respectively, of the set of objects with inferred distance $d$. The bias arises because objects with inferred distance $d$ come, with varying probability, from a range of true distances $r$ as a result of DI errors. They are constrained to lie within some vicinity of $d$, but their true distances are influenced by another factor wholly unrelated to the DI information: where along this line of sight galaxies are intrinsically likely to be found. This is quantified by the number of galaxies per unit distance $r^2 n(r)$, where $n(r)$ is the galaxy number density. If galaxies were in fact widely distributed along the line of sight ($n(r)$ roughly constant), then because there is more volume at larger distances (the $r^2$ factor), $d$ would *underestimate* the true distance. If on the other hand the common-distance model had been accurate in the first place, then regardless of the value of $d$ the true distance is the cluster distance $r_c$. Inferred distances $d < r_c$ would then necessarily underestimate, and $d > r_c$ necessarily overestimate, true distance. None of these considerations relates in any way to the selection criteria which define the sample.

When galaxies scatter from $r$ to $d$ because of DI errors, they "bring with them" the correct redshifts of their true positions. That is, DI errors do not affect redshifts, but only the distances at which they are evaluated and converted to peculiar velocities. Thus, if $\bar{r}_d$ (say) is the mean true distance of galaxies with inferred distance $d$, the mean redshift of these galaxies will to first order be $\langle cz(d) \rangle = \bar{r}_d + u_t(\bar{r}_d)$, where the subscript $t$ signifies *true* as opposed to inferred radial peculiar velocity[26]. From this redshift one would infer a peculiar velocity $u(d) = u_t(\bar{r}_d) + (\bar{r}_d - d)$. Thus, the distance bias ($d \neq \bar{r}_d$) translates directly into a peculiar velocity bias. In the case that the galaxies truly constitute a cluster (*i.e.*, $\bar{r}_d = r_c$ for all $d$), we see that $u > u_t$ when $d < r_c$, and $u < u_t$ when $d > r_c$. It follows that the inferred radial peculiar velocity field $u(d)$ would exhibit a compressional inflow into the cluster center, *even if there were no real peculiar velocities*. This last effect—spurious peculiar velocities associated with strong line of sight density gradients, of which a cluster is the most extreme case—is the most serious consequence of Malmquist bias. As density gradients generally are not known with great precision, Malmquist bias can be difficult to correct for reliably (§ 6.5.2).

### 6.4.3  Bias in a Field Sample

We used the example of a cluster to distinguish selection and Malmquist bias with maximum clarity. However, peculiar velocity analyses more often involve field galaxy samples. It is therefore important to realize that the two kinds of bias occur, for analogous reasons, in field sample analyses as well. In the cluster example, the Method II approach seemed "natural" in comparison with Method I. With a field sample, the choice of method is not clear cut. We illustrate these issues in what follows.

A Method II analysis of field galaxies usually entails the adoption of a model of the peculiar velocity field, $\mathbf{v}(\mathbf{r};\mathbf{a})$; the quantity $\mathbf{a}$ is a vector of free parameters which specifies the model. An important example is a velocity field obtained from Eq. (150) using redshift survey data, with $\beta$ as the only free parameter. Using the model, one obtains a distance $r$ for a galaxy of redshift $cz$ as the solution of the equation

$$r(cz;\mathbf{a}) = cz - \hat{\mathbf{r}} \cdot \mathbf{v}(\mathbf{r};\mathbf{a}) . \tag{163}$$

Such an expression of course assumes a unique redshift-distance mapping, which may not obtain in the vicinity of large mass concentrations (see § 5.9), an issue we will return to later. From Eq. (163) one can

---

[26] This last statement neglects second-order changes in the peculiar velocity field; cf. Willick 1991 for details.

predict the galaxy's apparent magnitude, given the value of its velocity width parameter $\eta$:

$$m(\eta, cz; A, b, \mathbf{a}) = M(\eta) + 5 \log r(cz; \mathbf{a}) ; \tag{164}$$

this predicted apparent magnitude depends, as indicated, not only on the velocity field parameter vector $\mathbf{a}$, but also on the TF parameters $A$ and $b$, whose determination might also be part of the analysis. One can then form a $\chi^2$ statistic of observed minus predicted apparent magnitudes:

$$\chi^2_{\text{II}}(A, b, \mathbf{a}) = \sum_{i=1}^{N} \left[ \frac{m_i - m(\eta_i, cz_i; A, b, \mathbf{a})}{\sigma} \right]^2 \tag{165}$$

(the individual object scatters are usually assumed equal in the simplest implementations of Method II). Minimization of this statistic might then be expected to yield the "best" values of the velocity field parameters $\mathbf{a}$ and (if necessary) the TF parameters $A$ and $b$.

A bit of reflection shows, however, that such an exercise will be subject to the same bias that affected the first TF analysis of cluster galaxies discussed in § 6.4.2. In particular, there is again the issue of "missing" galaxies, although in a more abstract sense than earlier. There is in fact only one galaxy at redshift $cz_i$, direction $\hat{\mathbf{r}}_i$, and width parameter $\eta_i$. However, the validity of the $\chi^2$-minimization procedure requires that this object be representative of the ensemble of all possible such objects. Because of the flux limit, however, it cannot be: if the predicted apparent magnitude $m(\eta_i, cz_i; A, b, \mathbf{a})$ is close to or fainter than the flux limit, the fainter members of this hypothetical ensemble cannot belong to the sample. As a result, the observed apparent magnitude $m_i$ is likely to be brighter than the predicted value; minimization of $\chi^2_{\text{II}}(A, b, \mathbf{a})$ consequently does not yield the correct values of the parameters on which it depends. A Method II analysis of a field sample is thus affected by selection bias, just as was the earlier one based on the assumption of equidistant cluster members. We note that although we have assumed the existence of a model peculiar velocity field, this is not fundamental to a Method II field sample analysis. For example, if pure Hubble flow were assumed ($\mathbf{v}(\mathbf{r}; \mathbf{a})$ vanishes identically), an exercise identical to that outlined above could be used just to calibrate the TF relation; its results also would be affected by selection bias.

A Method I analysis of the same sample would proceed differently. To make the comparison as direct as possible, we again suppose that the purpose is to fit a model velocity field $\mathbf{v}(\mathbf{r}; \mathbf{a})$. Now, however, there is no need to solve Eq. (163) to obtain a distance. Instead we use the inferred distance $d$ as the best indicator of true distance, and thus as the position at which we evaluate the velocity model. Inferred distances are given as described in the Method I cluster analysis (§ 6.4.2), and the TF relation is again assumed to have been independently calibrated [27]. Using the model velocity field we write a predicted redshift for a galaxy with inferred distance $d$ as

$$cz(d; \mathbf{a}) = d + \hat{\mathbf{r}} \cdot \mathbf{v}(\mathbf{d}; \mathbf{a}) \tag{166}$$

where $\mathbf{d} = d\,\hat{\mathbf{r}}$. This in turn motivates a $\chi^2$ statistic of the form

$$\chi^2_{\text{I}}(\mathbf{a}) = \sum_{i=1}^{N} \left[ \frac{cz_i - cz(d_i; \mathbf{a})}{\Delta_i} \right]^2 , \tag{167}$$

where $\Delta_i$ is a suitable measure of redshift and distance errors. Again one might suppose that minimization of this statistic leads to the best values of the velocity model parameters $\mathbf{a}$.

---

[27] It is not fundamental to our definition of Method I that the TF relation be calibrated *a priori*. However, unnecessary complications arise if the TF parameters are considered "free."

However, this naïve approach will lead to biased results. The validity of the $\chi^2$ minimization rests on the assumption that $cz(d_i; \mathbf{a})$ is an unbiased estimator of $cz_i$ when the velocity model is correct. As discussed in § 6.4.2, the object with inferred distance $d_i$ can come from a range of true distances with varying probability. Let the expected true distance of this object be $\bar{r}_{d_i}$. Its observed redshift then has a corresponding expected value $\langle cz_i \rangle = \bar{r}_{d_i} + u_t(\bar{r}_{d_i})$, where as in § 6.4.2 $u_t$ denotes the true radial peculiar velocity field and we neglect second order effects. From this it follows (compare with Eq. 166) that, even if the velocity model is correct, $\langle cz_i \rangle \neq cz(d_i, \mathbf{a})$ unless $\bar{r}_{d_i} = d_i$. But we know from our discussion of the Method I cluster analysis (§ 6.4.2) that this last condition does not hold in general; the relationship between $\bar{r}_{d_i}$ and $d_i$ depends on the details of the underlying density distribution $n(r)$, which is the essence of Malmquist bias. Thus $cz(d_i, \mathbf{a})$ is a biased measure of $cz_i$, and as a result the parameters which minimize $\chi^2_\mathrm{I}$ yield an incorrect velocity field. Again, while we have framed the problem in terms of a velocity model, Method I includes any analysis in which galaxies with similar inferred distances are assumed to be physically proximate and, therefore, to share a peculiar velocity vector. In particular, nonparametric methods, such as the POTENT reconstruction (Bertschinger & Dekel 1989; see § 7.5), which smooth in inferred-distance space, belong to Method I and suffer from Malmquist bias.

### 6.4.4 "Inverse" Distance Indicator Relations

There is another basic division between methods of peculiar velocity analysis, one which has sometimes been subsumed under the Method I/II dichotomy but in fact is entirely distinct. Until now we have considered only the "forward" form of DIs, in which the distance-independent quantity ($\eta$ in the case of TF) is thought of as predicting the distance-dependent quantity ($M$ in the case of TF). But we can just as easily "turn the relation on its side" and consider the TF relation as a predictor of velocity width given absolute magnitude[28]. We quantify this prediction as $\eta^0(M)$ (strictly speaking, the expectation value of $\eta$ given $M$), and call it the "inverse" TF relation. Like its forward counterpart $\eta^0(M)$ is well-described by a linear relation

$$\eta^0(M) = -e(M - D) ; \tag{168}$$

we denote its rms dispersion $\sigma_\eta$. Because of this finite scatter the forward and inverse forms are not mathematically inverse to one another, i.e., $e \neq b^{-1}$ and $D \neq A$. The forward and inverse forms of DIs are alternative representations of a single physical phenomenon; nonetheless, their statistical properties differ markedly.

We illustrate with the following example both this difference in bias properties, and why the forward/inverse and Method I/II distinctions have been conflated. Suppose that we wish to carry out a Method II analysis of the field galaxy sample of § 6.4.3 using the inverse TF relation. In that case we use Eq. (163) to predict the velocity width parameter $\eta$ of an object with redshift $cz$ and apparent magnitude $m$ as

$$\eta(m, cz; D, e, \mathbf{a}) = \eta^0 \left( m - 5 \log r(cz; \mathbf{a}) \right) ; \tag{169}$$

this equation motivates a new $\chi^2$ statistic

$$\chi^2_\mathrm{II}(D, e, \mathbf{a}) = \sum_{i=1}^{N} \left[ \frac{\eta_i - \eta(m_i, cz_i, ; D, e, \mathbf{a})}{\sigma_\eta} \right]^2 \tag{170}$$

---

[28] For pure standard candles, there is no distinction between forward and inverse methods; the material discussed in this section is only relevant for distance indicators which depend on a distance-independent observable such as line width.

apparently a perfect analogue to its forward counterpart, Eq. (165). The analogy breaks down, however, when one considers the bias properties of this inverse statistic. In the forward case bias arose because the sample galaxy with redshift $cz$ and velocity width parameter $\eta$ was not representative of the ensemble of all possible such objects; the presence of a flux limit meant that the faint end was excluded. In the inverse case the relevant question is whether the sample galaxy with redshift $cz$ and apparent magnitude $m$ is representative of all such objects. If we assume that sample selection is independent of the value of $\eta$ for a given $m$, then the answer to this question is evidently "yes," since $m$ must be brighter than the flux limit. There now are no "missing" galaxies, regardless of how close $m$ is to the flux limit, and $\eta(m, cz; D, e, \mathbf{a})$ is an unbiased estimator of $\eta_i$; minimization of $\chi^2_{\text{II}}(D, e, \mathbf{a})$ may thus be expected to yield essentially unbiased values of the free parameters of the model.

This felicitous property of inverse DIs was first noticed by Schechter (1980) in a study of the Virgocentric flow field, and was subsequently exploited to great advantage by Aaronson *et al.* (1982b) and by Faber & Burstein (1988). These studies enshrined the notion that the inverse relation was free of selection bias, and also cemented the association between Method II and the inverse form of DIs. They also, perhaps because of the catch-all nature of the term "Malmquist," led to the assertion (*e.g.*, Tully 1988) that inverse DIs are free of Malmquist bias. None of these conventional notions are accurate. The above example shows only that an inverse Method II analysis is unaffected by selection bias when sample selection is independent of velocity width. The paradigmatic form of sample selection, a photometric flux limit, which we have used here for illustrative purposes, is $\eta$-independent. However, as we discuss below (§ 6.5.4), in real TF samples selection criteria are more complex and may depend, usually weakly, on velocity width. In such cases, even an inverse Method II analysis will be subject to mild selection bias. Moreover, there is no inherent connection between inverse DIs and Method II; the essence of Method II is its *a priori* assignment of distances from redshift, after which either form of the DI can be employed.

Finally, because of the conventional association between Method II and inverse DIs, it has not generally been recognized that inverse DIs can be used in Method I analyses. We defer the details to § 6.5.5, and note here only that an *inverse inferred distance* is defined (in the case of TF) by the implicit relation $\eta^0(m - 5\log d) = \eta$. A consequence of this definition, as will be come clear, is the fact that inverse and forward inferred distances differ, in general, for a given galaxy. Nonetheless, the fundamental source of Malmquist bias in forward Method I analyses—that the likelihood an object will scatter from true distance $r$ to inferred distance $d$ depends on the density $n(r)$—is equally present in an inverse Method I analysis. While the use of forward vs. inverse affects its details, Malmquist bias inevitably accompanies a Method I analysis.

*6.4.5 The "Method Matrix" of Peculiar Velocity Analysis*

We provide here a succinct summary of the multiple and to some degree overlapping methods of peculiar velocity analysis in the form of a "method matrix." The notion of a matrix derives from the two-dimensional character of the methodology discussed above: peculiar velocity analyses can belong to either Method I or II, and can employ either the forward or inverse form of the relevant DI. The statistical biases present are specific to each case. Table 2 summarizes these ideas; two terms in that table not yet defined, the "Gould effect" and the "Landy-Szalay correction," are discussed in § 6.5.

We have defined Methods I and II and discussed the biases inherent in each; we have not yet indicated why we might choose one over the other. Method I has the advantage of being unburdened by preconceived notions about the relation between redshift and distance. Method II, which requires that we invoke at least some form of redshift-distance relation, is necessarily subject to our theoretical prejudice. Ideally, we would like to know where the galaxies really are, and only later compare these real-space positions with the observed redshifts. The most compelling motivation for this is the possibility of triple-valued zones in dense regions (Fig. 11), where infall patterns along the line-of-sight can result in widely separated galaxies having the same observed redshift. Since Method I considers redshift only *after* it has placed the galaxies at their DI-inferred distances, this presents no special problem. By contrast, Method II is inherently incapable of dealing with multivaluedness in the redshift-distance mapping. It assumes that redshift uniquely specifies distance and thus presents an overly smooth picture of a velocity field which in reality is rapidly changing. (In § 8.1.3 we will present a recent variant of Method II which circumvents this particular flaw.)

Table 2. "Method Matrix" of Peculiar Velocity Analysis

| DI type/Method Type | Method I<br>*DI-inferred distance best indicator of true distance* | Method II<br>*Redshift-space data best indicator of true distance* |
|---|---|---|
| Forward<br>*dist-dep (e.g. mag) predicted by dist-indep (e.g. $\eta$) quantity* | Malmquist bias (selection-independent unless $\exists$ "Gould effect") | Strong selection bias (depends on observational selection criteria) |
| Inverse<br>*dist-indep predicted by dist-dep quantity* | Malmquist bias (selection-dependent, Landy-Szalay possible) | Weak or no selection bias (bias present if selection related to dist-indep quantity) |

On the other hand, Method II has the advantage that its statistical aspects are comparatively well-constrained. If we suppose that the velocity field model $\mathbf{v}(\mathbf{r}; \mathbf{a})$ is not unreasonable, then the dominant errors in the assignment of distances (Eq. 163) come from "noise" in the velocity field, whose rms value in the field appears to be at most $\sim 200$ km s$^{-1}$ (§ 5.2.1). The dominant errors in Method I distance assignments come from DI scatter, and are therefore several times larger at interesting distances. Moreover, as we will see more clearly in the following section, we can correct for selection bias in terms of *knowable* information, namely, observational selection criteria. With inverse DIs we can realistically expect selection bias to be quite small. Correction for Malmquist bias, however, depends on density gradients along the line-of-sight, which are *not* knowable *a priori*. We can estimate such gradients from redshift data, but this involves additional assumptions concerning peculiar velocities (needed to map redshift- to real-space density) and smoothing methods. Working with inverse DIs gains us nothing in terms of Malmquist bias. In summary, both Method I and Method II have advantages and disadvantages. As we describe in Chapter 8, both are being used in contemporary peculiar velocity analyses. It should be pointed out, however, that many of the analyses we describe in Chapter 8 were done without proper attention to the biases we describe here, and of the differences between the different entries in the Method Matrix of Table 2. Indeed, there remains much work to be done in exploring all these methods; for example, there is no analysis beyond that of Landy & Szalay (1992) which exploits Method I with an inverse DI.

## 6.5 *Quantifying Statistical Bias*

In the last section we presented a qualitative outline of the main statistical biases affecting distance indicator and peculiar velocity analysis. In order to correct for these effects, the biases must be quantified and correction formulae and methods developed. This task has been the focus of a growing number of workers over the last decade or so. Sandage and coworkers (*e.g.*, Sandage, Tammann, & Yahil 1979; Sandage 1988; Federspiel, Sandage, & Tammann 1994) have provided useful, though mainly qualitative, suggestions for treating selection bias. A quantitative approach to the subject was pioneered by Teerikorpi, Bottinelli, and collaborators (*e.g.*, Teerikorpi 1984,1987; Bottinelli *et al.* 1986), and extended by Willick (1991,1994).

Malmquist bias (in the DI context) was first treated in detail by Lynden-Bell *et al.* (1988a); this work has been followed up and extended by Willick (1991,1995), Landy & Szalay (1992), and Gould (1993). The bias properties of inverse DIs have been studied by Schechter (1980), Aaronson *et al.* (1982b), Faber & Burstein (1988), Tully (1988), Han (1991), Roth (1993), Willick (1994,1995), and Triay, Lachièze-Rey, & Rauzy (1994). We will use the formalism developed by Willick (1991,1994,1995) who has advocated a comprehensive approach to the problem. We assume in what follows that the distance indicator in question is the TF relation, but the discussion is readily modified to treat comparable DIs such as $D_n$-$\sigma$.

The material discussed in this section remains controversial, and the formal mathematical approach we have adopted has not been accepted by all of the workers in the field. For example, Mathewson & Ford (1994) have argued that approaches like ours require idealized assumptions about sample selection that are invalid for realistic data sets. They instead separate their sample into "control" and "target" regions, using the former to calibrate the biases in the latter. We believe, however, that such qualitative or semiquantitative approaches are subject to their own caveats (are the control and target regions the same in their bias properties?), and run the risk of obscuring rather than illuminating the principal issues. Unlike much of the rest of the review, we have not attempted here to summarize all the analyses extant in the literature, but rather aim to give a comprehensive coverage of our own approach. This section can be skipped on a first reading.

### 6.5.1 *Forward DIs: Selection Bias*

We consider first the formalism appropriate to the forward relation; the inverse relation is treated in § 6.5.4. A starting point in the quantification of bias is the empirical fact that it is the TF *observables*, and not the inferred distances, which exhibit a roughly Gaussian distribution (*e.g.*, Willick *et al.* 1995c). Thus, the apparent magnitude of a galaxy with width parameter $\eta$ and true distance $r$ is normally distributed about a mean value $M(\eta) + 5\log r$ with rms dispersion $\sigma$, *when selection effects may be neglected.* We combine this Gaussian distribution with the *a priori* distributions of galaxy distances and width parameters, and with a function describing sample selection probability, to obtain the joint probability distribution of the observables $(m, \eta)$ and the true distance $r$ for a galaxy:

$$P(r, m, \eta) \propto r^2 \, n(r) \, S(m, \eta)$$
$$\times \phi(\eta) \frac{1}{\sqrt{2\pi}\,\sigma} \exp\left(-\frac{[m - (M(\eta) + \mu(r))]^2}{2\sigma^2}\right) , \tag{171}$$

where $\mu(r) = 5\log r$, $\phi(\eta)$ is the $\eta$-distribution function (similar to the luminosity function), and $n(r)$ is the galaxy number density along the line of sight in question. The quantity $S(m, \eta)$ is the *sample selection function*, defined as the probability an object with observables $(m, \eta)$ will be included in the sample irrespective of other considerations. Note that we express the selection function here in terms of the observables, rather than in terms of distance as we did in redshift survey analyses (§ 3.4). An idealized form of $S(m, \eta)$ describes a sample with a strict magnitude limit,

$$S(m, \eta) = \begin{cases} 1 & \text{if } m \leq m_\ell, \\ 0 & \text{otherwise}. \end{cases} \tag{172}$$

Selection and Malmquist bias effects are each implicit in Eq. (171). Which of the two comes into play depends, as already noted, on the assumptions underlying the method of data analysis; these assumptions determine which of several *conditional* probability distributions derivable from Eq. (171) are used. We first discuss selection bias; in § 6.5.2 we discuss Malmquist bias.

We know from § 6.4.1 that selection biases arise in Method II analyses, in which we take redshift-space information as the *a priori* distance indicator, either through the assumption of a common distance for cluster galaxies, or via a redshift–real space mapping such as Eq. (163). To the extent this mapping is a good one, we are in effect taking the true distance as "known"—dependent, perhaps, on model free parameters, but entirely independent of the TF observables.[29] In mathematical terms, this is equivalent to using probability distributions which are conditioned on the value of the true distance $r$. In the forward TF case, the relevant distribution is $P(m|\eta,r)$, which is readily derived from the joint probability distribution Eq. 171:

$$P(m|\eta,r) = \frac{P(r,m,\eta)}{\int_{-\infty}^{\infty} dm\, P(r,m,\eta)}$$

$$= \frac{S(m,\eta)\exp\left(-\frac{[m-(M(\eta)+\mu(r))]^2}{2\sigma^2}\right)}{\int_{-\infty}^{\infty} dm\, S(m,\eta)\exp\left(-\frac{[m-(M(\eta)+\mu(r))]^2}{2\sigma^2}\right)}\,. \tag{173}$$

Eq. (173) is the mathematical basis of a forward Method II analysis. The distance $r$ would be replaced by a redshift-space prediction, such as Eq. (163), in an actual implementation of Method II.

If sample selection were independent of apparent magnitude, $S(m,\eta)$ would drop out and Eq. (173) would reduce to the familiar normal distribution centered on the naïve expected apparent magnitude $M(\eta) + \mu(r)$. This is the condition that a simple $\chi^2$ minimization, such as that suggested by Eq. (165), be valid for determining values of any unknown model parameters. In a real study, however, sample selection almost always depends on apparent magnitude, although the dependence is generally indirect, as we explain below. For simplicity, we first suppose that in fact a strict magnitude limit, Eq. (172), applies. In that case Eq. (173) obviously does not reduce to a normal distribution centered on $M(\eta) + \mu(r)$. It is, rather, straightforward to show that the expected apparent magnitude is given by

$$E(m|\eta,r) = \int m P(m|\eta,r)\, dm = M(\eta) + \mu(r) - \sqrt{\frac{2}{\pi}}\sigma\frac{e^{-\mathcal{A}^2}}{1+\operatorname{erf}(\mathcal{A})}\,, \tag{174}$$

where

$$\mathcal{A}(\eta,r;m_\ell) \equiv \frac{m_\ell - (M(\eta)+\mu(r))}{\sqrt{2}\,\sigma}\,. \tag{175}$$

The last term on the right hand side of Eq. (174) is the apparent magnitude *bias*, the difference between actual and naïve expectation values. Its significance is determined by the parameter $\mathcal{A}(\eta,r;m_\ell)$, a dimensionless measure of how "close" the object is expected to be to the magnitude limit. For $\mathcal{A} \gtrsim 1$, the bias is small; in that case few galaxies are "missing" due to the magnitude limit. But for $\mathcal{A} \lesssim 0$, the hypothetical ensemble of galaxies of given $\eta$ and $r$ is strongly truncated by the magnitude limit, and the bias is comparable to or greater than the TF scatter. Note that the sense of the bias is that objects are typically *brighter* than the naïve TF prediction, as we saw in § 6.4.1. In any magnitude-limited sample, a large fraction of the objects are likely to be near the magnitude limit. Hence, the bias represented by Eq. (174) is a problematic issue in most Method II analyses [30].

---

[29] It is not, of course, literally correct that the true distance is specified in a Method II analysis. Errors in the velocity model being fit to the data, as well as small-scale "noise" in the actual velocity field, mean that $r$ is only approximately specified by redshift information. However, to first order this only increases the variance, but does not bias, a Method II solution. The formalism presented above is correct in this approximation, but the quantity $\sigma$ should be regarded as the quadrature sum of DI scatter and model errors. In § 8.1.3 we will describe a method which aims to go beyond this first order approximation and include the effect of model errors and velocity noise.

In real samples, the situation is more complicated. TF and $D_n$-$\sigma$ samples are drawn from catalogs culled from photographic plate material. Such catalogs (see § 3.1) typically list galaxy apparent magnitudes and/or angular diameters on a measurement system different from that used by the DI. The catalog might give, for example, blue-bandpass photographic magnitudes ($m_B$), while the TF study uses I band CCD magnitudes ($m_I$). Candidates for the TF sample might be required to satisfy $m_B \leq m_\ell$, but once thus selected, the allowed values of $m_I$ are unrestricted (the CCD photometry is far deeper than the catalog). Nonetheless, it is clear that the distribution of I band magnitudes must affected by this selection procedure, as $m_I$ is sure to be correlated with $m_B$. Analogous statements hold if the sample is limited by catalog diameter rather than magnitude.

Willick (1994) has shown how this problem of an "indirect" magnitude limit may be dealt with in several practical situations. Let us refer to the quantity determining sample inclusion as $\xi$, which might be a photographic magnitude or logarithmic diameter, and assume it is required to satisfy $\xi \geq \xi_\ell$. The selection probability for an object with TF observables $(m, \eta)$ is then given by

$$S(m, \eta) = P(\xi \geq \xi_\ell | m, \eta). \tag{176}$$

The quantity on the right hand side may be determined by first assuming that $P(\xi | m, \eta)$ is a Gaussian centered on a mean relation $\xi(m, \eta)$ and with rms dispersion $\sigma_\xi$. Both may be determined empirically from the data; Willick et al. (1995a,b) have shown that $\xi(m, \eta)$ is well modeled as linear in $m$ and $\eta$ when $\xi$ is a photographic apparent magnitude or logarithmic diameter. The resulting Gaussian probability distribution is then integrated over the allowed values of $\xi$:

$$S(m, \eta) = \frac{1}{\sqrt{2\pi}\,\sigma_\xi} \int_{\xi_\ell}^{\infty} d\xi \, \exp\left(-\frac{(\xi - \xi(m, \eta))^2}{2\sigma_\xi^2}\right). \tag{177}$$

This selection function may be substituted into Eq. (173), and the expected apparent magnitude, and thus the bias, may then be obtained in the usual way. The results, given by Willick (1994), are complicated and need not be reproduced here. We note only that, as in the case of a strict magnitude limit, the bias again depends on a "limit closeness parameter," given in the present case by

$$\mathcal{A}_\xi(\eta, r; \xi_\ell) \equiv \frac{\xi(M(\eta) + \mu(r), \eta) - \xi_\ell}{\sqrt{2}\,\sigma_\xi}, \tag{178}$$

and is also proportional to a "coupling" parameter $\beta_\xi \propto \sigma/\sigma_\xi$, which measures the tightness of the TF relation relative to that of the $\xi$–$(m, \eta)$ correlation. In particular, when the latter correlation is weak, the bias is small, since selection criteria only weakly constrain the apparent magnitude used in the TF analysis. In most real samples, however, the $\xi$–$(m, \eta)$ correlation is strong enough that even the bias due to "indirect" limits is significant. Another feature of an indirect limit is that, if $\xi$ is significantly correlated with $\eta$ as well as with $m$, sample selection probability is $\eta$-dependent; the consequences of this are discussed in § 6.5.4.

The formulae given by Willick (1994) for the magnitude and dispersion biases allow one to correct for selection bias. The basic idea is to iterate least-squares fits of the underlying model, using the parameter estimates from one iteration to correct the next for bias. "Corrected" apparent magnitudes $m^{(c)}$, defined as the observed magnitudes plus the computed bias, are used in the iterated fits. This procedure typically produces convergent parameter values in $\sim 10$ iterations, and has been used by Willick et al. (1995a,b) to calibrate TF relations for recent peculiar velocity analyses. An alternative to iteration is direct likelihood maximization, using individual object probability functions $P(m|\eta, r)$ given by Eq. (173).

---

[30] A related complication is that not only the expected apparent magnitude, but also its variance is biased, in the sense that the actual variance is strictly $< \sigma^2$. The causes of this effect, known as *dispersion bias*, were discussed in § 6.4.2. Quantitative expressions for dispersion bias are given by Willick (1994).

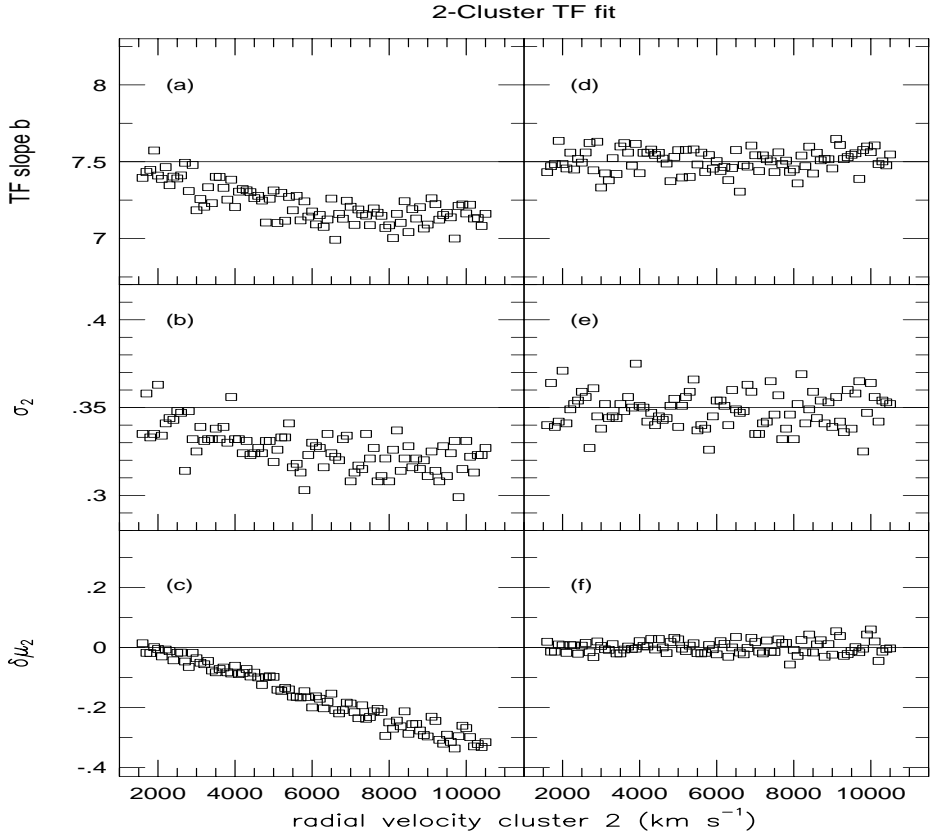

Fig. 14. The main effects of selection bias on a Method II TF analysis are shown in the left-hand panels; the same results following application of the iterative correction procedure (see text) are shown in the right-hand panels. Data sets consisting of 25 galaxies in each of two clusters were numerically simulated. The simulated galaxies obey a true TF relation with slope $b = 7.5$ and scatter $\sigma = 0.35$ mag, and are subject to a diameter limit. Each simulated data set consisted of one cluster at fixed distance $r_1 = 1500$ km s$^{-1}$, assumed known, and another at variable distance $r_2$ shown on the horizontal axis, which was taken as an unknown in the analysis. Each plotted point represents the mean of $\sim 25$ simulations within bins of 100 km s$^{-1}$. A TF relation was fitted to each data set; the free parameters were the zeropoint $A$ (not shown), slope $b$, and the distance modulus $\mu_2$ of the second cluster. The left hand panels exhibit the principal effects of selection bias: the fitted slope and scatter are too small, and the distance of the second cluster is underestimated. See text for details.

In Fig. 14 we illustrate the effects of selection bias on both DI calibration and distance determination using numerically simulated data. Each simulated data set consisted of two clusters with 25 galaxies each. The distance to the first cluster was fixed at $r_1 = 1500$ km s$^{-1}$, which was assumed known to the "observer." The second cluster was stepped out in distance, as shown on the horizontal axis; its distance $r_2$ was assumed unknown to the "observer." Determination of its distance modulus $\mu_2$ was assumed to be one goal of the analysis; the other was determination of the TF parameters $A$ and $b$. The simulated data obey a TF relation with slope $b = 7.5$ and scatter $\sigma = 0.35$ mag; sample selection was imposed by requiring galaxies to have photographic diameters $\geq 1'$. These diameters were generated from the TF observables $(m, \eta)$ using empirical relations from real data sets (Willick et al. 1995ab).

For each simulated data set the TF parameters $A$ and $b$, and the distance modulus of the second cluster $\mu_2$, were solved for using standard least-squares techniques. An rms scatter about the fit was computed separately in each cluster. The results of these least-squares fits are shown in the left-hand panels of Fig. 14. As the second cluster moves farther away, so that an ever-larger fraction of its member galaxies are near the diameter limit, the bias increases: the top and middle panels show that the fitted slope and rms scatter (in the second cluster) become progressively smaller as the second cluster moves farther away. (The fitted slope levels out at large distances because it is strongly constrained by data in the first cluster, which is close enough to be nearly unbiased.) The bottom panel shows the error in the least-squares determination of the distance modulus of the second cluster, $\delta\mu_2 = \mu_2 - 5\log r_2$. This error is in the sense that the distance is *underestimated*, and grows fractionally worse with increasing distance. The right hand panel of Fig. 14 show the same quantities following application of the iterative bias correction scheme discussed above. As can be seen, the iterative method yields unbiased final values of the slope, scatter, and distance of the second cluster.

### 6.5.2 Forward DIs: Malmquist Bias

Malmquist bias arises in Method I analyses, in which we first assign distances directly from the DI, and consider redshift information only afterward. As discussed above, the large scatter of DIs such as TF or $D_n$-$\sigma$ guarantees that objects with inferred distance $d$ come, with varying probability, from a wide range of true distances. The probability distribution of true distance $r$ given inferred distance $d$ is thus the mathematical basis for treating Malmquist bias.

The simplest way to arrive at $P(r|d)$ is first to consider $P(r, \mu(d))$, which we obtain from Eq. (171) as follows:

$$P(r, \mu(d)) = \int_{-\infty}^{\infty} d\eta \int_{-\infty}^{\infty} dm \, P(r, m, \eta) \times \delta\left(m - [M(\eta) + \mu(d)]\right)$$

$$\propto r^2 n(r) \frac{1}{\sqrt{2\pi}\,\sigma} \exp\left(-\frac{(\mu(r) - \mu(d))^2}{2\sigma^2}\right) \mathcal{S}(d). \tag{179}$$

In Eq. (179) $\delta$ is the Dirac delta function, and we have defined

$$\mathcal{S}(d) \equiv \int d\eta \, \phi(\eta) \, S(M(\eta) + \mu(d), \eta). \tag{180}$$

Converting to natural logarithms we derive

$$P(r, d) = P(r, \mu(d)) \frac{\partial \mu}{\partial d} \propto r^2 n(r) \frac{1}{\sqrt{2\pi}\,\Delta} \exp\left(-\frac{[\ln r/d]^2}{2\Delta^2}\right) \frac{\mathcal{S}(d)}{d}, \tag{181}$$

where $\Delta \equiv \left(\frac{\ln 10}{5}\right)\sigma \simeq 0.46\,\sigma$ is a measure of the fractional distance uncertainty of the DI. The desired *conditional* probability is then given by

$$P(r|d) = \frac{P(r,d)}{\int_0^\infty dr\, P(r,d)} = \frac{r^2 n(r) \exp\left(-\frac{[\ln r/d]^2}{2\Delta^2}\right)}{\int_0^\infty dr\, r^2 n(r) \exp\left(-\frac{[\ln r/d]^2}{2\Delta^2}\right)}. \tag{182}$$

We see that $S(m, \eta)$ has dropped out entirely in Eq. (182); selection bias does not play a role in forward Method I analyses. This may appear odd at first glance; one would think that because selection depletes the sample more at $r>d$ than at $r<d$, it must affect the ratio of near and far objects at inferred distance $d$. The reason it does not may be understood heuristically as follows. Objects at $r<d$ must be intrinsically fainter than expected for their velocity width ($M>M(\eta)$) in order to scatter *out* to $d$; objects at $r>d$ must be brighter than expected ($M<M(\eta)$) in order to scatter *in* to $d$. Thus, although the sample as a whole is progressively depleted with distance, the set of objects with inferred distance $d$ is atypical: its closer members are unusually faint, and its more distant members are unusually bright. This discrepancy exactly balances the overall decrease in sampling density with distance. It is important to note, however, that this argument is specific to the forward form of the DI (we show how inverse DIs differ in this respect in § 6.5.5) and to the absence of an explicit distance-dependence in the sample selection function (in § 6.5.3 we demonstrate how such a dependence might arise).

Although independent of observational selection, forward Malmquist bias is nonetheless a complicated effect. It is best illustrated by considering the expectation value of $r$ given $d$:

$$E(r|d) = \frac{\int_0^\infty r P(r|d)\,\mathrm{d}r}{\int_0^\infty P(r|d)\,\mathrm{d}r} = d\,\frac{\int_0^\infty x^3 n(dx)\,\mathrm{e}^{-\frac{(\ln x)^2}{2\Delta^2}}\,\mathrm{d}x}{\int_0^\infty x^2 n(dx)\,\mathrm{e}^{-\frac{(\ln x)^2}{2\Delta^2}}\,\mathrm{d}x}, \quad (183)$$

where $x = r/d$. Eq. (183) shows that, in general $E(r|d) \neq d$; that is, the inferred distance is biased. It also shows that Malmquist bias depends on the number density of galaxies all along the line of sight, and thus is a nonlocal effect.

The simplest (though usually unrealistic) example of Malmquist bias occurs when density gradients are negligible on the scale of distance errors (*i.e.*, $\frac{n'}{n} \ll (d\Delta)^{-1}$). In that case, the density term drops out of Eq. (183). The integrals are then easily evaluated with the result

$$E(r|d) = d\,\mathrm{e}^{\frac{7}{2}\Delta^2}. \quad (184)$$

Thus even in the case of uniform density $E(r|d) \neq d$. This effect, *uniform density Malmquist bias*, is due to the increasing volume element at larger distances: there are more galaxies at $r>d$ than at $r<d$, and thus $E(r|d) > d$; for a typical distance indicator ($\Delta \sim 0.2$) it is a $\sim 15\%$ effect. More generally, density gradients cannot be neglected, and can either reinforce or work against the volume effect. This is most easily seen in the approximation that $n(r)$ is slowly varying on the scale of DI errors. Expanding $n(r)$ to first order in Eq. (183) then yields

$$E(r|d) = d\left[1 + \left(\frac{7}{2} + \gamma(d)\right)\Delta^2\right] \quad (185)$$

where

$$\gamma(d) \equiv \left.\frac{\mathrm{d}\ln n(r)}{\mathrm{d}\ln r}\right|_{r=d}, \quad (186)$$

where we have further assumed $\Delta^2 \ll 1$. Eqs. (184) and (185) were first derived in the extragalactic context by Lynden-Bell *et al.* (1988a; cf., Malmquist 1920).

The slowly-varying approximation Eq. (185) illustrates in a general way the effect of density gradients on the statistical properties of inferred distances. This effect is also known as *inhomogeneous Malmquist bias* or IHM. When $\gamma(d) > 0$, IHM reinforces the uniform density bias; when $\gamma(d) < 0$ it works against it. While the essence of IHM is shown by the slowly-varying density approximation, Eq. (185) is in fact a rather poor approximation to the true IHM for the density fields seen in the real universe (see below). A proper calculation of IHM requires the full numerical integration indicated by Eq. (183).

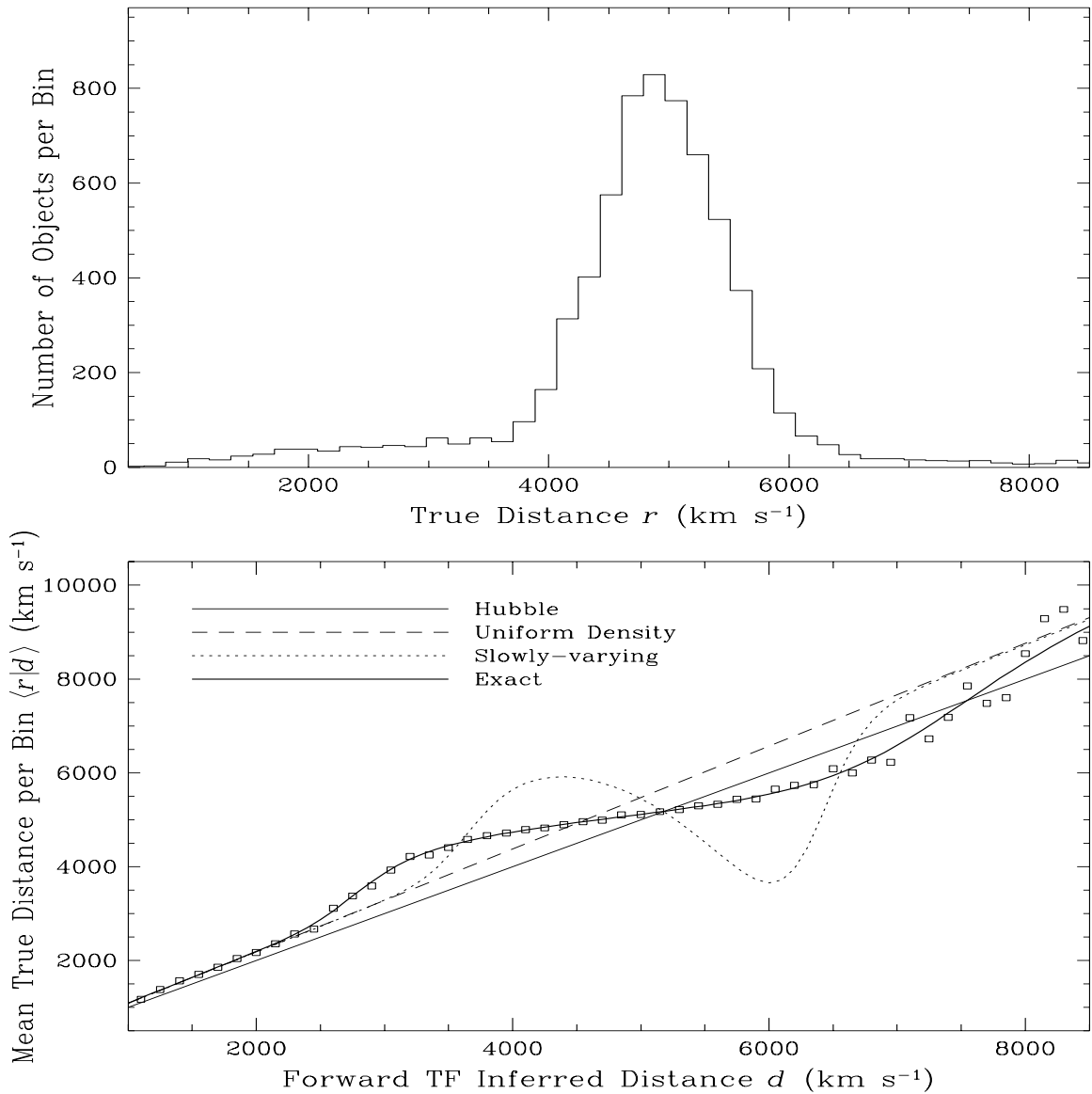

Fig. 15. Malmquist bias induced by a density peak at a distance of 5000 km s$^{-1}$. Simulated galaxies were generated along a line of sight with a Gaussian perturbation of full width $\sim 1200$ km s$^{-1}$ and maximum overdensity $\delta_{\rm max} = 25$, atop a uniform distribution. The upper panel shows the distribution of simulated galaxies as a function of their true distance. The lower panel plots, as open squares, average true distance vs. inferred distance for objects in 150 km s$^{-1}$ wide bins. The meaning of the various lines drawn through the points is indicated in the panel.

These points are illustrated using simulated data in Fig. 14, which shows the Malmquist bias associated with a very strong density perturbation. Galaxies were simulated along a line of sight in which the density is given by $n_0 + n_p(r)$, where $n_p(r)$ is a Gaussian perturbation centered at 5000 km s$^{-1}$, with full width $\sim 1200$ km s$^{-1}$ and maximum overdensity $\delta_{\rm max} = 25$. Galaxies were assumed to be moving with Hubble flow; there are no true peculiar velocities in this simulation. The number of galaxies per unit true distance along the line of sight is shown in the upper panel; the effect of the density peak on the distribution is clear. The simulated galaxies were required to have diameters $\geq 1.2'$, although this diameter limit does not affect the character of Malmquist bias, for reasons described above. The simulated galaxies obey a TF relation with scatter $\sigma = 0.35$ mag; the "observer" of the simulation assigned inferred distances $d$ with a properly calibrated TF relation. In the lower panel the galaxies have been binned according to the values of their inferred distances. The average true distances of the galaxies in each bin are plotted as open squares.

Nowhere do the simulated data follow the Hubble line. Very far from the density peak ($d \lesssim 2500$ km s$^{-1}$ and $d \gtrsim 7500$ km s$^{-1}$) the simulated data follow the uniform density Malmquist bias prediction Eq. (184). However, within $\sim 2000$ km s$^{-1}$ of the peak, the data depart strongly from the uniform density prediction; this is the IHM effect. The sense is that in the foreground of the peak distances are underestimated (the squares lie above the long dashed line) and in the background they are overestimated. The amplitude of these effects is 500–1000 km s$^{-1}$. The short dashed line is the prediction of the slowly varying approximation Eq. (185), using the exact density gradient. It is apparent that this approximation is extremely poor; this is because the density varies rapidly on the scale of distance errors. The heavy solid line is obtained from the exact formula Eq. (183) by numerical integration of the true density field, and correctly predicts the bias at all distances.

Fig. 15 gives further insight into an issue touched upon in § 6.4.2, namely, why IHM can lead to spurious peculiar velocities. Suppose that the galaxies depicted in the figure followed the Hubble flow. Their redshift vs. inferred distance diagram would then closely resemble the lower panel. The "S-wave" seen in the figure—measured with respect to the uniform density prediction—would be interpreted as peculiar velocities if IHM were neglected. In particular, motions of many hundreds of km s$^{-1}$ toward the density peak would be seen in both the foreground and the background. Such motions are what is expected from gravitational instability, but in this example they are merely an artifact of Malmquist bias. More generally, if gravitational instability is correct, there are real motions but IHM will cause the observed motions to be larger. *This bias mimics the effect of having more mass in the density enhancement than actually exists.* It follows that, if not properly accounted for, IHM will result in an overestimate of $\beta$ (*e.g.*, Eq. 149).

*6.5.3 Distance-Dependent Selection Functions: The "Gould Effect"*

An important feature of the Malmquist bias equations in § 6.5.2 was their independence of observational selection. This greatly simplifies their application in peculiar velocity analyses; in particular, it makes the uniform density correction trivial. However, as was first pointed out by Gould (1993), under certain circumstances this desirable attribute of Malmquist bias might break down. The origin of this effect is best understood in terms of the selection function, which we have assumed to depend *explicitly* only on the observables $(m, \eta)$, not on distance. As a result, in the steps leading from Eq. (179) to Eq. (182) $S(m, \eta)$ naturally cancelled out. Had the selection function actually been of the form $S(m, \eta, r)$, it would not have dropped out, and the result would have been a selection-dependent equation for Malmquist bias.

To see how distance-dependent selection might arise, suppose that selection is based on a blue apparent magnitude which must satisfy $m_B \leq m_\ell$, while the TF analysis involves a distinct magnitude $m$. Now, instead of empirically relating $m_B$ to the TF observables and deriving $S(m, \eta)$ according to Eq. (176), let us consider the TF relation $M_B(\eta)$ satisfied by the absolute magnitude corresponding to $m_B$. We can then relate the probability that $m_B \leq m_\ell$, for a galaxy of width parameter $\eta$ and distance $r$, to a probability involving the absolute magnitude $M_B$:

$$P(m_B \leq m_\ell | \eta, r) = P(M_B \leq m_\ell - \mu(r) | \eta)$$
$$= \int_{-\infty}^{m_\ell - \mu(r)} \frac{1}{\sqrt{2\pi}\sigma_B} e^{-\frac{[M_B - M_B(\eta)]^2}{2\sigma_B^2}} \, dM_B \,, \qquad (187)$$

where we have assumed a normally distributed blue TF relation with scatter $\sigma_B$. Eq. (187) is apparently nothing but the sample selection function as we have defined it. However, it is independent of the TF magnitude $m$, and explicitly dependent on $r$: rather than writing it $S(m, \eta)$ we must write it $S(\eta, r)$.

How can this derivation be reconciled with Eq. (176), whose premises appear equally valid but which yields a qualitatively different result? The answer is quite subtle, and we refer the reader to Willick (1995) for a detailed explanation. There it is shown that Eq. (176) is valid to the extent that the residuals of the TF relations $M_B(\eta)$ and $M(\eta)$ are strongly correlated, as is normally the case. Eq. (187) is valid in the opposite limit of uncorrelated residuals. A rigorous derivation of sample selection must take this effect into account; the general result is a selection function $S(m, \eta, r)$, with the strength of the $r$-dependence inversely related to the degree of correlation between the two TF relations.

There are two methods to determine whether the Gould effect is present is real samples. The first is to directly examine correlations between the $M_B(\eta)$ and $M(\eta)$ TF residuals. The second is to test whether $m_B$ is statistically best described as a function of $m$ and $\eta$, of $\eta$ and $r$, or of all three. In the first of the three cases, there is no Gould effect; otherwise it is present at some level. Willick et al. (1995a,b) used the second test (with $r$ modeled by redshift) to show that the effect is significant for samples selected on the ESO angular diameter (e.g., Mathewson et al. 1992b). Peculiar velocity analyses using such samples (e.g., Hudson et al. 1995) rely heavily on redshift-space grouping to minimize the effect.

*6.5.4  Inverse DIs: Selection Bias*

A formalism quite analogous to that developed above may be presented for inverse DIs; we assume, as usual, that the DI in question is TF. The central element of this formalism is the Gaussian distribution of observed width parameters about the expectation value $\eta^0(m - \mu(r))$ when selection effects are neglected. Adding in the remaining elements we arrive at the joint distribution of $m$, $\eta$, and $r$ written in terms of the inverse TF relation:

$$P(r, m, \eta) \propto r^2 \, n(r) \, S(m, \eta)$$
$$\times \Phi(m - \mu(r)) \frac{1}{\sqrt{2\pi}\,\sigma_\eta} \exp\left( -\frac{[\eta - \eta^0(m - \mu(r))]^2}{2\sigma_\eta^2} \right) . \quad (188)$$

We note that, in contrast to its forward analogue Eq. (171), this equation involves the *luminosity function* $\Phi(M)$.

Both selection and Malmquist bias are implicit in Eq. (188). As before the two are distinguished by the assumptions of data analysis. In a Method II analysis we (implicitly) condition on the true distance (see § 6.5.1). The probability distribution which underlies such an analysis is thus $P(\eta|m, r)$, obtained from Eq. (188) according to

$$P(\eta|m, r) = \frac{P(r, m, \eta)}{\int_{-\infty}^{\infty} d\eta \, P(r, m, \eta)}$$
$$= \frac{S(m, \eta) \exp\left(-\frac{[\eta - \eta^0(m - \mu(r))]^2}{2\sigma_\eta^2}\right)}{\int_{-\infty}^{\infty} d\eta \, S(m, \eta) \exp\left(-\frac{[\eta - \eta^0(m - \mu(r))]^2}{2\sigma_\eta^2}\right)} . \quad (189)$$

This equation is the inverse counterpart of Eq. (173). In the forward case the condition for selection bias to vanish was that $S(m, \eta)$ be independent of $m$, which is almost always unrealistic. Now the condition is that $S(m, \eta)$ be independent of $\eta$. We can expect that, to a good approximation, this condition will frequently hold. To the extent that it does, Eq. (189) reduces to a Gaussian distribution of $\eta$ about $\eta^0(m - \mu(r))$. It follows that a standard $\chi^2$ minimization procedure will yield unbiased values of model parameters, as discussed in § 6.4.4.

However, the assumption that inverse Method II analyses are free of selection bias is subject to several caveats which are often overlooked:

(i) As noted in § 6.5.1, using different photometric quantities for sample selection than for DI analysis can result in an $\eta$-dependence of the function $S(m, \eta)$. Willick (1994) has quantified and described correction techniques for this effect.

(ii) *Direct* dependence of sample selection on velocity width can occur in studies in which $\eta$ is measured from the H I 21 cm line. Signal-to-noise ratio considerations can mean that large velocity widths more likely to be undetected at low 21 cm flux levels (Roth 1993). Willick *et al.* (1995a) have attempted to estimate this effect for the Mould *et al.* (1991) TF sample, and found it small but non-negligible.

(iii) The slope of the inverse relation—the quantity $e$ in Eq. (168)—is affected by the accuracy of the redshift-distance model used in calibrating the relation (Willick 1991). This fitting bias does not affect the forward relation, in which a poor model increases scatter but does not bias the slope.

(iv) Finally and perhaps most importantly, while inverse DIs can avoid selection bias in the case of velocity width-independent selection, they have no special advantage over forward DIs in terms of susceptibility to Malmquist bias. Indeed, inverse Malmquist bias is inherently more complicated than forward, as we describe in the following section.

### 6.5.5 *Inverse DIs: Malmquist Bias*

An inverse Method I analysis is fully analogous to a forward Method I analysis, except that the inferred distances are obtained by solving the equation $\eta^0(m - 5\log d) = \eta$ implicitly for $d$. For the linear inverse TF relation Eq. (168), this gives $d = 10^{0.2[m-(D-\eta/e)]}$, which may be compared with the forward inferred distance $d = 10^{0.2[m-(A-b\eta)]}$. Although we have used the symbol $d$ for both the forward and inverse inferred distance, the two are not equal in general; the non-equality of the forward and inverse TF parameters (§ 6.4.4) guarantees this. In fact, inverse DIs are *steeper* than their forward counterparts when plotted in the usual way ($M$ vs. $\eta$). Inverse inferred distances are thus larger for luminous, and smaller for faint, galaxies. Since luminous objects dominate a sample at large redshifts, it is clear that a redshift vs. inferred distance diagram will look different depending on whether the forward or inverse relation has been used, with quite different biases in the two cases. Let us now quantify these ideas.

If we begin with Eq. (188) and go through a series of steps analogous to those which took us from Eq. (171) through Eq. (182), we obtain the following for the probability distribution of the true distance of a galaxy with inverse inferred distance $d$:

$$P(r|d) = \frac{r^2 n(r) s(r) \exp\left(-\frac{[\ln r/d]^2}{2\Delta^2}\right)}{\int_0^\infty r^2 n(r) s(r) \exp\left(-\frac{[\ln r/d]^2}{2\Delta^2}\right)\, dr}. \tag{190}$$

Here $\Delta \equiv \sigma_\eta/e$, the effective fractional inverse distance error, is roughly but not precisely equal to its forward counterpart. Derivation of Eq. (190) required the assumption of a linear inverse TF, and also of a selection function at most weakly dependent upon $\eta$ (details are given in Willick 1995). With these (realistic) assumptions Eq. (190) resembles its forward analogue Eq. (182), but with a crucial difference: the presence of the *real space selection function*

$$s(r) \equiv \int_{-\infty}^{\infty} dm\, \Phi(m - \mu(r))\, S(m, \eta^0[m - \mu(r)]), \tag{191}$$

the probability that a sample object lies at true distance $r$ irrespective of the observables $(m, \eta)$.

Eq. (190) shows that the inverse Malmquist effect quantitatively resembles the forward, with the exception that now the product $n(r)s(r)$ replaces $n(r)$ alone. The differences between inverse and forward Malmquist bias are most clearly illustrated by considering the uniform density case, with the further assumption that $s(r)$ varies slowly on scales $d\Delta$. By analogy with Eq. (185) it is then easy to derive the

result

$$E(r|d) = d \left[ 1 + \left( \frac{7}{2} + \gamma_s(d) \right) \Delta^2 \right] , \qquad (192)$$

where

$$\gamma_s(d) = \left. \frac{d \ln s(r)}{d \ln r} \right|_{r=d} . \qquad (193)$$

We see again that, like their forward counterparts, inverse inferred distances are biased: $E(r|d) \neq d$. However, in contrast with the forward case, the effect *is not independent of sample selection or the luminosity function.* Even in the case of uniform density, it is not simple to obtain a Malmquist correction for an inverse distance. One must not only accurately model the selection function $S(m, \eta)$, but integrate it with respect to a believable luminosity function $\Phi(M)$. These difficulties are only compounded by the realistic requirement of estimating $n(r)$ in order to account for IHM. The net effect of these considerations is that inverse DIs are less suitable than forward DIs for Method I analyses. This point must be kept in mind when one speaks of the "unbiased" character of inverse DIs. The quantity $s(r)$ might be modeled by the observables $s(z)$ or $s(d)$, and thus perhaps be dealt with easily. There is, however, a method which, at least in principle, permits effective use of inverse inferred distances, as we discuss in the next section.

### 6.5.6 The Method of Landy and Szalay

A severe obstacle to correcting for IHM is the requirement of knowing the number density $n(r)$ along any given line of sight. One way around this obstacle is to estimate $n(r)$ from redshift survey information. This has been done, for example, in comparisons of the *IRAS* density field with the POTENT mass reconstruction (Dekel *et al.* 1993; Hudson 1994a; Hudson *et al.* 1995). Although reasonable, this procedure has unsatisfactory features, such as uncertainties introduced by smoothing and the need for an assumed peculiar velocity field to convert redshift- to real-space density. One would like, if possible, to derive the IHM correction self-consistently from the DI data themselves. Such a method was outlined by Landy & Szalay (1992;LS), and independently by Willick (1991). The LS technique relies on the mathematical relationship between $E(r|d)$ and a quantity which is in principle observable: the number of objects per unit inferred distance.

Let us define $f(d)$ as the number of objects, per unit solid angle along a particular line of sight, per unit inferred distance. If the galaxy distribution is densely enough sampled, one can imagine measuring $f(d)$ by binning objects in inferred distance. The resulting histogram representation of $f(d)$ is proportional to the probability, $P(d)$, that an object along the given line of sight has inferred distance $d$. This probability may in turn be calculated by integrating $P(r, d)$, given by Eq. (181) in the forward case and derivable from Eq. (188) in the inverse case, over the true distance $r$:

$$f(d) \propto P(d) = \int_0^\infty P(r, d) \, dr . \qquad (194)$$

At this point, the expressions for the forward and cases diverge, and we present separately the result for each:

$$f_{\text{forw}}(d) \propto \frac{S(d)}{d} \int_0^\infty r^2 n(r) \frac{1}{\sqrt{2\pi}\,\Delta} \exp\left( -\frac{[\ln r/d]^2}{2\Delta^2} \right) \, dr , \qquad (195)$$

and

$$f_{\text{inv}}(d) \propto \frac{1}{d} \int_0^\infty r^2 n(r) s(r) \frac{1}{\sqrt{2\pi}\,\Delta} \exp\left(-\frac{[\ln r/d]^2}{2\Delta^2}\right)\,dr\,. \tag{196}$$

In the forward case, the quantity $\mathcal{S}(d)$, which dropped out of the conditional expectation $E(r|d)$, is present here in the expression for $f(d)$. In the inverse case, however, $f(d)$ contains a dependence on selection identical to that seen in $E(r|d)$. Thus, it is only for the inverse case that the relationship between $f(d)$ and $E(r|d)$ will be straightforward (cf., Teerikorpi 1993), and we consider only the inverse case (now without a subscript) in what follows.

Taking as usual $x = r/d$ we have

$$f(d) \propto d^2 \int_0^\infty x^2 n(dx) s(dx)\,e^{-\frac{(\ln x)^2}{2\Delta^2}}\,dx \tag{197}$$

for inverse distances. To relate this to $E(r|d)$ we rely on the mathematical identity (e.g., LS or Willick 1995) that

$$\int_0^\infty x^3 g(dx)\,e^{-\frac{(\ln x)^2}{2\Delta^2}}\,dx = e^{\frac{7}{2}\Delta^2} \int_0^\infty x^2 g(d\,e^{\Delta^2} x)\,e^{-\frac{(\ln x)^2}{2\Delta^2}}\,dx \tag{198}$$

where $g(r)$ is an arbitrary function. Combining Eqs. (183), (197), and (198) one readily obtains

$$E(r|d) = d\,e^{\frac{3}{2}\Delta^2}\,\frac{f(d\,e^{\Delta^2})}{f(d)}\,. \tag{199}$$

Eq. (199) is the basis of the Landy-Szalay method of Malmquist bias correction. It says that we can obtain an unbiased estimate of the true distance, given an inverse inferred distance $d$, from the observable distribution $f(d)$.

The LS method is attractive for several reasons. First, as already noted, it enables us to correct for IHM using the DI data alone. Second, because LS involves inverse DI distances, it is based on a DI relation whose *calibration* was very nearly free of selection bias. In this sense, inverse DIs can be viewed as (nearly) bias-free, but only when the LS method is used. Finally, while we have not made this explicit in the above derivation, the LS method is applicable even when the Gould effect is present. This is because even when the selection function is of the form $S(m,\eta,r)$, one arrives at the expression Eq. 190.

With this rosy view of LS must go several realistic caveats. Accurate determination of the quantity $f(d)$ presents a number of technical problems (Willick 1995). Basically, the issue is one of resolution: one needs a sufficient number of objects per bin that the histogram representation of $f(d)$ is not overwhelmed by shot noise; to obtain this number one must make bigger bins in both angle and distance. The larger the bin, of course, the less accurately one determines the actual IHM correction. For these reasons, LS has yet to be implemented for a real sample. An effort is presently underway (Dekel et al. 1995) to do so for the large TF samples currently available (§ 7.2). It remains unclear at this time whether implementation of the LS method will constitute a major advance in dealing with Malmquist bias.

# 7  Statistical Measures of the Velocity Field

We begin this chapter with a discussion of bulk flows in § 7.1; this has been the major theme of much of the work in peculiar velocities in the last decade. § 7.2 summarizes the history of peculiar velocity work with a discussion of the effort needed to put on a common basis the various samples available. The rest of the chapter discusses various statistics measurable from the observed flow field. The velocity correlation function and the Cosmic Mach Number are discussed in § 7.3 and § 7.4, respectively. § 7.5 is devoted to derivations of the full three-dimensional velocity field from its radial component, with emphasis on the POTENT technique of Bertschinger & Dekel (1989), and the results that have come from this work.

## 7.1 A History of Observations of Large-Scale Flow

Since the mid-1970's coherent departures from uniform Hubble flow[31] have been observed on ever larger scales. A comprehensive historical review of the subject through $\sim 1989$ is given by Burstein (1990). Here we briefly outline this history from $\sim 1976$ to the present, with an emphasis on recent years.

When one speaks of a coherent bulk flow, one requires a velocity frame of reference. Prior to the mid-1970's, the most logical choice for such a frame was that defined by the barycenter of the Local Group of galaxies (LG), the bulk of whose mass is found in the Milky Way and the Andromeda galaxy M31. Our own motion with respect to such a frame was thought to arise from LG dynamics and the rotation of the Milky Way; the LG barycenter itself was considered at rest with respect to the Hubble expansion. This reasoning changed following the discovery of the CMB dipole anisotropy in 1976 (see § 5.7). The CMB dipole was and is most naturally interpreted as due to the motion of the LG with respect to the rest frame of the totality of mass within our observable universe, in the direction $l = 276°$, $b = 30°$, and with amplitude 627 km s$^{-1}$. With this finding it became clear that significant peculiar velocities exist in the Universe: unless the LG is atypical, peculiar motions of hundreds of km s$^{-1}$ could no longer be considered unusual. The CMB dipole also suggested a natural reference frame for the analysis of peculiar motions.

### 7.1.1 Early Work

The first claimed detection of large-scale streaming was that of Rubin et al. (1976ab), who assumed that giant Sc spiral galaxies are standard candles. They studied a sample of 96 such galaxies in the redshift range 3500-6500 km s$^{-1}$, and found a bulk flow relative to the Local Group frame of $\sim 600$ km s$^{-1}$ in the direction $l = 160°$, $b = -10°$. The Rubin et al. result provoked a great deal of skepticism in the astronomical community, despite the (nearly simultaneous) discovery of the CMB dipole. The direction of the Rubin et al. bulk flow was nearly orthogonal to the LG velocity vector, making the relationship of the two measurements difficult to understand. Indeed, the measurement of Rubin et al. to this day has neither been confirmed nor entirely refuted, but instead has simply been supplanted by more modern data based on better distance indicators.

In the late 1970s and early 1980s, peculiar velocity work focused on the detection of infall to the Virgo cluster, which lies near the North Galactic Pole at $l \simeq 270°$, $b \simeq 75°$, and is the nearest large cluster to the Local Group. A number of workers using TF, Faber-Jackson, and other techniques estimated the amplitude of this motion at the Local Group to be in the range $\sim 150$–400 km s$^{-1}$ (e.g., Schechter 1980; Aaronson et al. 1982a; de Vaucouleurs & Peters 1981; Tonry & Davis 1981a,b; Hart & Davies 1982; Dressler 1984). Even if the largest of these estimates were correct, the misalignment between the Virgo direction and the CMB dipole meant that Virgocentric infall could not be the sole cause of the LG motion (Davis & Peebles 1983a; de Vaucouleurs & Peters 1984; Sandage & Tammann 1984; Yahil 1985). This led to suggestions that the remaining part of the LG motion resulted from the pull of the relatively nearby Hydra-Centaurus supercluster complex, which lies almost near the apex of the microwave dipole at a redshift $cz \simeq 3000$ km s$^{-1}$ (Shaya 1984; Tammann & Sandage 1985; Davies & Stavely-Smith 1985). Lilje, Yahil, & Jones (1986) studied residuals from the Virgocentric flow solution of Aaronson et al. (1982a) and concluded they were best fit by a quadrupolar term arising from the pull of Hydra-Centaurus. These suggestions were consistent with the generally accepted view that peculiar motions arose from relatively "local" (scales $\lesssim 5000$ km s$^{-1}$) mass density perturbations.

---

[31] On large scales, these flows are referred to as as "large-scale streaming motions," "large-scale flow," "bulk flow," among other terms. We use all these terms interchangeably here.

*7.1.2  1986–1990: The "Great Attractor"*

Confirmation of this view appeared to come from a long term study by the Aaronson group. Using IRTF distances to ten clusters in the 4000–10,000 km s$^{-1}$ redshift range, Aaronson et al. (1986) concluded that these clusters exhibited no net motion relative to the CMB [32]. They argued, as a corollary, that the LG motion had to be entirely generated by mass fluctuations within $\sim 5000$ km s$^{-1}$. The picture that prevailed in the mid-1980's was thus one in which the Hubble flow was unperturbed on scales $\gtrsim 5000$ km s$^{-1}$. But within several years the paradigm had changed radically, due mainly to the work of the "7 Samurai" (7S). Applying the $D_n$-$\sigma$ relation to an all-sky sample of over 400 bright ($m_B \leq 13$ mag, $\langle cz \rangle \simeq 3000$ km s$^{-1}$) elliptical galaxies, the group reported a bulk flow of amplitude $600 \pm 100$ km s$^{-1}$ relative to the CMB frame, in the direction $l = 312 \pm 11°$, $b = 6 \pm 10°$ (Dressler et al. 1987a). They emphasized that galaxies in the Hydra-Centaurus concentration participated in, and therefore could not be the source of, the observed flow. They concluded that, contrary to the conventional wisdom, the greater part of the LG motion relative to the CMB had to be generated on scales $\gtrsim 5000$ km s$^{-1}$.

This result shook up the cosmology community considerably. Its significance lay not so much in the amplitude as in the coherence scale of the flow. Indeed, as initially reported by Dressler et al. (1987a), the true scale of the bulk flow was unconstrained, potentially larger than the effective limit ($\sim 6000$ km s$^{-1}$) of their sample. It was quickly realized (e.g., Vittorio, Juszkiewicz, & Davis 1986) that a bulk flow on this scale was inconsistent with essentially all of the then-popular scenarios for large-scale structure formation, including the standard cold and hot dark matter models. The power spectrum $P(k)$ for any of those models did not contain sufficient power on large scales to generate flows of amplitude $\sim 600$ km s$^{-1}$ on scales $\gtrsim 5000$ km s$^{-1}$ (Eq. 40). Kaiser (1988) showed that the situation for cosmological models was not as dire as Vittorio et al. (1986) implied, as the effective depth of the 7S sample was not as large as had been assumed. Indeed, the fitted bulk flow was not inconsistent with the standard CDM model with high enough normalization. However, in this case, the small-scale velocity dispersion is larger than that observed; Bertschinger & Juszkiewicz (1988) point out that there is no normalization of standard CDM which can simultaneously match the 7S bulk flow and the observed small velocity dispersion (cf., § 7.4).

The controversial nature of the 7S result was at least partially alleviated when the group undertook a reinterpretation of their data in Lynden-Bell et al. (1988a, hereafter LB88). They showed that much of the signal for the elliptical streaming motion was provided by galaxies within $\sim 2000$ km s$^{-1}$; if they confined their analysis to objects at distances $\gtrsim 3000$ km s$^{-1}$, bulk flow was only marginally detected. Moreover, in a conical region within $\sim 60°$ of the apex of the apparent bulk flow, the derived amplitude was nearly 1000 km s$^{-1}$. They interpreted this as due to infall into an attracting point. They fit their data using a Method I approach (§ 6.4) to a model in which the flow is generated primarily by a spherically symmetric density perturbation they called the "Great Attractor" (GA) (cf., § 4). In addition to the GA distance and flow amplitude, which were treated as free parameters, the model of LB88 included a Virgo infall motion with a fixed amplitude of 250 km s$^{-1}$ at the position of the LG. Their maximum-likelihood fits placed the GA at a distance of $4350 \pm 350$ km s$^{-1}$ in the direction $l = 307°$, $b = 9°$, indicated that it induced an infall velocity of $570 \pm 60$ km s$^{-1}$ at the position of the LG, and attributed to it a mass of $\sim 5.4 \times 10^{16} M_\odot$.

A further elaboration of the GA model was provided by Faber & Burstein (1988, FB88), who combined TF data for nearby spirals from Aaronson et al. (1982b) with the 7S elliptical data. FB88 fitted the combined data set to a model in which the local velocity field was fully described by infall motions into the GA and Virgo, but with the parameters characterizing each motion allowed to vary independently (see also Han & Mould 1990). They obtained consistent results using Method I and Method II analyses. Two important conclusions resulted from this work. First, the nearby TF spirals ($cz \lesssim 2500$ km s$^{-1}$) mapped out a velocity field similar to that of the ellipticals. Second, the effect of Virgo was found to be very small in comparison with the tidal effect of the GA: the infall motion due to Virgo alone at the position of the LG had an amplitude of $\sim 100$ km s$^{-1}$. This contrasted strongly with earlier estimates (*e.g.*, Aaronson *et al.* 1982a) which had found a Virgocentric motion of $\sim 300$ km s$^{-1}$ and higher (§ 7.1, § 8.1.1). FB88 verified that the discrepancy between their Virgo flow solution and that of Aaronson *et al.* was due solely to the shear caused by the GA in the model (cf., Lilje *et al.* 1986)[33].

---

[32] Willick (unpublished) has shown, however, that the limited sky coverage of the Aaronson et al. sample means that its ruling out of large-scale bulk flows is not definitive.

Apparent confirmation of the GA picture came from two directions. First, three newly-completed redshift surveys (da Costa *et al.* 1988; Dressler 1988; Strauss & Davis 1988) showed that there was in fact a very strong excess number of galaxies in the direction of the proposed GA, peaking at $cz \sim 4000$ km s$^{-1}$ (cf., Fig. 6). The second source of support for the GA model came from further peculiar velocity studies of the region. Aaronson *et al.* (1989) extended their infrared TF cluster study to the Hydra, Centaurus, and other neighboring clusters and confirmed large positive peculiar velocities for these clusters (see also Mould *et al.* 1991). Dressler & Faber (1990a,b) studied field galaxies in the GA region using both the TF (using CCD photometry and optical linewidths) and $D_n$-$\sigma$ methods. Probing deeper than earlier studies, they claimed to detect "back-side infall" into the GA, *i.e.*, negative peculiar motions of galaxies at distances $\gtrsim 5000$ km s$^{-1}$. Such motions would have to exist if the positive velocities detected in the foreground were indeed generated by the GA, and the GA was an isolated structure. The detection of backside infall remains controversial (cf., Burstein *et al.* 1990; Roth 1993; Mathewson & Ford 1994), given the large and uncertain correction for inhomogeneous Malmquist bias. In addition, it is clear that the velocity field within the Centaurus cluster is quite complex; it consists of two pieces with substantially different distances and peculiar velocities (e.g., Lucey, Curry, & Dickens 1986; Burstein *et al.* 1990).

*7.1.3 1990-1994: Very Large-Scale Streaming?*

Several independent peculiar velocity surveys came to fruition around 1990, with results that challenged the consensus which had emerged around the GA picture. Willick (1990,1991; hereafter W91) reported an R band CCD TF study of 155 cluster spirals from the sample of Aaronson *et al.* (1986), and of 326 field spirals in the Perseus-Pisces supercluster region (PP). One of the more prominent structures in the nearby universe (Fig. 6), PP consists of a dense filament of galaxies at redshift $\sim 5000$ km s$^{-1}$, centered on $l \sim 110°$, $b \sim -30°$. W91 obtained velocity widths for PP galaxies from the published 21 cm data of Giovanelli & Haynes (1985,1989) and Giovanelli *et al.* (1986). The W91 PP sample was concentrated in a relatively narrow ($\sim 15°$) strip in galactic latitude, stretching over $\sim 90°$ in longitude. PP lies nearly opposite on the sky from the GA at a similar distance, but was not well sampled in the 7S study. The PP region thus represented a significant uncharted volume in the velocity field as mapped by the elliptical data.

W91 used the cluster galaxies to calibrate the R band TF relation, assuming (following the finding of the Aaronson group) that the Northern Sky clusters were at rest in the CMB. When applied to the PP galaxies, the TF data revealed a coherent motion directed toward the LG, with a mean amplitude of $\sim 350$ km s$^{-1}$. Superposed on the coherent streaming were compressional motions presumably due to the overdense PP filament, but the bulk motion was the most prominent feature of the data. Because of the location of PP, the streaming motion was directed not only toward the LG, but also approximately toward the GA. The amplitude of this streaming was much larger than what was predicted at PP ($\sim 150$ km s$^{-1}$) by the FB88 GA model. W91 suggested that the residual motion might be due to very large-scale, low amplitude density fluctuations, but noted that at the largest distances in his sample ($\sim 10,000$ km s$^{-1}$), objects appeared to be at rest in the CMB frame.

---

[33] The very small Virgo infall amplitude found by FB88 remains controversial. Tonry *et al.* (1992) have applied the SBF technique to local ellipticals and found an infall velocity of $340 \pm 80$ km s$^{-1}$ at the LG. Work on the subject will undoubtedly continue in the coming years.

The small solid angle covered by the W91 PP sample raised the question of whether the PP inflow extended over a larger region of the sky. Confirmation that it did was provided by the R band CCD TF survey of Courteau and Faber (Courteau 1992; Courteau et al. 1993; CF), who obtained data for 321 galaxies distributed over the Northern sky using a dilute sampling approach. CF used optical H$\alpha$ rotation curves to measure velocity widths. Because of their large sky coverage and very uniform selection criteria, CF were able to compare different regions of the sky reliably. Well away from PP their data suggested a relatively quiescent velocity field, as indicated by the small scatter of the TF relation when distances were modeled by pure Hubble flow. However, within the PP region—now defined to cover an area several times larger than the W91 sample—the scatter was larger and the residuals systematically offset from Hubble flow. The implied velocity was $\sim 300$–$350$ km s$^{-1}$ toward the LG, in agreement with W91. Further confirmation of a bulk motion came from Han & Mould (1992), who used the I band CCD TF relation to study five clusters embedded in the PP supercluster. They found PP to be moving towards the LG with amplitude $\sim 400$ km s$^{-1}$.

Han, Mould and collaborators also obtained I band TF data for 16 further clusters distributed over much of the sky (Han 1992; Mould et al. 1991). This sample of clusters largely overlapped those studied using the IRTF by Aaronson et al. (1986,1989), but the I band relation was considered more accurate. Han & Mould (1992) fitted the combined data for 21 clusters to two models: a large-scale bulk streaming similar to the original model proposed by the 7S group (Dressler et al. 1987a); and a "bi-infall" model in which the local universe is dominated by competing gravitational pulls of the GA and Perseus-Pisces. They found that both models were consistent with the data.

Connolly (1993) measured Tully-Fisher parameters for a sample of 218 Sc galaxies in both Galactic hemispheres, with mean redshift of 5500 km s$^{-1}$. He fits the data to a dipole and finds good agreement with the analysis of a much larger sample by Courteau et al. (1993).

The largest single contribution of Tully-Fisher data in recent years has been that of Mathewson and collaborators (Mathewson et al. 1992ab, hereafter MAT92; Mathewson & Ford 1994). Using I band CCD photometry and a combination of H I and optical velocity widths from several telescopes in Australia, MAT92 obtained TF data for 1355 Southern sky galaxies. Roughly half of these were in the steradian around the GA region as defined by the 7S group, and another half, thought of as a control group, were outside. The sample was constructed to be roughly complete to an ESO diameter limit of 1.7$'$. Most objects in the sample have redshifts $\lesssim 4000$ km s$^{-1}$, but a substantial number have $5000 \lesssim cz \lesssim 7000$ km s$^{-1}$. MAT92 calibrated their TF relation relative to $\sim 40$ galaxies with TF data from Aaronson et al. (1989), Willick (1991), and Han & Mould (1992).

The redshift vs. TF distance diagram for galaxies in the "control" region exhibited pure Hubble flow in the mean. Within the GA region, MAT92 confirmed the large ($\sim 1000$ km s$^{-1}$) outflowing motions in the GA foreground that had been detected in the 7S elliptical data. However, the spirals did not appear to trace a velocity field which fell to zero at the distance ($\sim 4500$ km s$^{-1}$) of the nominal Great Attractor. Instead, the data indicated that the GA itself (i.e., galaxies in the redshift range $\sim 4000$–$5000$ km s$^{-1}$) possessed a mean positive peculiar velocity of $\sim 600$ km s$^{-1}$. Furthermore, the MAT92 data did not exhibit the "backside infall" which Dressler & Faber (1990ab) had claimed to detect: even at distances $\gtrsim 5000$ km s$^{-1}$ a return to the Hubble line was not apparent. MAT92 argued that their data, in combination with those of Willick (1990), suggested a vast bulk flow in the Supergalactic plane on scales well in excess of 10,000 km s$^{-1}$.

However, a recent reanalysis of the MAT92 data, as well as further data from the Mathewson group itself, have not lent support to this hypothesis. MAT92 did not attempt a careful correction for selection or Malmquist biases; in addition, their zeropoint calibration used only a small number of independent comparisons. Using a careful zeropointing technique which ties together the distance scale of several TF samples, Willick et al. (1995abc) have recalibrated the TF relations used by MAT92 as well as other TF samples (see § 7.2). Preliminary analyses of the recalibrated data (Courteau et al. 1993; Faber et al. 1993, 1994) have derived modest ($\sim 300$–$350$ km s$^{-1}$) bulk flows in the GA direction, as well as evidence of a return to the Hubble line at the distance of the GA. The backside infall, however, remains undetected in these recent studies. Very recently, Mathewson & Ford (1994) have enlarged their Southern sky sample by $\sim 1000$ galaxies. The new sample is selected to an ESO diameter of $1'$ and thus probes to greater distances ($cz \lesssim 10,000$ km s$^{-1}$). Mathewson & Ford measure a bulk flow in the GA direction of $\sim 370$ km s$^{-1}$ for the combined sample, in good agreement with Courteau et al. (1993), and a return to the Hubble line at $cz \simeq 7000$ km s$^{-1}$. The new data have not been reanalyzed using the techniques of Willick et al.; this will be carried out in the near future.

### 7.1.4 The Lauer-Postman Result

The most dramatic apparent detection of very large-scale streaming to date comes from the work of Lauer & Postman (1994, hereafter LP). Using the BCG $L$-$\alpha$ relation (§ 6.3.2), LP obtained redshift-independent distances to a volume-limited sample of 119 Abell clusters in the redshift range $cz < 15,000$ km s$^{-1}$. The typical distance error from the method was estimated by LP as $\sim 16\%$, so that the peculiar velocity error for a typical sample cluster is $\sim 2000$ km s$^{-1}$, much too large to estimate individual cluster peculiar velocities. However, the data do allow the determination of a mean bulk velocity vector for the sample.

LP defined an "Abell Cluster Inertial Frame," or ACIF, and measured the velocity of the LG $\mathbf{L}$ relative to it. They solved for $\mathbf{L}$ by minimizing the $\chi^2$-like quantity [34]

$$E = \sum_{i=1}^{N} \frac{(2-\alpha_i)^2}{z_i^2} \left[ cz_i \left( 10^{\frac{0.4 \Delta M(\alpha_i)}{2-\alpha_i}} - 1 \right) - \mathbf{L} \cdot \hat{\mathbf{g}}_i \right]^2 \qquad (200)$$

where:
 (i) $\alpha_i$ is the $\alpha$-value for galaxy $i$;
 (ii) $z_i$ its redshift in the LG frame;
 (iii) $\Delta M(\alpha_i)$ is its residual (in magnitudes) from the $L$-$\alpha$ relation;
 (iv) $\hat{\mathbf{g}}_i$ is a unit vector toward the galaxy; and
 (v) $\mathbf{L}$ is the motion of the LG with respect to the ACIF.
The minimization is carried out with respect to the parameters specifying the $L$-$\alpha$ relation (which determine the quantity $\Delta M(\alpha)$), and the three components of $\mathbf{L}$. The quantity $(2 - \alpha)$ which appears both in the weighting term and in the exponent arises from the fact that both $L$ and $\alpha$ must be measured within *metric* apertures; as a result the effective scatter of the relation is actually given by $2\sigma_M/(2-\alpha)$ where $\sigma_M = 0.24$ mag is the scatter about the relation. The $z^{-2}$ term gives greatest weight to the nearby clusters, although LP show the solution is insensitive to the redshift weighting.

LP found that the LG motion relative to the ACIF was in the direction $l = 220°$, $b = -28°$ (total angular uncertainty $\pm 27°$). That the LG would have a significant motion relative to an inertial rest frame is of course no surprise; we already know that the LG moves at $\sim 600$ km s$^{-1}$ relative to the CMB. However, the latter motion is in the direction $l = 276°$, $b = 30°$ (with negligible angular uncertainty). Consequently, according to LP the ACIF differs significantly from the frame defined by the CMB dipole. Stated another way, if the CMB defines the true cosmic rest frame, the LP data indicate that the Abell clusters out to 15,000 km s$^{-1}$ possess a net peculiar motion of 689 ± 178 km s$^{-1}$ in the direction $l = 343°$, $b = +52$ (total angular uncertainty ± 23°). Aware of the controversial nature of their result, LP performed a suite of tests aimed at identifying systematic effects in their analysis. Among other things, they considered the effects of cutting their sample at low and high redshifts, and at low and high values of $\alpha$, of removing prominent superclusters (such as the Hydra-Centaurus complex) from their sample, of using a linear rather than a quadratic $L$-$\alpha$ relation, and of removing galaxies with very large residuals. They found that none of these tests produced significant changes in their solution. They did identify one non-negligible systematic effect called "geometry bias," which arises due to coupling between the dipole moment of the sky distribution of the sample and the distance indicator relation. LP calibrated and corrected for this effect using Monte Carlo experiments; in practice, it was rather small ($< 100$ km s$^{-1}$) since their sample has excellent sky coverage.

---

[34] Colless (1995) derives a more accurate version of Eq. (200); his reanalysis of their data gives results differing only slightly from those of Lauer & Postman (1994).

One could take the view that a $\sim 700$ km s$^{-1}$ bulk motion of a volume $\sim 30,000$ km s$^{-1}$ in diameter is highly implausible, and attribute it instead to an incorrect choice of cosmic rest frame. One would then be obliged to jettison the conventional view that the CMB dipole is kinematic in origin; possible explanations of a *cosmological* origin for the CMB dipole have been suggested by Gunn (1988), Paczyński & Piran (1990), and Turner (1991). While the possibility that the CMB does not define a cosmic rest frame cannot be excluded, such a view has become increasingly difficult to uphold in light of the discovery of small but measurable CMB anisotropies by the COBE satellite (Smoot *et al.* 1992). These COBE measurements have corroborated most theoretical ideas about the nature of the CMB.

Alternatively, if one accepts the kinematic origin of the CMB dipole and the resultant CMB-defined rest frame, the LP data indicate a $\sim 5\%$ perturbation to the Hubble flow on a 15,000 km s$^{-1}$ scale. This is difficult to reconcile with bulk flow observations from the TF and $D_n$-$\sigma$ relations. Courteau *et al.* (1993) used the Mark III dataset (§ 7.2) to estimate the bulk flow within 6000 km s$^{-1}$ to have an amplitude of 360 ± 40 km s$^{-1}$ in a nearly orthogonal direction, toward $l \simeq 300°$, $b \simeq +10°$. It is difficult to understand how the amplitude of the bulk flow could grow with scale in a universe which approaches uniformity in the galaxy distribution in the large, especially given the apparent convergence of the dipole in the galaxy distribution between 6000 and 15,000 km s$^{-1}$(§ 5.7).

The LP result also poses severe problems for a number of currently popular theories for the growth of large-scale structure. Strauss *et al.* (1995) have generated mock LP-like BCG samples in N-body simulations based on six different cosmogonic scenarios, including CDM, HDM, and PIB models. These mock samples were constructed to mimic all the observational errors in the actual LP sample. Strauss *et al.* found that no more than 2–5% of the realizations of any given model exhibited an apparent flow as large as that measured by LP. Similar results were found by Feldman & Watkins (1994) using an analytic approach. Jaffe & Kaiser (1995) suggest comparing not just the bulk flow, but the full sample of peculiar velocities, to the predictions of various models. They find better agreement between the LP data and models than do Strauss *et al.* and Feldman & Watkins. However, as Jaffe & Kaiser ignore the correlations between the measured peculiar velocities of different clusters, which arise from the self-calibrating nature of the LP sample, the validity of their analysis is uncertain.

The very careful treatment by LP makes it unlikely that the observed flow will "go away" with a future reanalysis of the data. Rather, it seems likely that future measurements on comparably large scales (§ 9.7) will determine the ultimate evaluation of LP, as was the case with the Rubin *et al.* (1976ab) result. If LP is confirmed, it will need to be reconciled with the discrepant flows seen on smaller scales, and theoretical models are likely to require significant revision.

*7.1.5 Bulk Flows: A Summary*

The myriad of bulk flow analyses we have discussed here have used a wide variety of samples, methodologies and techniques, and have been carried out during a time when our knowledge of the properties of distance indicators, and the various biases which plague them, was evolving rapidly (as indeed it still is). The bulk flows measured by different samples necessarily probe different volumes of space; a direct comparison between them is problematic, and we have not attempted one in this review. Nevertheless, there is a rough consensus between the different workers in the field as to the nature of the bulk flow within 6000 km s$^{-1}$ (cf., the review by Postman 1995), both in its amplitude and in its direction. Even the Lauer-Postman bulk flow, which is on appreciably larger scales than the others, agrees very roughly in direction. The surprising thing about this result is the fact that it also agrees roughly in amplitude, on such a large scale.

One of our aims in measuring bulk flows is to compare with theoretical predictions, as given for a given model by Eq. (40). This equation refers to the *volume-weighted* average of the three-dimensional velocity of particles within a spherical volume, which can differ substantially from the *galaxy- and error-weighted* average of radial velocities that is usually measured. Indeed, as Juszkiewicz et al. (1990) and Strauss et al. (1995) point out, one can have a bulk flow within a volume even in the absence of density fluctuations on larger scales, if the center of mass of the volume does not coincide with its geometric center. There are a number of such subtleties which are only now being appreciated in this game; further theoretical analyses along the lines of Kaiser (1988; 1991) and Feldman & Watkins (1994) are needed.

Of the samples discussed here, only that of Lauer & Postman (1994) approached being volume-limited (cf., Roth 1993, 1994). However, one can measure an effective volume-limited bulk flow if one weights the galaxies appropriately. The POTENT technique (to be discussed in detail in § 7.5) does exactly that, and Bertschinger et al. (1990) quote volume-limited bulk flows measured within a series of spheres centered on the Local Group. More recent results using POTENT applied to a more complete data set by Dekel et al. (1995) show a bulk flow of 400 km s$^{-1}$ within a sphere of radius 6000 km s$^{-1}$ sphere, which is in fact consistent with standard CDM normalized to COBE. Again, the Lauer-Postman result, which probes appreciably larger scales, is in disagreement with this and all other viable models; we suspect that measurements of bulk flows on large scales will continue to be a very active area of research for at least the next five years.

## 7.2 Homogeneous Peculiar Velocity Catalogs

Tully-Fisher and $D_n$-$\sigma$ data have been collected, at an increasing rate, over the past fifteen years. There now exists a large number of independent data sets potentially useful in peculiar velocity surveys. This trend will accelerate in the coming years, as most ordinary spiral and elliptical galaxies out to $\sim 10{,}000$ km s$^{-1}$ acquire quality photometric and spectroscopic measurements. There is and will be a need for uniform catalogs containing the TF and $D_n$-$\sigma$ observable data, and the corresponding estimated distances and peculiar velocities. Distances obtained using the newer DIs such as the PNLF, SBF, BCG, and SNe methods will eventually be cataloged as well.

The greatest problem in constructing a peculiar velocity catalog is ensuring the *homogeneity* of the separate samples. The originally published inferred distances to galaxies found in two or more independent data sets are almost sure to differ systematically, for two basic reasons. First, observers measure the relevant quantities—apparent magnitudes, diameters, and internal velocity widths—differently; they not only use different observational methods, but apply different corrections (*e.g.* for internal extinction) to the raw observables. As a result, there are as many TF relations (say) as there are data acquisition and reduction techniques. Nonetheless, a TF relation derived for one data set is sometimes heedlessly applied to another. Second, observers determine the appropriate TF relation (say) from their data in different ways. Some fit an inverse relation, some a forward relation; in each case, the treatment (if any) of selection bias can differ. A naïvely constructed peculiar velocity catalog, in which inferred distances are simply appropriated from the published literature, will thus be plagued with systematic errors within and between data sets.

Until recently, attempts to produce uniform peculiar velocity catalogs from the literature have been largely the work of D. Burstein, a member of the 7S collaboration. Two early versions known as the Mark I and Mark II, were distributed privately to the community by electronic mail in 1987 and 1989 respectively. Burstein tabulated the $D_n$-$\sigma$ elliptical galaxy data of the 7S group, Lucey & Carter (1988) and Dressler & Faber (1990a), as well as the infrared TF spiral data of the Aaronson group. Burstein was able to make direct comparisons of the $D_n$-$\sigma$ data from the different groups. However, it was difficult to guarantee that the spiral and elliptical data were on the same system; comparison of TF and $D_n$-$\sigma$ *group* distances was used to make small adjustments, but these comparisons suffered from the inevitable uncertainty associated with the spatial segregation of spiral and elliptical galaxies. Although not optimal, the Mark II catalog played an important role in some of the major statistical studies of peculiar velocity in recent years (e.g., Bertschinger et al. 1990; Dekel et al. 1993).

With the advent of new large samples of peculiar velocity data, there has been a strong need to extend Burstein's effort, particularly in view of the more complete understanding of bias effects acquired in recent years (§ 6.5). A group led by one of the present authors (J.W.) has recently completed this task. The basic methodology is described in Willick et al. (1995ab); the *Mark III Catalog of Galaxy Peculiar Velocities* is presented in Willick et al. 1995c, and is available for electronic distribution as well. The Mark III consists of the CCD TF samples of Han, Mould, and collaborators (Han 1991, 1992; Han & Mould 1992; Mould et al. 1991,1993; collectively, HM); Willick (1991, W91); Courteau and Faber (Courteau 1992; CF); and Mathewson et al. 1992b; and of the infrared TF sample of Aaronson et al. (1982b), as recently reanalyzed by Tormen & Burstein (1994, 1995), who corrected previously unrecognized systematic errors in the diameter system used by the Aaronson group (see § 6.1.1). The Mark III catalog will also include expanded versions of the elliptical $D_n$-$\sigma$ samples found in the Mark II.

Several basic principles guide the construction of the Mark III catalog. The first is the adoption of a uniform set of rules for correcting the raw observable quantities, in particular apparent magnitude and velocity width, for the effects of extinction, inclination, and redshift. Internal extinction corrections have been a source of particular controversy in the past (see Burstein et al. 1995 for a review); the Mark III compilation adopts bandpass-specific internal extinctions which minimize TF scatter. The second principle is to re-derive a Tully-Fisher relation for each sample separately, taking full account of and correcting for selection biases in the manner suggested by Willick (1994; cf. § 6.5). An important result of this re-derivation is that the TF relations in the Mark III compilation can differ markedly from those of the original authors. Third, the zeropoints (the constant $A$ in Eq. 152) of the individual TF relations are mutually adjusted to yield the maximum agreement in the inferred distances for several hundred galaxies which are common to two or more individual samples. Finally, the global catalog zeropoint is determined by assuming that in the sample as a whole, the volume-weighted radial peculiar velocity component vanishes in the mean.

A representation of the Mark III velocity field is shown in Fig. 16, which shows the measured radial peculiar velocities of all galaxies within 22.5° of the Supergalactic plane. The point is drawn at the measured distance of the galaxy, while the line is drawn to its redshift, in the CMB rest frame. Positive peculiar velocities are drawn with solid points and solid lines, while negative peculiar velocities use open points and dashed lines. When possible, galaxies are grouped in order to decrease the errors and reduce the Malmquist bias; points representing groups of more than three galaxies are drawn somewhat larger. The zone of avoidance is apparent as the missing wedge out of the middle of the figure. Compare this figure with the *IRAS* density field of Fig. 6; we have labeled the major structures as we did in that figure. The bulk flow into the GA is apparent, as is the coherence of the flow even back to the Pisces-Perseus supercluster. It is difficult, however, to draw quantitative conclusions from this figure alone about bulk flows or infalls into specific structures. In the remainder of this chapter, we discuss various statistical analyses of peculiar velocity data, culminating in reconstruction methods of the full three-dimensional velocity field in § 7.5.

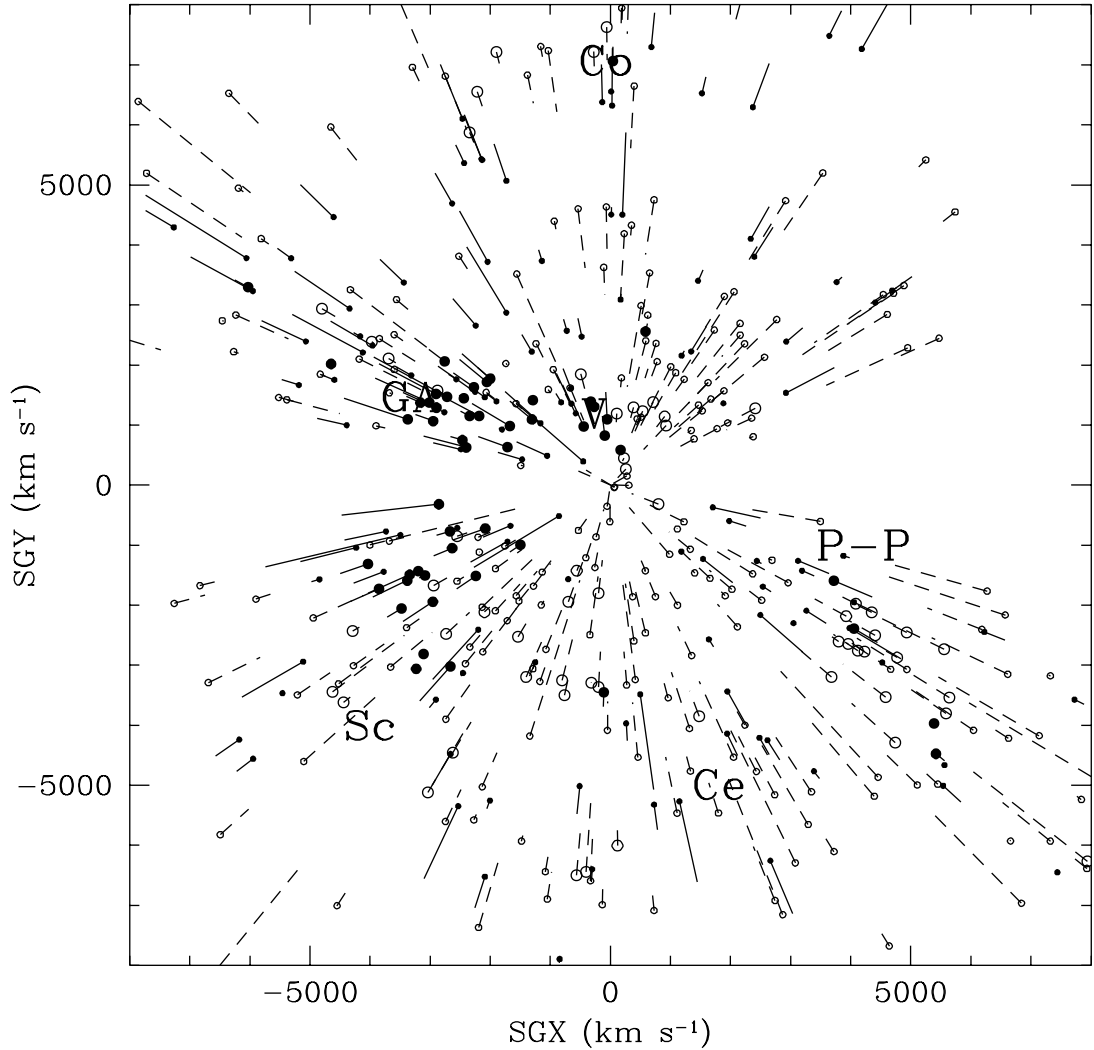

Fig. 16. The Mark III peculiar velocities of all galaxies within 22.5° of the Supergalactic plane. The point is drawn at the measured distance of the galaxy, while the line is drawn to its redshift, in the CMB rest frame. Positive peculiar velocities are drawn with solid points and solid lines, while negative peculiar velocities use open points and dashed lines. Points representing groups of more than three galaxies are drawn somewhat larger.

*7.3 Velocity Correlation Function*

One of the most striking features of the observed large-scale velocity field is its coherence. One way to quantify this is by fitting bulk flows to the data (§ 7.1). Another approach was suggested by Górski (1988): the correlation function of the velocity field. Górski defines the correlation function between the $i$th and $j$th Cartesian coordinate of the velocity field as:

$$\Psi_{ij}(\mathbf{r}) \equiv \langle v_i(\mathbf{x}) v_j(\mathbf{x} + \mathbf{r}) \rangle = \Psi_\perp(r)\delta_{ij} + \left[\Psi_\|(r) - \Psi_\perp(r)\right] \hat{r}_i \hat{r}_j, \tag{201}$$

where $\Psi_\perp$ and $\Psi_\|$ are the transverse and radial correlation functions. The second equality holds if the velocity field is homogeneous and isotropic. If the velocity field is derivable from a potential, then these

two are not independent:

$$\Psi_\|(r) = \frac{\mathrm{d}[r\Psi_\perp(r)]}{\mathrm{d}r}. \tag{202}$$

In linear perturbation theory, one can derive simple expressions for these quantities:

$$\Psi_\perp(r) = \frac{H_0^2 f(\Omega_0)^2}{2\pi^2} \int P(k) \frac{j_1(kr)}{kr}\,\mathrm{d}k, \tag{203}$$

and

$$\Psi_\|(r) = \frac{H_0^2 f(\Omega_0)^2}{2\pi^2} \int P(k) \left(j_0(kr) - 2\frac{j_1(kr)}{kr}\right)\,\mathrm{d}k; \tag{204}$$

compare with Eq. (40). Of course, we observe only the radial component of the velocity field. Górski et al. (1989) and Groth, Juszkiewicz, & Ostriker (1989) have suggested methods to determine the velocity correlation function from observational data. The first of these papers defines a quantity $\psi$ from a sample with radial peculiar velocities $u_i$:

$$\psi(r) \equiv \frac{\sum_{\text{pairs }(r)} u_1 u_2 \cos\theta_{12}}{\sum_{\text{pairs }(r)} \cos^2\theta_{12}}, \tag{205}$$

where the sum is over all pairs of galaxies with separations between $r$ and $r + \mathrm{d}r$, and $\theta_{12}$ is the angle between galaxies 1 and 2 on the sky. A uniformly selected full-sky sample exhibiting a bulk flow of amplitude $v$ will have a velocity correlation function $\psi_1 = v^2/3$. The second equality of Eq. (201) implies that $\psi(r)$ is a linear combination of $\Psi_\|(r)$ and $\Psi_\perp(r)$, with coefficients depending on the spatial distribution of galaxies in the sample.

The velocity correlation function at zero lag is a measure of the root-mean-square velocity dispersion of galaxies, while the scale on which it drops to zero is a measure of the coherence length of the velocity field. Górski et al. (1989) looked at two datasets: the spiral galaxies of Aaronson et al. (1982) and the elliptical galaxies of the 7 Samurai. The correlation function $\psi$ of both drop to zero at separations of 2000 km s$^{-1}$. The amplitude of the elliptical galaxy dataset has a much larger amplitude than that of the spirals, although this amplitude was not robust: deleting a small number of galaxies in the Great Attractor region caused the amplitude to drop by more than a factor of two. These results were compared to $N$-body simulations of CDM and PBI models. The CDM models fit the observed correlation length well, and required a normalization $\sigma_8 > 0.5$ to match the amplitude. The PBI models tended to show *more* coherence than is seen in the real data.

Groth et al. (1989) used the linear relation between $\psi$ and $\Psi_\perp, \Psi_\|$ to solve for the latter. They emphasized the fact that in the frame comoving with the bulk flow of the Aaronson et al. (1982) data, the velocity correlations were essentially zero; the flow was very cold. This is consistent with the small value of the pairwise velocity dispersion deduced from redshift surveys (§ 5.2.1). Seen in this light, the correlation length of 2000 km s$^{-1}$ reported by Górski et al. (1989) is probably at least partly an edge effect due to a finite sample. It is time to revisit the velocity correlation function now that the data samples have improved and expanded. More work is needed to characterize the effects of survey geometry, and in particular, peculiar velocity errors, on this statistic. It has the potential to place strong constraints on cosmological models, once these various effects are understood better.

Juszkiewicz & Yahil (1989) point out that the comparison of the velocity and the spatial correlation function yields an estimate of $\beta$; in particular, in linear theory:

$$\left\langle [\mathbf{v}(\mathbf{r}_1) - \mathbf{v}(\mathbf{r}_2)]^2 \right\rangle = 2[\Psi_{ii}(0) - \Psi_{ii}(r_{12})] = \frac{\beta^2 H_0^2}{2\pi} \int_0^{r_{12}} \frac{J_3(x)}{x^2} \, dx. \tag{206}$$

Note that the velocity correlations in this form on scale $r$ depend only on the the correlation functions on scales smaller than $r$. Thus accurate measurements of the velocity correlation function and the spatial correlation function have the potential to yield a measurement of $\beta$. Moreover, the extent to which the two sides of Eq. (206) agree with one another as a function of $r$ on linear scales is a test of gravitational instability theory. Unfortunately, existing data are not yet at the stage to allow this test to be done.

## 7.4 The Cosmic Mach Number

Ostriker & Suto (1990), struck by the coldness of the velocity field observed in the Aaronson *et al.* (1982) data, suggested a new statistic to quantify this coldness: the Cosmic Mach Number. The Mach number in standard usage is the ratio of the flow velocity in some medium to the sound velocity in that medium. In the cosmological context, the equivalent of the sound velocity is the small-scale velocity dispersion of galaxies. Following Eq. (40), the characteristic bulk velocities in a given cosmological model measure the large-scale component of the power spectrum, while the small-scale velocity dispersion depends on the power on small scales. Their ratio is thus independent of the *amplitude* of the power spectrum (at least in linear theory), and is a diagnostic of its *shape*. Strauss, Cen, & Ostriker (1993; cf. Suto, Cen, & Ostriker 1992) fit bulk flows to observed datasets; all components of the velocity field on smaller scales were attributed to incoherent small-scale peculiar velocities. Subtracting off the estimated errors in quadrature from the rms of the residuals allowed them to define a small-scale velocity dispersion, and thus a Mach number. Strauss *et al.* (1992) compared the Mach number results from three different datasets to the distribution of Mach numbers observed in Monte-Carlo simulations of the observational data. The Aaronson *et al.* (1982) sample, with its relatively small errors, gave the strongest constraints on models; standard CDM was ruled out at the 95% confidence level by this statistic. This is largely a consequence of the fact that it greatly over-predicts the velocity dispersion on small scales, as was discussed in § 5.2.1. Other models with less power on small scales relative to large, including tilted CDM and HDM, fared much better by this statistic.

## 7.5 Reconstructing the Three-Dimensional Velocity Field

The bulk flows discussed in § 7.1 represent one quantitative statistic that can be extracted from observations of peculiar velocities. One would like a method to characterize all the information available in the velocity field. In particular, given gravitational instability theory (Eq. 30 or quasi-linear extensions thereof), the observed velocity field gives a measure of the gravitating density field, independent of any assumptions about the relative distribution of galaxies and dark matter. Bertschinger & Dekel (1989) have developed a technique they call *POTENT* (cf. Dekel 1994 for a review) which starts from the basic assumption that the observed velocity field is derivable from a potential $\Phi(\mathbf{r})$ such that

$$\mathbf{v}(\mathbf{r}) = -\nabla \Phi. \tag{207}$$

This is valid to the extent that the velocity field is curl-free; Kelvin's circulation theorem implies that vorticity is generated only in regions of shell-crossing. Moreover, initial vorticity decays in an expanding universe just as do initial peculiar velocities. Thus we expect that at the present time, if we smooth on large enough scales, the vorticity is likely to be negligible. In this case, the radial component of the velocity field (which is all that is observable) determines the full three-dimensional velocity field. If $u(r, \theta, \phi)$ is the observed radial velocity field, then the potential (normalized to zero at the origin) is given by:

$$\Phi(\mathbf{r}) = -\int_0^r u(r', \theta, \phi) \, dr'; \tag{208}$$

differentiation via Eq. (207) then yields the full three-dimensional velocity field. Indeed, there is no reason to restrict the integration in Eq. (208) to radial rays; Simmons et al. (1994) discuss optimal integration paths for recovering the potential.

Given the three-dimensional velocity field, linear theory gives a simple relation to the density field (Eq. 30). The POTENT method as currently implemented uses a non-linear generalization of this following Nusser et al. (1991):

$$\delta_P(\mathbf{r}) = f(\Omega_0) \left( \left\| \mathbf{I} - f(\Omega_0)^{-1} \frac{\partial \mathbf{v}}{\partial \mathbf{r}} \right\| - 1 \right), \tag{209}$$

where $\mathbf{I}$ is the unit matrix; Eq. (209) reduces to $\nabla \cdot \mathbf{v}$ in the linear limit.

Of course, we do not observe a radial velocity field $u(\mathbf{r})$; we have noisy data for a non-uniform and sparsely sampled set of galaxies. Thus another crucial part of the POTENT technique involves turning the data available into a continuous radial velocity field. This smoothing has several features:
 (i) It allows the radial velocity field to be defined at every point in space;
 (ii) It smoothes over small-scale vorticity in the velocity field due to shell-crossing;
(iii) It smoothes over very non-linear effects that are not modeled by Eq. (209);
(iv) It averages together the very noisy individual peculiar velocity measurements.
Dekel, Bertschinger, & Faber (1990) use a tensor window function smoothing that takes into account the radial nature of the observed velocity field. There are three particularly pernicious sources of error in the resulting smoothed velocity field:
 (i) Statistical noise in the velocity field due to the errors in the individual peculiar velocities. One minimizes this noise by weighting each galaxy by the inverse square errors;
 (ii) Malmquist bias, both homogeneous and inhomogeneous, resulting from the peculiar velocity errors (§ 6.5.2); POTENT uses peculiar velocities derived via Method I applied to forward DIs;
(iii) Sampling gradient bias, due to the inhomogeneous sampling of the velocity field within a smoothing window. This is minimized using equal volume weighting.
Note that the effects of items (i) and (iii) are minimized with different weightings; one can never minimize both simultaneously. In practice, the effects of these sources of noise and biases for any given dataset and smoothing scheme are calculated using extensive Monte-Carlo simulations. First results of the POTENT technique are presented in Bertschinger et al. (1990), using the Mark II data. The resulting density field, smoothed with a 1200 km s$^{-1}$ Gaussian window, clearly shows the Great Attractor, and the void in the foreground of the Pisces-Perseus supercluster. The smoothed velocity field and resulting density field as gained from the Mark III data (Dekel et al. 1995) are shown in Fig. 17.

The POTENT method and results have been used for a number of other studies, now using the Mark III data described in § 7.2. We have already referred to its use to trace the density field at low Galactic latitudes (§ 3.8), and to compare the velocity fields as traced by elliptical and spiral galaxies separately (§ 6.2). Seljak & Bertschinger (1994) have used a maximum-likelihood method to fit the amplitude of fluctuations in the Mark II POTENT density field, assuming a given power spectrum. The analysis is complicated by the fact that data points are coupled not just by real correlations, but also by noise, necessitating a full Monte-Carlo approach to the covariance matrix. A summary of their results for a range

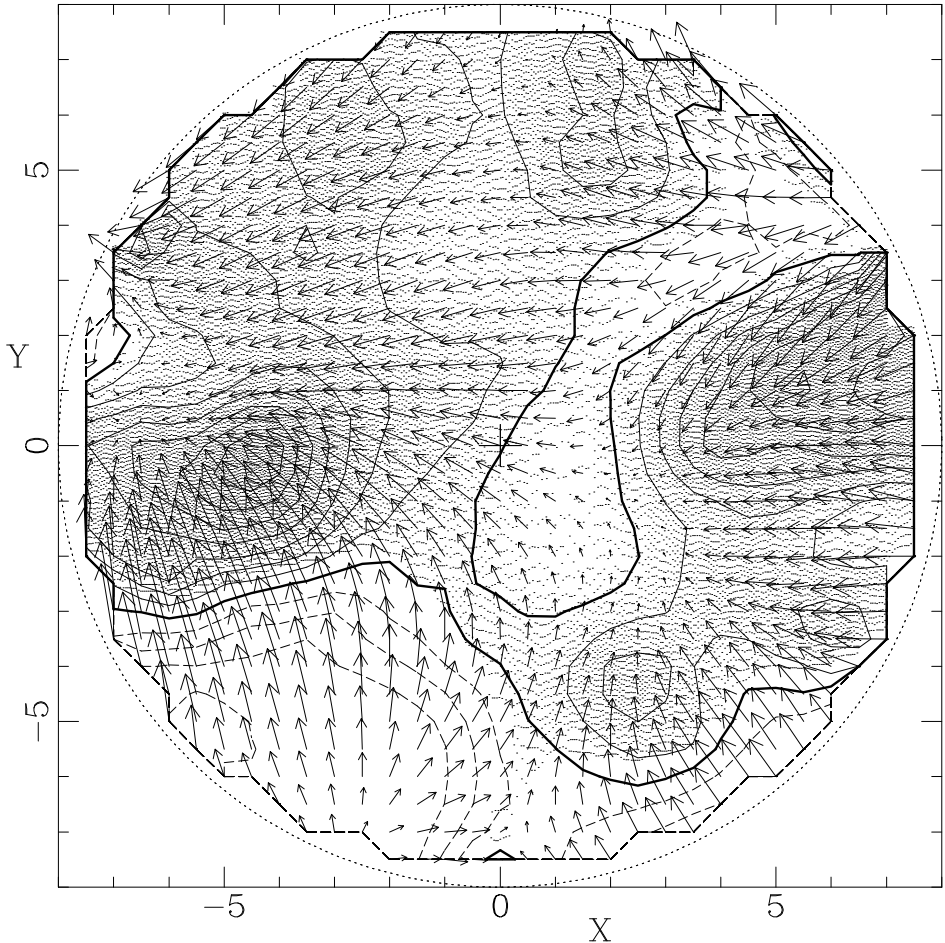

Fig. 17. The smoothed velocity field and resultant density field from the Mark III data, in the Supergalactic Plane. The smoothing used was a 1200 km s$^{-1}$ Gaussian. The arrows give the X-Y components of the three-dimensional velocity field, while the contours are of $\delta$ assuming $\Omega_0 = b = 1$, spaced at intervals of 0.2.

of plausible CDM-like models is

$$\sigma_{8,v}\Omega_0^{0.6} = 1.3^{+0.4}_{-0.3}, \qquad (210)$$

where $\sigma_{8,v}$ is the rms value of $\nabla \cdot \mathbf{v}$ in spheres of radius 8 $h^{-1}$ Mpc. This result is in good agreement with standard CDM, normalized to the COBE quadrupole.

*7.5.1  The Initial Density Distribution Function*

In § 5.2.2, we described a method developed by Nusser & Dekel (1993) which uses the Zel'dovich approximation to reconstruct the initial density distribution function from the eigenvalues of the spatial derivatives of the velocity field. It can be shown that this reconstruction is independent of $\Omega_0$ when reconstructing from the velocity field *predicted* from a redshift survey, while the reconstruction from the *observed* velocity field is $\Omega_0$-dependent; in particular, the shape of the derived initial density distribution function will depend on the value of $\Omega_0$ used. Nusser & Dekel (1993) carried out a preliminary analysis of the Mark II POTENT velocity field using this technique, and found that they needed values of $\Omega_0$ close to unity in order to match the Gaussian distribution function seen in the reconstruction from the density field. That is, for small values of $\Omega_0$, the distribution function reconstructed from the observed density and velocity fields did not agree at all. Nusser & Dekel (1993) used this approach to rule out $\Omega_0 < 0.3$ models at the 4-6 $\sigma$ confidence level; the data are consistent with $\Omega_0 = 1$. Further tests are needed, however, to test whether very low $\Omega_0$ models, in which the derived errors become large, can be ruled out as well. The Mark III data should give much superior results.

*7.5.2 Higher-Order Moments of the Velocity Field*

In linear perturbation theory with Gaussian initial conditions, we saw that the density field showed a Gaussian distribution. Given the direct proportionality between $\delta$ and $\theta \equiv \nabla \cdot \mathbf{v}$ (Eq. 30), $\theta$ will also have a Gaussian distribution. However, just as the distribution of $\delta$ develops skewness in second-order perturbation theory, so does $\theta$. Bernardeau (1994ab; cf. Bouchet et al. 1994) calculates the higher-order moments of $\theta$ in perturbation theory for top-hat smoothing (cf. Bernardeau et al. 1994 and Łokas et al. 1994 for Gaussian smoothing), and shows that the ratio of the skewness to the variance squared is given by:

$$S_{3\theta} = -\frac{1}{\Omega_0^{0.6}} \left[ \frac{26}{7} - (\gamma_1 + 3) \right] \tag{211}$$

(compare with Eq. 124; like that result, this is valid only in the range $-3 \leq \gamma_1 < 1$). Unlike the case of the density field, the skewness of the velocity field is strongly dependent on $\Omega_0$. Unlike the density-velocity comparisons discussed below, galaxy biasing does not affect the results (at least for equal volume weighting). Bernardeau et al. (1994) have made a tentative measurement of the skewness of the POTENT density field from the Mark III data, and conclude that the data are consistent with $\Omega_0 = 1$, with $\Omega_0 = 0.3$ ruled out at the 2 $\sigma$ level. Because the skewness is so heavily weighted by the tails of the distribution, and the substantial sources of errors and biases in POTENT will strongly affect these tails, this result remains tentative, and will require extensive Monte-Carlo simulations to test it thoroughly. Higher-order moments remain unmeasurable from current data.

*7.5.3 Voids in the Reconstructed Density Field*

The dimensionless density field $\delta$ has a firm lower limit, $-1$, corresponding to the absence of matter. In linear theory, $\delta$ is proportional to the divergence of the velocity field $\theta$, which says that $\theta$ also has a lower limit, depending only on the proportionality constant $f(\Omega_0)$. Comparison with lowest observed point in the POTENT maps thus puts a lower limit on $f(\Omega_0)$, again independent of galaxy biasing. Dekel & Rees (1994) have applied this idea to the Sculptor Void seen in the Mark III POTENT data (cf., Fig. 6), and find that $\Omega_0 > 0.3$ at the 2.4 $\sigma$ level. The systematics of this method have not yet been properly treated, however. In particular, as there is a strong correlation between $\delta_{\text{galaxies}}$ and $\theta$, the voids that one wants to use for this test are in those regions where there are fewest galaxies, and thus the noise in the POTENT maps are highest. In addition, these are regions which potentially suffer from strong inhomogeneous Malmquist bias, precisely because the galaxy density field shows strong gradients in voids.

*7.5.4 Other Approaches to Reconstructing the Velocity Field*

The POTENT approach can be thought of as a parameterized fit to the velocity field, in which the velocity field in each smoothing volume is fit to a bulk flow (plus shear terms; cf. the discussion in Dekel et al. 1994). An alternative approach is to expand the velocity field in Fourier modes. Kaiser & Stebbins (1991) and Stebbins (1994) use a Bayesian approach, regularizing their solution for the Fourier coefficients by assuming that the velocities are drawn from a Gaussian distribution with a given power spectrum.

Indeed, this regularization is equivalent to applying a Wiener filter to the data, and thus has the same feature as we saw above in § 3.7: the derived density field goes to zero in regions of poor data. The results they get are a strong function of the power spectrum assumed; we await a detailed exposition of their technique in refereed journals.

As a Method I technique (Table 3) the POTENT method has the serious drawback that it assumes an *a priori* distance indicator relation, calibrated independently of the data in question. A small error in the slope of the Tully-Fisher relation, for example, will cause systematic errors in the derived velocity field. An alternative approach is to solve simultaneously for the parameters of the distance indicator relation and the velocity field, in a Method II approach. In the first paper that takes into account all the selection effects in such a problem, Han & Mould (1990) fit the Aaronson *et al.* (1982) peculiar velocity data to a model involving infall into the Virgo cluster and the Great Attractor (cf., Faber & Burstein 1988). We describe a generalization of their technique in § 8.1.3 using the *IRAS* predicted velocity field.

Nusser & Davis (1994b) suggest an expansion of the radial velocity field in spherical harmonics and radial spherical Bessel functions. They find linear combinations of the basis functions that are orthonormal at the positions of the galaxies for which data exist, allowing them to find an analytic solution for the Tully-Fisher parameters and the coefficients of these orthonormalized functions *as measured in redshift space* by minimizing the scatter in the inverse Tully-Fisher relation (thereby eliminating selection bias). Small-scale noise and triple-valued zones eliminate the one-to-one mapping between real space and redshift space, causing a bias in the derived velocity field, although this seems to be a small effect with real data. This offers an alternative method to smooth peculiar velocity data, and may be the ideal way to compare with the predicted velocity field from redshift surveys using the Nusser & Davis (1994a) approach (cf., Davis & Nusser 1995). The applicability of the derived velocity field to points other than those where data exist remains unclear, and the method is limited to data characterized by a single distance indicator relation (or at least is analytic only in this limit). These problems are not fatal by any means, and this method holds great promise.

# 8 Comparing the Density and Velocity Fields

In this penultimate chapter, we bring the results of the previous chapters together with a discussion of analyses involving both the velocity and density fields. This can be done either by predicting the velocity field from redshift surveys using Eq. (150), and comparing with the observed velocities (§ 8.1), or predicting the density field from peculiar velocity surveys using Eq. (149), and comparing with the observed redshift surveys (§ 8.2).

## 8.1 *Comparison via the Velocity Field*

### 8.1.1 *Cluster Infall Models*

Much of the motivation for measuring the velocity field has been to compare it to models of what is expected given the density field. Our historical review of our gradual understanding of the nature of the large-scale flow field focussed on measurements of bulk flows, but in fact much of the motivation of the early work was for measurements of cluster infall. A spherically symmetric cluster embedded in a homogeneous medium induces a spherically symmetric radial velocity field. In linear theory, the cluster infall velocity is simply given by (Eq. 33) :

$$v(r) = \frac{H_0 \Omega_0^{0.6}}{3} r \bar{\delta}(r), \qquad (212)$$

where $r$ is the distance to the center of the cluster, and $\bar{\delta}(r)$ is the mean overdensity within $r$. In fact, for a top-hat initial density perturbation (i.e., a spherically symmetric overdensity that is constant in amplitude out to some given radius), the evolution can be calculated exactly (Silk 1974, 1977; Schechter 1980; Bertschinger 1985ab; Regös & Geller 1989) by writing the evolution of the perturbation and the background density as separate isotropic expanding or contracting bodies, and matching the boundary conditions at the edge of the tophat. For the case of an open universe with an initial mean tophat overdensity $\bar{\delta}_i > 1$ at a time when the Hubble constant was $H_i$ and the density parameter was $\Omega_i$, the radial velocity field at time $t$ is

$$v(r) = \frac{r}{t} \frac{\sinh\theta(\sinh\theta - \theta)}{(\cosh\theta - 1)^2}, \qquad (213)$$

where $\theta$ is defined implicitly by the equation:

$$\sinh\theta - \theta = \frac{2tH_i\Omega_i^{1/2}\left[\bar{\delta}_i - (\Omega_i^{-1} - 1)\right]^{3/2}}{1 + \bar{\delta}_i}. \qquad (214)$$

Similar equations can be written down for a closed universe. Much of the early work on interpreting results of velocity fields concentrated on fitting these formulae to the infall around the Virgo cluster (Tonry & Davis 1981ab; Aaronson *et al.* 1982b; Davis *et al.* 1982; Tully & Shaya 1984; Tammann & Sandage 1985; Gudehus 1989; cf., the review of Davis & Peebles 1983a). There has been a great deal of controversy in the literature about the amplitude of the cluster infall detected, with characteristic numbers at the Local Group ranging from 100 km s$^{-1}$ to 450 km s$^{-1}$ (§ 7.1); this, together with the uncertainty in the overdensity of the Virgo cluster itself in galaxies (Sandage, Tammann, & Yahil 1979; Davis *et al.* 1982; Strauss *et al.* 1992a), has meant that values of $\beta$ determined from Virgocentric infall have been equally uncertain. Bushouse *et al.* (1985) and Villumsen & Davis (1986) used $N$-body models to test the ability of this method to constrain $\Omega_0$, and concluded that it works to the extent that one's peculiar velocity data surrounds $4\pi$ steradians of the cluster, otherwise, shear motions from more distant mass concentrations can strongly bias the results.

In the meantime, studies of the Virgo cluster have shown that the approximation of it as an isolated spherically symmetric cluster is less and less applicable. It has been known for years that it shows appreciable substructure on the sky; accurate distances to Virgo galaxies have shown that it probably has appreciable depth (at least in spiral galaxies), which may contribute to some of the controversy as to its distance, and the intrinsic scatter of the Tully-Fisher relation (Pierce & Tully 1988; Fukugita, Okamura, & Yasuda 1993). Moreover, the velocity field around it is affected by other, more distant, mass concentrations and voids; in particular, Lilje *et al.* (1986) demonstrated the presence of a tidal field in the Aaronson *et al.* (1982a) data from what we now interpret as the Great Attractor.

Spherically symmetric cluster infall models are starting to be applied to more distant clusters. Kaiser (1987) and Regös & Geller (1989) showed that cluster infall causes characteristic caustics in the redshift-space maps of galaxies around clusters, although the structure in the galaxy distribution in the field in which the cluster is embedded can make these caustics difficult to identify. Careful measurement of these caustics has the potential to yield the linear cluster infall velocity, which, together with the measured overdensity of galaxies in the clusters, will yield $\beta$. However, it is not yet clear how practical this approach is given the complicated effects of intrinsic small-scale structure in the galaxy distribution outside the clusters themselves.

There have been a variety of attempts to go beyond the single cluster model for the velocity field by invoking two or more clusters (e.g., Faber & Burstein 1988; Han & Mould 1990; Rowan-Robinson *et al.* 1990), or even to fit the flow field around a void (Bothun *et al.* 1992). But given the availability of redshift surveys covering much of the sky, we can trace out the full density field at every point in space (at some modest smoothing length) in the local universe, and compare the resulting predicted velocity field (§ 5.9) with observations.

### 8.1.2 Unparameterized Velocity Field Models

In this section and the next, we describe the most direct comparisons between peculiar velocity and redshift surveys. Linear theory gives a relation between galaxy density and peculiar velocity (Eq. 33), which can be used to derive a velocity field from a redshift survey (§ 5.9). The resulting velocity field can be compared point-by-point with measured peculiar velocities in a Method I analysis (usually using the forward DI relations); the slope of the resulting scatter plot is thus in principle a measure of $\beta$. This process is actually somewhat subtle: sampling the predicted peculiar velocity field at the measured distance of each galaxy gives biased results, due to the substantial errors in the distances. Rather, one should calculate the predicted peculiar velocities given the redshifts to each object, by inverting the predicted redshift-distance diagram along each line of sight. This is subject to the ambiguities of triple-valued zones (Fig. 11). Furthermore, because the self-consistent velocity field from the redshift survey predicts the peculiar velocity of the Local Group, this comparison is best made in the frame in which the Local Group is at rest. This will give different results from a comparison in the frame in which the CMB shows no dipole, because the *IRAS* velocity field does not exactly match the peculiar velocity of the Local Group (§ 5.7). Much of the early work in this field was done before a proper understanding of selection and Malmquist biases were at hand (§ 6.5), making these results somewhat suspect.

Strauss (1989) compared the *IRAS* 1.936 Jy predicted velocity field with the Mark II peculiar velocity data. A strong correlation between observed and peculiar velocities was seen, and the slope was consistent with $\beta = 0.8$. Similar results were found by minimizing the scatter in the inverse Tully-Fisher relation for the Aaronson *et al.* (1982a) data in a Method II analysis. However, the error in the derived $\beta$ was not properly quantified; nor for that matter was it demonstrated that the scatter was consistent with the observational errors.

Kaiser *et al.* (1991) used a Method I approach to compare the velocity field from the QDOT redshift survey with the Mark II data. The QDOT density field (and therefore predicted velocity field) was corrected for redshift space distortions not by iterations, as in § 5.9, but rather by applying a correction to the smoothed density field at each point taken from Kaiser (1987) (cf, Eq. 91). The resulting predicted peculiar velocity field was compared to the Mark II data, binned on the same grid used to define the density field; least-squares fits yielded a slope $\beta = 0.86 \pm 0.14$.

A similar approach was taken by Hudson (1994b) who used the density field of optically selected galaxies to obtain a predicted peculiar velocity field to compare with the Mark II data. He used the techniques of Hudson (1994a) to correct the data for inhomogeneous Malmquist bias, assuming the galaxies for which peculiar velocities were measured to be drawn from the same density distribution as in his maps. He also included in his models a bulk flow from scales beyond those surveyed. He concluded from his peculiar velocity scatter plots that $\beta = 0.50 \pm 0.06$, with an additional bulk flow of 405 km s$^{-1}$ towards $l = 292°$, $b = +7°$. However, this derived bulk flow is almost in the Galactic plane, in the direction of the Great Attractor, and thus may be due to overdensities not surveyed by Hudson's sample. On the other hand, this bulk flow is roughly consistent with that observed from the Mark III data, so it may indeed represent flows on scales larger than his sample.

Shaya, Tully, & Pierce (1992) also compared peculiar velocities (in this case, from a combination of the Aaronson *et al.* (1982a) data and their own TF data; Tully, Shaya, & Pierce 1992) with a redshift survey, namely the catalog of galaxies within 3000 km s$^{-1}$ compiled by Tully (1987b). They used luminosity rather than number weighting, and carried out an elaborate analysis which includes components to the density field clustered on a variety of length scales. They concluded that $\Omega_0$ associated with galaxies is only 0.1, clustered on 1 $h^{-1}$ Mpc scales. However, their modeling of the effects of mass concentrations in the Zone of Avoidance, and beyond 3000 km s$^{-1}$, is simplistic, and their resulting predicted peculiar velocity field only bears qualitative resemblance to that measured. A more recent analysis is presented by Shaya, Peebles, & Tully (1994), using Peebles' (1989) variational technique to extend the linear theory relation between the density and velocity fields. They conclude that $\beta_{optical} < 0.4$, but emphasize that their analysis is still in progress, and awaits improved Tully-Fisher data.

Roth (1993; 1994) carried out a Tully-Fisher survey in the $I$ band of 91 galaxies selected from a volume-limited subset of galaxies within 4000 km s$^{-1}$ from the 1.936 Jy *IRAS* redshift survey, and minimized the scatter of the forward Tully-Fisher relation as a function of $\beta$ in the *IRAS* velocity field model, using a Method II approach. Extensive Monte-Carlo simulations demonstrated that this method gives an unbiased estimate of $\beta$; he found $\beta \sim 0.6$. Unfortunately, systematic errors in the line-width data, and the dominance of the triple-valued zone around the Virgo cluster (which is prominent in the sample) mean that the systematic errors associated with this result are large. Schlegel (1995) is extending the survey to contain $\sim 250$ galaxies with accurately measured line widths from H$\alpha$ rotation curves, and with more uniform sky coverage; this dataset promises to give tighter constraints on $\beta$.

Finally, Nusser & Davis (1994a) compared the predicted dipole moment of their multipole expansion of the *IRAS* velocity and density field (cf., Eq. 145) with that measured from the POTENT map. They show that the dipole of a shell *as measured in the Local Group frame* depends only on the density field interior to that shell, making this a semi-local comparison. They conclude $\beta = 0.6 \pm 0.2$, although the error was estimated by eye from their plots. This is a promising way to proceed, especially with their more sophisticated technique for determining the multipole moments of the measured velocity field (§ 7.5.4).

### 8.1.3 Method II$^+$

Common to the handful of Method II velocity comparisons discussed above is the assumption of a unique redshift–distance mapping, as required in a Method II analysis (§ 6.4.3). In the real world, however, a distance cannot be unambiguously assigned from an observed redshift—even when the peculiar velocity model is "correct." In what follows, we explain why this is so, and describe a maximum likelihood method to overcome the problems that result.

One can distinguish two contributions to a galaxy's peculiar velocity. The first is what is usually meant by the peculiar velocity "field." It has a coherence length of a few Mpc or greater, is due to perturbations in the linear or quasi-linear regime, and is predictable from an analysis of density fluctuations. The second is what is loosely referred to as velocity "noise." It has zero coherence length, arises from strongly nonlinear processes, and is unpredictable except in a statistical sense. We label the coherent part $\mathbf{v}(\mathbf{r})$, and describe the random part in terms of an rms radial velocity dispersion $\sigma_v$. Each can separately invalidate the assumption of a unique redshift-distance mapping, as follows:

(i) Method II assumes that the distance $r$ to a galaxy is the "crossing point" in the redshift-distance diagram, given implicitly by

$$r = cz - \hat{\mathbf{r}} \cdot \mathbf{v}(\mathbf{r}) \qquad (215)$$

(cf. § 6.4.3). There is, however, no guarantee that this equation will have only one solution. When line of sight gradients in $\mathbf{v}(\mathbf{r})$ are of order unity, there can be three or more crossing points for given $cz$. Such regions are generically called "triple-valued zones" (cf., Fig. 11).

(ii) Even if Eq. 215 has a unique solution $r$, this solution will differ from the true distance because of velocity noise. In essence, $cz$ is a random realization of the "redshift field" $r + \hat{\mathbf{r}} \cdot \mathbf{v}(\mathbf{r})$, and only defines the crossing point to accuracy $\sim \sigma_v$.

The situations just described are summarized in Fig. 11, which shows a triple-valued zone around the Virgo cluster. The inherent uncertainty due to velocity noise is indicated with the scatter of points.

Because of these effects, Method II is subject to biases over and above selection bias (§ 6.4). Willick *et al.* (1995d) have developed a modified form of Method II which neutralizes these biases by explicitly allowing for non-uniqueness in the redshift-distance mapping. The basic idea is to derive correct probability distributions of observable quantities, taking into account the complexities of the redshift-distance relation, and then to maximize likelihood over the entire data set. This approach shares features of both Methods I and II (§ 6.4.3), but is closer in spirit to the latter, and will accordingly be called "Method II$^+$" in what follows.

We assume the goal is to fit a model peculiar velocity field $\mathbf{v}(\mathbf{r};\mathbf{a})$, where $\mathbf{a}$ is a vector of free parameters, and adopt the useful abbreviation

$$u(r;\mathbf{a}) = \hat{\mathbf{r}} \cdot \mathbf{v}(\mathbf{r};\mathbf{a}) \tag{216}$$

for its radial component. The central element of Method II$^+$ is a description of the redshift-distance relation in terms of a Gaussian probability distribution:

$$P(cz|r) = \frac{1}{\sqrt{2\pi}\,\sigma_v} \exp\left(-\frac{[cz - (r + u(r;\mathbf{a}))]^2}{2\sigma_v^2}\right) . \tag{217}$$

A related probability function is given as a function of $r$ in the lower half of Fig. 11 (that curve shows the probability distribution in Eq. (144), which differs from $P(cz|r)$ by the additional factors of $n(\mathbf{r})r^2\Phi(4\pi r^2 f_{\min})$). Three subtleties of Eq. (217) deserve mention. First, $\sigma_v$ is not merely the true velocity noise, whose value is thought to be $\sim 150$ km s$^{-1}$ (e.g., Groth *et al.* 1989), but rather its convolution with two additional effects: redshift measurement errors and velocity model errors. The former are small (typically $\sim 50$ km s$^{-1}$) but not entirely negligible. The latter, which reflect the finite accuracy of our predictions, can be estimated from N-body simulations (*e.g.*, Fisher *et al.* 1994d) and are of order 200 km s$^{-1}$. Second, $\sigma_v$ is not necessarily constant; both the true velocity noise and model errors are larger in dense regions. Third, whereas true velocity noise is incoherent, model prediction errors are not. Contributions to $\mathbf{v}(\mathbf{r})$ arising on scales too small to be included in the model, but which unlike true noise have spatial coherence, manifest themselves as coherent prediction errors. In what follows, we will neglect these subtleties and treat $\sigma_v$ as a spatial constant of order 200 km s$^{-1}$ whose value may be held fixed or treated as a free parameter in the likelihood analysis.

We quantified Method II selection bias (§ 6.5) based on the probability distribution $P(m,\eta,r)$, arguing that we could in effect treat redshift as distance. Using Eq. (217), we may now write down the joint distribution of the TF observables, distance, *and* redshift:

$$P(m,\eta,r,cz) = P(m,\eta|r) \times P(cz|r) \times P(r) , \tag{218}$$

where we have assumed that the TF observables and redshift couple only via their mutual dependence on true distance. The observables in a redshift-distance sample are $m$, $\eta$, and $cz$. Their distribution is obtained by integration:

$$P(m,\eta,cz) = \int_0^\infty P(m,\eta|r)$$
$$\times \frac{1}{\sqrt{2\pi}\,\sigma_v} \exp\left(-\frac{[cz - (r + u(r;\mathbf{a}))]^2}{2\sigma_v^2}\right) n(r)r^2 \, dr . \tag{219}$$

Eq. (219) gives the likelihood of a data point in a redshift-distance sample, valid for arbitrary $\mathbf{v}(\mathbf{r};\mathbf{a})$ and $\sigma_v$. Method II$^+$ consists of maximizing the product of the likelihood (Eq. 219) over the peculiar velocity sample, with respect to the parameters of the velocity field model, the parameters of the TF relation (including its scatter), and $\sigma_v$.

The overall likelihood depends on TF probability evaluated not only at the crossing point(s), but over a range of distances roughly characterized by

$$|cz - r - u(r; \mathbf{a})| \sim \sigma_v \,. \tag{220}$$

This likelihood will differ from its Method II counterpart to the extent that the length scale on which $P(m, \eta, r)$ varies is comparable to or smaller than the interval defined by Eq. (220). The former scale is given by $\sim \Delta d$, where $\Delta$ is the TF fractional distance error and $d = 10^{0.2[m-M(\eta)]}$ is the inferred distance (§ 6.5.2). For galaxies beyond 3000 km s$^{-1}$, $\Delta d \gtrsim 600$ km s$^{-1}$. This is considerably *larger* than the range given by Eq. (220) for typical $\sigma_v$, outside triple valued or flat zones. In these circumstances Methods II and II$^+$ differ little. Indeed, it is easy to see that—again away from triple-valued or flat zones—Method II$^+$ reduces exactly to Method II (Eq. 165) in the limit $\sigma_v \to 0$. However, Method II$^+$ represents a substantial correction to classical Method II at small ($d \lesssim 2000$ km s$^{-1}$) distances, in triple-valued or flat zones, or when $\sigma_v$ becomes anomalously large. Method II$^+$ is therefore necessary for rigorous analysis of the very local universe and, in particular, of the Local Supercluster region, where small distances and triple-valuedness are often combined.

Application of Method II$^+$ requires two additional steps. First, one must decide whether to use the forward or inverse form of the TF relation, and thus whether Eq. (171) or Eq. (188) is used for $P(m, \eta, r)$ in Eq. (219). As in Method II (§ 6.4), the inverse method is advantageous when sample selection is independent of velocity width[35], though with Method II$^+$ this choice introduces a new uncertainty discussed below. Second, one does not apply Eq. (219) directly, but instead derives from it suitable *conditional* probabilities: $P(m|\eta, cz)$ in the forward case, $P(\eta|m, cz)$ in the inverse. These are obtained from Eq. (219) as follows:

$$P(m|\eta, cz) = P(m, \eta, cz) \Big/ \int_{-\infty}^{\infty} \mathrm{d}m \, P(m, \eta, cz) =$$

$$\frac{\int_0^\infty \mathrm{d}r \, r^2 n(r) \, P(cz|r) \, S(m, \eta, r) \exp\left(-\frac{[m-(M(\eta)+\mu(r))]^2}{2\sigma^2}\right)}{\int_0^\infty \mathrm{d}r \, r^2 n(r) \, P(cz|r) \int_{-\infty}^{\infty} \mathrm{d}m \, S(m, \eta, r) \exp\left(-\frac{[m-(M(\eta)+\mu(r))]^2}{2\sigma^2}\right)} \tag{221}$$

and

$$P(\eta|m, cz) = P(m, \eta, cz) \Big/ \int_{-\infty}^{\infty} \mathrm{d}\eta \, P(m, \eta, cz) =$$

$$\frac{\int_0^\infty \mathrm{d}r \, r^2 n(r) \, \Phi(m - \mu(r)) \, P(cz|r) \, S(m, \eta, r) \exp\left(-\frac{[\eta - \eta^0(m-\mu(r))]^2}{2\sigma_\eta^2}\right)}{\int_0^\infty \mathrm{d}r \, r^2 n(r) \, \Phi(m - \mu(r)) \, P(cz|r) \int_{-\infty}^{\infty} \mathrm{d}\eta \, S(m, \eta, r) \exp\left(-\frac{[\eta - \eta^0(m-\mu(r))]^2}{2\sigma_\eta^2}\right)} \,, \tag{222}$$

where we have restored the compact notation $P(cz|r)$ (Eq. 217), reversed the order of integration in the denominators, and allowed for an explicit $r$-dependence of sample selection (which in fact exists for some of the Mark III samples; cf. § 6.5.3). The integrals over $m$ and $\eta$ in the denominators of Eqs. (221) and (222) can be done analytically for simple forms of $S$ (Willick 1994), so two-dimensional integrations are not required. The chief advantage of conditional probabilities is that they are less sensitive than $P(m, \eta, cz)$ to the precision with which the number density $n(r)$ is modeled. As $n(r)$ appears in both numerator and denominator, it only weakly affects the conditional distributions provided it varies slowly compared with $P(cz|r)$ or the TF probability term, as will generally be the case. This is important, since it is the accuracy of our velocity model, not our density model, with which we are mainly concerned. Still, Method II$^+$ (unlike Method II) does require a density model, and in that sense resembles a Method I analysis. Finally, note that the velocity width distribution function $\phi(\eta)$ (§ 6.5) strictly cancels out of $P(m|\eta, cz)$, while the luminosity function $\Phi(M)$ (§ 6.5.4) does not cancel out of $P(\eta|m, cz)$ (although the results are insensitive to the luminosity function, as it again appears in numerator and denominator).

---

[35] In principle, Method II$^+$ rigorously incorporates selection effects into either the forward or inverse formalism. However, with real data characterization of sample selection is often subject to uncertainty (Willick *et al.* 1995ab). Its relatively smaller susceptibility to selection bias thus remains a virtue of the inverse approach.

Willick et al. (1995d) used Method II$^+$ to analyze data from the Mark III peculiar velocity catalog (§ 7.2). They limited their analysis to the 900 galaxies in the TF samples of Mathewson et al. (1992b) and Aaronson et al. (1982b), which densely sample the local region, and to redshifts $\leq 3500$ km s$^{-1}$. The forward method (Eq. 221) was used, and the TF parameters for each sample (slope, zeropoint, and scatter) were allowed to vary in the search for maximum likelihood, rather than fixing them at their values determined in the Mark III analysis (§ 7.2).

The *IRAS* 1.2 Jy predicted peculiar velocity field was taken as the model to be fitted to the data. To first order this model velocity field depends only on $\beta$, although the analysis was carried out for two different smoothing lengths, and with and without nonlinear effects included. Using the *IRAS* galaxy density field for the quantity $n(r)$ that appears in Eqs. (221) and (222) assumes that *IRAS* galaxies are distributed like the Mark III sample objects. This assumption may not be correct in detail but is expected to have a minimal effect on our conditional analysis. The forward Method II$^+$ analysis was carried out in terms of a quantity $\Lambda$ defined by

$$\Lambda = -2 \sum_i \ln P(m_i | \eta_i, cz_i) \tag{223}$$

where the sum runs over all galaxies used in the comparison. An analogous statistic using the inverse relation is discussed by Willick et al. and gives consistent results.

We summarize these results in Fig. 18, which shows $\Lambda$ vs. $\beta$ curves for four realizations of the *IRAS* fields: 300 and 500 km s$^{-1}$ gaussian smoothing lengths, each with and without nonlinear corrections, following Nusser et al. (1991), and setting $b = 1$. The "best" values of $\beta$ occur at the minima of the curves; they range from $\beta \simeq 0.48$ for 300 km s$^{-1}$ smoothing linear to $\beta \simeq 0.65$ for 500 km s$^{-1}$ smoothing nonlinear. The *statistical* error at 95% confidence level associated with these numbers is $\sim 0.15$, as determined by deviations of the likelihood function from its minimum. Note that there are systematic effects acting as well: a larger smoothing length leads to a larger $\beta$, and nonlinear corrections yield larger $\beta$ for a given smoothing length. These results are to be expected: both increasing the smoothing scale and making nonlinear corrections *decrease* the amplitude of predicted peculiar velocities, and thus are qualitatively similar to decreasing $\beta$. For the nonlinear fields, the predicted velocities now depend separately on $\Omega_0$ and $b$; the curves here were obtained for the case $b = 1$, i.e., $\beta = \Omega_0^{0.6}$.

Which of the four *IRAS* predicted velocity fields shown should ideally be used in estimating $\beta$? In Method II$^+$ we are comparing the predicted velocity field with unsmoothed data. The field obtained from 300 km s$^{-1}$ smoothing thus gives rise to the most valid comparison here. At smaller smoothing lengths the nonlinear corrections cease to be valid. Nonlinear effects must be present, as the overdensities at 300 km s$^{-1}$ smoothing can be considerably in excess of unity. However, Fig. 18 shows that the nonlinear curve has formally smaller likelihood than the linear curve. The reason for this is not well understood at present. A compromise value is $\beta \simeq 0.55$, roughly midway between the linear and nonlinear minima. It is not clear how to combine the systematic and statistical errors we have identified; for now we conservatively put the 95% confidence level at $\pm 0.25$.

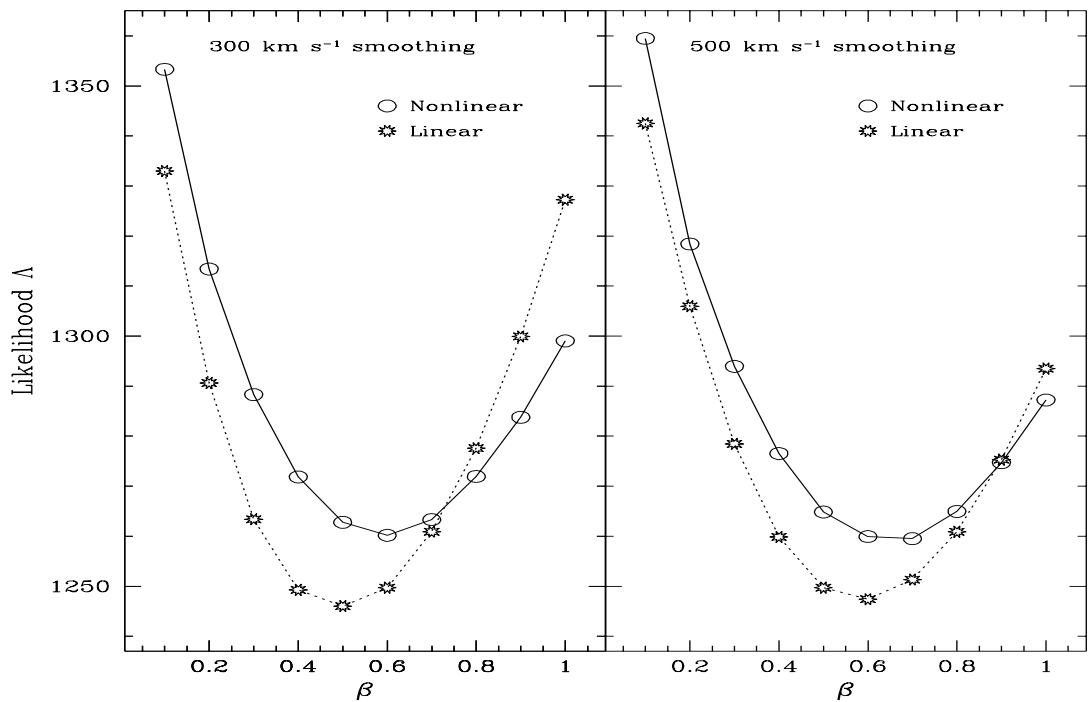

Fig. 18. Likelihood $\Lambda$ plotted as a function of $\beta$ for four realizations of the *IRAS* predicted peculiar velocity and density fields. Likelihoods have been computed using a Method II$^+$ comparison with a subset of the Mark III peculiar velocity catalog (see text for details). The left hand panel depicts results in which a 300 km s$^{-1}$ smoothing length was used in the *IRAS* velocity and density field reconstruction; a 500 km s$^{-1}$ smoothing length was used in the right hand panel. Each panel shows the results for linear (dotted line) and nonlinear (solid line) predictions. The latter were obtained by holding the bias factor fixed at $b = 1$.

As mentioned above the value of the rms velocity noise $\sigma_v$ was treated as a free parameter, its final value determined by maximizing likelihood for each $\beta$. Willick *et al.* found that $\sigma_v \simeq 150$–$160$ km s$^{-1}$ for the likelihood-maximizing values of $\beta$. This value is remarkably small, in view of the fact that the *IRAS* predictions themselves are thought to have rms errors of 100–200 km s$^{-1}$, as discussed above; the implication would appear to be that the true noise is $\lesssim 100$ km s$^{-1}$. It is unlikely that the velocity field is in reality that cold. It is likely instead that by neglecting correlated model prediction errors (see above), the likelihood analysis ends up underestimating $\sigma_v$. Future implementations of both Method II$^+$ (and Method II, which similarly assumes uncorrelated residuals) will need to address this issue.

*8.2 Comparison via the Density Field*

The reconstruction of the mass density field using the POTENT method begs a comparison with the galaxy density field as observed in redshift surveys. Indeed, to the extent that the two fields are proportional to one another, their ratio gives a measure of $\beta$ via Eq. (149). Dekel *et al.* (1993) carried out this comparison, using the Mark II POTENT maps of Bertschinger *et al.* (1990) and the *IRAS* 1.936 Jy redshift survey. The first, and most striking result from this comparison, is that the POTENT and IRAS density field show qualitative agreement. Given the noise and sparseness in the peculiar velocity data, the POTENT map has much greater noise than does the *IRAS* map, and therefore the region in which the comparison of the two can be made is limited. Nevertheless, both show the Great Attractor and the

void in front of the Perseus-Pisces supercluster. With 1200 km s$^{-1}$ Gaussian smoothing, there are $\sim 10$ independent volumes within which the comparison of the two density fields can be made. A scatter plot of the two shows a strong correlation. In the absence of biases, the slope of the regression would be an estimate of $\beta$. However, as discussed in § 7.5, the POTENT density field is subject to a number of biases. The most severe of these in this context is Sampling Gradient bias, with inhomogeneous Malmquist bias taking a close second. One can quantify the first by sampling the *IRAS* predicted velocity field at the positions of the Mark II galaxies, and running the results through the POTENT machinery; comparing the resulting density field to the input density field yields a regression slope of 0.65, substantially different from unity. Given this fact, Dekel *et al.* used an elaborate maximum likelihood technique to quantify the agreement between the POTENT and *IRAS* density fields. For a given value of $\beta$ (actually given values of $\Omega_0$ and $b$; Eq. (209) rather than Eq. (149) is used throughout), they create mock Mark II datasets given the *IRAS* predicted peculiar velocity field (via the method described in § 5.9), noise is added, and the results are fed into POTENT. Note that these simulations suffer from sampling gradient bias exactly as does the real POTENT data. The slope and scatter of the regression between the original *IRAS* map and the mock POTENT map are recorded, and the two-dimensional distribution of these two quantities is calculated for 100 such realizations. Elliptical fits to this distribution allows them to calculate the probability that the observed slope and scatter of the regression is consistent with the model assumed (i.e., the *IRAS* predicted peculiar velocity field for the given values of $\Omega_0$ and $b$ are consistent with the observed peculiar velocity field, given the errors). This process is then repeated for a grid of values of $\Omega_0$ and $b$.

The conclusions from this work are as follows:
(i) There exist values of $\Omega_0$ and $b$ for which the likelihood that the model is is consistent with the data is high. That is, the observed scatter is consistent with the assumed errors and the assumption that *IRAS* galaxies are at least a biased tracer of the density field that gives rise to the observed peculiar velocities.
(ii) The tightest constraint is on $\beta$, for which the likelihood curve gives $\beta = 1.28^{+0.75}_{-0.59}$ at 95% confidence.
(iii) Not surprisingly, the constraints on $\Omega_0$ and $b$ separately are quite a bit weaker. Indeed, the likelihood contour levels do not close at large $\Omega_0$ for a given $\beta$, meaning that the data are consistent with purely linear theory in which no second-order effects exist. The data are *inconsistent* with highly non-linear conditions (i.e., very small $\Omega_0$ for a given $\beta$), but this is the regime in which the non-linear approximation used starts to break down anyway.

May we then conclude that the gravitational instability picture has been proven? After all, we have seen consistency in the data with one of its strongest predictions, namely Eq. (30). However, as Babul *et al.* (1994) emphasize, Eq. (30) can be derived directly from the first of Eqs. (21), the continuity equation alone; only the constant of proportionality comes from gravity. Indeed, $\delta$ and $\nabla \cdot \mathbf{v}$ are proportional for any model in which galaxies are a linearly biased tracer of the mass, and for which the time-averaged acceleration is proportional to the final acceleration. They show analytically and with the aid of simulations that a range of models with non-gravitational forces exhibit correlations between $\delta$ and $\nabla \cdot \mathbf{v}$ at least as strong as that seen in the POTENT-*IRAS* comparison. Thus a proof of the gravitational instability picture will require ruling out these alternative models by other means.

Work is in progress as this review is being written, to update the POTENT results using the Mark III peculiar velocity compilation. The Mark III data have been corrected for inhomogeneous Malmquist bias, assuming that the *IRAS* density field is that of the galaxies of the Mark III sample. More importantly, however, the systematic errors in the overlap between datasets have been minimized, and the volume surveyed well with the Mark III data is such that the *IRAS*-POTENT comparison can be done over four times as many data points as before. In addition, the more complete sampling means that the sampling gradient bias is smaller than with the Mark II data, by roughly a factor of two in the mean. The left-hand panel of Fig. 19 shows a preliminary version of the POTENT density field in the Supergalactic plane using 1200 km s$^{-1}$ Gaussian smoothing, taken from Fig. 17. The right hand panel shows the density field of the *IRAS* 1.2 Jy survey at the same smoothing. The qualitative agreement is remarkable. In the Supergalactic plane, both maps show the Great Attractor, the Perseus-Pisces Supercluster, the Coma-A1367 Supercluster, as well as voids between Coma and Perseus, and South of the Great Attractor (the Sculptor Void). Work is ongoing to quantify the differences between the two maps, and to put exact error bars on the derived $\beta$.

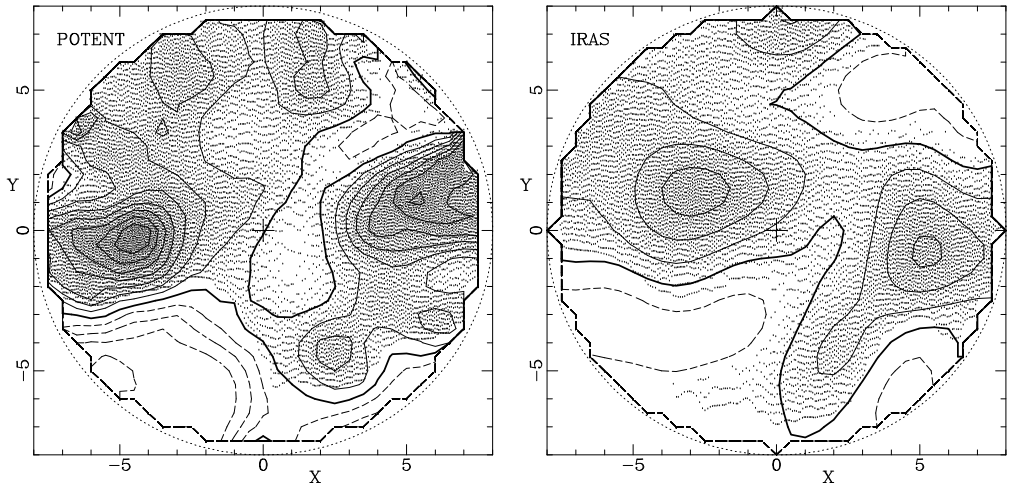

Fig. 19. The left panel is the POTENT density field $\nabla \cdot \mathbf{v}$ in the Supergalactic plane from the Mark III peculiar velocity data. The right panel is the independently determined density field of *IRAS* galaxies. The smoothing in both panels is 1200 km s$^{-1}$. The axes are labeled in 1000 km s$^{-1}$. The Local Group sits at the center of each panel.

Hudson *et al.* (1995) have carried out a comparison of the Mark III POTENT results with the optical galaxy density field of Hudson (1993a,b). They also find good agreement. between the two density fields; a less elaborate analysis than that of Dekel *et al.* (1993) shows $\beta = 0.74 \pm 0.13$.

## 9 Discussion

In this brief concluding chapter, we summarize what it is that has been learned from redshift and peculiar velocity surveys, and put the results into the context of the larger field of observational cosmology.

*9.1 The Initial Power Spectrum*

We have put constraints on the power spectrum directly from measurements of the distribution of galaxies (§ 5.3). The fact that power spectra derived from redshift surveys in different areas of the sky, using different samples, agree, implies that it is meaningful to define a power spectrum in the first place. That is, we do not live in a simple fractal universe, in which a mean density depends on the scale on which it is measured. Moreover, our samples are starting to become big enough that our statistical measures are not completely dominated by sampling fluctuations, at least for measures probing relatively small scales (50 $h^{-1}$ Mpc and smaller). This is not to say that improvements in the statistical errors in the power spectrum on these scales are not needed!

The redshift survey data strongly rule out the standard CDM model, and are fit much better by a $\Gamma = 0.20 - 0.30$ model (Fig. 9). This is in accord with analyses of the small-scale velocity dispersion of galaxies, their large-scale angular clustering, observations of bulk flows on large scales, as well as constraints from the CMB fluctuations (Efstathiou et al. 1992; Kamionkowski & Spergel 1994; Kamionkowski, Spergel, & Sugiyama 1994), the distribution and mass spectrum of clusters (Bahcall & Cen 1992; 1993), and a host of other constraints. Unfortunately, we cannot conclude from this that the dark matter problem is solved. A number of the different suggested power spectra are degenerate over the scales probed by redshift surveys, with very different implications for the nature of the dark matter (compare the range of Fig. 9 to that of Fig. 2). In particular, the data we have presented are also consistent with the Mixed Dark Matter model and the Tilted Cold Dark Matter model, which are two of the more popular models being discussed.

The Mixed Dark Matter model has a long history, starting with the idea that adiabatic damping of the power spectrum on small scales by baryons will cause a turn-down in the power spectrum (Silk 1968; Dekel 1981); its recent incarnation is in terms of a mix of hot and cold dark matter (Schaefer, Shafi, & Stecker 1989; Schaefer & Shafi 1992; Taylor & Rowan-Robinson 1992; Davis, Summers, & Schlegel 1992; Klypin et al. 1993). In particular, the hot dark matter suppresses the power spectrum on small scales, decreasing the small-scale velocity dispersion relative to standard CDM, and increasing the amount of large-scale power for a given normalization on small scales. However, the model perhaps suppresses small-scale power overly much: galaxies cannot form on these small scales until very late, which is difficult to reconcile with observations of galaxies and quasars at very high redshifts (Cen & Ostriker 1994; cf., Efstathiou & Rees 1988 for a similar critique of standard CDM).

The tilted CDM model was suggested simultaneously by a number of workers (Cen et al. 1992; Lidsey & Coles 1992; Lucchin, Matarrese & Mollerach 1992; Liddle, Lyth & Sutherland 1992; Adams et al. 1993; Cen & Ostriker 1993). It also increases the amount of power on large scales relative to small, in this case by changing the slope of the primordial power spectrum. However, Muciaccia et al. (1993) showed that standard CDM is preferred over tilted models when velocity field data and CMB fluctuations are taken into account as well.

There are a number of further constraints on the power spectrum, some of which we've only touched upon on this review. The most important of these is the fluctuations in the CMB. On the largest scales, the fluctuations as detected by COBE appear to be consistent with a primordial power spectrum of index $n = 1$ (the inflationary prediction) (Górski et al. 1994), with an amplitude that matches well that predicted from the $\Gamma = 0.2$ CDM model (Fig. 9). Observations of the CMB fluctuations on sub-COBE scales (a few degrees) are just beginning to yield reproducible results; because more than the Sachs-Wolfe effect is operating on these smaller scales, these data can potentially yield information on $\Omega_0$, the density parameter in baryons, as well as the power spectrum (Hu & Sugiyama 1994). In addition, these probe scales now being reached by the largest peculiar velocity and redshift surveys, spanning much of the gap seen in Fig. 2. In this regard, the bulk flow results of Lauer & Postman (1994) (§ 7.1.4) remain unexplained in the context of models for large-scale structure. It is vitally important that this result be checked, as a number of workers are now doing (§ 9.7). The comparison between large-scale flows and CMB fluctuations has the potential to check gravitational instability theory directly, independent of the power spectrum (Juszkiewicz, Górski, & Silk 1987; Tegmark, Bunn & Hu 1994).

Constraints can be put on the power spectrum from observations of galaxies at high redshift. The amount of small-scale power determines at what epoch galaxies will form; measures of the age of galaxies and their evolution thus tell us something about the power spectrum. Balancing the need for enough small-scale power to allow galaxies to form early, as seems to be required by observations of high-redshift galaxies and quasars, against the requirement of not too much small-scale power, in order to restrict the small-scale velocity dispersion (§ 5.2.1) has not yet been self-consistently done for any model. Similarly, observations of clusters of galaxies and their evolution also have the power to constrain cosmological models (eg., Peebles, Daly, & Juszkiewicz 1989). Finally, detection of the evolution of clustering in the universe would be a tremendously important observation. In an open universe, clustering ceases to grow when $\Omega$ deviates significantly from unity (§ 2.2); unambiguous detection of this effect would be a sensitive measure of $\Omega_0$.

Perhaps the most dramatic constraint one could imagine on the power spectrum would be the laboratory detection of dark matter (Primack et al. 1988). It would be a tremendous triumph of our theoretical framework if a dark matter particle were discovered with properties consistent with the best-fit power spectrum from astronomical data. The HDM model has been out of favor for some time, given its unphysically late formation epoch for galaxies (e.g., White, Frenk, & Davis 1983), but if a definite non-zero mass for the muon neutrino were measured appropriate to close the universe, HDM models would certainly enjoy a resurgence of popularity!

### 9.2 The Distribution Function of the Initial Fluctuations

All the tests we have described for the random-phase hypothesis have yielded positive results; there is no direct evidence for non-Gaussian fluctuations in the initial density field. Perhaps the strongest such claim comes from the direct measure of the distribution function of initial fluctuations as found by the time machine of Nusser & Dekel (1993). Similar conclusions are found in analyses of the COBE CMB fluctuations (e.g., Hinshaw et al. 1994). However, as we emphasized in § 5.4, this by no means allows us to conclude that all non-Gaussian models are dead. Each of the tests described in § 5.4 refer to a specific smoothing scale, and a model that is non-Gaussian on a given scale need not be so on another. What is needed is a systematic test of each non-Gaussian model proposed against the various observational constraints. This has been done to a certain extent for models of cosmic strings (Bennett, Stebbins, & Bouchet 1992) and texture models (Pen, Spergel, & Turok 1994), mostly in the context of non-Gaussian signatures in CMB fluctuations.

### 9.3 The Gravitational Instability Paradigm

The results we have presented here are all consistent with the gravitational instability picture. In particular, we have seen that there exist physically plausible power spectra which can simultaneously match the observed large-scale distribution of galaxies, large-scale flows, and the CMB fluctuations[36]. When redshift surveys began to reveal extensive structures such as giant voids and the Great Wall, many questioned whether this could be explained in the context of gravitational instability theory. However, simulations by Weinberg & Gunn (1990), and Park (1990), as well as arguments based on the Zel'dovich approximation (Shandarin & Zel'dovich 1989), showed that these structures were not unexpected given gravitational instability and plausible models for the power spectrum.

The most direct test of gravitational instability comes from the comparison of peculiar velocity and redshift surveys. Dekel et al. (1993) in particular claim that the Mark II peculiar velocity data are consistent with the velocity field predicted from the distribution of *IRAS* galaxies, and gravitational instability theory. However, this is not a proof; Babul et al. (1994) demonstrate models with velocities due to non-gravitational forces (in particular, large-scale explosion models) that the Dekel et al. tests would not rule out.

We have not discussed features of the velocity field on small scales, smaller than are resolved by the POTENT-*IRAS* comparison, with its 1200 km s$^{-1}$ Gaussian smoothing. Burstein (1990) presents evidence that the very local velocity field, as measured with the Aaronson *et al.* (1982a) TF data, differ qualitatively from the *IRAS* predictions, a conclusion that continues to hold with the Tormen & Burstein (1995) reanalysis of the Aaronson *et al.* data. Similarly, the complete lack of an infall signature in the spiral galaxies around the Coma cluster found by Bernstein *et al.* (1994) is worrisome, and remains unexplained. Understanding these results remains a task for the future.

---

[36] An obvious exception to this statement is the Lauer-Postman (1994) bulk flow; if it is confirmed by further observations, we may find ourselves questioning the gravitational instability paradigm.

*9.4 The Value of $\Omega_0$*

Gravitational instability theory has given us a tool to measure the cosmological density parameter, by comparing peculiar velocities with the density distribution (Eq. 30), although in most applications, galaxy biasing means that we constrain only $\beta \equiv \Omega_0^{0.6}/b$. In Table 3, we summarize the various constraints on $\Omega_0$ that we have discussed in this review.

Is there some consensus in the literature as to the value of $\beta$? One way to assess this, at least qualitatively, is to plot each of the determinations in Table 3 as a series of Gaussians of unit integral, with means and standard deviations given by the numbers in the table. For simplicity, asymmetric error bars have been symmetrized, and those determinations without quoted error bars are not included. Determinations based on *IRAS* and optical samples are plotted with different symbol types. We now simply add the Gaussians together, to yield the two heavy curves in the plots. Note that this procedure tends to give lower weight to those determinations with more realistic (i.e., larger) error bars. No attempt has been made to assess the relative quality of these different determinations. Note also that because many of these determinations are from common datasets, they are not independent. Thus this form of qualitative summary gives an unprejudiced view of literature of determinations of $\beta$ from redshift and peculiar velocity surveys. The heavy curves have a mean of 0.78 and standard deviation of 0.33 (*IRAS*) and a mean of 0.71 and standard deviation of 0.25 (optical). These values are actually in quite close agreement, although that seems more coincidental than anything else, given the large spread of individual determinations.

The community is clearly not quite ready to settle on a single value for $\beta$ for the *IRAS* galaxies. The determinations range from 0.45 (Fisher *et al.* 1994b) to 1.28 (Dekel *et al.* 1993, although the latter is likely to come down slightly with the Mark III data; Dekel, private communication). This is reflected in the large standard deviation, larger than any individual determination, and the flat top to the heavy curve in Fig. 20. The optical $\beta$ shows a smaller spread, perhaps simply because there are fewer individual determinations of it. The odd man out is the determination of $\Omega_0$ by Shaya *et al.* (1994), although their determination is heavily affected by their modeling of the background density within 3000 km s$^{-1}$ and the density field beyond there. Moreover, their work remains in flux (compare with Shaya *et al.* 1992) and it is not clear where their final results will lie.

Most of the references in Table 3 are very recent, and we have not done a thorough job of reviewing the earlier literature, especially on Virgocentric infall. However, the common impression that estimates of $\Omega_0$ have taken a dramatic upturn in recent years is wrong. Davis *et al.* (1980) used observations of Virgocentric flow to find $\beta = 0.6 \pm 0.1$, in good agreement with the values for optical galaxies here. The value from the Cosmic Virial Theorem from Davis & Peebles (1983b) is difficult to interpret in terms of a biasing model, but corresponds to $\beta = 0.4$ for an unbiased model.

*9.5 The Relative Distribution of Galaxies and Mass*

We have very few handles on the biasing parameter independent of $\beta$. One approach has been to assume a model for the power spectrum, normalize it to the COBE fluctuations, and then compare the results predicted for the galaxy fluctuations at 8 $h^{-1}$ Mpc with observations. This approach is by definition model-dependent; for standard CDM, one finds that optical galaxies are unbiased and that *IRAS* galaxies are anti-biased, while a model like $\Gamma = 0.2$ CDM gives a normalization that leaves the *IRAS* galaxies unbiased.

Table 3: Constraints on $\Omega_0$ and $\beta$

| Quantity | Value | Comments | Reference |
|---|---|---|---|
| *Redshift Space Anisotropy* | | | |
| $\beta_{CfA}$ | $0.53 \pm 0.15$ | $\xi(s)$ vs. $w(\theta)$ | Fry & Gaztañaga 1993 |
| $\beta_{SSRS}$ | $1.10 \pm 0.16$ | $\xi(s)$ vs. $w(\theta)$ | Fry & Gaztañaga 1993 |
| $\beta_{IRAS}$ | $0.84 \pm 0.45$ | $\xi(s)$ vs. $w(\theta)$ | Fry & Gaztañaga 1993 |
| $\beta_{IRAS}$ | $1.0 \pm 0.2$ | $\xi(s)$ vs. $w(\theta)$ | Peacock & Dodds 1994 |
| $\beta_{optical}$ | $0.77 \pm 0.15$ | $\xi(s)$ vs. $w(\theta)$ | Peacock & Dodds 1994 |
| $\beta_{IRAS}$ | $0.69^{+0.28}_{-0.24}$ | Angular moments of $\xi$ | Hamilton 1993a |
| $\beta_{IRAS}$ | $0.45^{+0.27}_{-0.18}$ | Angular moments of $\xi$ | Fisher et al. 1994b |
| $\beta_{IRAS}$ | $0.52 \pm 0.14$ | Angular moments of $P(k)$ | Cole et al. 1995 |
| $\beta_{IRAS}$ | $0.94 \pm 0.17$ | Spherical Harmonics | Fisher et al. 1994c |
| $\beta_{IRAS}$ | $1.1 \pm 0.3$ | Spherical Harmonics | Heavens & Taylor 1994 |
| $\Omega_0$ | $0.2^{+0.1}_{-0.07}$ | Cosmic Virial Theorem | Davis & Peebles 1983b |
| *Acceleration on Local Group* | | | |
| $\beta_{optical}$ | $0.38 - 0.74$ | Infall into Virgo | Davis & Huchra 1982 |
| $\beta_{IRAS}$ | $0.82 \pm 0.15$ | QDOT | Rowan-Robinson et al. 1990 |
| $\beta_{IRAS}$ | $0.55^{+0.20}_{-0.12}$ | IRAS 1.2 Jy | Strauss et al. 1992c |
| $\beta_{optical}$ | $0.80^{+0.21}_{-0.13}$ | | Hudson 1993b |
| *Analyses of Velocity Field* | | | |
| $\sigma_{8,v}\Omega_0^{0.6}$ | $1.3^{+0.4}_{-0.3}$ | Amplitude of fluctuations | Seljak & Bertschinger 1994 |
| $\Omega_0$ | $1.0$ | Gaussianity of $\nabla \cdot \mathbf{v}$ | Nusser & Dekel 1993 |
| $\Omega_0$ | $1.0$ | Skewness of $\nabla \cdot \mathbf{v}$ | Bernardeau et al. 1994 |
| $\Omega_0$ | $> 0.3$ | Voids in $\nabla \cdot \mathbf{v}$ | Dekel & Rees 1994 |
| *Comparison of Velocity and Density Fields* | | | |
| $\beta_{IRAS}$ | $0.8$ | Velocity Scatterplot | Strauss 1989 |
| $\beta_{IRAS}$ | $0.86 \pm 0.14$ | Velocity Scatterplot | Kaiser et al. 1991 |
| $\beta_{optical}$ | $0.50 \pm 0.06$ | Velocity Scatterplot | Hudson 1994b |
| $\beta_{optical}$ | $< 0.4$ | Variational Method | Shaya et al. 1994 |
| $\beta_{IRAS}$ | $0.6$ | Scatter in TF diagram | Roth 1994 |
| $\beta_{IRAS}$ | $0.6$ | Differential Dipole | Nusser & Davis 1994a |
| $\beta_{IRAS}$ | $1.28^{+0.38}_{-0.30}$ | $\delta$ vs $\nabla \cdot \mathbf{v}$ | Dekel et al. 1993 |
| $\beta_{IRAS}$ | $0.55 \pm 0.13$ | Likelihood analysis | Willick et al. 1995d |
| $\beta_{optical}$ | $0.74 \pm 0.13$ | $\delta$ vs $\nabla \cdot \mathbf{v}$ | Hudson et al. 1995 |

Alternatively, one can constrain biasing by looking for non-linear effects to break the degeneracy between $\Omega_0$ and $b$. The skewness is one such effect. Fry & Gaztañaga (1993; 1994) and Frieman & Gaztañaga (1994) claim that the beautiful agreement between the measured higher-order moments of the APM counts-in-cells with that predicted given the power spectrum, implies that biasing of optical galaxies is very weak, and to the extent that there is biasing, that it is local. Dekel *et al.* (1993) attempted to look for non-linear effects in the *IRAS*-POTENT comparison; they could only show that the data are inconsistent with very strong non-linearities, thereby ruling out very small values of $b$.

Finally, one can look for relative biasing of different types of galaxies, as we described in § 5.10. The effects are subtle: outside of clusters, there are no two populations of galaxies known that have qualitatively different large-scale distributions. The lack of such effects have motivated several workers (Valls-Gabaud *et al.* 1989; Peebles 1993) to argue that biasing cannot be acting at all. But differential effects are seen between galaxies of different luminosities and morphological types. It is time for a detailed comparison of these observed effects with hydrodynamic simulations, in order to see what constraints these put on general biasing schemes.

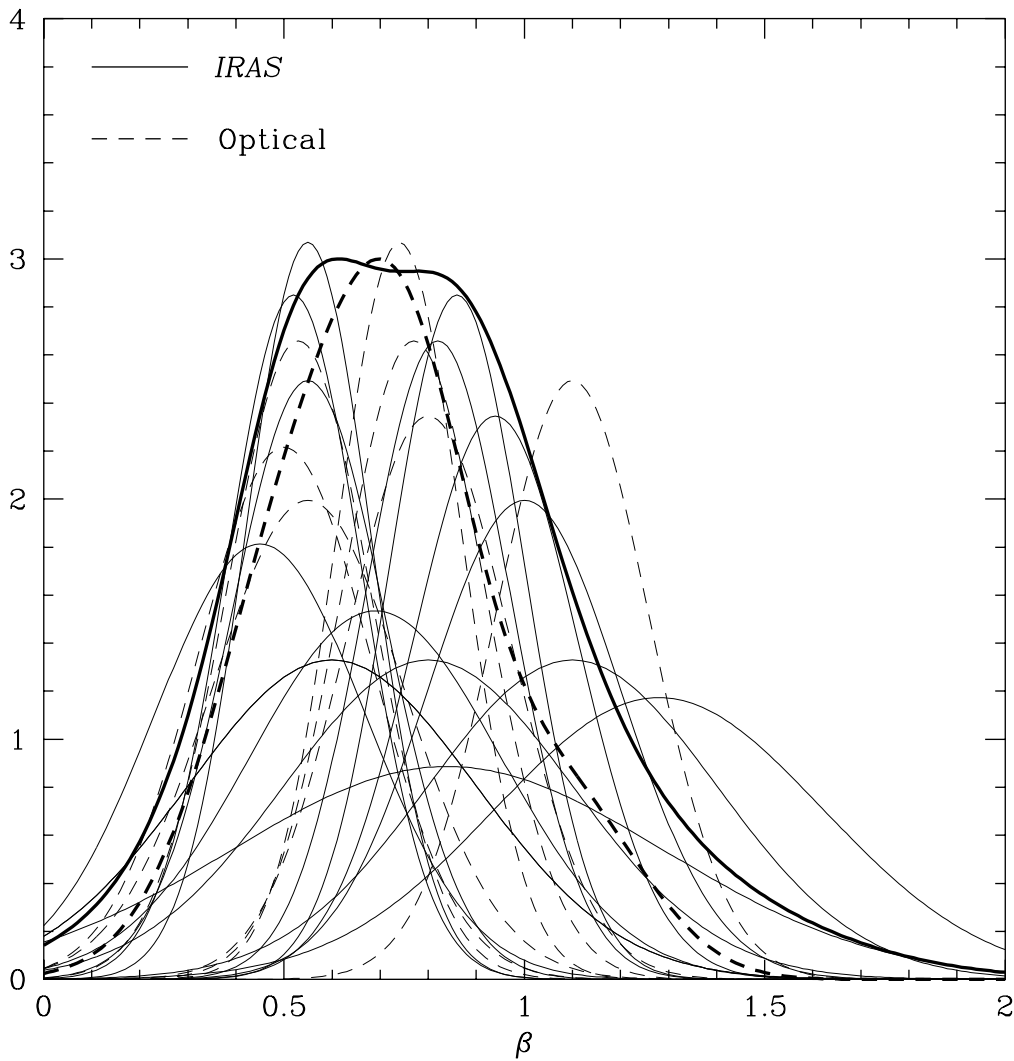

Fig. 20. The distribution of determinations of $\beta$ discussed in this review. Each determination is shown as a Gaussian of unit integral with mean and standard deviation given by the values listed in Table 3. *IRAS* and optical determinations are plotted with different line types. The (renormalized) sums of all curves (optical and *IRAS* separately) are shown as the heavy curves.

In any case, the consensus of the community is that biasing is relatively weak; few authors are arguing for $b > 1.5$ these days. This is quite a contrast to a decade ago, when the idea of biasing was first introduced; values of $b = 2.5$ or higher were popular (e.g., Davis *et al.* 1985). Thus we conclude that $\beta^{5/3} < \Omega_0 < 2\beta^{5/3}$; the results of Table 3 are still consistent with values in the range $\Omega_0 = 0.3$ to $\Omega_0 = 1$. It has been quite popular in recent years to argue for the lower value, given the coincidence with the value of $\Omega_0$ needed to match the $\Gamma = 0.25$ value preferred by the power spectrum (Coles & Ellis 1994).

## 9.6 Is the Big Bang Model Right?

One tests the Big Bang model with redshift and peculiar velocity data only to the extent that they give results which can be fit into our grander picture of the evolution of the universe, with input from all the subjects we did not discuss: observations of distant galaxies and quasars, measurements of individual galaxy properties, abundances of the light elements, and so on. We should point out one serious problem which we see on the horizon. The data we have discussed point towards a value of $\Omega_0$ close to unity, implying an age of the universe given roughly by $t_0 = 2/3H_0^{-1}$ (Eq. 17). With recent determinations of the Hubble Constant of the order of 80 km s$^{-1}$Mpc$^{-1}$ (Jacoby et al. 1992; Pierce et al. 1994; Freedman et al. 1994), this gives an age of 8 billion years, less than half the currently accepted ages of the oldest globular clusters (e.g., Chaboyer, Sarajedini, & Demarque 1992). Note that this would be a problem even if $\Omega_0 \to 0$, for which $t_0 = 1/H_0 = 12$ billion years. We may find ourselves invoking theoretically awkward models in which $\Omega_0 \approx \Omega_\Lambda \lesssim 1$. In any case, the next few years should be very exciting, as we come to grips with this rather basic problem.

## 9.7 The Future

We conclude this review with a quick discussion of the various on-going and planned redshift surveys and peculiar velocity surveys of which we are aware. As these become available, we can look forward to applying the statistics developed so far to vastly superior datasets; moreover, these will allow us to do analyses of much more subtle statistics.

There are a number of large-scale peculiar velocity surveys in progress. Giovanelli, Haynes, and collaborators are doing a Tully-Fisher survey of Sc I galaxies from the Northern sky drawn from the UGC catalog, together with calibrating galaxies drawn from a number of clusters. They have data for roughly 800 galaxies. In the meantime, Mathewson & Ford (1994) have extended their Tully-Fisher survey in the Southern Hemisphere to smaller diameters, as reported in § 7.1.3; their sample now includes a total of 2473 galaxies.

At higher redshift, a team of eight astronomers started by three of the original 7 Samurai (Burstein, Davies, and Wegner), has extended the 7 Samurai $D_n$-$\sigma$ survey of elliptical galaxies to a further $\sim 500$ galaxies in clusters at redshifts $\sim 10,000$ km s$^{-1}$ (Colless et al. 1993).

Several groups are attempting to check the large-scale bulk flow measured by Lauer & Postman (1994). The same authors, in collaboration with Strauss, are in the process of extending the survey to include the BCG's of all Abell clusters to $z = 0.08$, a total of over 600 clusters. They expect to complete the gathering of the data by mid-1996. Fruchter & Moore are measuring distances to the same clusters as the original Lauer & Postman (1994) dataset by fitting Schechter functions to the luminosity distributions in the clusters. In a complementary effort, Willick is measuring accurate distances to 15 clusters around the sky at redshifts of $\approx 10,000$ km s$^{-1}$, using TF and $D_n$-$\sigma$ distances to spirals and ellipticals in each cluster. Finally, Hudson, Davies, Lucey, and Baggley are measuring $D_n$-$\sigma$ parameters of 6-10 ellipticals in each Lauer-Postman cluster with redshift less than 12,000 km s$^{-1}$. This will result in distance errors of $\approx 8\%$ per cluster.

There are two major new redshift survey projects in preparation. A British collaboration led by Ellis plans to measure redshifts for 250,000 galaxies to $b_J = 19.7$ selected from the APM galaxy catalog in a series of fields in the Southern Sky, using the 2dF 400-fiber spectrograph on the Anglo-Australian telescope (Gray et al. 1992). The survey geometry consists of two long strips in the Fall and Spring skies, plus 100 randomly placed fields of 2° diameter, totaling 0.53 ster. The principal motivation is to measure the large-scale power spectrum of the galaxy distribution, redshift space distortions to constrain $\Omega_0$, and evolutionary effects.

The Sloan Digital Sky Survey (SDSS) will use a dedicated 2.5m telescope to survey 3 ster around the Northern Galactic Cap with CCD's in five photometric colors. A multi-object spectrograph with 640 fibers will be used to carry out a flux-limited redshift survey of galaxies to roughly $R = 18.0$. Over five years, this survey will measure redshifts for $\sim 10^6$ galaxies, with a median redshift of $\approx 31,000$ km s$^{-1}$. The survey will see first light in the second half of 1995. Details may be found in Gunn & Knapp (1993), and Gunn & Weinberg (1995). The SDSS is one of the few large-scale surveys in in which the photometric data from which the redshift galaxy sample will be selected is obtained as part of the survey itself. The use of CCD data and careful calibration guarantees that it will be the best calibrated of these surveys. It does not go as deep as the 2dF survey mentioned above, but covers much more area.

Thus we look forward to tremendous growth in the quantity and quality of both peculiar velocity and redshift data. We set forth a series of questions in the beginning of this review (§ 2.5) which we hoped to address with the data available. We have reviewed the analyses that have been done with redshift and peculiar velocity surveys to answer these questions. However, as we have summarized in this concluding chapter, there are few of these questions for which we now have definitive answers. Indeed, most of the quantities we hope to measure are known to within a factor of two at best, and more often only within an order of magnitude. We expect that the next decade will be a period of intense activity in this branch of observational cosmology, during which superior data and a more complete understanding of the theoretical issues will allow us to make observational cosmology a precision science; there is no doubt qualitatively new science to be discovered when we measure the power spectrum on large scales, the value of $\Omega_0$, the bias parameter of different galaxy types, and many other quantities, to 10% accuracy.

## Acknowledgement

We thank Alan Dressler and Sandra Faber for comments and suggestions on parts of the text. Avishai Dekel supplied two of the figures. Karl Fisher, Mike Hudson, and David Weinberg read through the entire paper and made many valuable comments; in addition, we received useful comments and suggestions from Yehuda Hoffmann, David Burstein, Roman Juszkiewicz, and an anonymous referee. Maggie Best helped tremendously in the compilation of the references. JAW thanks his collaborators on the Mark III project for permission to discuss aspects of this work prior to publication. MAS acknowledges the support of the WM Keck Foundation during the writing of this review.